# REGENERATIVE MEDICINE FOR TENDON/LIGAMENT INJURIES: DE NOVO EQUINE TENDON/LIGAMENT NEOTISSUE GENERATION AND APPLICATION

A Dissertation

Submitted to the Graduate Faculty of the
Louisiana State University and
Agricultural and Mechanical College
in partial fulfillment of the
requirements for the degree of
Doctor of Philosophy

in

Department of Veterinary Clinical Sciences

by
Takashi Taguchi
B.Sc., Osaka Prefecture University, 2008
M.Sc., Western University of Health Sciences, 2017
August 2023

# Acknowledgements

This dissertation is dedicated to all who supported me throughout the Ph.D. program. My mentor Dr. Mandi J. Lopez always supported me to explore many new ideas that led to groundbreaking discoveries and provided me an environment where I made lifetime friends and colleagues. My colleagues at Laboratory of Equine and Comparative Orthopedic Research, Dr. Qingqiu Yang, Rita Aoun, Catherine Takawira, and others were all my source of inspiration. We grew together and supported each other, and now they are my sisters whom I will provide support unconditionally. I would like to also express my appreciation to collaborators, especially Dr. Mustajab H. Mirza, Dr. Francisco J. Morales-Yniguez, Dr. Frank M. Andrew, and Dr. Nathalie Rademacher, for their selfless contribution to bring studies to fruition. The committee members, Dr. Craig M. Hart, Dr. Michelle Osborn, Dr. Kevin F. Hoffseth, and Dr. Javier Nevarez were all instructive in guiding my studies. All the projects were also not available without the support from School of Veterinary Medicine. Especially, Michael L. Keowen, Dawn Kelley, Rouchelle Gage, Kendra Shultz, and Thaya Stoufflet always gave me help without questions being asked.

Alongside the Ph.D. program, I received countless support from my family, mentors, and friends. My family was always supportive and a guide for my both personal and professional growth. Mentors from Department of Experimental Statistics, especially Dr. Brian D. Marx, Dr. BeiBei Guo, Dr. Bin Li, and Dr. Luis A. Escobar were my true inspiration for their strict professionalism and excellence in the field. Mentors from my previous educations and training were also always supportive, and provided me a guidance whenever asked. My friends from a wide range of disciplines are now lifetime colleagues, and we grow together by exchanging ideas and experiences.

# Table of Contents

# Abstract

Tendon and ligament injuries are debilitating conditions across species. Poor regenerative capacities of these tissues limit restoration of original functions. The first study evaluated the effect of cellular administration on tendon/ligament injuries in horses using meta-analysis. The cellular administration was effective in restoring ultrasonographic echogenicity and increasing vascularity during early phase of healing. Additionally, it improved microstructural organization of healed tissue in terms of cellularity and fiber alignment. However, the study did not support its use for increasing rate of return to performance, expression/deposition of tendon-specific genes/proteins, or mechanical properties.

The findings led to the second study that engineered implantable *de novo* tendon neotissue using equine adipose-derived multipotent stromal cells and collagen type I. Neotendon cultured in tenogenic medium using custom-designed bioreactor contained highly proliferative tenoblast-like cells that matured and progressively deposited abundant extracellular matrix over 21-day culture period, forming embryonic tendon-like tissue, whereas neotissue cultured in stromal medium contained non-proliferative sparse spherical cells loosely attached to template throughout the culture period.

The neotendon was evaluated for its biocompatibility and therapeutic potential in the third study using immunocompetent and immunocompromised rat bilateral calcaneal tendon elongation model. Neotendon was demonstrated to be biocompatible in immunocompetent rats, and pre-implantation differentiation was essential to maintain phenotype, as neotissue cultured in stromal medium formed amorphous tissue. And the role immune cells play on aberrant tissue formation of non-differentiated neotissue was suggested to be critical.

The fourth study investigated the therapeutic effects of neotendon in surgically-induced non-terminal equine accessory ligament of deep digital flexor tendon injury model. The model rendered minimum morbidity with impaired limb use until 9 weeks post-injury. Lesion recapitulated naturally-occurring injuries ultrasonographically and clinically. Implanted neotendon led to formation of pre-mature tendon tissue that integrated with native tissue within lesion, demonstrating robust healing potential.

Collectively, the work in this dissertation systematically analyzed the effects of cellular therapies to treat tendon/ligament injuries and identified the limitations. Following development of novel neotendon represents a potential breakthrough to overcome these limitations. Further evaluation of neotendon on its therapeutic efficacies is likely to contribute to improved healing of tendon and ligament in equine and other species.

# Chapter 1. Meta-analysis of the Effects of Adult Tissue-derived Cell Therapy on Tendinopathy and Ligamentopathy in the Horse

## 1.1. Introduction

Tendinopathy and desmitis are chronic and disabling conditions to humans that affect as high as 2% of the general population.[1] The fraction of the affected population is even higher in people doing physical work (as much as 11% )[2] or those involved with athletics 80%.[3] In the United States alone, up to 15 million musculoskeletal injuries associated with tendon and ligament injuries have been reported annually, and the aggregate total cost from loss of productivity and burden to the healthcare system can exceed nearly $ 200 billion annually.[4-6] The treatment regimens for tendon and ligament injuries are selected based on symptoms and stages of the healing process, and include anti-inflammatory drugs, eccentric exercises, extracorporeal shock wave therapy (ESWT), as well as surgical release or reconstruction.[7] Among these, the most common intervention for long-term therapy is eccentric exercise,[8] demonstrating relative lack of long-term efficacy among currently available pharmacological interventions. This can be attributed to the inherently low regenerative potential of tendons and ligaments. Furthermore, aging populations increase risk of impaired healing potentially due to reduced collagen synthesis as well as responsiveness to mechanical stimulus of tenocytes.[9,10] As a result, although depending on the type of injury and activity levels of patients, re-injury can occur in up to 18% of affected populations.[11]

Similarly in equine athletes, tendinopathy and desmitis comprise a large majority of musculoskeletal injuries that are responsible for up to 72% of lost training days and 14% of early retirements.[12-14] Superficial digital flexor tendinopathy and suspensory ligament (SL) desmitis are the most common, comprising 46% of all limb injuries.[15,16] The predominant type of tendon and ligament injury varies among disciplines, but virtually all equine companions can be impacted. Strain induced injuries are common in the equine suspensory apparatus including the SL, superficial digital flexor tendon (SDFT), and deep digital flexor tendon (DDFT).[17] Many acute and chronic tendon and ligament lesions are thought to result from focal accumulation of microtrauma and poorly organized repair tissue that can coalesce into large lesions and predispose to spontaneous rupture in many species.[18] Treatments vary widely and can range from rest with anti-inflammatory drugs, cold therapy and pressure bandaging to intralesional injections of various therapeutics and extracorporeal shock wave therapy.[19] However, poor or abnormal tissue repair contributes to a reinjury rate in horses as high as 67% within 2 years.[20,21] Therefore, intralesional regenerative therapies such as platelet rich plasma, stem cells, and genetic material have been applied with variable success.[22] Short-term outcomes of these treatments are favorable. To date, there is no single, gold standard to promote healing of ligament and tendon lesions.

Autologous tenocyte implantation (ATI) is one mechanism to deliver endogenous cells to the site of tendon or ligament injury in adult animals and humans.[23-25] Animal studies demonstrated ATI combined with biomaterials augment tendon regeneration by immunomodulatory effects and preventive effects of ectopic chondroid formation.[26,27] In a human clinical prospective study, ATI has also been shown to alleviate pain associated with gluteal tendinopathy over 24 months period.[28] However, the therapy is limited by few harvest sites and harvest morbidity, which limits its wide use. Among other cell types, administration of exogenous adult multipotent stromal cells (MSCs) is more accessible and reported to augment natural healing in human tendinopathy, as well as in naturally-occurring and experimentally-induced equine tendon and ligament injuries.[29-32] Results are mixed, however, in part due to differences among cell isolates, lesions, individual healing capacity, and low engraftment of exogenous cells (< 0.001%).[30,33] Further, there is evidence that an inflammatory environment may impede differentiation of MSCs, and the cells may assume an abnormal phenotype leading to unwanted side effects.[34,35] Hence, it is of utmost importance to inform clinicians with a best available evidence-based guidance with respect to the effects of cellular therapies on tendon and ligament injuries.

To date, several regenerative therapies have been investigated in meta-analysis for their efficacy on tendinopathy. One widely utilized regenerative interventions is platelet-rich plasma (PRP) due to its ease of preparation and FDA approval. The application of PRP combined with arthroscopic rotator cuff (RC) repair was determined to be effective in reducing retear rate and improving several short-term pain scores.[36] The effects were also evident in mid- to long-term pain alleviation when used as conservative treatment without surgical interventions.[37] And PRP was effective in reducing long-term pain in ligament injury as well.[38] In contrast to PRP application, clinical studies investigating the effects of cellular therapies on tendinopathy and desmitis are limited, as are meta-analysis studies, largely because cellular therapies are still in the preclinical stage. Despite this paucity, the effect of MSCs application on tendon disorder has been evaluated in a meta-analysis of human clinical data. According to the study, the treatment with MSCs improved all the aspects analyzed: pain, functional scores, radiological parameters (magnetic resonance image or ultrasonography), and arthroscopic findings.[32] And pain relief was cell dose-dependent. In addition to humans and canines, equines also commonly develop naturally-occurring tendinopathy. The horse has been utilized in multiple randomized-controlled studies to investigate the efficacy of cellular therapies on tendon or ligament injuries, making horse studies an ideal resource for large scale meta-analysis. Additionally, investigations on equine tendinopathy and desmitis allow extensive analysis of repaired tissue upon terminal endpoint, which is not feasible in humans. Collectively, systematic meta-analysis conducted on equine studies will add tremendous value to the existing knowledge of human tendinopathy/desmitis therapies.

In this study, systematic meta-analysis was conducted to investigate the effects of cellular therapies on treating both naturally-occurring and experimentally-induced equine tendinopathy and desmitis. The hypotheses tested were: 1) cellular therapies are effective in increasing long-term rate of return to previous level of performance; 2) the effects of cellular therapies include microstructural improvement of repaired tissue as well as increased tenogenic/ligamentogenic gene expression and protein synthesis; yet 3) cellular therapies fail to restore original mechanical properties of injured tissues. Due to current regulatory restrictions on the use of pluripotent stem cells in clinical applications, cellular therapies included in the study were of primary isolates of adult tissues, such as MSCs, tenogenically-induced MSCs, and tenocytes.

## 1.2. Materials and Methods

Literature were screened, data extracted, and crosschecks performed in accordance with the recommendations published in the PRISMA statement.[39]

### 1.2.1. Literature Search

A comprehensive literature search was carried out with PubMed, and Web of Science from January 2001 to December 2022. No language restrictions were applied. Key words used in the database included "equine", "horse", "tendon", "stem", and "cells". Randomized controlled trials were identified using the refined search function in the databases, if available. Articles were also directly searched in veterinary journals (Veterinary Surgery, Journal of Veterinary Internal Medicine, American Journal of Veterinary Research, Equine Veterinary Education, Equine Veterinary Journal, Journal of Veterinary Emergency Critical Care, and Journal of American Veterinary Medical Association) using the same key words.

### 1.2.2. Study Selection Criteria

Studies were identified by the following inclusion criteria: (1) a clinical study including randomized controlled trials, cohort studies, case–control studies, or case reports and case series; (2) prospective studies evaluating the efficacy of MSCs on surgically or chemically induced lesion in tendon or ligament of horses; (3) retrospective studies evaluating the efficacy of MSCs on naturally-occurring lesion in tendon or ligament of horses; (4) application of undifferentiated or tenogenic MSCs of autologous or allogenic origin; and (5) studies reporting a quantitative outcome of interest.

Exclusion criteria were the following: (1) secondary analyses of published data, review papers or expert opinions; (2) individual case reports without control cases; (3) studies not reporting a quantitative outcome of interest or this information not being available after contact with the authors; (4) studies with a quantitative outcome of interest unavailable in English; and (5) application of a pluripotent stem cells including embryonic stem cells or induced pluripotent stem cells.

### 1.2.3. Data Collection and Extraction

Data collected included authors' names, journal name, title of journal article, year of publication, volume, and page numbers, study design, injured limb, location of lesion, type of lesion, type of treatment, type of control, timing of treatment, location of treatment, delivery of treatment, range or mean of age of horses, range or mean of weight of horses, breed of horses, sex of horses, discipline of horses, number of horses in treatment group, number of horses in control group, applicability of pairing limbs between control and treatment in a same horse, maximum follow-up period, and primary outcome measures for treatment and control groups.

The quantitative outcome measures extracted for this study were: (1) clinical performance evaluation (number of starts, palpation score, and lameness score); (2) ultrasound (US) imaging and magnetic resonance imaging (MRI) evaluation; (3) histological score evaluation; (4) biochemical (gene expression and protein quantification) evaluation; (5) biomechanical evaluation; and (6) number of horses returned to soundness. The quantitative outcome measures were extracted from the records directly or estimated from the graphs. When standard error of the mean (SEM) or interquartile range (IR) were reported instead of standard deviation (SD), SD was calculated based on the sample size or IR was divided by 1.35.[40]

### 1.2.4. Statistical Analysis

The results were expressed as the odds ratio (OR) for the categorical variables and as the weighted mean difference (MD) for the continuous variables, with 95% confidence intervals (CIs). Weighted standardized mean differences (SMD) were used when functional outcomes where different scales are used but with the same direction, such as scoring systems representing abnormal findings with low score and normal findings with high score. Because of differences in type of cells used for treatment, a subgroup analysis was performed on the basis of cell types (ASCs, BMSCs, and tenogenic MSCs). For each outcome analyzed, a minimum of 2 studies were required to interpret the result. Studies were excluded from the analysis when both groups were not included in the study or SD was 0.

Statistical analysis was performed using RevMan 5.3 software, which was provided by the Cochrane Collaboration (https://training.cochrane.org/online-learning/core-software-cochrane-reviews/revman/revman-5-download). A Cochran Q test was used to test for evidence of statistical heterogeneity. Heterogeneity between studies was evaluated via I-squared statistic and p value. When heterogeneity was significant ($p < 0.05$ or I-squared $> 50\%$), a random effects model was adopted to pool the results. Then, $\chi^2$ tests were applied to identify statistical differences of pooled estimates between groups, the effect measure was the adjusted odds ratio (OR) with 95% confidence intervals (CIs), and p < 0.05 was defined as statistically significant. The OR was calculated using the Mantel-Haenszel method without zero-cell corrections, SMD calculated using inverse variance method.

### 1.2.5. Quality Assessment of the Studies

The following domains were assessed: (1) randomization process, (2) deviations from intended interventions, (3) missing outcome data, (4) outcome measurement, (5) selection of the reported result, and (6) overall bias. Risk-of-bias assessment was conducted by two independent reviewers according to Cochrane Handbook for Systematic Reviews of Interventions.[40] Funnel plot was not used to test publication bias because of the limited number (< 10) of studies included in each analysis

### 1.3. Results

### 1.3.1. Description of Included Studies

The described search strategy identified a total of 737 peer reviewed publications and conference abstracts. Among those, 148 duplicates were identified and excluded. A total of 589 records were screened by reviewing titles and abstracts and excluded according to the exclusion criteria. Records with no English version available (n = 148), no standard error or deviation reported (n = 2), conference papers without quantitative outcomes of interest (n = 73), review papers (n = 71), case reports without a control group (n = 26), non-horse clinical studies (n = 376), no treatment efficacy evaluation reported (n = 12), and treatment with ESCs (n = 1) were excluded (Fig 1.1). After the full application of inclusion criteria, a total of 20 relevant sources remained. (Table 1.1 and 1.2).

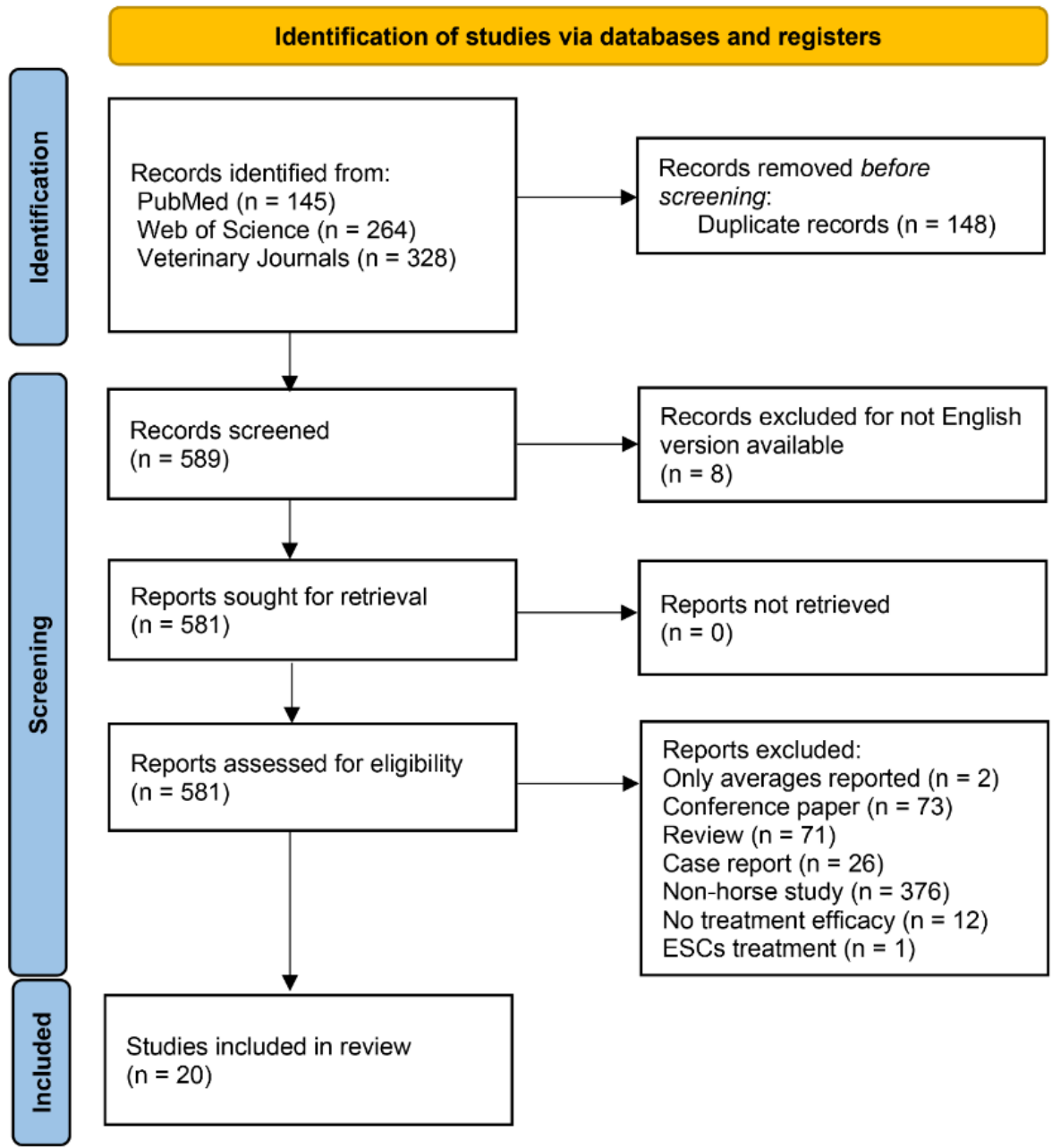

Figure 1.1. PRISMA flow diagram of the systematic literature search process.

Table 1.1. Randomized controlled trial and prospective cohort study characteristics and outcomes.

| Study Characteristics | Treatment Outcomes | Control Outcomes |
|---|---|---|
| Ahrberg *et al*[41], 2018<br>Design: RCT<br>Breed: Standard<br>Sex: Male (3), Female (3)<br>Discipline: Trot Racing<br>Tx: 6 Horses (Fore, Hind)<br>Ctrl: 6 Horses (Fore, Hind)<br>Limb: Fore & Hind<br>Mean Age: 6 (3 − 10)<br>Mean Weight: (400 − 550)<br>Lesion Location: SDFT<br>Lesion Creation: Collagenase<br>Tx: Autologous ASCs (10 million) +<br>Autologous Serum (1 ml)<br>Ctrl: Autologous Serum (1 ml)<br>Tx Timing: 3 wk post-injury<br>Tx Location: Intralesional<br>Paired Limb: Yes<br>Follow-up: 24 wk | <u>Clinical Outcomes</u><br>Palpation Score: $5.77 \pm 2.78$<br>Lameness Score: $8.11 \pm 1.50$<br><br><u>Ultrasound Outcomes</u><br>Lesion CSA: $5.5 \pm 8.4\%$<br>Lesion Score: $1.70 \pm 1.17$<br><br><u>MRI Outcomes</u><br>Lesion Volume: $83.64 \pm 334.57$ mm$^3$<br>Lesion Signal Intensity: $340.74 \pm 151.85$<br><br><u>Histology Outcomes</u><br>Macroscopic Score: $6.35 \pm 5.85$<br>Total HE Score: $3.93 \pm 1.64$<br>Fuchsin Staining: $50.66 \pm 32.31\%$<br>Healthy Crimp: $55.46 \pm 22.27\%$<br>Collagen I IHC Score: $1.41 \pm 0.48$<br>Nuclei: $33.37 \pm 27.86\%$<br>Erythrocytes: $1.94 \pm 2.36\%$<br><br><u>Biochemical Outcomes</u><br>*COL1A2*: $5.3 \pm 7.0$ x $10^3$<br>*COL3A1*: $1.8 \pm 3.2$ x $10^3$<br>*Dcn*: $1.16 \pm 0.91$<br>*Scx*: $0.015 \pm 0.024$<br>*Tnc*: $0.00011 \pm 0.00024$<br>*Opn*: $0.42 \pm 1.14$ | <u>Clinical Outcomes</u><br>Palpation Score: $5.02 \pm 1.16$<br>Lameness Score: $8.11 \pm 1.50$<br><br><u>Ultrasound Outcomes</u><br>Lesion CSA: $3.25 \pm 3.5\%$<br>Lesion Score: $1.59 \pm 2.30$<br><br><u>MRI Outcomes</u><br>MRI Lesion Volume: $27.88 \pm 66.91$ mm$^3$<br>MRI Lesion Intensity: $485.19 \pm 259.26$<br><br><u>Histology Outcomes</u><br>Macroscopic Score: $4.99 \pm 3.52$<br>Total HE Score: $4.59 \pm 4.69$<br>Fuchsin Staining: $51.09 \pm 34.93\%$<br>Healthy Crimp: $58.08 \pm 30.57\%$<br>Collagen I IHC Score: $1.43 \pm 0.67$<br>Nuclei: $45.61 \pm 20.51\%$<br>Erythrocytes: $1.47 \pm 1.78\%$<br><br><u>Biochemical Outcomes</u><br>*COL1A2*: $35 \pm 110$ x $10^3$<br>*COL3A1*: $0.11 \pm 0.16$<br>*Dcn*: $1.72 \pm 1.89$<br>*Scx*: $0.018 \pm 0.025$<br>*Tnc*: $0.00013 \pm 0.00017$<br>*Opn*: $0.49 \pm 1.06$ |

(table cont'd)

| Study Characteristics | Treatment Outcomes | Control Outcomes |
|---|---|---|
| Caniglia *et al*[42], 2012<br>Design: RCT<br>Tx: 6 Horses; Ctrl: 6 Horses<br>Limb: Fore<br>Lesion Location: SDFT<br>Lesion Creation: Surgery<br>Tx: Autologous BMSCs (10 million)<br>+ Autologous Citrated BM (2 ml)<br>Ctrl: Autologous Citrated BM (2 ml)<br>Tx Timing: 4 wk post-injury<br>Tx Location: Intralesional (3 – 5 sites)<br>Paired Limb: Yes<br>Follow-up: 12 wk | <u>Electron Microscope Outcomes</u><br>Collagen Fibril Index: $0.30 \pm 0.07$<br>Mass Average Diameter: $76.89 \pm 14.89$ | <u>Electron Microscope Outcomes</u><br>Collagen Fibril Index: $0.33 \pm 0.04$<br>Mass Average Diameter: $89.71 \pm 26.90$ |
| Carvalho *et al*[43], 2013<br>Design: RCT<br>Breed: Mixed<br>Sex: Male (1), Female (7)<br>Tx: 4 Horses; Ctrl: 4 Horses<br>Limb: Fore<br>Mean Age: (3 – 4.5)<br>Lesion Location: SDFT<br>Lesion Creation: Collagenase<br>Tx: Autologous ASCs (10 million)<br>+ Platelet Concentrate (1 ml)<br>Ctrl: PBS (1 ml)<br>Tx Timing: 2 wk post-injury<br>Tx Location: Intralesional<br>Tx Delivery: Needle (21-G)<br>Follow-up: 16 wk | <u>Ultrasound Outcomes</u><br>Lesion area/tendon area: $6.081 \pm 2.16\%$<br>Lesion Area: $3.25 \pm 0.94$ mm$^2$<br>Power Doppler Score (2 wk): $7.53 \pm 1.98$<br><br><u>Histology Outcomes</u><br>Overall Score: $14.75 \pm 5.96$<br>Inflammatory Cell Infiltrate: $2 \pm 2$ | <u>Ultrasound Outcomes</u><br>Lesion area/tendon area: $14.46 \pm 2.70\%$<br>Lesion area: $11.49 \pm 1.89$ mm$^2$<br>Power Doppler Score (2 wk): $4.99 \pm 1.88$<br><br><u>Histology Outcomes</u><br>Overall Score: $22.66 \pm 5.02$<br>Inflammatory Cell Infiltrate: $3.66 \pm 1.14$ |

(table cont'd)

| Study Characteristics | Treatment Outcomes | Control Outcomes |
|---|---|---|
| Conze *et al*[44], 2014<br>Design: RCT<br>Breed: Warmblood (8), Standard (1)<br>Sex: Gelding (2), Female (7)<br>Tx: 9 Horses; Ctrl: 9 Horses<br>Limb: Fore<br>Mean Age: 4 (3 – 6)<br>Mean Weight: $545 \pm 34$<br>Lesion Location: SDFT<br>Lesion Creation: Surgery<br>Tx: Autologous ASCs (10 million) + Autologous Serum (0.5 ml)<br>Ctrl: Autologous Serum (0.5 ml)<br>Tx Timing: 2 wk post-injury<br>Tx Location: Intralesional<br>Paired Limb: Yes<br>Tx Delivery: Needle (22-G)<br>Follow-up: 22 wk | <u>Ultrasound Outcomes</u><br>Color Doppler Density (2 wk): $28.7 \pm 19.8$<br><br><u>Histology Outcomes</u><br>Vessel number (HE): $1226.8 \pm 54.8$<br>Vessel number (VIII total): $2285 \pm 21.1$<br>Vessel number (VIII small vessel): $1606.2 \pm 84.3$<br>Vessel number (VIII large vessel): $674.5 \pm 46.4$ | <u>Ultrasound Outcomes</u><br>Color Doppler Density (2 wk): $11.3 \pm 10.4$<br><br><u>Histology Outcomes</u><br>Vessel number (HE): $1011.8 \pm 84.3$<br>Vessel number (VIII total): $1627.12 \pm 29.66$<br>Vessel number (VIII small vessel): $1216.1 \pm 101.7$<br>Vessel number (VIII large vessel): $394.1 \pm 50.9$ |

(table cont'd)

| Study Characteristics | Treatment Outcomes | Control Outcomes |
| --- | --- | --- |
| Crovace *et al*[45], 2010<br>Design: RCT<br>Breed: Standard<br>Sex: Male<br>Tx: 6 Horses; Ctrl: 6 Horses<br>Limb: Fore & Hind<br>Mean Age: 4<br>Mean Weight: 522 ± 30<br>Lesion Location: SDFT<br>Lesion Creation: Collagenase<br>Tx: Autologous BMSCs (~ 6 million)<br>+ Fibrin Glue (~ 4 ml)<br>Ctrl: Fibrin Glue<br>Tx Timing: 3 wk post-injury<br>Tx Location: Intralesional<br>Tx Delivery: Needle (23-G)<br>Follow-up: 21 wk | <u>Histology Outcomes</u><br>Fiber Orientation Score: 3.0 ± 0.0<br>MNC Infiltration Score: 1.6 ± 0.5<br>Collagen I Expression: 2.6 ± 0.5<br>Collagen III Expression: 1.2 ± 0.4<br>COMP Expression: 2.2 ± 0.4<br>CD34 Expression: 1.0 ± 0.7 | <u>Histology Outcomes</u><br>Fiber Orientation Score: 0.5 ± 0.7<br>MNC Infiltration Score: 0.5 ± 0.7<br>Collagen I Expression: 1.0 ± 0.0<br>Collagen III Expression: 3.0 ± 0.0<br>COMP Expression: 0.5 ± 0.7<br>CD34 Expression: 0.0 ± 0.0 |

(table cont'd)

| Study Characteristics | Treatment Outcomes | Control Outcomes |
|---|---|---|
| DePuydt *et al*[46], 2021<br>Design: RCT<br>Breed: Warmblood<br>Sex: Gelding (4), Female (4)<br>Tx: 8 Horses; Ctrl: 8 Horses<br>Limb: Fore<br>Mean Age: (3 − 12)<br>Lesion Location: SDFT<br>Lesion Creation: Surgery<br>Tx: Allogenic Tenogenic MSCs<br>Ctrl: Saline (1 ml)<br>Tx Timing: 7 d post-injury<br>Tx Location: Intralesional (1 sites)<br>Tx Delivery: Needle<br>Paired Limb: Yes<br>Follow-up: 112 d | <u>Ultrasound Outcomes</u><br>APT: $0.63 \pm 0.05$ cm<br>Echogenicity Score: $0.6250 \pm 0.5175$<br>Fiber Alignment Score: $0.1250 \pm 0.3536$<br>UTC Type I: $73.96 \pm 10.00$<br>UTC Type II: $12.21 \pm 6.14$<br>UTC Type III: $6.35 \pm 4.25$<br>UTC Type IV: $7.50 \pm 5.93$<br><br><u>Histology Outcomes</u><br>Fiber structure: $0.6 \pm 0.9$<br>Fiber arrangement: $0.6 \pm 0.9$<br>Roundness of the nuclei: $1.6 \pm 0.9$<br>Regional variations in cellularity: $1.6 \pm 0.9$<br>Vascularity: $1.8 \pm 0.5$<br>Collagen stainability: $2.1 \pm 0.8$<br>Glycosaminoglycan content: $2.8 \pm 0.5$<br>Presence of inflammatory cells: $0.6 \pm 1.1$<br>Distribution COLI: $83.44 \pm 5.92\%$<br>Distribution COLIII: $0.53 \pm 0.33\%$<br>Distribution VWF: $8.74 \pm 0.62\%$<br>Distribution SMA: $0.46 \pm 0.29\%$ | <u>Ultrasound Outcomes</u><br>APT: $0.74 \pm 0.05$ cm<br>Echogenicity Score: $1.875 \pm 0.6409$<br>Fiber Alignment Score: $1.875 \pm 0.6409$<br>UTC Type I: $40.71 \pm 16.34$<br>UTC Type II: $12.81 \pm 7.63$<br>UTC Type III: $24.58 \pm 10.86$<br>UTC Type IV: $21.91 \pm 13.12$<br><br><u>Histology Outcomes</u><br>Fiber structure: $0.0 \pm 0.0$<br>Fiber arrangement: $0.1 \pm 0.4$<br>Roundness of the nuclei: $1.3 \pm 0.5$<br>Regional variations in cellularity: $1.9 \pm 0.8$<br>Vascularity: $1.4 \pm 0.5$<br>Collagen stainability: $1.6 \pm 0.9$<br>Glycosaminoglycan content: $2.4 \pm 0.7$<br>Presence of inflammatory cells: $0.5 \pm 1.1$<br>Distribution COLI: $49.73 \pm 6.91\%$<br>Distribution COLIII: $10.58 \pm 2.33\%$<br>Distribution VWF: $1.21 \pm 0.25\%$<br>Distribution SMA: $9.22 \pm 2.18\%$ |

(table cont'd)

| Study Characteristics | Treatment Outcomes | Control Outcomes |
|---|---|---|
| Durgam *et al*[47], 2016<br>Design: RCT<br>Tx: 8 Horses; Ctrl: 8 Horses<br>Limb: Fore<br>Mean Age: (2 − 4)<br>Lesion Location: SDFT<br>Lesion Creation: Collagenase<br>Tx: Autologous TDPCs (5 million)<br>+ PBS (0.15 ml)<br>Ctrl: PBS (0.15 ml)<br>Tx Timing: 4 wk post-injury<br>Tx Location: Intralesional (2 sites)<br>Paired Limb: Yes<br>Follow-up: 12 wk | <u>Histology Outcomes</u><br>CFA (SHG): 105.67 ± 8.3°<br><br><u>Biochemical Outcomes</u><br>*COLI*: 7.48 ± 4.24<br>*COLIII*: 12.81 ± 9.76<br>*COMP*: 3.13 ± 1.7<br>*Tenomodulin*: 23.32 ± 19.94<br>Collagen: 0.2371 ± 0.12 µg/mg<br>Glycosaminoglycan: 0.4496 ± 0.086 µg/mg<br>DNA: 0.2564 ± 0.092 µg/mg<br><br><u>Biomechanical Outcomes</u><br>CSA: 2.29 ± 0.2 cm2<br>Yield stress: 19.75 ± 16.62 MPa<br>Maximum stress: 20.375 ± 19.45 MPa<br>Elastic modulus: 229.29 ± 84.85 MPa<br>Stiffness: 10200 ± 1544.321 N/cm | <u>Histology Outcomes</u><br>CFA (SHG): 114.7 ± 16.2°<br><br><u>Biochemical Outcomes</u><br>*COLI*: 9.76 ± 8.2<br>*COLIII*: 20.79 ± 24.78<br>*COMP*: 3.17 ± 2.1<br>*Tenomodulin*: 18.23 ± 16.24<br>Collagen: 0.2398 ± 0.083 µg/mg<br>Glycosaminoglycan: 0.4344 ± 0.085 µg/mg<br>DNA: 0.2694 ± 0.089 µg/mg<br><br><u>Biomechanical Outcomes</u><br>CSA: 2.57 ± 0.28 cm2<br>Yield stress: 10 ± 2.47 MPa<br>Maximum stress: 11 ± 3.18 MPa<br>Elastic modulus: 147.86 ± 60.61 MPa<br>Stiffness: 9600 ± 1221.881 N/cm |

(table cont'd)

| Study Characteristics | Treatment Outcomes | Control Outcomes |
|---|---|---|
| Garbin *et al*[48], 2019<br>Design: RCT<br>Sex: Male (1), Female (5)<br>Tx: 6 Horses; Ctrl: 6 Horses<br>Limb: Fore & Hind<br>Mean Age: $10.5 \pm 3.5$<br>Mean Weight: $375 \pm 75$<br>Lesion Location: SL<br>Lesion Creation: Surgery<br>Tx: Autologous ASCs (1 million)<br>+ Saline (0.8 ml)<br>Ctrl: Saline (0.8 ml)<br>Tx Timing: 2 d post-injury<br>Tx Location: Intralesional<br>Paired Limb: Yes<br>Follow-up: 6 d | <u>Histology Outcomes</u><br>Birefringence Score: $1.807 \pm 0.478$ | <u>Histology Outcomes</u><br>Birefringence Score: $1.221 \pm 0.335$ |

(table cont'd)

| Study Characteristics | Treatment Outcomes | Control Outcomes |
|---|---|---|
| Geburek *et al*[30], 2017<br>Design: RCT<br>Tx: 9 Horses; Ctrl: 9 Horses<br>Limb: Fore<br>Mean Age: 4 (3 - 6)<br>Mean Weight: 545 (498 – 607)<br>Lesion Location: SDFT<br>Lesion Creation: Surgery<br>Tx: Autologous ASCs (10 million)<br>+ AIS (1 ml)<br>Ctrl: AIS (1 ml)<br>Tx Timing: 2 wk post-injury<br>Tx Location: Intralesional (2 sites)<br>Tx Delivery: Needle (22-G)<br>Paired Limb: Yes<br>Follow-up: 24 wk | <u>Ultrasound Outcomes</u><br>SDFT CSA: $7.9294 \pm 1.0409$ cm$^2$<br>Total Fibre Alignment Score: $3.07 \pm 0.72$<br><br><u>Histology Outcomes</u><br>Fibre Structure: $8.09 \pm 1.62$<br>Fibre Alignment: $8.25 \pm 1.62$<br>Morphology of Tenocyte Nuclei: $8.41 \pm 1.94$<br>Variation in Cell Density: $10.36 \pm 2.75$<br>Vascularization: $8.58 \pm 3.07$<br>Structural Integrity: $15.70 \pm 2.75$<br>Metabolic Activity: $26.86 \pm 6.31$<br>Total Score: $43.20 \pm 8.90$<br><br><u>Biochemical Outcomes</u><br>Total Collagen: $510 \pm 76$ µg/mg<br>Glycosaminoglycan: $21.38 \pm 11.44$ µg/mg<br>DNA: $3.91 \pm 0.96$ µg/mg<br>Hyp: $67 \pm 1$ µg/mg<br>HP: $0.196 \pm 0.025$ mol/mol col<br>LP: $0.016 \pm 0.005$ mol/mol col<br>HLys: $6.90 \pm 1.76$ mol/mol col<br><br><u>Biomechanical Outcomes</u><br>Stress at Failure: $2.32 \pm 5.33$ Mpa<br>Modulus of Elasticity: $46.34 \pm 52.39$ Mpa | <u>Ultrasound Outcomes</u><br>SDFT CSA: $7.7063 \pm 0.7881$ cm$^2$<br>Total Fibre Alignment Score: $2.69 \pm 0.65$<br><br><u>Histology Outcomes</u><br>Fibre Structure: $7.77 \pm 1.62$<br>Fibre Alignment: $8.09 \pm 1.62$<br>Morphology of Tenocyte Nuclei: $7.61 \pm 1.62$<br>Variation in Cell Density: $9.06 \pm 2.59$<br>Vascularization: $6.96 \pm 2.75$<br>Structural Integrity: $15.86 \pm 3.56$<br>Metabolic Activity: $23.79 \pm 6.47$<br>Total Score: $39.32 \pm 9.71$<br><br><u>Biochemical Outcomes</u><br>Total Collagen: $469 \pm 93$ µg/mg<br>Glycosaminoglycan: $26.80 \pm 10.26$ µg/mg<br>DNA: $4.24 \pm 1.31$ µg/mg<br>Hyp: $62 \pm 12$ µg/mg<br>HP: $0.184 \pm 0.026$ mol/mol col<br>LP: $0.017 \pm 0.007$ mol/mol col<br>HLys: $7.39 \pm 2.37$ mol/mol col<br><br><u>Biomechanical Outcomes</u><br>Stress at Failure: $5 \pm 1.90$ Mpa<br>Modulus of Elasticity: $75.61 \pm 29.80$ Mpa |

(table cont'd)

| Study Characteristics | Treatment Outcomes | Control Outcomes |
|---|---|---|
| Nixon *et al*[49], 2008<br>Design: RCT<br>Tx: 4 Horses<br>Ctrl: 4 Horses; Limb: Fore<br>Mean Age: (2 - 6)<br>Lesion Location: SDFT<br>Lesion Creation: Collagenase<br>Tx: Autologous ADNCs (13.83 ± 3.41 million)<br>+ PBS (0.6 ml) per site<br>Ctrl: PBS (0.6 ml) per site<br>Tx Timing: 1 wk post-injury<br>Tx Location: Intralesional (3 sites)<br>Tx Delivery: Needle (22-G)<br>Paired Limb: No<br>Follow-up: 6 wk | <u>Histology Outcomes</u><br>Overall Healing Score: 24.8 ± 2.1<br>Fiber Organization Score: 1.79 ± 0.6<br><br><u>Biochemical Outcomes</u><br>*COL1*: 11.98 ± 1.95<br>*COL3*: 15.38 ± 2.79<br>*Dcn*: 7.62 ± 3.27<br>*COMP*: 2.33 ± 0.64<br>Collagen: 612.12 ± 34.55 µg/mg<br>Glycosaminoglycan: 20.76 ± 7.76 µg/mg<br>DNA: 1.73 ± 0.27 µg/mg | <u>Histology Outcomes</u><br>Overall Healing Score: 33.5 ± 2.8<br>Fiber Organization Score: 2.79 ± 0.4<br><br><u>Biochemical Outcomes</u><br>*COL1*: 11.78 ± 3.73<br>*COL3*: 14.70 ± 4.62<br>*Dcn*: 9.57 ± 2.30<br>*COMP*: 1.10 ± 0.48<br>Collagen: 617.40 ± 34.55 µg/mg<br>Glycosaminoglycan: 21.52 ± 7.25 µg/mg<br>DNA: 2.09 ± 0.46 µg/mg |

(table cont'd)

| Study Characteristics | Treatment Outcomes | Control Outcomes |
|---|---|---|
| Romero *et al*[50], 2017<br>Design: RCT<br>Breed: Cross<br>Sex: Gelding<br>Tx: 6 Horses; Ctrl: 6 Horses<br>Limb: Fore<br>Mean Age: (5 - 8)<br>Lesion Location: SDFT<br>Lesion Creation: Surgery<br>Tx: Autologous BMSCs (20 million)<br>+ LRS (7 ml)<br>Ctrl: LRS (7 ml)<br>Tx Timing: 1 wk post-injury<br>Tx Location: Intralesional<br>Tx Delivery: Needle (18-G)<br>Follow-up: 45 wk | <u>Ultrasound Outcomes</u><br>Tendon Echogenicity Score: $2.16 \pm 0.375$<br>Fiber Pattern Score: $4.92 \pm 0.75$<br>CSA: $18.78 \pm 3.75\%$<br><br><u>Histology Outcomes</u><br>Cell Number: $12.2 \pm 3.33$<br>Collagen Orientation: $5.67 \pm 2.78$<br>Ground Substance: $13.1 \pm 4.78$<br>Tenocyte Morphology: $12.2 \pm 4.56$<br> Vascularity: $11.11 \pm 2.11$<br><br><u>Biochemical Outcomes</u><br>*COLI*: $3.25 \pm 0.94$<br>*COMP*: $0.34 \pm 0.35$<br>*Dcn*: $1.33 \pm 0.54$<br>*COLIII*: $0.16 \pm 0.17$<br>*ACAN*: $0.5 \pm 0.14$<br>*Tnc*: $1.36 \pm 0.61$<br>*MMP-3*: $0.235 \pm 0.094$<br>*Scx*: $0.13 \pm 0.11$<br>*Tnmd*: $0.64 \pm 0.82$ | <u>Ultrasound Outcomes</u><br>Tendon Echogenicity Score: $3.75 \pm 1.08$<br>Fiber Pattern Score: $5.58 \pm 0.52$<br>CSA: $32.5 \pm 7.5\%$<br><br><u>Histology Outcomes</u><br>Cell Number: $17.3 \pm 2.22$<br>Collagen Orientation: $16.8 \pm 4$<br>Ground Substance: $24.2 \pm 4.33$<br>Tenocyte Morphology: $19.8 \pm 3.33$<br>Vascularity: $16.7 \pm 4.67$<br><br><u>Biochemical Outcomes</u><br>*COLI*: $1.19 \pm 2.04$<br>*COMP*: $1.29 \pm 1.25$<br>*Dcn*: $0.87 \pm 0.87$<br>*COLIII*: $0.33 \pm 0.63$<br>*ACAN*: $3 \pm 3.27$<br>*Tnc*: $0.49 \pm 0.24$<br>*MMP-3*: $0.125 \pm 0.12$<br>*Scx*: $0.18 \pm 0.12$<br>*Tnmd*: $1.1 \pm 0.95$ |

(table cont'd)

| Study Characteristics | Treatment Outcomes | Control Outcomes |
|---|---|---|
| Schnabel *et al*[51], 2009<br>Design: RCT<br>Sex: Male (5), Female (7)<br>Tx: 6 Horses; Ctrl: 6 Horses<br>Limb: Fore<br>Mean Age: (2 - 5)<br>Lesion Location: SDFT<br>Lesion Creation: Collagenase<br>Tx: Autologous BMSCs (10 million)<br>+ PBS (1 ml)<br>Ctrl: PBS (1 ml)<br>Tx Timing: 5 d post-injury<br>Paired Limb: Yes<br>Follow-up: 8 wk | <u>Histology Outcomes</u><br>Cell shape: $2.5 \pm 0.54$<br>Cell density: $2.5 \pm 0.54$<br>Free hemorrhage: $1.58 \pm 0.81$<br> Neo-vascularization: $2.33 \pm 0.51$<br>Perivascular cuffing: $1.58 \pm 0.73$<br>Collagen linearity: $2 \pm 0$<br>Collagen uniformity: $2 \pm 0$<br>Polarized crimping: $2.33 \pm 0.81$<br>Epitenon thickening: $2.5 \pm 0.54$<br>COLI immunohistochemistry: $2.58 \pm 0.37$<br>Cumulative score: $21.92 \pm 4.07$<br><br><u>Biochemical Outcomes</u><br>*COL1*: $56.8 \pm 17.15$<br>*COL3*: $320 \pm 110.23$<br>*COMP*: $9.3 \pm 5.14$<br>*IGF-1*: $1.77 \times 10^3 \pm 0.28$<br>*ADAMTS-4*: $0.0348 \pm 0.026$<br>*MMP-13*: $0.0191 \pm 0.0073$<br>*MMP-3*: $0.0022 \pm 0.0022$<br>Collagen: $330.96 \pm 50.83$ μg/mg<br>Glycosaminoglycan: $12.79 \pm 3.31$ μg/mg<br>DNA: $1.55 \pm 0.93$ μg/mg<br><br><u>Biomechanical Outcomes</u><br>Modulus: $61.4 \pm 15.5$ ksi | <u>Histology Outcomes</u><br>Cell shape: $2.75 \pm 0.42$<br>Cell density: $3 \pm 0$<br>Free hemorrhage: $2.25 \pm 0.42$<br> Neo-vascularization: $2.75 \pm 0.42$<br>Perivascular cuffing: $2.08 \pm 0.2$<br>Collagen linearity: $2.5 \pm 0.54$<br>Collagen uniformity: $2.67 \pm 0.51$<br>Polarized crimping: $3 \pm 0$<br>Epitenon thickening: $3 \pm 0$<br>COLI immunohistochemistry: $3.17 \pm 0.76$<br>Cumulative score: $27.17 \pm 1.13$<br><br><u>Biochemical Outcomes</u><br>*COL1*: $67 \pm 18.86$<br>*COL3*: $352 \pm 83.28$<br>*COMP*: $11.2 \pm 4.16$<br>*IGF-1*: $0.163 \pm 0.064$<br>*ADAMTS-4*: $0.02 \pm 0.0069$<br>*MMP-13*: $0.0281 \pm 0.0199$<br>*MMP-3*: $0.0018 \pm 0.0012$<br>Collagen: $298.67 \pm 43.48$ μg/mg<br>Glycosaminoglycan: $10.70 \pm 5.49$ μg/mg<br>DNA: $1.85 \pm 1.1$ μg/mg<br><br><u>Biomechanical Outcomes</u><br>Modulus: $38.9 \pm 27.0$ ksi |

(table cont'd)

| Study Characteristics | Treatment Outcomes | Control Outcomes |
|---|---|---|
| Renzi *et al*[52], 2013<br>Design: PCS<br>Breed: Thoroughbred<br>Sex: Gelding<br>Discipline: Steeplechase<br>Tx: 21 Horses; Ctrl: 12 Horses<br>Limb: Fore<br>Mean Age: 5.2 (2 - 8)<br>Lesion Location: SDFT (15), DDFT (8), SL (10)<br>Lesion Creation: Natural Injury<br>Tx: Autologous BMSCs<br>+ Autologous PRP (2 million/ml: 2 - 6 ml)<br>Ctrl: Pin Firing<br>Tx Timing: ≈ 20 d post-injury<br>Tx Location: Intralesional<br>Tx Delivery: Needle (22-G)<br>Paired Limb: No<br>Follow-up: ~ 15 mo | <u>Clinical Outcomes</u><br>Return to Soundness: 13 | <u>Clinical Outcomes</u><br>Return to Soundness: 3 |

(table cont'd)

| Study Characteristics | Treatment Outcomes | Control Outcomes |
|---|---|---|
| Rivera *et al*[53], 2020<br>Design: PCS<br>Breed: Holsteriner<br>Discipline: Racehorse<br>Tx: 5 Horses; Ctrl: 5 Horses<br>Limb: Fore<br>Lesion Location: SDFT<br>Lesion Creation: Natural Injury<br>Tx: Autologous ASCs (0.6 million)<br>+ PBS (0.6 ml)<br>Ctrl: Conventional Therapy<br>Tx Location: Intralesional<br>Tx Delivery: Needle (22-G)<br>Paired Limb: No<br>Follow-up: 16 wk | <u>Ultrasound Outcomes</u><br>Scar Length: $26.7 \pm 3.33\%$ | <u>Ultrasound Outcomes</u><br>Scar Length: $83.92 \pm 17.1\%$ |

(table cont'd)

| Study Characteristics | Treatment Outcomes | Control Outcomes |
|---|---|---|
| Smith *et al*[29], 2013<br>Design: PCS<br>Breed: TB, TB Cross<br>Sex: Gelding<br>Tx: 6 Horses; Ctrl: 6 Horses<br>Limb: Fore<br>Mean Age: 7.8 ± 3.0 (5 - 15)<br>Lesion Location: SDFT<br>Lesion Creation: Natural Injury<br>Tx: Autologous BMSCs (10 million)<br>+ Autologous Citrated BM (2 ml)<br>Ctrl: Saline (2 ml)<br>Tx Timing: ~ 2 mo post-injury<br>Tx Location: Intralesional (2 – 4 sites)<br>Tx Delivery: Needle (19-G)<br>Follow-up: 24 wk | Ultrasound Outcomes<br>SDFT CSA 1.79 ± 1.04 $cm^2$<br>SDFT 24 wk/Initial: 90.29 ± 12.35%<br><br>Histology Outcomes<br>Organization Score: 19.10 ± 4.02<br>Crimp Score: 0 ± 0.73<br>Cellularity Score: 18.58 ± 6.56<br>Vascularity Score: 11.82 ± 4.98<br><br>Biochemical Outcomes<br>DNA: 2.00 ± 1.04 µg/mg<br>Hydroxyproline: 83.68 ± 37.92 µg/mg<br>sGAG: 13.00 ± 10.95 µg/mg<br>TLF: 1846.7 ± 197.9 units/mg<br>Active MMP-13: 0.79 ± 0.22 µg/mg<br><br>Biomechanical Outcomes<br>Stiffness: 1146.34 ± 600.73 N%<br>Modulus: 607.05 ± 414.55 $N/cm^2$ | Ultrasound Outcomes<br>SDFT CSA: 4.01 ± 0.54 $cm^2$<br>SDFT CSA 24 wk/Initial: 113.34 ± 37.60%<br><br>Histology Outcomes<br>Organization Score: 34.46 ± 3.47<br>Crimp Score: 2 ± 0<br>Cellularity Score: 38.94 ± 4.72<br>Vascularity Score: 31.09 ± 14.01<br><br>Biochemical Outcomes<br>DNA: 3.70 ± 1.45 µg/mg<br>Hydroxyproline: 99.70 ± 37.58 µg/mg<br>sGAG: 36.14 ± 12.40 µg/mg<br>TLF: 1645.4 ± 181.3 units/mg<br>Active MMP-13: 1.63 ± 0.32 µg/mg<br><br>Biomechanical Outcomes<br>Stiffness: 1524.39 ± 135.49 N%<br>Modulus: 384.15 ± 122.97 $N/cm^2$ |

Age is presented as mean ± standard deviation (range) years. Weight is presented presented as mean ± standard deviation (range) kg. Outcomes included are those at the endpoint of follow-up period, unless stated otherwise in parenthesis. Quantitative outcomes are presented as mean ± standard deviation. Gene (italic letter) expression is presented as fold change normalized to unaffected region. Tx: treatment; Ctrl: control; RCT: randomized controlled trial; PCS: prospective cohort study; TB: thoroughbred; SDFT: superficial digital flexor tendon; DDFT: deep digital flexor tendon; SL: suspensory ligament; ASCs: adipose-derived stromal multipotent cells; BMSCs: bone marrow-derived multipotent stromal cells; MSCs: multipotent stromal cells; CSA: cross sectional area; MRI: magnetic resonance imaging; HE: hematoxylin and eosin staining; MNC: mononuclear cells; APT: anterior posterior thickness; UTC: ultrasound tissue source characterization; CFA: collagen fiber alignment; SHG: second harmonic generation microscopy; AIS: autologous inactivated serum; VIII: von Willebrand factor III; TLF: tissue-linked fluorescence.

Table 1.2. Retrospective case series characteristics and outcomes.

| Study Characteristics | Intervention | Treatment Outcomes | Control Outcomes |
|---|---|---|---|
| Hawkins *et al*[54], 2022<br>Design: RCS<br>Limb: Fore, Hind<br>Lesion Location: SML<br>Lesion Creation: Natural Injury | Tx: 2 Horses; Ctrl: 4 Horses<br>Tx: Autologous BMSCs (10 million)<br>+ Tenoscopic Debridement<br>Ctrl: Tenoscopic Debridement<br>Tx Timing: 14 - 16 d post-BM aspiration | Return to Soundness: 1 | Return to Soundness: 2 |
| Marfe *et al*[55], 2012<br>Design: RCS<br>Sex: Male (5), Female (1)<br>Discipline: Racehorse (3)<br>Mean Age: (10 - 20)<br>Lesion Location: SDFT<br>Lesion Creation: Natural Injury | Tx: 3 Horses; Ctrl: 3 Horses<br>Tx: Autologous CD90$^+$ BDSCs<br>+ PBS<br>Ctrl: Conventional Therapy<br>Tx Location: Intralesional +<br>Intravenous<br>Follow-up: 3 yr | Return to Soundness: 3 | Return to Soundness: 0 |
| Murphy *et al*[56], 2021<br>Design: RCS<br>Breed: TB, Standard<br>Discipline: Racehorse<br>Limb: Fore<br>Lesion Location: SDFT<br>Lesion Creation: Natural Injury | Tx: 39 Horses; Ctrl: 38 Horses<br>Tx: ALSDFT Desmotomy<br>+ Autologous BM Aspirate<br>Ctrl: Conventional Therapy<br>Tx Location: Intralesional ($\geq$ 1 site)<br>Tx Delivery: Needle (19-G)<br>Follow-up: 12 mo | Return to Soundness: 32<br>Number of Starts: $14.8 \pm 2.8$ | Return to Soundness: 21<br>Number of Starts: $9.1 \pm 2.8$ |
| Pacini *et al*[57], 2007<br>Design: RCS<br>Sex: Male (20), Female (6)<br>Mean Age: (2 - 15)<br>Lesion Location: SDFT<br>Lesion Creation: Natural Injury | Tx: 11 Horses; Ctrl: 15 Horses<br>Tx: Autologous BMSCs (0.6 - 31.2 million) + Autologous Serum (1.5 ml)<br>Ctrl: Conventional Therapy<br>Tx Location: Intralesional<br>Tx Delivery: Needle<br>Follow-up: ~ 12 mo | Return to Soundness: 9 | Return to Soundness: 0 |

(table cont'd)

| Study Characteristics | Intervention | Treatment Outcomes | Control Outcomes |
|---|---|---|---|
| Van Loon *et al*[31], 2014<br>Design: RCS<br>Breed: Warmblood<br>Sex: Male, Gelding, Female<br>Mean Age: 9.9 ± 3.5<br>Discipline: Dressage, Show Jumping<br>Limb: Fore, Hind<br>Lesion Location: SDFT, SL, DDFT, ALDDFT<br>Lesion Creation: Natural Injury | Tx: 52 Horses; Ctrl: 3 Horses<br>Tx: Allogenic UCB-MSCs (~ 10 million)<br>Ctrl: Conventional Therapy<br>Tx Location: Intralesional (≥ 1 site, 1 – 2 times)<br>Tx Delivery: Needle (23-G)<br>Follow-up: ≥ 6 mo | Return to Soundness: 40 | Return to Soundness: 2 |

Age is presented as mean ± standard deviation (range) years. Weight is presented presented as mean ± standard deviation (range) kg. Outcomes included are those at the endpoint of follow-up period, unless stated otherwise in parenthesis. Tx: treatment; Ctrl: control; RCS: retrospective case series; TB: thoroughbred; SML: sesamoidean ligament; SDFT: superficial digital flexor tendon; DDFT: deep digital flexor tendon; SL: suspensory ligament; ALDDFT: accessory ligament of deep digital flexor tendon; ALSDFT: accessory ligament of the superficial digital flexor tendon; BMSCs: bone marrow-derived multipotent stromal cells; BDCSs: blood-derived stem cells; BM: bone marrow; UCB-MSCs: umbilical cord blood-derived multipotent stromal cells.

### 1.3.2. Rate of Returning to Soundness after Injury

The rate of returning to soundness was extracted from a prospective cohort study and 5 retrospective case series,[31,52,54-57] which included 128 horses in the stem cell group and 75 in the control group. A total of 98 (76.6%) horses treated with stem cell, whereas 28 (37.3%) horses treated with non-stem cell returned to soundness (Fig 1.2). The overall heterogeneity across the subgroups was low ($I^2$ = 29%, p = 0.22), which suggested the use of the fixed effect model was appropriate. The rate of returning to soundness was higher in the stem cell group than in the control group (OR = 5.30, 95% CI [2.65, 10.62], p < 0.0001).

### 1.3.3. Ultrasound Evaluation of Injured Tendon and Ligament

The percentage of ultrasonographic lesion cross sectional area (CSA) at the endpoint of the follow-up period was extracted from 3 randomized controlled trials,[41,43,50] which included 22 horses each in both the stem cell group and the control group (Fig 1.3). The overall heterogeneity across was high ($I^2$ = 88%, p = 0.0002), which suggested use of the random effect model was appropriate. The percentage of lesion CSA was not different between the stem cell and control groups (MD = -6.47, 95% CI [-14.74, 1.81], p = 0.13).

The echogenicity score[58,59] of lesions at the endpoint of the follow-up period was extracted from 2 randomized controlled trials,[46,50] which included 14 horses each in both the stem cell group and the control group (Fig 1.4). The overall heterogeneity across was low ($I^2$ = 0%, p = 0.83), which suggested use of the fixed effect model was appropriate. The echogenicity score of lesions was lower, indicating less hyperechoic and more normal appearance, in the stem cell group than the control group (SMD = -1.94, 95% CI [-2.89, 0.98], p < 0.0001).

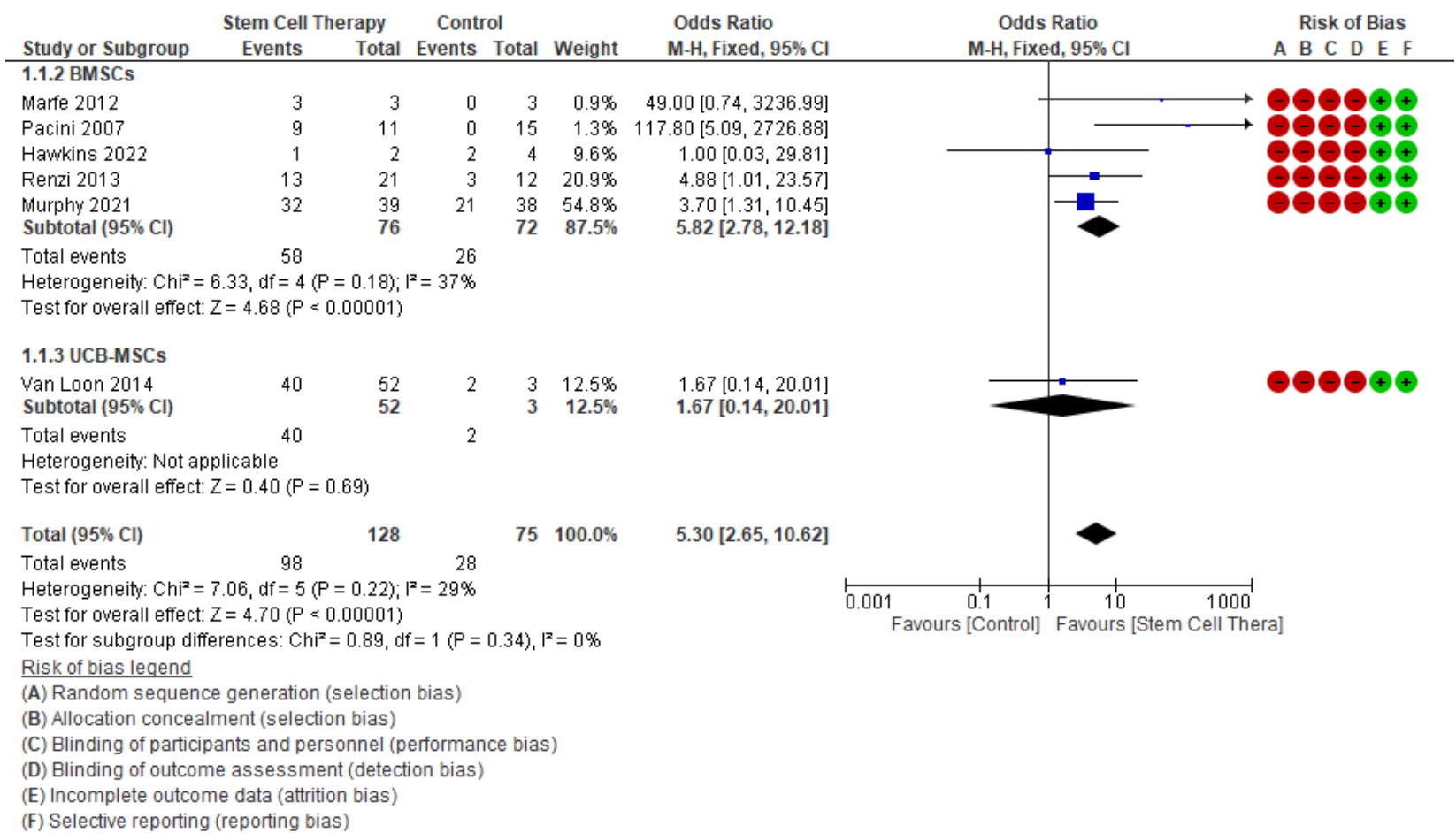

Figure 1.2. Forest plot of the effect of stem cell treatment on the odds of return to soundness with a fixed effects model. Odds ratio (OR) = 1 indicates no difference between treatment and control, whereas an OR > 1 indicates that stem cell treatment is associated with an increased frequency of return to soundness/non-return to soundness. Confidence intervals (95% CI) that overlap an OR of 1 suggest a lack of association between the treatment and outcome.

The fiber alignment score[58-61] of lesions at the endpoint of the follow-up period was extracted from 3 randomized controlled trials,[30,46,50] which included 23 horses each in both the stem cell group and the control group (Fig 1.5). The overall heterogeneity across was low ($I^2 = 87\%$, p = 0.0004), which suggested use of the random effect model was appropriate. Fiber alignment score of lesions was not different between the stem cell and control groups (SMD = -1.12, 95% CI [-3.12, 0.88], p = 0.27).

Gross transactional area of SDFT measured as CSA[29,30] or anterior-posterior thickness[46] at the endpoint of the follow-up period was extracted from 2 randomized controlled trials and a prospective cohort study, which included 23 horses each in both the stem cell group and the control group (Fig 1.6). The overall heterogeneity across was high ($I^2 = 84\%$, p = 0.002), which suggested use of the random effect model was appropriate. Gross transactional area of SDFT was not different between the stem cell and control groups (SMD = -1.36, 95% CI [-3.18, 0.46], p = 0.14).

Vascularity of SDFT measured as power doppler score[43] or color doppler density[44] at the early phase of healing (2 week post-injury) was extracted from 2 randomized controlled trials, which included 13 horses each in both the stem cell group and the control group (Fig 1.7). The overall heterogeneity across was low ($I^2 = 0\%$, p = 0.92), which suggested use of the fixed effect model was appropriate. More abundant vascularity of the SDFT represented by the higher value of power doppler score and color doppler density was observed in the stem cell group as compared to the control group (SMD = 1.07, 95% CI [0.23, 1.92], p = 0.01).

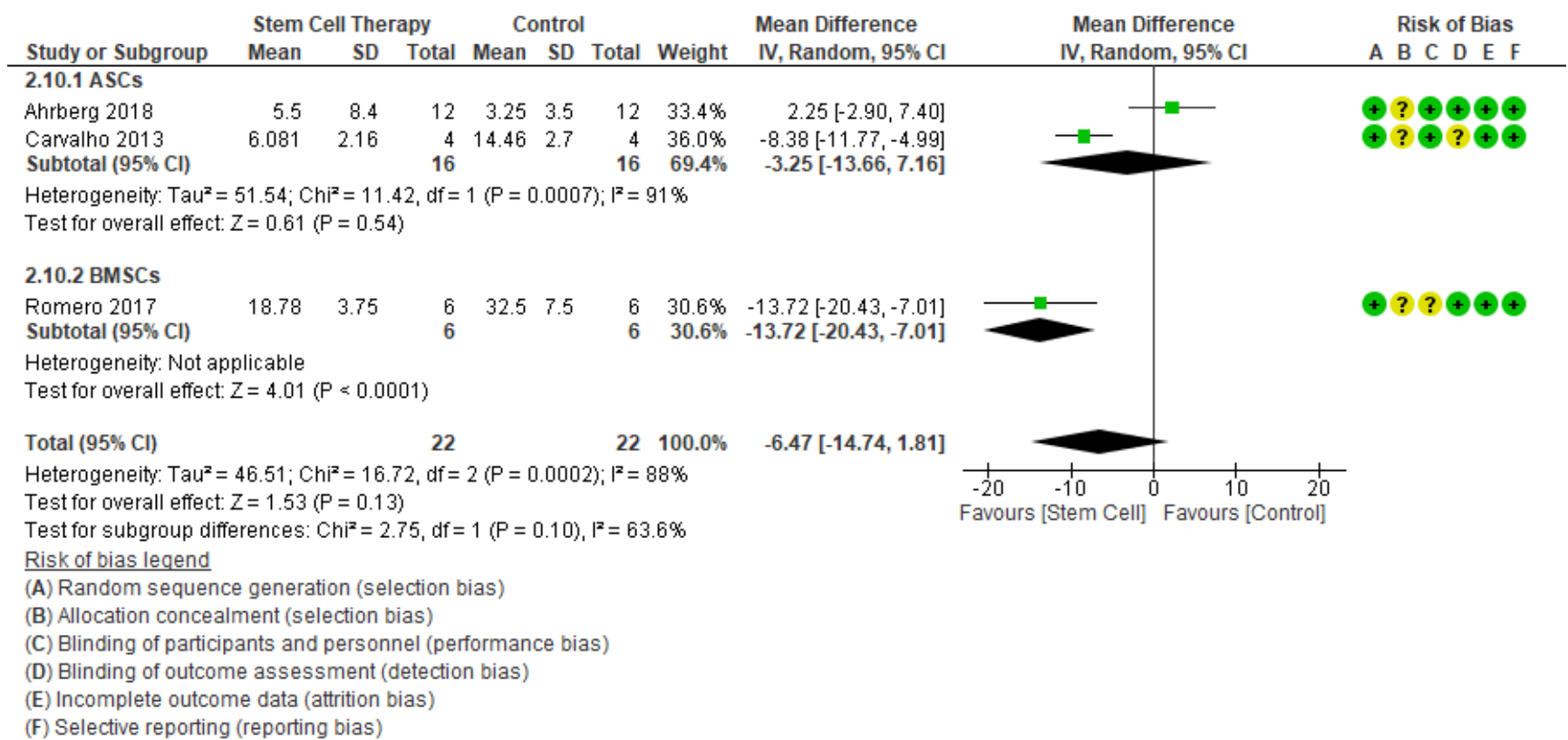

Figure 1.3. Forest plot for the ultrasonography lesion CSA percentage.

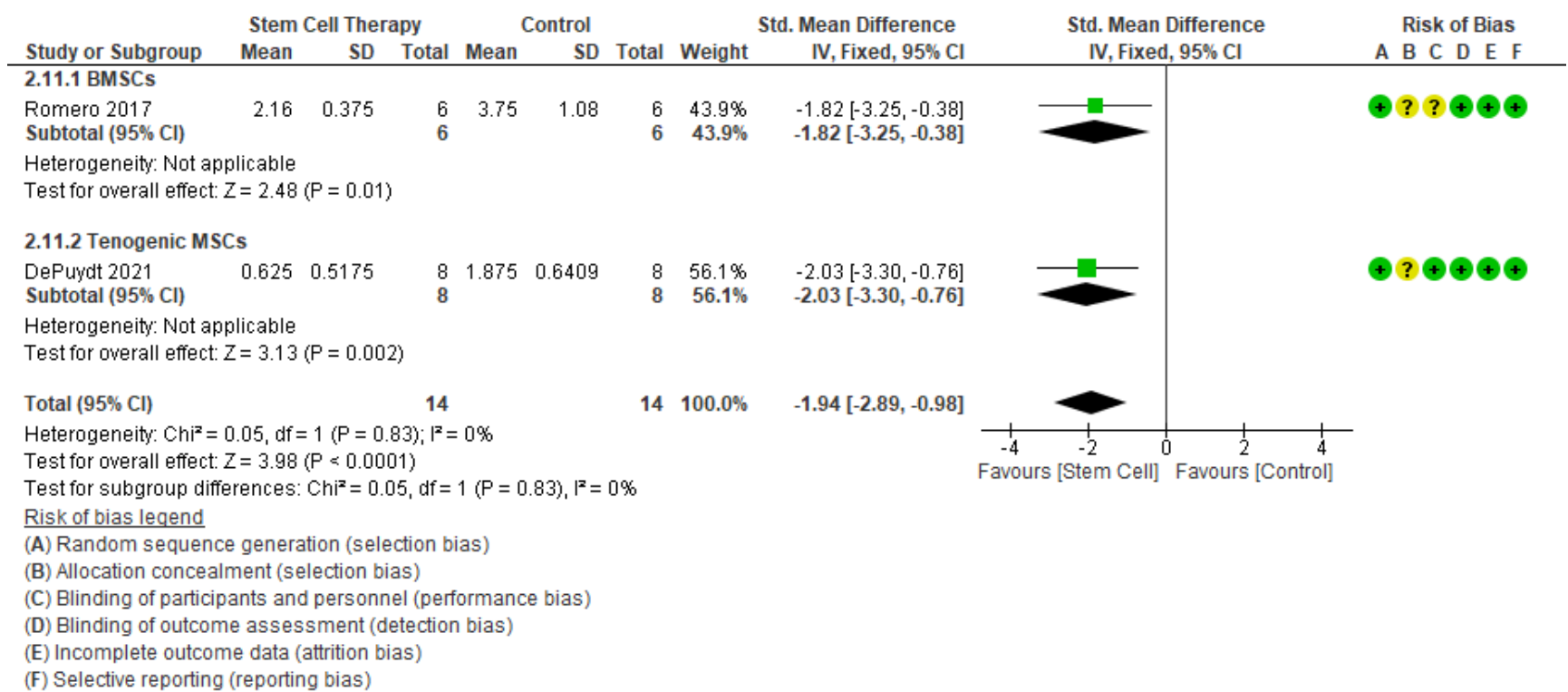

Figure 1.4. Forest plot for the ultrasonography echogenicity score. Lower score indicates normal echogenicity and higher score indicates abnormal echogenicity.

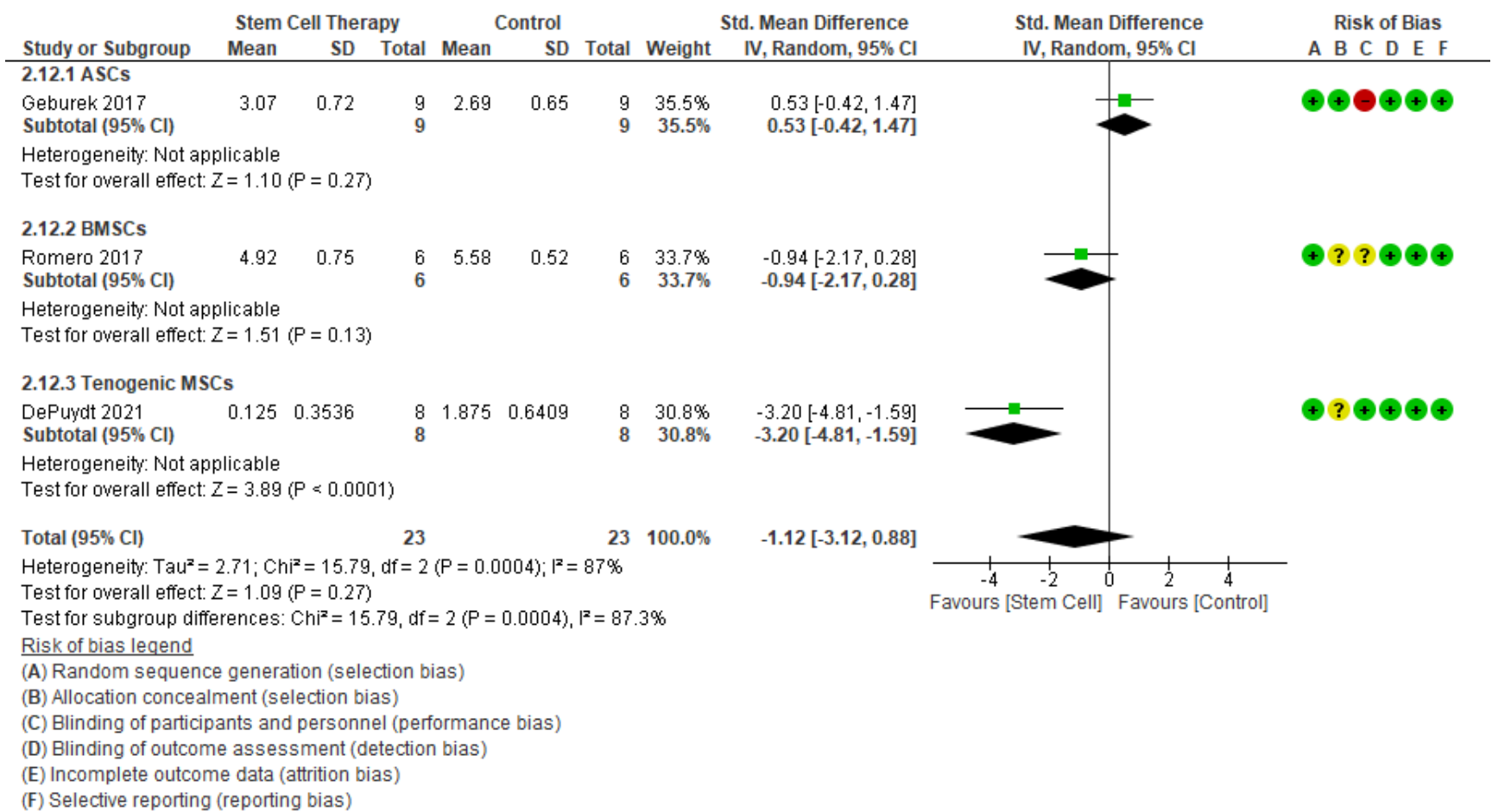

Figure 1.5. Forest plot for the ultrasonography fiber alignment score. Lower score indicates normal fiber alignment and higher score indicates abnormal fiber alignment.

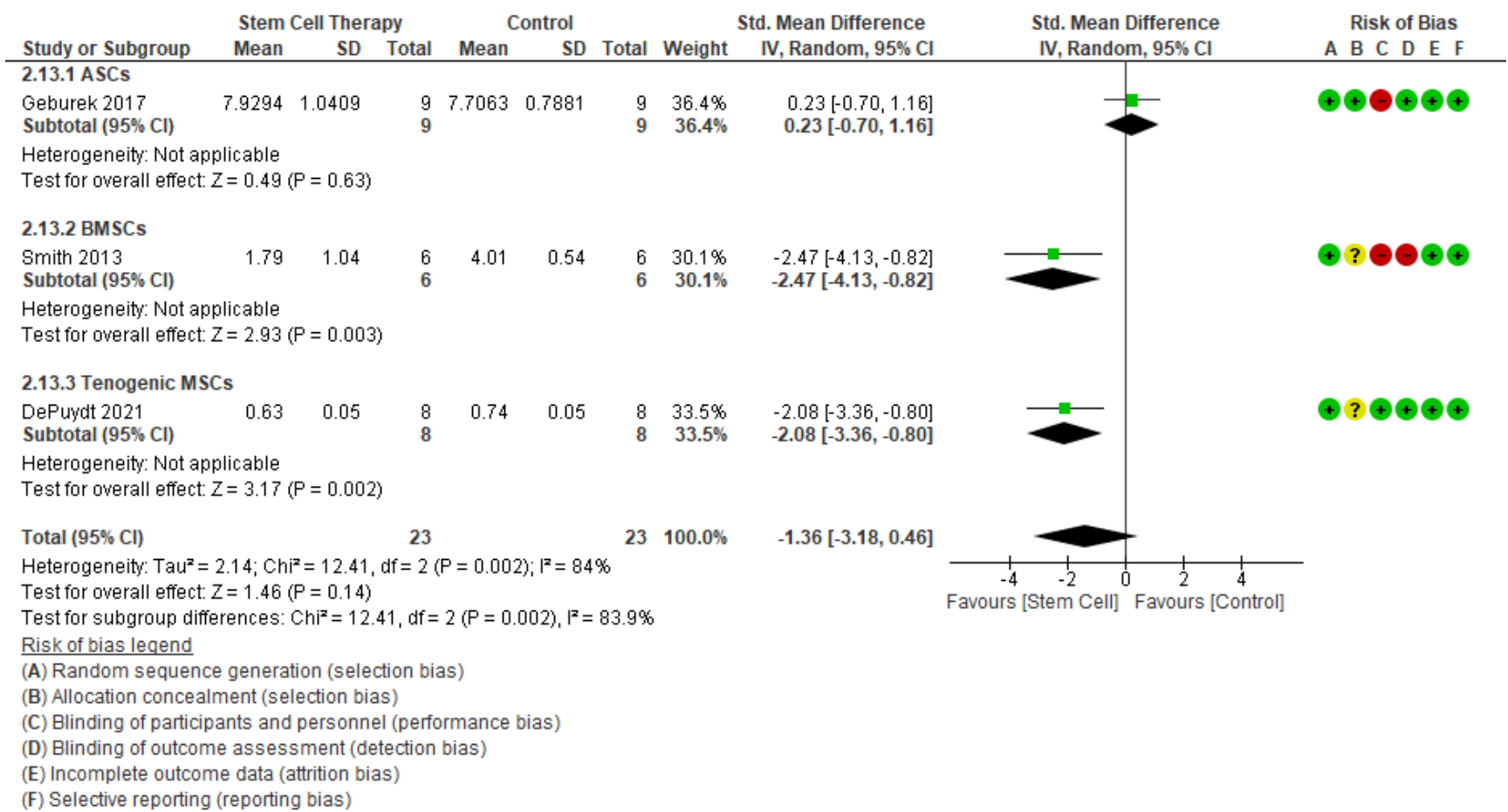

Figure 1.6. Forest plot for the ultrasonography SDFT CSA and anterior-posterior thickness.

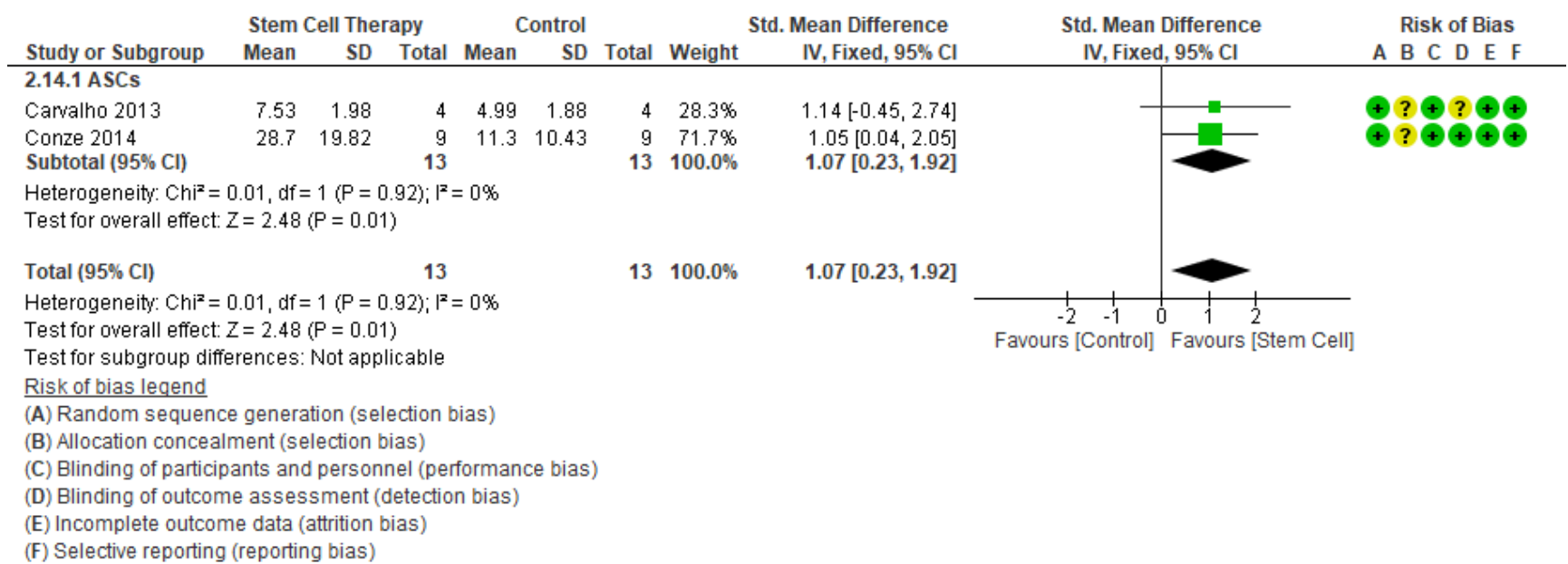

Figure 1.7. Forest plot for the ultrasonography color doppler.

### 1.3.4. Gene Expression of Injured Tendon and Ligament

Expression of the *scleraxis* (*Scx*) gene, expressed as fold change relative to housekeeping genes ($2^{-\Delta Ct}$)[41,50], at the endpoint of the follow-up period was extracted from 2 randomized controlled trials, which included 18 horses each in both the stem cell group and the control group (Fig 1.8). SMD was compared because although both studies applied the same $2^{-\Delta Ct}$ method, different housekeeping genes were used. The overall heterogeneity across was low ($I^2 = 0\%$, p = 0.69), which suggested use of the fixed effect model was appropriate. The expression of *Scx* was not different between stem cell and control groups (SMD = -0.21, 95% CI [-0.87, 0.45], p = 0.53).

Expression of *tenomodulin* (*Tnmd*) gene, expressed as fold change relative to housekeeping genes ($2^{-\Delta Ct}$)[50] or fold change relative to housekeeping gene and normal tendon within each horse ($2^{-\Delta\Delta Ct}$)[47], at the endpoint of the follow-up period was extracted from 2 randomized controlled trials, which included 14 horses each in both the stem cell group and the control group (Fig 1.9). SMD was compared because both studies applied different methods to report gene expression level. The overall heterogeneity across was low ($I^2 = 0\%$, p = 0.34), which suggested use of the fixed effect model was appropriate. The expression of *Tnmd* was not different between stem cell and control groups (SMD = -0.05, 95% CI [-0.80, 0.70], p = 0.90).

Expression of *tenascin-C* (*Tnc*) gene, expressed as fold change relative to housekeeping genes ($2^{-\Delta Ct}$)[41,50], at the endpoint of the follow-up period was extracted from 2 randomized controlled trials, which included 18 horses each in both the stem cell group and the control group (Fig 1.10). SMD was compared because although both studies applied same $2^{-\Delta Ct}$ method, different housekeeping genes were used. The overall heterogeneity across was high ($I^2 = 79\%$, p = 0.03), which suggested use of the random effect model was appropriate. The expression of *Tnc* was not different between stem cell and control groups (SMD = 0.72, 95% CI [-1.06, 2.50], p = 0.43).

Expression of *matrix metalloproteinase 3* (*MMP-3*) gene, expressed as fold change relative to housekeeping genes ($2^{-\Delta Ct}$)[50] or total copy number obtained from a standard curve and normalized to 18S rRNA expression[51], at the endpoint of the follow-up period was extracted from 2 randomized controlled trials, which included 12 horses each in both the stem cell group and the control group (Fig 1.11). SMD was compared because both studies applied different methods to report gene expression level. The overall heterogeneity across was low ($I^2 = 0\%$, p = 0.39), which suggested use of the fixed effect model was appropriate. The expression of *MMP-3* was not different between stem cell and control groups (SMD = 0.55, 95% CI [-0.28, 1.38], p = 0.20).

Expression of *collagen 1* (*ColI*) gene, expressed as fold change relative to housekeeping genes ($2^{-\Delta Ct}$)[41,50] or fold change relative to housekeeping gene and normal tendon within each horse ($2^{-\Delta\Delta Ct}$)[47] or total copy number obtained from a standard curve and normalized to 18S rRNA expression[49,51], at the endpoint of the follow-up period was extracted from 5 randomized controlled trials, which included 36 horses each in both the stem cell group and the control group (Fig 1.12). SMD was compared because the studies applied different methods to report gene expression level. The overall heterogeneity across was low ($I^2 = 23\%$, p = 0.27), which suggested use of the fixed effect model was appropriate. The expression of *ColI* was not different between stem cell and control groups (SMD = -0.12, 95% CI [-0.59, 0.36], p = 0.63).

Expression of *collagen 3* (*ColIII*) gene, expressed as fold change relative to housekeeping genes ($2^{-\Delta Ct}$)[41,50] or fold change relative to housekeeping gene and normal tendon within each horse ($2^{-\Delta\Delta Ct}$)[47] or total copy number obtained from a standard curve and normalized to 18S rRNA expression[49,51], at the endpoint of the follow-up period was extracted from 5 randomized controlled trials, which included 36 horses each in both the stem cell group and the control group (Fig 1.13). SMD was compared because the studies applied different methods to report gene expression level. The overall heterogeneity across was low ($I^2 = 0\%$, p = 0.41), which suggested use of the fixed effect model was appropriate. The expression of *ColIII* was not different between stem cell and control groups (SMD = -0.20, 95% CI [-0.67, 0.28], p = 0.42).

Expression of *cartilage oligomeric matrix protein* (*COMP*) gene, expressed as fold change relative to housekeeping genes ($2^{-\Delta Ct}$)[50] or fold change relative to housekeeping gene and normal tendon within each horse ($2^{-\Delta\Delta Ct}$)[47] or total copy number obtained from a standard curve and normalized to 18S rRNA expression[49,51], at the endpoint of the follow-up period was extracted from 4 randomized controlled trials, which included 24 horses each in both the stem cell group and the control group (Fig 1.14). SMD was compared because the studies applied different methods to report gene expression level. The overall heterogeneity across was high ($I^2 = 52\%$, p = 0.10), which suggested use of the random effect model was appropriate. The expression of *COMP* was not different between stem cell and control groups (SMD = -0.06, 95% CI [-0.97, 0.85], p = 0.89).

Expression of *decorin* (*Dcn*) gene, expressed as fold change relative to housekeeping genes ($2^{-\Delta Ct}$)[41,50] or total copy number obtained from a standard curve and normalized to 18S rRNA expression[49], at the endpoint of follow-up period was extracted from 3 randomized controlled trials, which included 22 horses each in both the stem cell group and the control group (Fig 1.15). SMD was compared because the studies applied different methods to report gene expression level. The overall heterogeneity across was low ($I^2 = 8\%$, p = 0.34), which suggested use of the fixed effect model was appropriate. The expression of *Dcn* was not different between stem cell and control groups (SMD = -0.15, 95% CI [-0.75, 0.45], p = 0.62).

### 1.3.5. Compositional Analysis of Injured Tendon and Ligament

The amount of DNA in injured tendon or ligament, measured by bisbenzimide staining and expressed as μg/mg dry weight tendon[29,30,49,51] or μg/mg wet weight tendon[47], at the endpoint of the follow-up period was extracted from 4 randomized controlled trials and a prospective cohort study, which included 33 horses each in both the stem cell group and the control group (Fig 1.16). SMD was compared because, although all studies measured DNA with bisbenzimide, the initial tissue processing differs. The overall heterogeneity across was low ($I^2 = 0\%$, p = 0.67), which suggested use of the fixed effect model was appropriate. The amount of DNA was not different between stem cell and control groups (SMD = -0.45, 95% CI [-0.94, 0.05], p = 0.08).

The amount of glycosaminoglycan (GAG) in injured tendon or ligament, measured by dimethylmethylene blue (DMMB) dye staining and expressed as μg/mg dry weight tendon[29,30,49,51] or μg/mg wet weight tendon[47], at the endpoint of the follow-up period was extracted from 4 randomized controlled trials and a prospective cohort study, which included 33 horses each in both the stem cell group and the control group (Fig 1.17). SMD was compared because, although all studies measured GAG with DMMB, the initial tissue processing differs. The overall heterogeneity across was low ($I^2 = 42\%$, p = 0.14), which suggested use of the fixed effect model was appropriate. The amount of GAG was not different between stem cell and control groups (SMD = -0.25, 95% CI [-0.75, 0.26], p = 0.34).

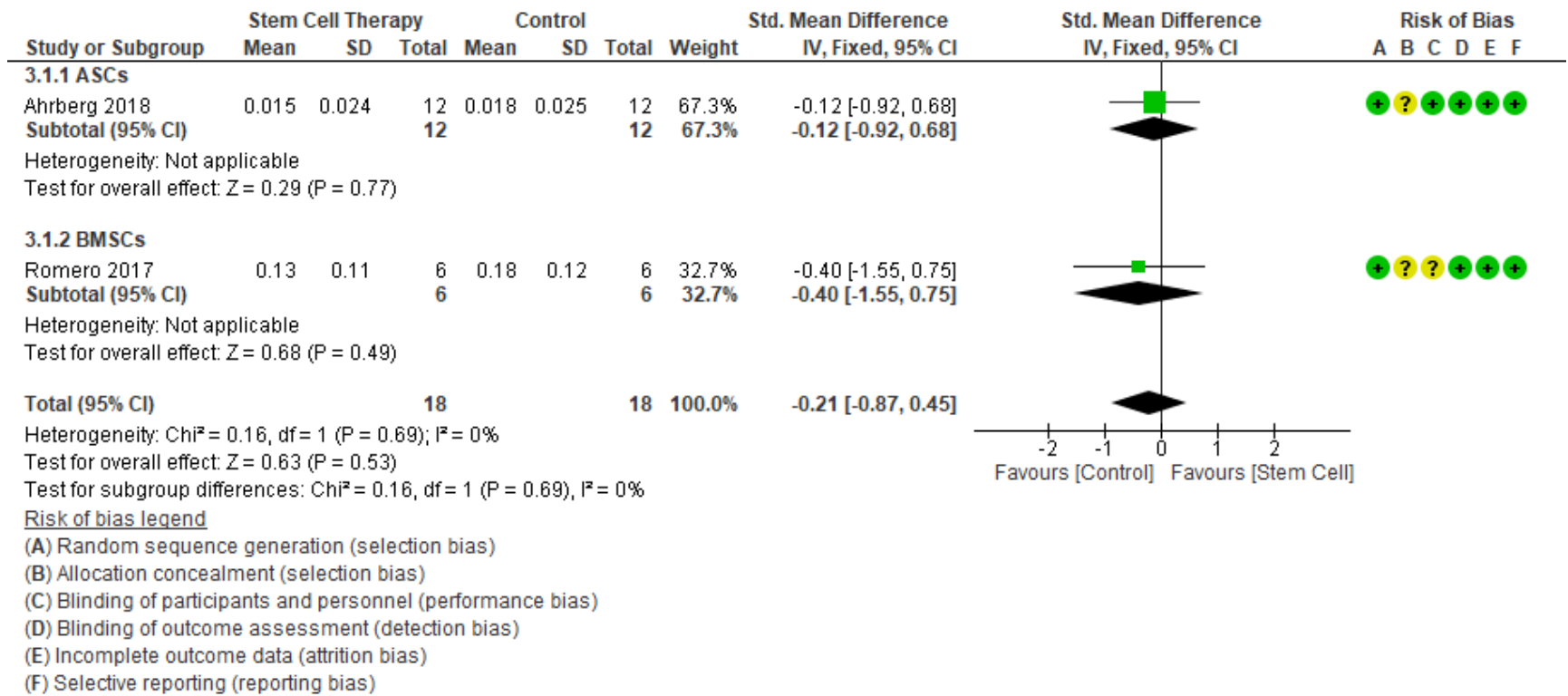

Figure 1.8. Forest plot for *Scx* gene expression.

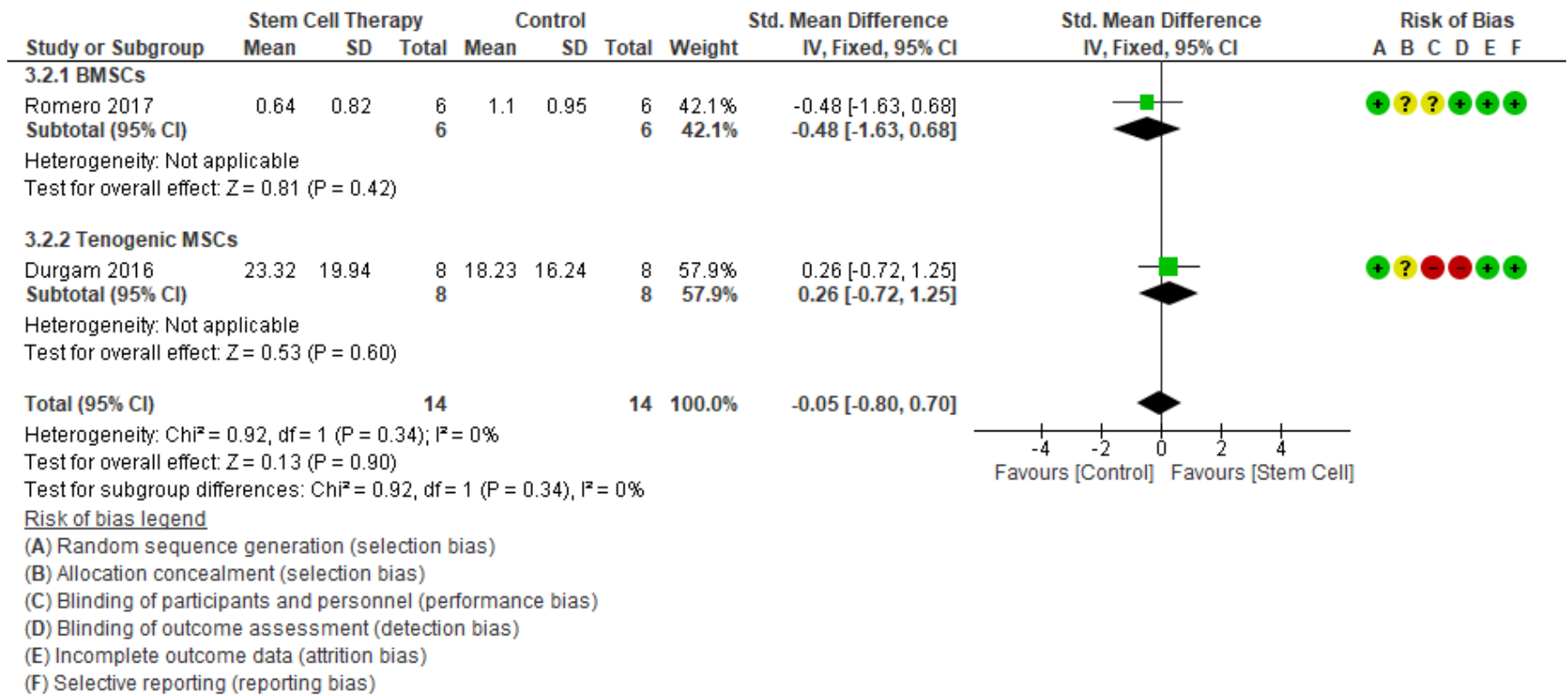

Figure 1.9. Forest plot for *Tnmd* gene expression.

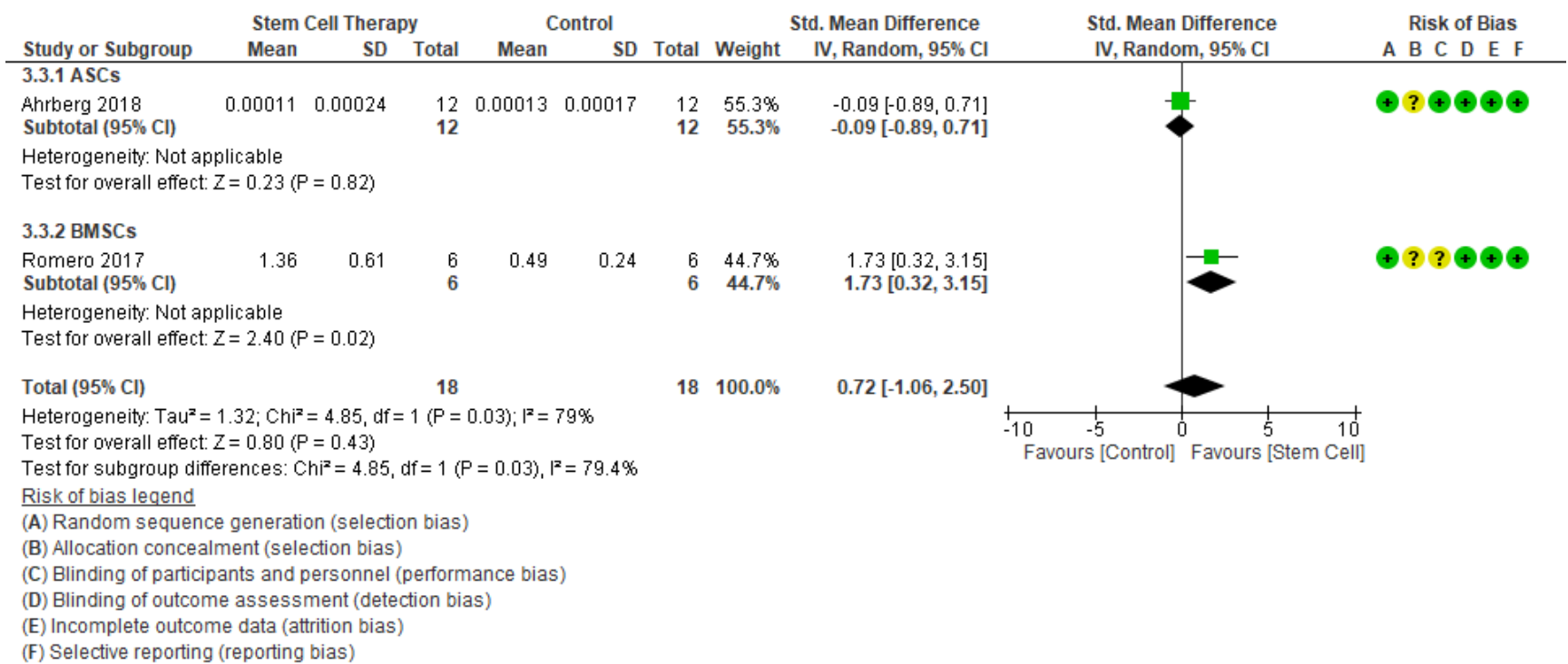

Figure 1.10. Forest plot for *Tnc* gene expression.

|  | Stem Cell Therapy | | | Control | | | | Std. Mean Difference | Std. Mean Difference | Risk of Bias |
|---|---|---|---|---|---|---|---|---|---|---|
| Study or Subgroup | Mean | SD | Total | Mean | SD | Total | Weight | IV, Fixed, 95% CI | IV, Fixed, 95% CI | A  B  C  D  E  F |
| **3.4.1 BMSCs** | | | | | | | | | | |
| Romero 2017 | 0.235 | 0.094 | 6 | 0.125 | 0.12 | 6 | 46.4% | 0.94 [-0.28, 2.16] | | + ? ? + + + |
| Schnabel 2009 | 0.0022 | 0.0022 | 6 | 0.0018 | 0.0012 | 6 | 53.6% | 0.21 [-0.93, 1.34] | | + ? ? + + + |
| **Subtotal (95% CI)** | | | **12** | | | **12** | **100.0%** | **0.55 [-0.28, 1.38]** | | |

Heterogeneity: Chi² = 0.74, df = 1 (P = 0.39); I² = 0%
Test for overall effect: Z = 1.29 (P = 0.20)

| **Total (95% CI)** | | | **12** | | | **12** | **100.0%** | **0.55 [-0.28, 1.38]** | | |

Heterogeneity: Chi² = 0.74, df = 1 (P = 0.39); I² = 0%
Test for overall effect: Z = 1.29 (P = 0.20)
Test for subgroup differences: Not applicable

Risk of bias legend
(A) Random sequence generation (selection bias)
(B) Allocation concealment (selection bias)
(C) Blinding of participants and personnel (performance bias)
(D) Blinding of outcome assessment (detection bias)
(E) Incomplete outcome data (attrition bias)
(F) Selective reporting (reporting bias)

Figure 1.11. Forest plot for *MMP-3* gene expression.

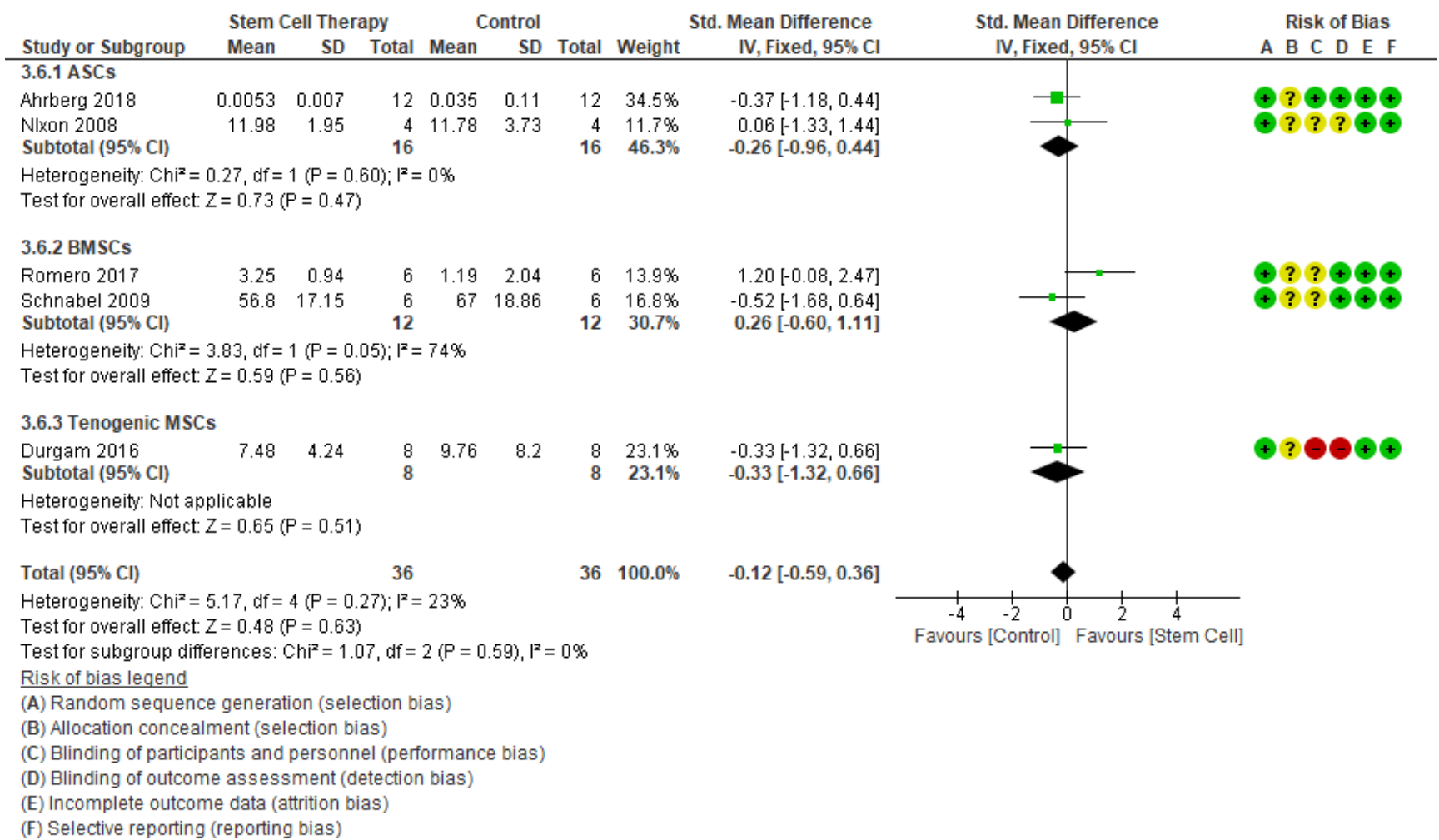

Figure 1.12. Forest plot for *ColI* gene expression.

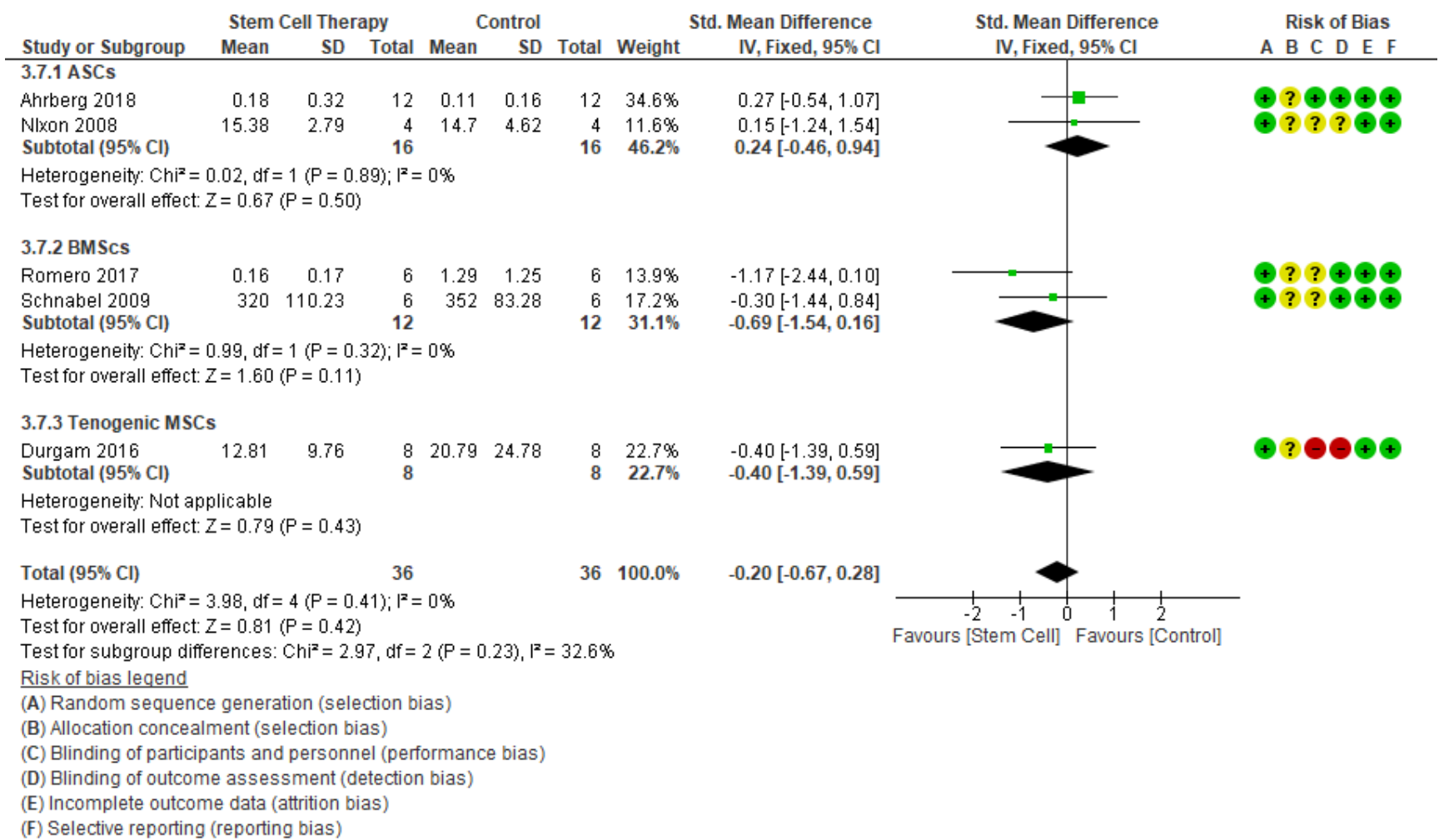

Figure 1.13. Forest plot for *ColIII* gene expression.

|  | Stem Cell Therapy | | | Control | | | | Std. Mean Difference | Std. Mean Difference | Risk of Bias |
| Study or Subgroup | Mean | SD | Total | Mean | SD | Total | Weight | IV, Random, 95% CI | IV, Random, 95% CI | A B C D E F |
| **3.8.1 ASCs** | | | | | | | | | | |
| Nixon 2008 | 2.33 | 0.64 | 4 | 1.1 | 0.48 | 4 | 15.6% | 1.89 [-0.01, 3.79] | | + ? ? ? + + |
| Subtotal (95% CI) | | | 4 | | | 4 | 15.6% | 1.89 [-0.01, 3.79] | | |
| Heterogeneity: Not applicable | | | | | | | | | | |
| Test for overall effect: Z = 1.95 (P = 0.05) | | | | | | | | | | |
| **3.8.2 BMSCs** | | | | | | | | | | |
| Romero 2017 | 0.34 | 0.35 | 6 | 1.29 | 1.25 | 6 | 25.9% | -0.96 [-2.18, 0.27] | | + ? ? + + + |
| Schnabel 2009 | 9.3 | 5.14 | 6 | 11.2 | 4.16 | 6 | 27.5% | -0.38 [-1.52, 0.77] | | + ? ? + + + |
| Subtotal (95% CI) | | | 12 | | | 12 | 53.3% | -0.65 [-1.48, 0.19] | | |
| Heterogeneity: Tau² = 0.00; Chi² = 0.46, df = 1 (P = 0.50); I² = 0% | | | | | | | | | | |
| Test for overall effect: Z = 1.51 (P = 0.13) | | | | | | | | | | |
| **3.8.3 Tenogenic MSCs** | | | | | | | | | | |
| Durgam 2016 | 3.13 | 1.7 | 8 | 3.17 | 2.1 | 8 | 31.1% | -0.02 [-1.00, 0.96] | | + ? ● ● + + |
| Subtotal (95% CI) | | | 8 | | | 8 | 31.1% | -0.02 [-1.00, 0.96] | | |
| Heterogeneity: Not applicable | | | | | | | | | | |
| Test for overall effect: Z = 0.04 (P = 0.97) | | | | | | | | | | |
| **Total (95% CI)** | | | 24 | | | 24 | 100.0% | -0.06 [-0.97, 0.85] | | |
| Heterogeneity: Tau² = 0.44; Chi² = 6.31, df = 3 (P = 0.10); I² = 52% | | | | | | | | | | |
| Test for overall effect: Z = 0.13 (P = 0.89) | | | | | | | | | | |
| Test for subgroup differences: Chi² = 5.85, df = 2 (P = 0.05), I² = 65.8% | | | | | | | | | | |

Favours [Control]   Favours [Stem Cell]

Risk of bias legend
(A) Random sequence generation (selection bias)
(B) Allocation concealment (selection bias)
(C) Blinding of participants and personnel (performance bias)
(D) Blinding of outcome assessment (detection bias)
(E) Incomplete outcome data (attrition bias)
(F) Selective reporting (reporting bias)

Figure 1.14. Forest plot for *COMP* gene expression.

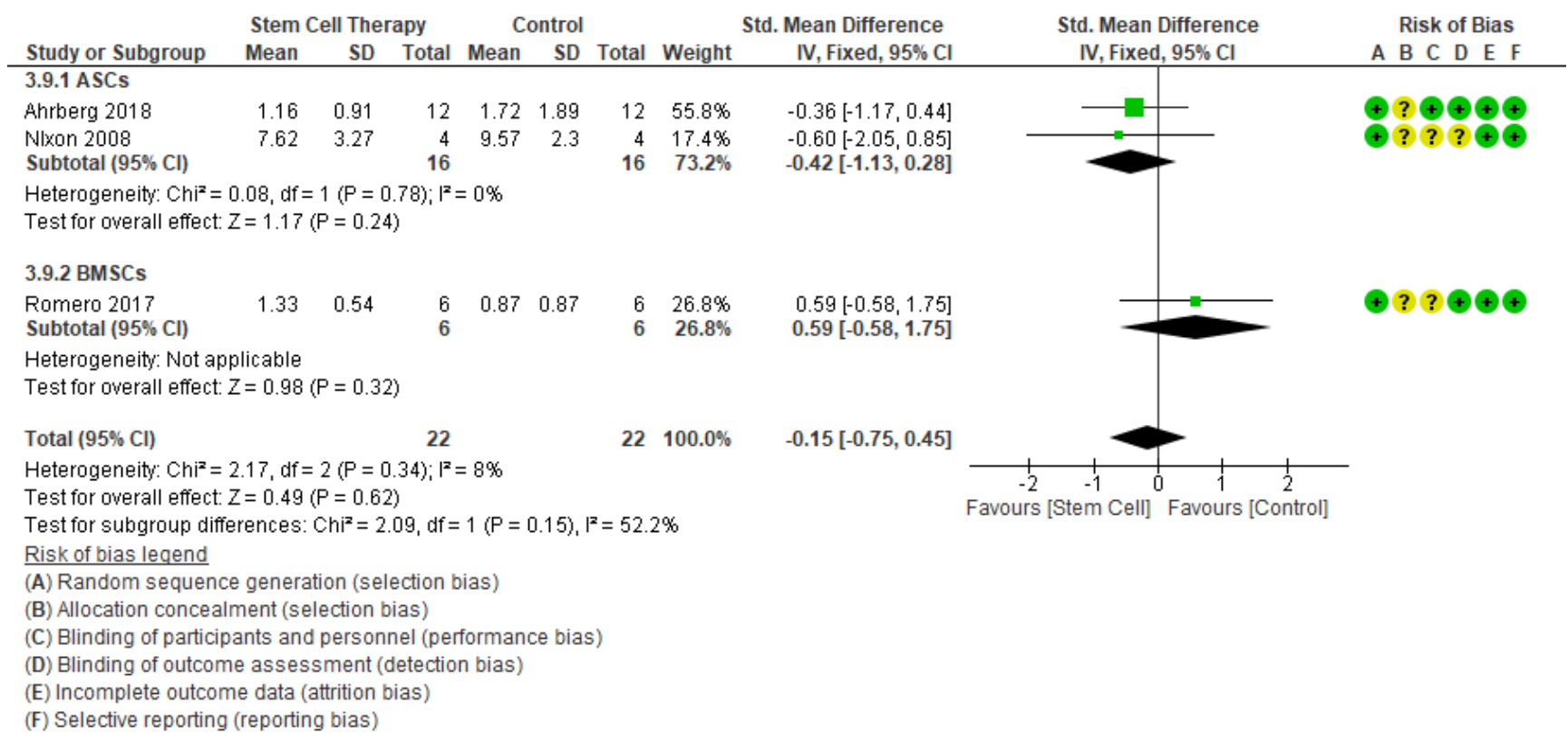

Figure 1.15. Forest plot for *Dcn* gene expression.

The amount of total collagen in injured tendon or ligament, measured by mass spectrometry quantification of hydroxyproline[30] or picrosirius red staining[47,49,51] and expressed as μg/mg dry weight tendon[30,49,51] or μg/mg wet weight tendon[47], at the endpoint of the follow-up period was extracted from 4 randomized controlled trials, which included 27 horses each in both the stem cell group and the control group (Fig 1.18). SMD was compared because quantification methods and initial tissue processing methods differ among the studies. The overall heterogeneity across was low ($I^2 = 0\%$, p = 0.76), which suggested use of the fixed effect model was appropriate. The amount of collagen was not different between stem cell and control groups (SMD = 0.26, 95% CI [-0.28, 0.80], p = 0.35).

Another measure of total collagen in injured tendon or ligament, measured by mass spectrometry quantification of hydroxyproline[30] or colorimetric quantification of hydroxyproline[29] and expressed as μg/mg dry weight tendon, at the endpoint of the follow-up period was extracted from a randomized controlled trial and a prospective cohort study, which included 15 horses each in both the stem cell group and the control group (Fig 1.19). SMD was compared because quantification methods differ among the studies. The overall heterogeneity across was low ($I^2 = 36\%$, p = 0.21), which suggested use of the fixed effect model was appropriate. The amount of collagen was not different between stem cell and control groups (SMD = 0.17, 95% CI [-0.56, 0.90], p = 0.64).

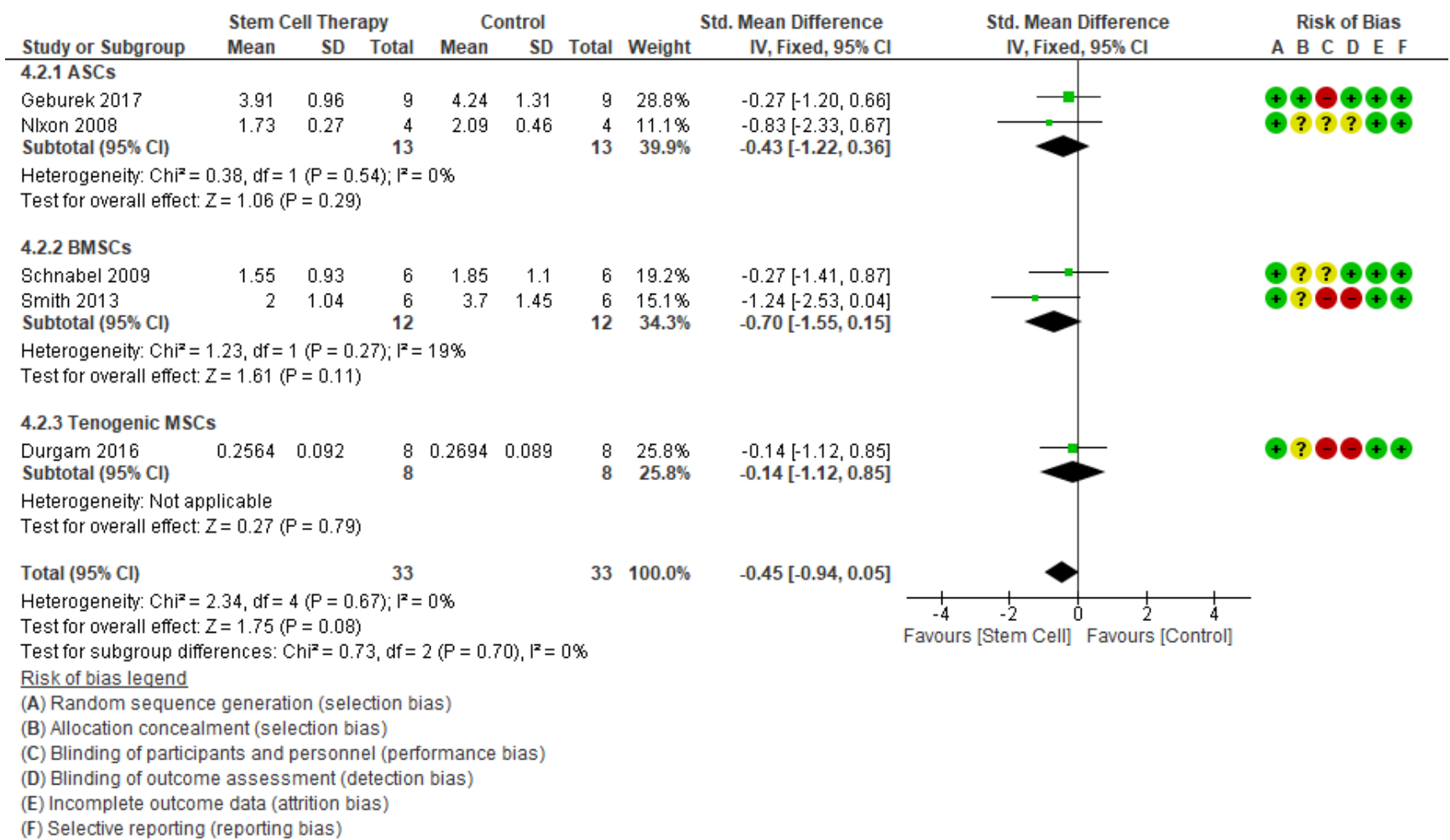

Figure 1.16. Forest plot for DNA contents.

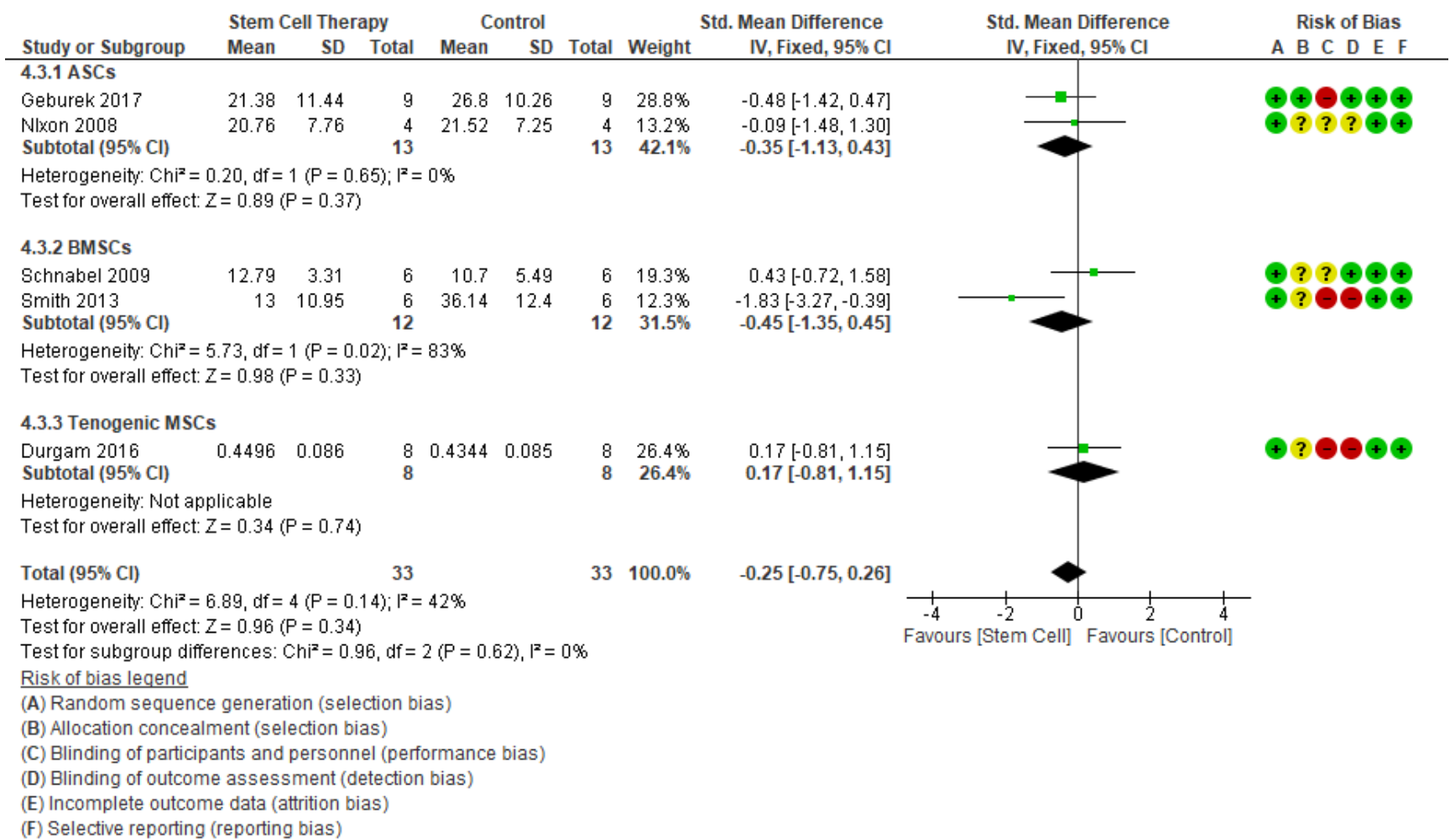

Figure 1.17. Forest plot for GAG contents.

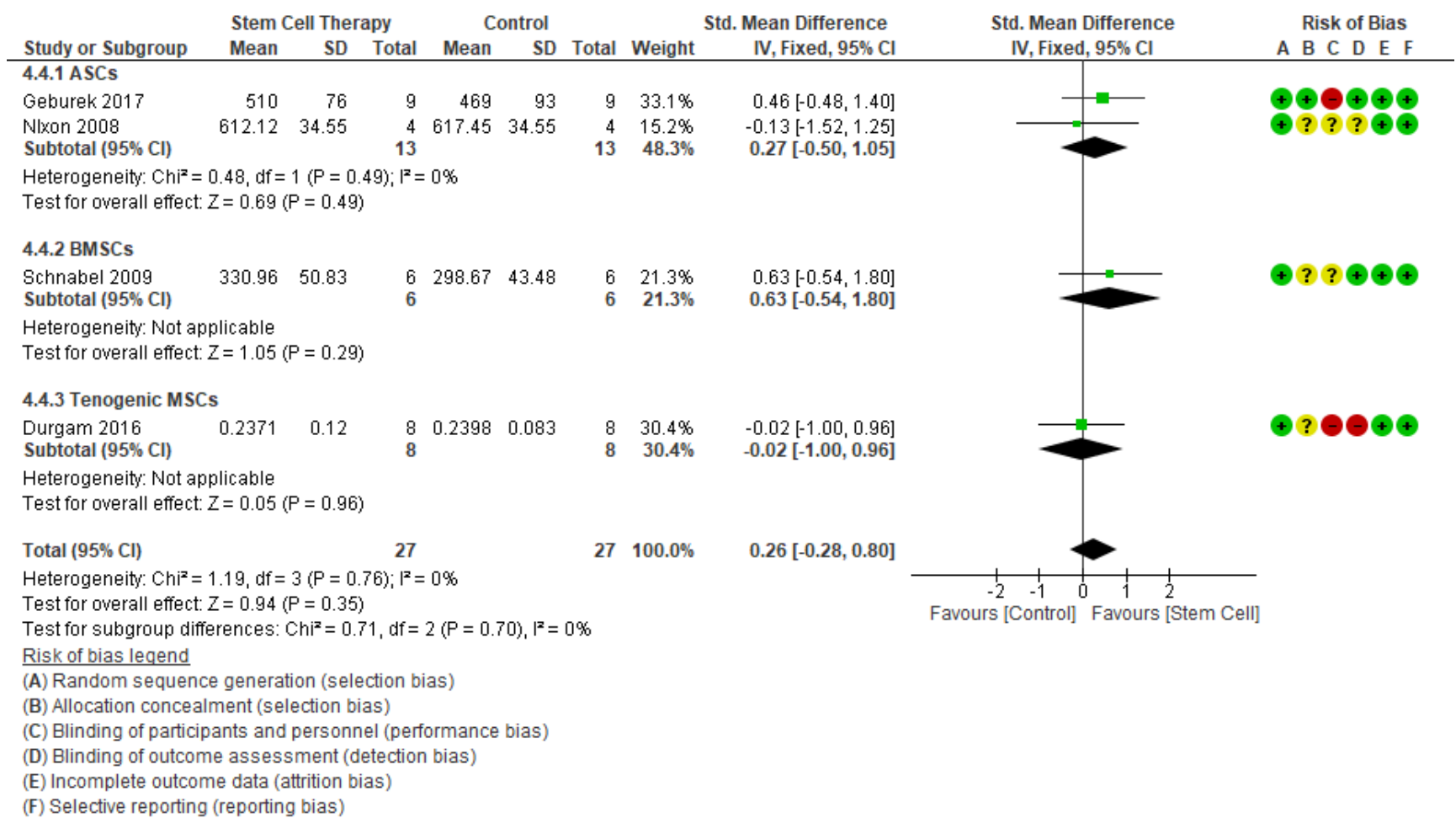

Figure 1.18. Forest plot for collagen contents.

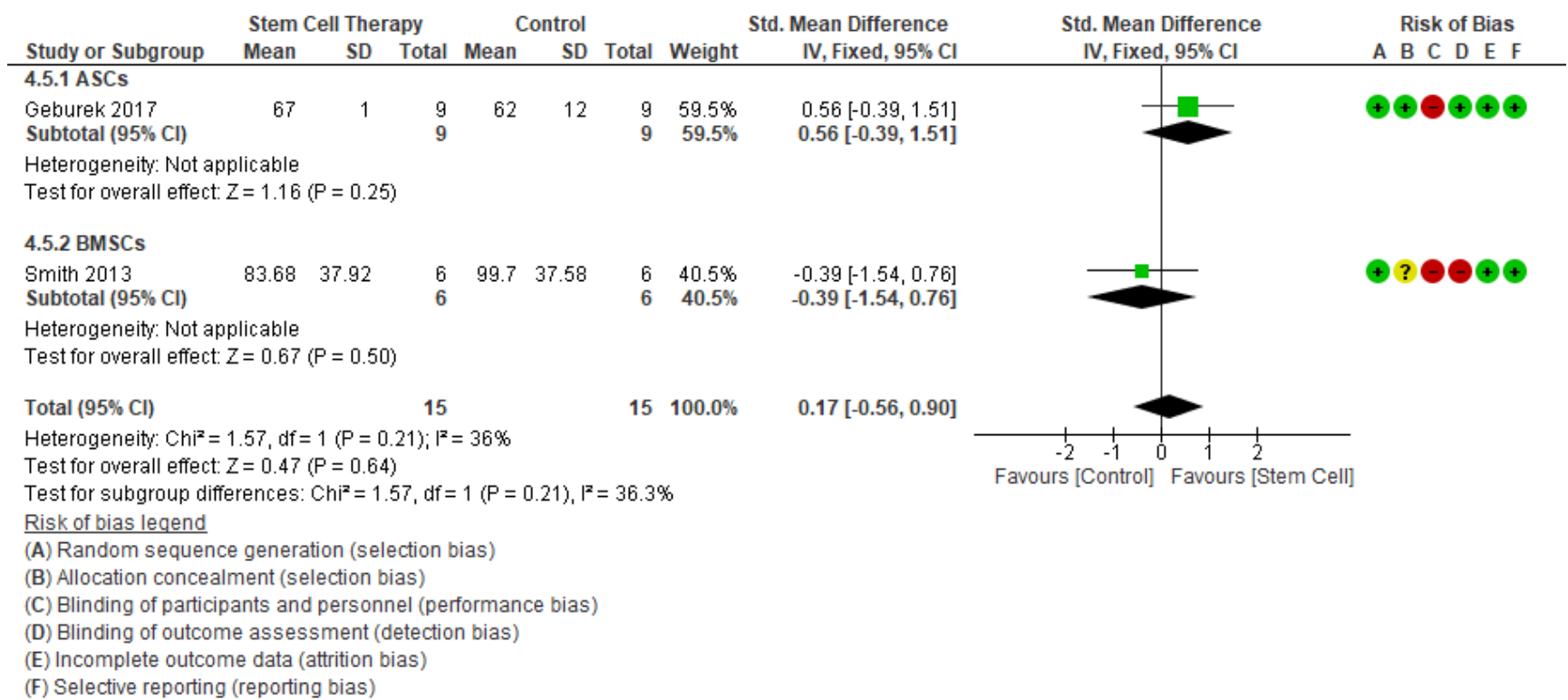

Figure 1.19. Forest plot for hydroxyproline contents.

### 1.3.6. Histological Analysis of Injured Tendon and Ligament

Histological evaluation of cellularity in injured tendon or ligament, measured as a percentage of 4,6-diamidino-2-phenylindole (DAPI) staining per total area of each section[41] or scores using established scoring rubrics[58,59], at the endpoint of the follow-up period was extracted from 3 randomized controlled trials[41,50,51] and a prospective cohort study[29], of which one study had a standard deviation (SD) = 0 and was excluded from the overall effect analysis, resulting in a total of 24 horses each in both the stem cell group and the control group (Fig 1.20). SMD was compared because cellularity evaluation methods and scoring systems varied among studies. The overall heterogeneity across was high ($I^2 = 74\%$, p = 0.02), which suggested use of the random effect model was appropriate. The histological cellularity in injured tendon or ligament was lower in the stem cell treatment group than in the control group (SMD = -1.60, 95% CI [-3.11, -0.10], p = 0.04).

Histological evaluation of tenocyte morphology in injured tendon or ligament, measured as scores using established scoring rubrics[58-61], at the endpoint of the follow-up period was extracted from 4 randomized controlled trials[30,46,50,51], which included 29 horses each in both the stem cell group and the control group (Fig 1.21). SMD was compared because cellularity evaluation scoring systems varied among studies. The overall heterogeneity across was high ($I^2 = 61\%$, p = 0.05), which suggested use of the random effect model was appropriate. The histological tenocyte morphology was not different between stem cell and control groups (SMD = -0.24, 95% CI [-1.13, 0.65], p = 0.59).

Histological evaluation of variation in cell density in injured tendon or ligament, measured as scores using established scoring rubrics[58,60,61], at the endpoint of the follow-up period was extracted from 2 randomized controlled trials[30,46], which included 17 horses each in both the stem cell group and the control group (Fig 1.22). SMD was compared because variation in cell density evaluation scoring systems varied among studies. The overall heterogeneity across was low ($I^2 = 24\%$, p = 0.25), which suggested use of the fixed effect model was appropriate. The histological variation in cell density was not different between stem cell and control groups (SMD = 0.09, 95% CI [-0.60, 0.77], p = 0.81).

Histological evaluation of collagen type I (COLI) in injured tendon or ligament, measured as a distribution percentage[46] or scores using novel scoring rubrics[41,45] following immunohistochemical staining, at the endpoint of the follow-up period was extracted from 3 randomized controlled trials[41,45,46], of which one study had SD = 0 and was excluded from the overall effect analysis, resulting in a total of 20 horses each in both the stem cell group and the control group (Fig 1.23). SMD was compared because COLI distribution evaluation methods and scoring systems varied among studies. The overall heterogeneity across was high ($I^2 = 94\%$, p < 0.0001), which suggested use of the random effect model was appropriate. The histological COLI distribution was not different between stem cell and control groups (SMD = 2.35, 95% CI [-2.53, 7.23], p = 0.35).

Histological evaluation of collagen type III (COLIII) in injured tendon or ligament, measured as a distribution percentage[46] or scores using novel scoring rubrics[45] following immunohistochemical staining, at the endpoint of the follow-up period was extracted from 2 randomized controlled trials[45,46], of which one study had SD = 0 and was excluded from the overall effect analysis, resulting in a total of 8 horses each in both the stem cell group and the control group (Fig 1.24). Since only a single study was included in the analysis, MD was compared, heterogeneity assessment was not performed, and the fixed effect model was applied.

The histological COLIII distribution in injured tendon or ligament was lower in the stem cell treatment group than in the control group (MD = -10.05, 95% CI [-11.68, -8.42], p < 0.00001).

Histological evaluation of fiber structure in injured tendon or ligament, measured as scores using established scoring rubrics[58,60,61], at the endpoint of the follow-up period was extracted from 2 randomized controlled trials[30,46], of which one study had SD = 0 and was excluded from the overall effect analysis, resulting in total of 9 horses each in both the stem cell group and the control group (Fig 1.25). Since only a single study was included in the analysis, MD was compared, heterogeneity assessment was not performed, and the fixed effect model was applied. The histological fiber structure was not different between stem cell and control groups (MD = 0.32, 95% CI [-1.18, 1.82], p = 0.68).

Histological evaluation of fiber alignment in injured tendon or ligament, measured as scores using established and novel scoring rubrics[49,58-61] or the mean orientation of the collagen fibers obtained from a second harmonic generation (SHG) microscope with deviation from 90° representing more random fiber alignment[47], at the endpoint of the follow-up period was extracted from 6 randomized controlled trials[30,46,47,49-51] and a prospective cohort study[29], of which one study had SD = 0 and was excluded from overall effect analysis, resulting in a total of 42 horses each in both the stem cell group and the control group (Fig 1.26). SMD was compared because fiber alignment evaluation methods and scoring systems varied among studies. The overall heterogeneity across was high ($I^2 = 71\%$, p = 0.87), which suggested use of the random effect model was appropriate. The histological fiber alignment score was lower in the stem cell treatment group than in the control group (SMD = -1.31, 95% CI [-2.32, -0.30], p = 0.01).

Histological evaluation of crimp in injured tendon or ligament, measured as scores using established scoring rubrics[58,59], at the endpoint of the follow-up period was extracted from a randomized controlled trial[51] and a prospective cohort study[29], all of which had SD = 0 and were excluded from the overall effect analysis (Fig 1.27).

Histological evaluation of vascularity in injured tendon or ligament, measured as scores using established scoring rubrics[58-61] or percentage of erythrocyte autofluorescence per total section area[41] or total number of vessels after immunohistochemical staining for von Willebrand factor VIII[44], at the endpoint of the follow-up period was extracted from 5 randomized controlled trials[30,44,46,50,51] and a prospective cohort study[29], which included 56 horses each in both the stem cell group and the control group (Fig 1.28). SMD was compared because vascularity evaluation methods and scoring systems varied among studies. The overall heterogeneity across was high ($I^2 = 81\%$, $p < 0.0001$), which suggested use of the random effect model was appropriate. The histological vascularity score was not different between stem cell and control groups (SMD = 0.08, 95% CI [-0.89, 1.04], p = 0.88).

Histological evaluation of inflammatory cell infiltrate in injured tendon or ligament, measured as scores using established and novel scoring rubrics[45,49,58,59], at the endpoint of the follow-up period was extracted from 4 randomized controlled trials[43,45,46,51], which included 24 horses each in both the stem cell group and the control group (Fig 1.29). SMD was compared because inflammatory cell infiltrate evaluation scoring systems varied among studies. The overall heterogeneity across was high ($I^2 = 66\%$, p = 0.03), which suggested use of the random effect model was appropriate. The histological inflammatory cell infiltrate score was not different between stem cell and control groups (SMD = 0.00, 95% CI [-1.08, 1.07], p = 1.00).

Histological evaluation of glycosaminoglycan (GAG) deposition in injured tendon or ligament, measured as scores using established and novel scoring rubrics[58,59], at the endpoint of the follow-up period was extracted from 2 randomized controlled trials[46,50], which included 14 horses each in both the stem cell group and the control group (Fig 1.30). SMD was compared because GAG deposition evaluation scoring systems varied among studies. The overall heterogeneity across was high ($I^2$ = 89%, p = 0.003), which suggested use of the random effect model was appropriate. The histological GAG deposition was not different between stem cell and control groups (SMD = -0.75, 95% CI [-3.55, 2.06], p = 0.60).

Histological evaluation of morphological and structural quality in injured tendon or ligament, measured as total scores using established and novel scoring rubrics[45,49,59-61], at the endpoint of the follow-up period was extracted from 5 randomized controlled trials[30,41,43,49,51], which included 35 horses each in both the stem cell group and the control group (Fig 1.31). SMD was compared because histological evaluation scoring systems varied among studies. The overall heterogeneity across was high ($I^2$ = 65%, p = 0.02), which suggested use of the random effect model was appropriate. The histological morphological and structural quality was not different between stem cell and control groups (SMD = -0.80, 95% CI [-1.76, 0.17], p = 0.11).

### 1.3.7. Mechanical Properties of Injured Tendon and Ligament

Mechanical maximum stress at failure in injured tendon or ligament at the endpoint of the follow-up period was extracted from 2 randomized controlled trials[30,47], which included a total of 17 horses each in both the stem cell group and the control group (Fig 1.32). SMD was compared because, although both studies reported stress using the same unit (MPa), specimen sizes and specific experimental conditions differed. The overall heterogeneity across was high ($I^2$ = 69%, p = 0.07), which suggested use of the random effect model was appropriate. The mechanical maximum stress in injured tendon or ligament was not different between stem cell and control groups (SMD = -0.01, 95% CI [-1.26, 1.24], p = 0.98).

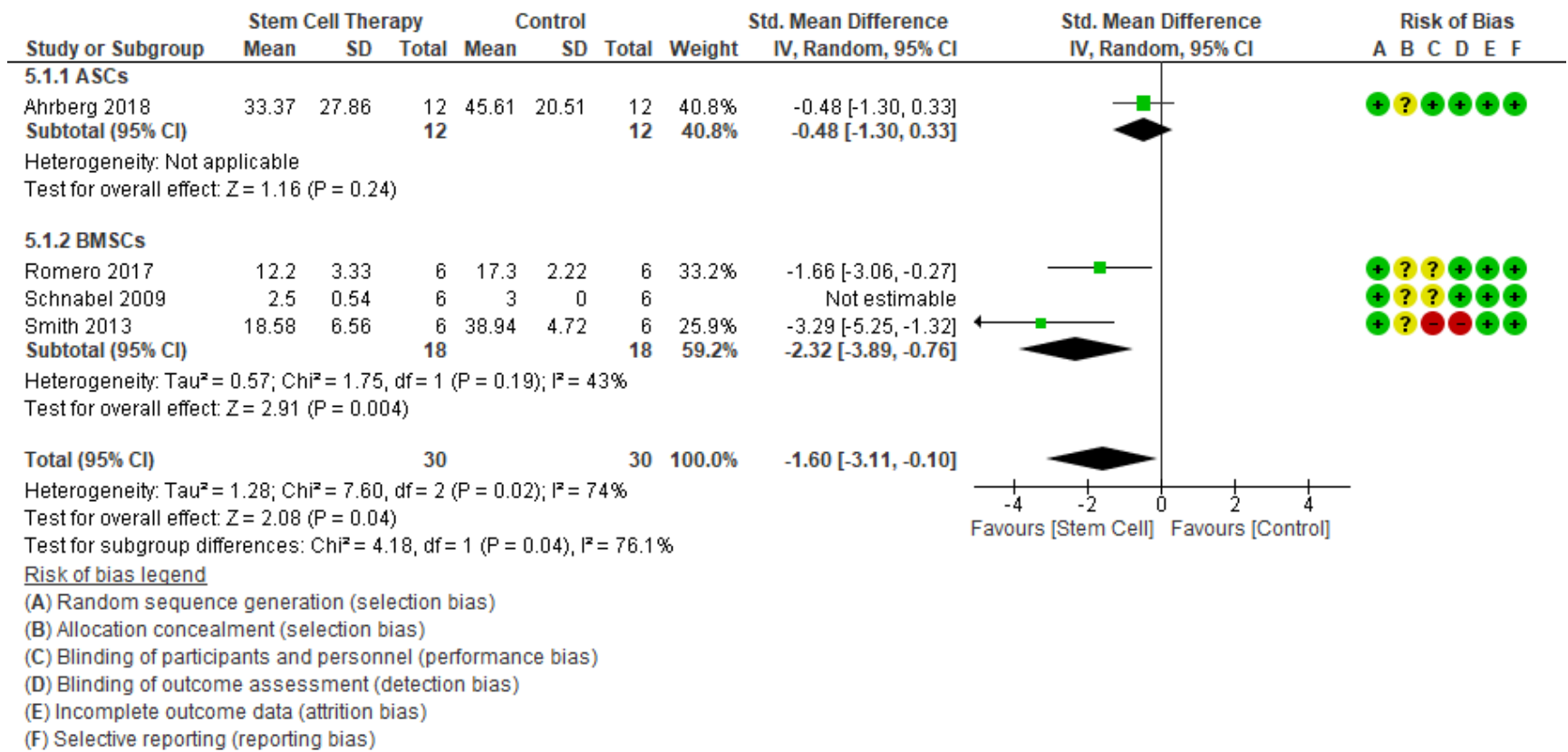

Figure 1.20. Forest plot for histological cellularity. The lower value indicates lower cellularity and normal appearance, while the higher value indicates higher cellularity and abnormal appearance.

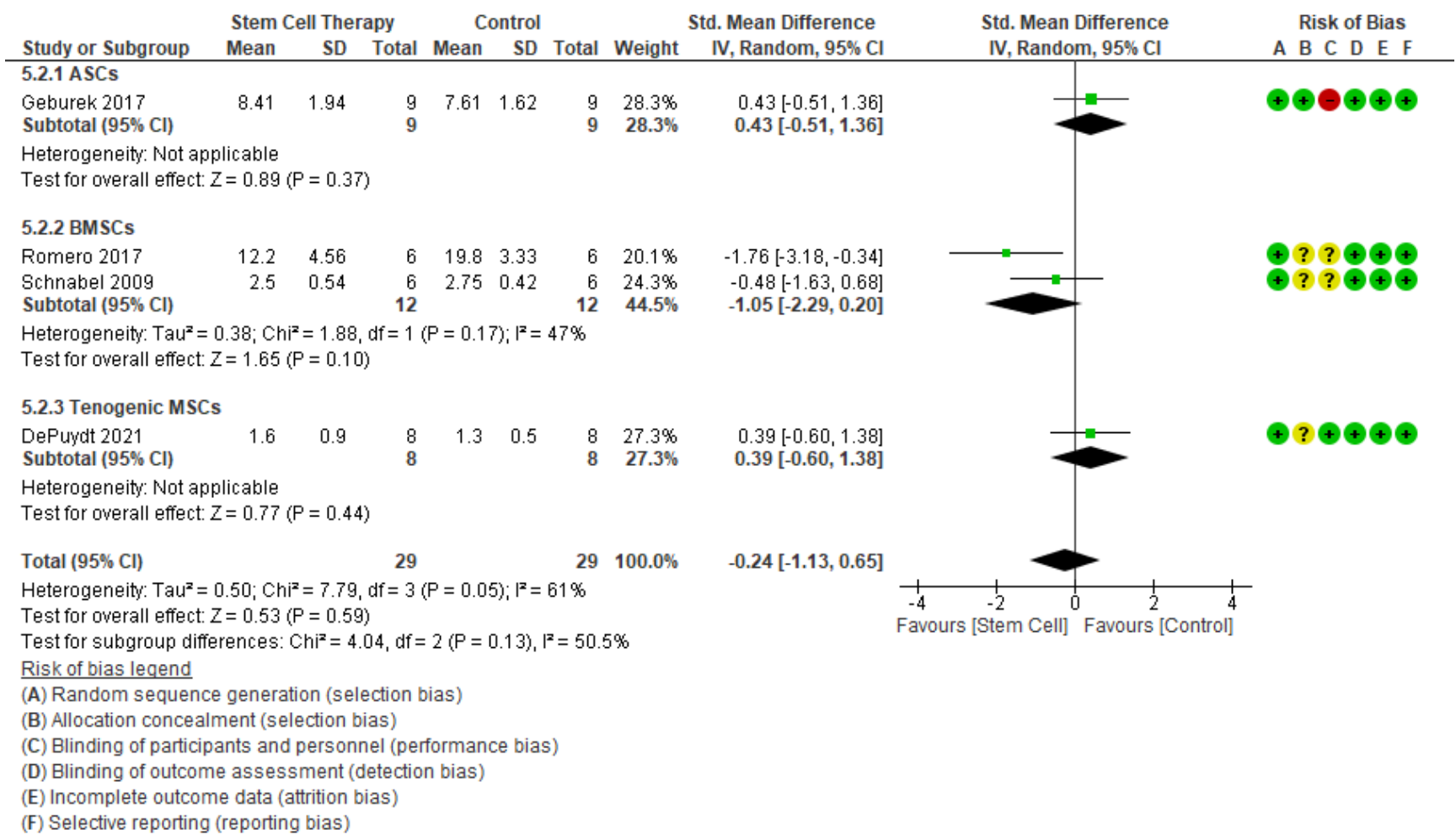

Figure 1.21. Forest plot for histological tenocyte morphology score. The lower value indicates normal morphology, while the higher value indicates abnormal morphology.

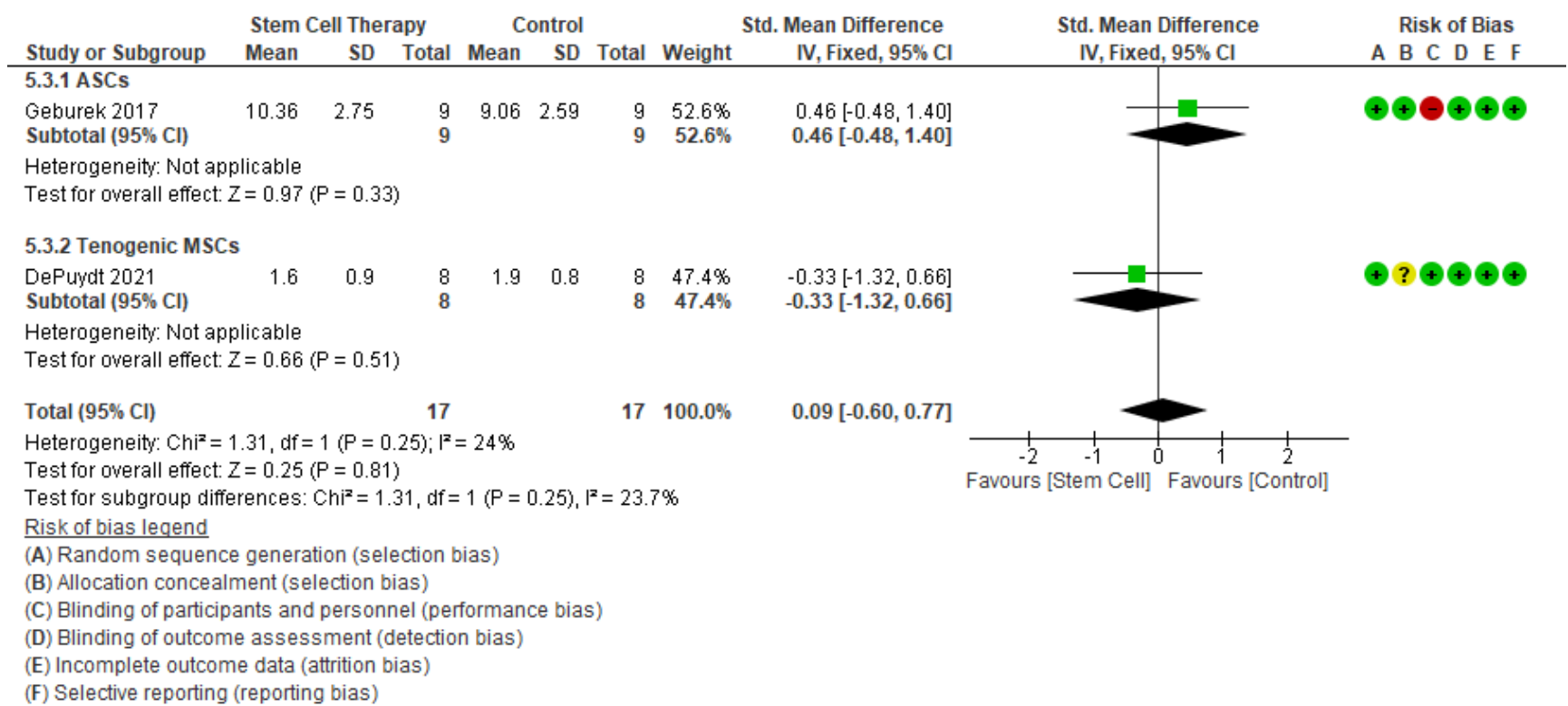

Figure 1.22. Forest plot for histological variation in cell density score. The lower value indicates normal uniform cell density, while the higher value indicates abnormal non-uniform cell density.

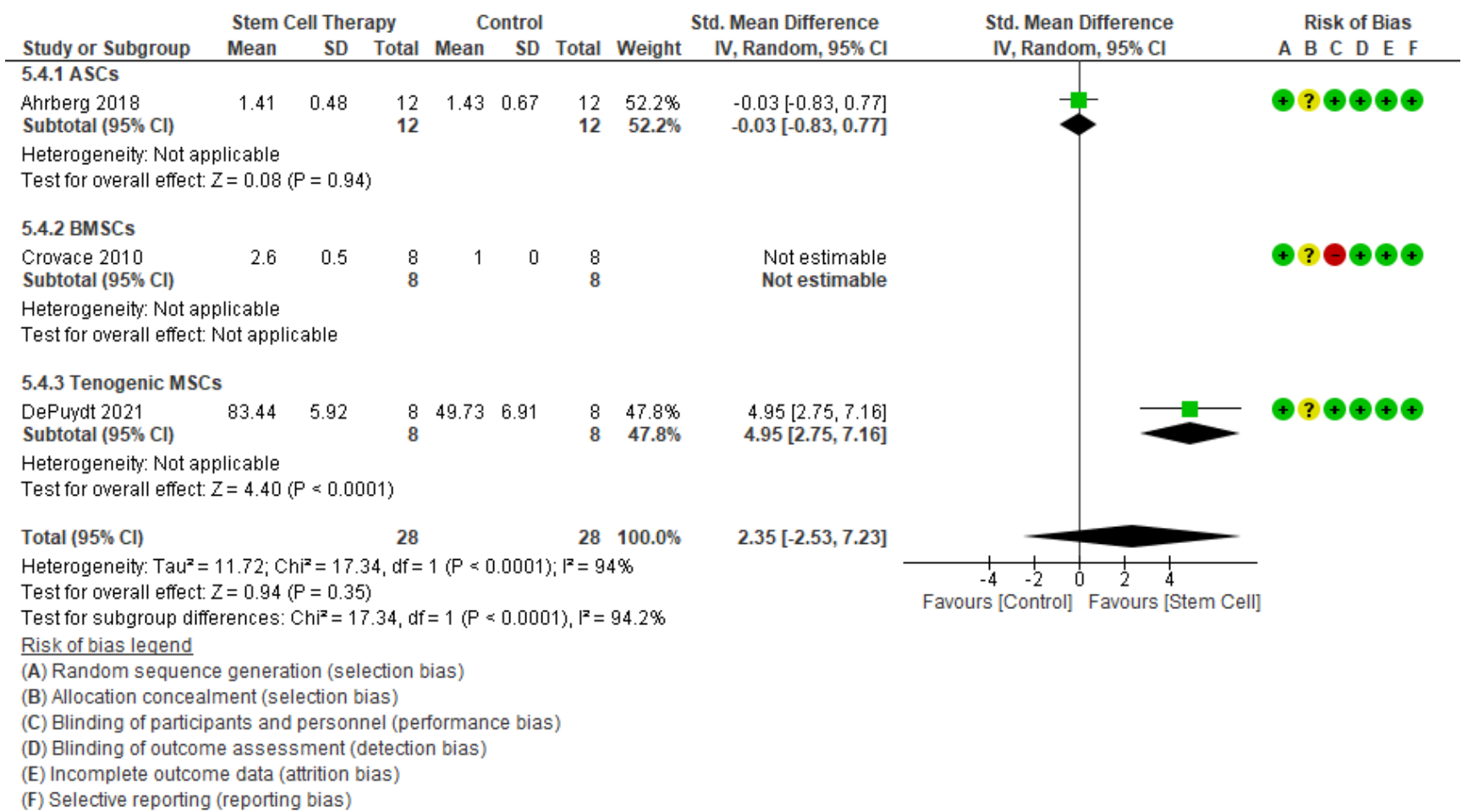

Figure 1.23. Forest plot for histological collagen type I distribution. The lower value indicates less abundant collagen type I distribution, while the higher value indicates more abundant collagen type I distribution.

| Study or Subgroup | Stem Cell Therapy | | | Control | | | Weight | Mean Difference IV, Fixed, 95% CI | Mean Difference IV, Fixed, 95% CI | Risk of Bias A B C D E F |
|---|---|---|---|---|---|---|---|---|---|---|
| | Mean | SD | Total | Mean | SD | Total | | | | |
| **5.5.1 BMSCs** | | | | | | | | | | |
| Crovace 2010 | 1.2 | 0.4 | 6 | 3 | 0 | 6 | | Not estimable | | + ? ● + + + |
| **Subtotal (95% CI)** | | | 6 | | | 6 | | Not estimable | | |
| Heterogeneity: Not applicable | | | | | | | | | | |
| Test for overall effect: Not applicable | | | | | | | | | | |
| | | | | | | | | | | |
| **5.5.2 Tenogenic MSCs** | | | | | | | | | | |
| DePuydt 2021 | 0.53 | 0.33 | 8 | 10.58 | 2.33 | 8 | 100.0% | -10.05 [-11.68, -8.42] | | + ? + + + + |
| **Subtotal (95% CI)** | | | 8 | | | 8 | 100.0% | -10.05 [-11.68, -8.42] | | |
| Heterogeneity: Not applicable | | | | | | | | | | |
| Test for overall effect: Z = 12.08 (P < 0.00001) | | | | | | | | | | |
| | | | | | | | | | | |
| **Total (95% CI)** | | | 14 | | | 14 | 100.0% | -10.05 [-11.68, -8.42] | | |
| Heterogeneity: Not applicable | | | | | | | | | | |
| Test for overall effect: Z = 12.08 (P < 0.00001) | | | | | | | | | | |
| Test for subgroup differences: Not applicable | | | | | | | | | | |

Favours [Stem Cell]    Favours [Control]

Risk of bias legend
(A) Random sequence generation (selection bias)
(B) Allocation concealment (selection bias)
(C) Blinding of participants and personnel (performance bias)
(D) Blinding of outcome assessment (detection bias)
(E) Incomplete outcome data (attrition bias)
(F) Selective reporting (reporting bias)

Figure 1.24. Forest plot for histological collagen type III distribution. The lower value indicates less abundant collagen type III distribution, while the higher value indicates more abundant collagen type III distribution.

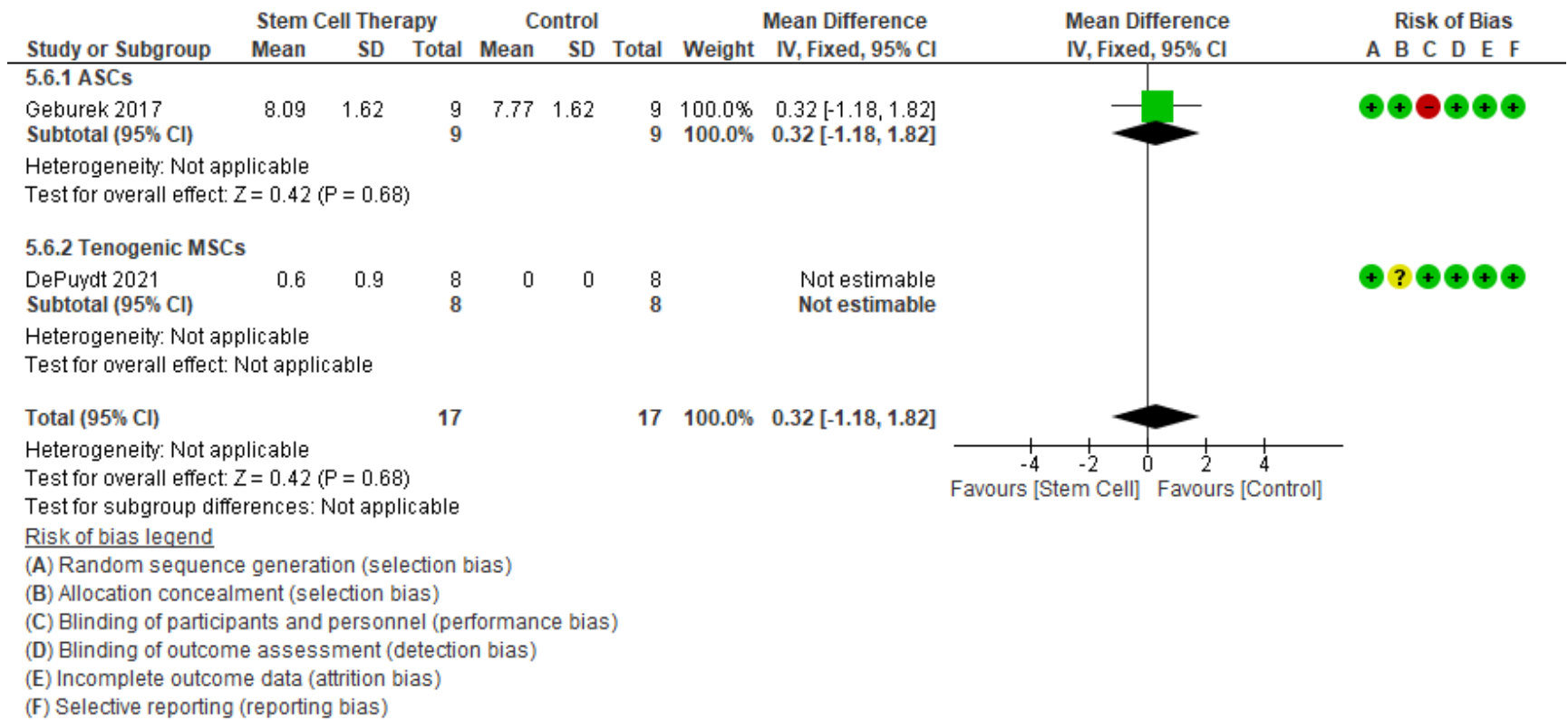

Figure 1.25. Forest plot for histological fiber structure score. The lower value indicates normal fiber structure, while the higher value indicates abnormal fiber structure.

| Study or Subgroup | Stem Cell Therapy Mean | SD | Total | Control Mean | SD | Total | Weight | Std. Mean Difference IV, Random, 95% CI | Std. Mean Difference IV, Random, 95% CI | Risk of Bias A B C D E F |
|---|---|---|---|---|---|---|---|---|---|---|
| **5.7.1 ASCs** | | | | | | | | | | |
| Geburek 2017 | 8.25 | 1.62 | 9 | 8.09 | 1.62 | 9 | 20.7% | 0.09 [-0.83, 1.02] | | |
| Nixon 2008 | 1.79 | 0.6 | 4 | 2.79 | 0.4 | 4 | 13.9% | -1.71 [-3.52, 0.11] | | |
| Subtotal (95% CI) | | | 13 | | | 13 | 34.6% | -0.63 [-2.36, 1.10] | | |
| Heterogeneity: Tau² = 1.08; Chi² = 3.00, df = 1 (P = 0.08); I² = 67% | | | | | | | | | | |
| Test for overall effect: Z = 0.71 (P = 0.48) | | | | | | | | | | |
| | | | | | | | | | | |
| **5.7.2 BMSCs** | | | | | | | | | | |
| Romero 2017 | 5.67 | 2.78 | 6 | 16.8 | 4 | 6 | 13.7% | -2.98 [-4.83, -1.14] | | |
| Schnabel 2009 | 2 | 0 | 6 | 2.5 | 0.54 | 6 | | Not estimable | | |
| Smith 2013 | 19.1 | 4.02 | 6 | 34.46 | 3.47 | 6 | 11.7% | -3.78 [-5.94, -1.61] | | |
| Subtotal (95% CI) | | | 18 | | | 18 | 25.4% | -3.32 [-4.72, -1.91] | | |
| Heterogeneity: Tau² = 0.00; Chi² = 0.30, df = 1 (P = 0.58); I² = 0% | | | | | | | | | | |
| Test for overall effect: Z = 4.63 (P < 0.00001) | | | | | | | | | | |
| | | | | | | | | | | |
| **5.7.3 Tenogenic MSCs** | | | | | | | | | | |
| DePuydt 2021 | 0.6 | 0.9 | 8 | 1 | 0.4 | 8 | 20.1% | -0.54 [-1.55, 0.46] | | |
| Durgam 2016 | 105.67 | 8.3 | 8 | 114.7 | 16.2 | 8 | 20.0% | -0.66 [-1.68, 0.35] | | |
| Subtotal (95% CI) | | | 16 | | | 16 | 40.1% | -0.60 [-1.32, 0.11] | | |
| Heterogeneity: Tau² = 0.00; Chi² = 0.03, df = 1 (P = 0.87); I² = 0% | | | | | | | | | | |
| Test for overall effect: Z = 1.65 (P = 0.10) | | | | | | | | | | |
| | | | | | | | | | | |
| **Total (95% CI)** | | | 47 | | | 47 | 100.0% | -1.31 [-2.32, -0.30] | | |
| Heterogeneity: Tau² = 1.07; Chi² = 17.51, df = 5 (P = 0.004); I² = 71% | | | | | | | | | | |
| Test for overall effect: Z = 2.53 (P = 0.01) | | | | | | | | | | |
| Test for subgroup differences: Chi² = 11.74, df = 2 (P = 0.003), I² = 83.0% | | | | | | | | | | |

Favours [Stem Cell]   Favours [Control]

Risk of bias legend
(A) Random sequence generation (selection bias)
(B) Allocation concealment (selection bias)
(C) Blinding of participants and personnel (performance bias)
(D) Blinding of outcome assessment (detection bias)
(E) Incomplete outcome data (attrition bias)
(F) Selective reporting (reporting bias)

Figure 1.26. Forest plot for histological fiber alignment score. The lower value indicates normal fiber structure, while the higher value indicates abnormal fiber alignment.

| Study or Subgroup | Stem Cell Therapy | | | Control | | | Weight | Std. Mean Difference IV, Random, 95% CI | Std. Mean Difference IV, Random, 95% CI | Risk of Bias A B C D E F |
|---|---|---|---|---|---|---|---|---|---|---|
| | Mean | SD | Total | Mean | SD | Total | | | | |
| **5.15.1 BMSCs** | | | | | | | | | | |
| Schnabel 2009 | 2.33 | 0.81 | 6 | 3 | 0 | 6 | | Not estimable | | |
| Smith 2013 | 0 | 0.73 | 6 | 2 | 0 | 6 | | Not estimable | | |
| **Subtotal (95% CI)** | | | **12** | | | **12** | | **Not estimable** | | |
| Heterogeneity: Not applicable | | | | | | | | | | |
| Test for overall effect: Not applicable | | | | | | | | | | |
| | | | | | | | | | | |
| **Total (95% CI)** | | | **12** | | | **12** | | **Not estimable** | | |

Heterogeneity: Not applicable
Test for overall effect: Not applicable
Test for subgroup differences: Not applicable

Risk of bias legend
(A) Random sequence generation (selection bias)
(B) Allocation concealment (selection bias)
(C) Blinding of participants and personnel (performance bias)
(D) Blinding of outcome assessment (detection bias)
(E) Incomplete outcome data (attrition bias)
(F) Selective reporting (reporting bias)

Figure 1.27. Forest plot for histological crimp score. The lower value indicates normal crimp, while the higher value indicates abnormal crimp.

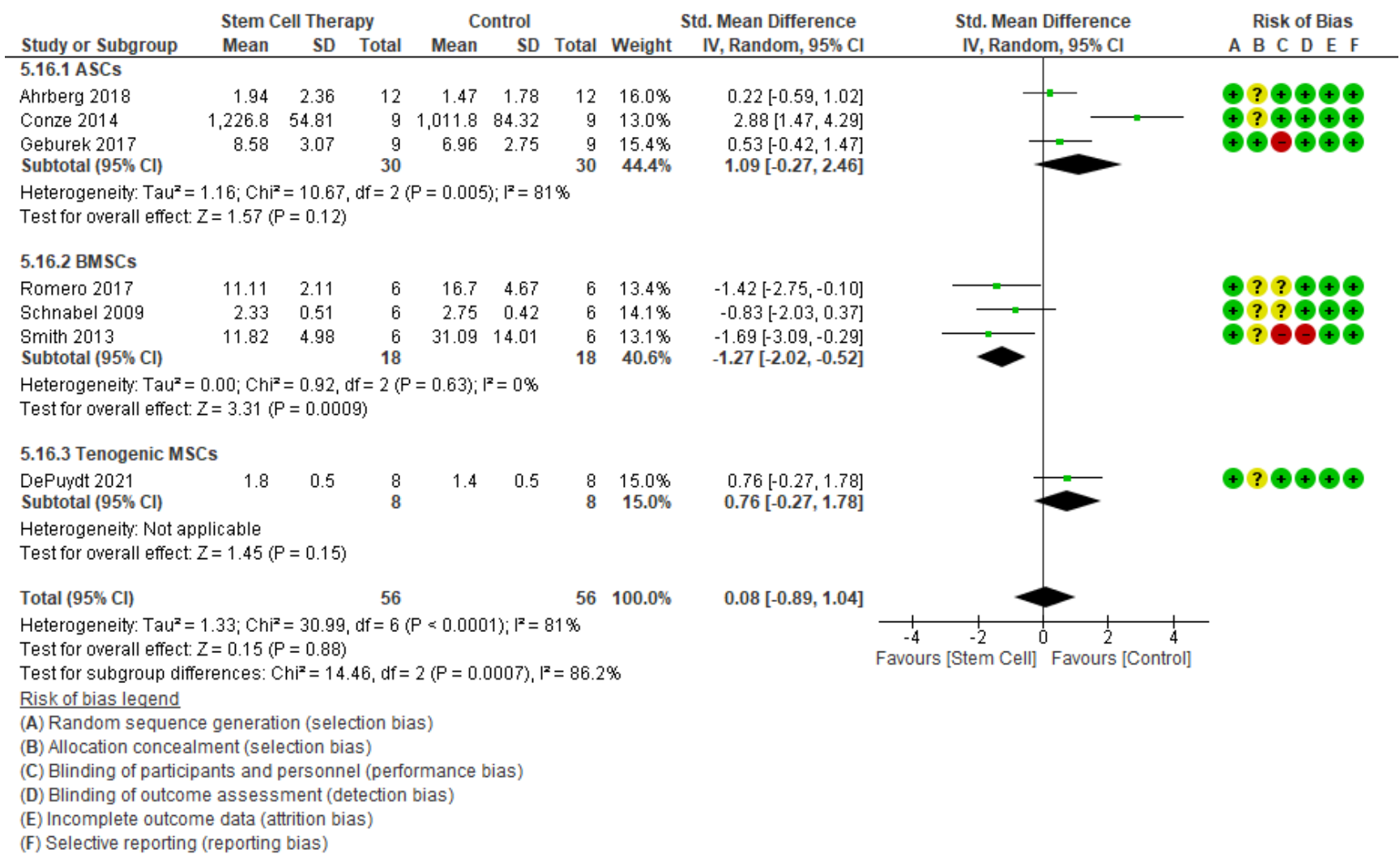

Figure 1.28. Forest plot for histological vascularity. The lower value indicates normal appearance with less vascularity, while the higher value indicates abnormal appearance with more vascularity.

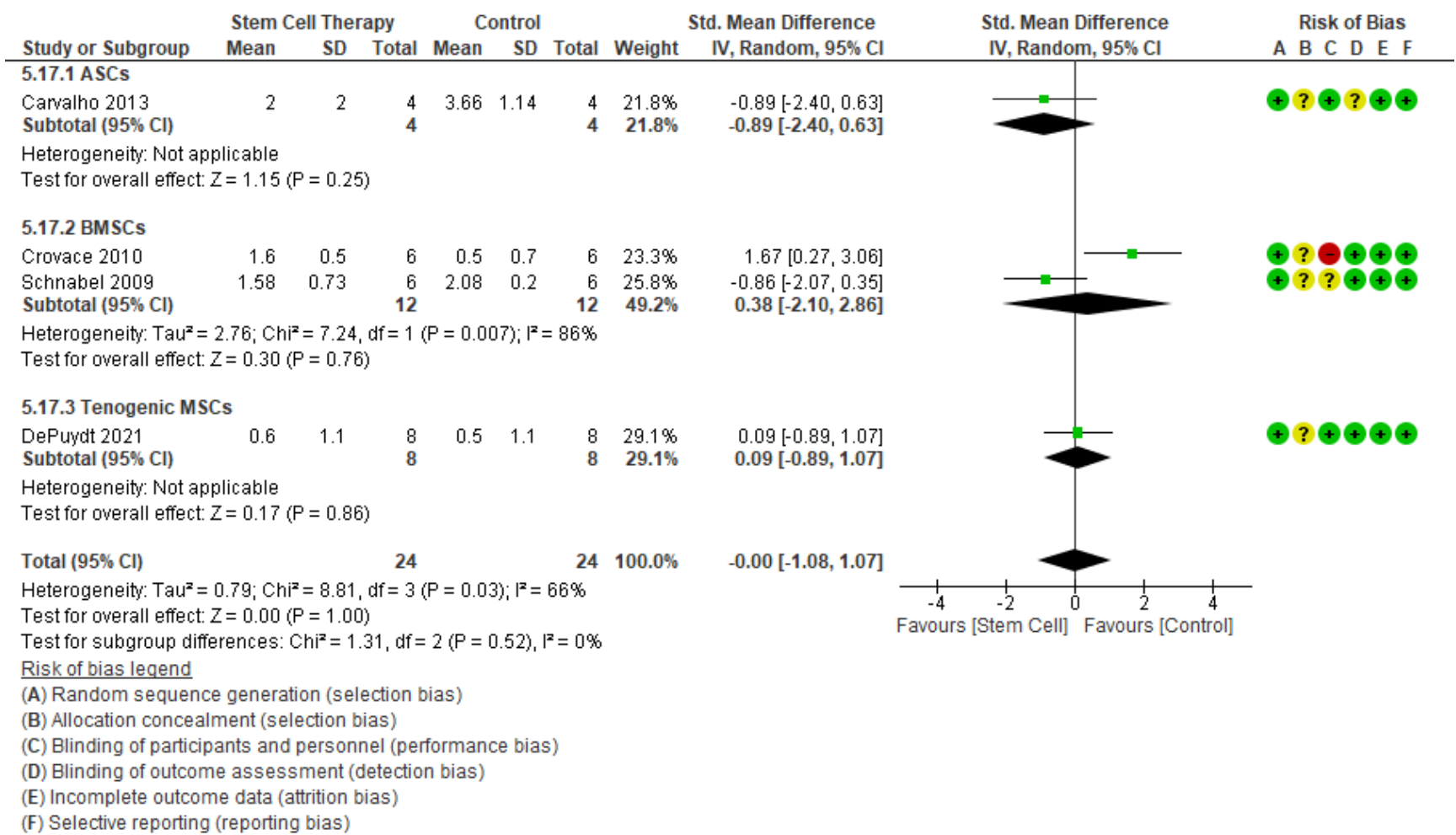

Figure 1.29. Forest plot for histological inflammatory cell infiltrate score. The lower value indicates normal appearance with less inflammatory cell infiltrate, while the higher value indicates abnormal appearance with more inflammatory cell infiltrate.

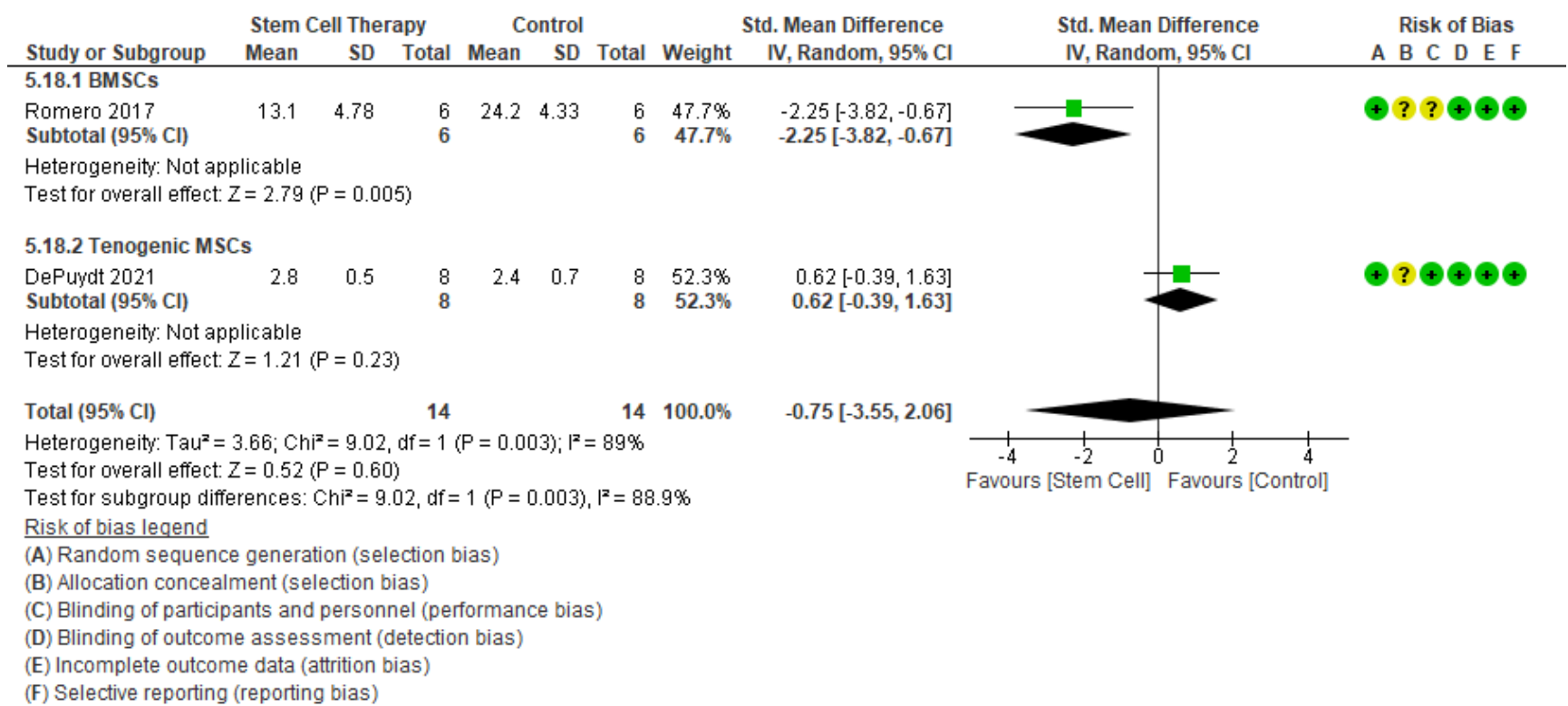

Figure 1.30. Forest plot for histological GAG. The lower value indicates normal appearance with lower presence of GAG, while the higher value indicates abnormal appearance with higher presence of GAG.

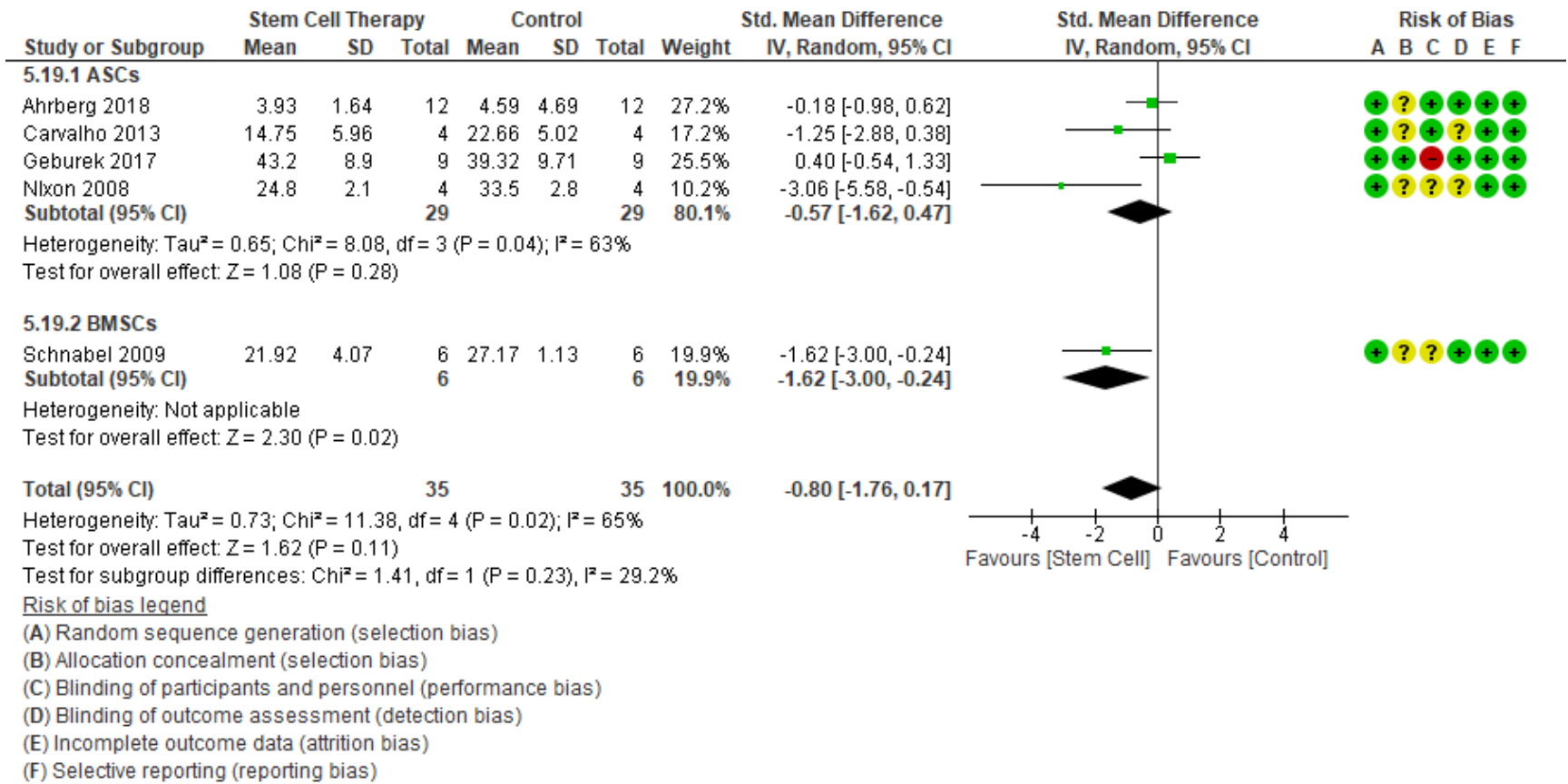

Figure 1.31. Forest plot for histological total score. The lower value indicates normal appearance similar to native tissue, while the higher value indicates abnormal appearance similar to damaged tissue.

Mechanical stiffness in injured tendon or ligament at the endpoint of the follow-up period was extracted from randomized controlled trials[47] and a prospective cohort study[29], which included a total of 14 horses each in both the stem cell group and the control group (Fig 1.33). SMD was compared because, although both studies reported stiffness as macroscopic mechanical property expressed in load over strain, specimen sizes and specific experimental conditions as well as reported units differed (N/cm and N%). The overall heterogeneity across was high ($I^2$ = 57%, p = 0.13), which suggested use of the random effect model was appropriate. The mechanical stiffness in injured tendon or ligament was not different between stem cell and control groups (SMD = -0.15, 95% CI [-1.33, 1.03], p = 0.80).

Mechanical elastic modulus in injured tendon or ligament at the endpoint of follow-up period was extracted from 3 randomized controlled trials[30,47,51] and a prospective cohort study[29], which included a total of 29 horses each in both the stem cell group and the control group (Fig 1.34). SMD was compared because, although all studies reported elastic modulus as microscopic structural property, specimen sizes and specific experimental conditions as well as reported units (MPa, ksi, and $N/cm^2$) differed. The overall heterogeneity across was high ($I^2$ = 57%, p = 0.07), which suggested use of the random effect model was appropriate. The mechanical elastic modulus in injured tendon or ligament was not different between stem cell and control groups (SMD = 0.46, 95% CI [-0.38, 1.30], p = 0.28).

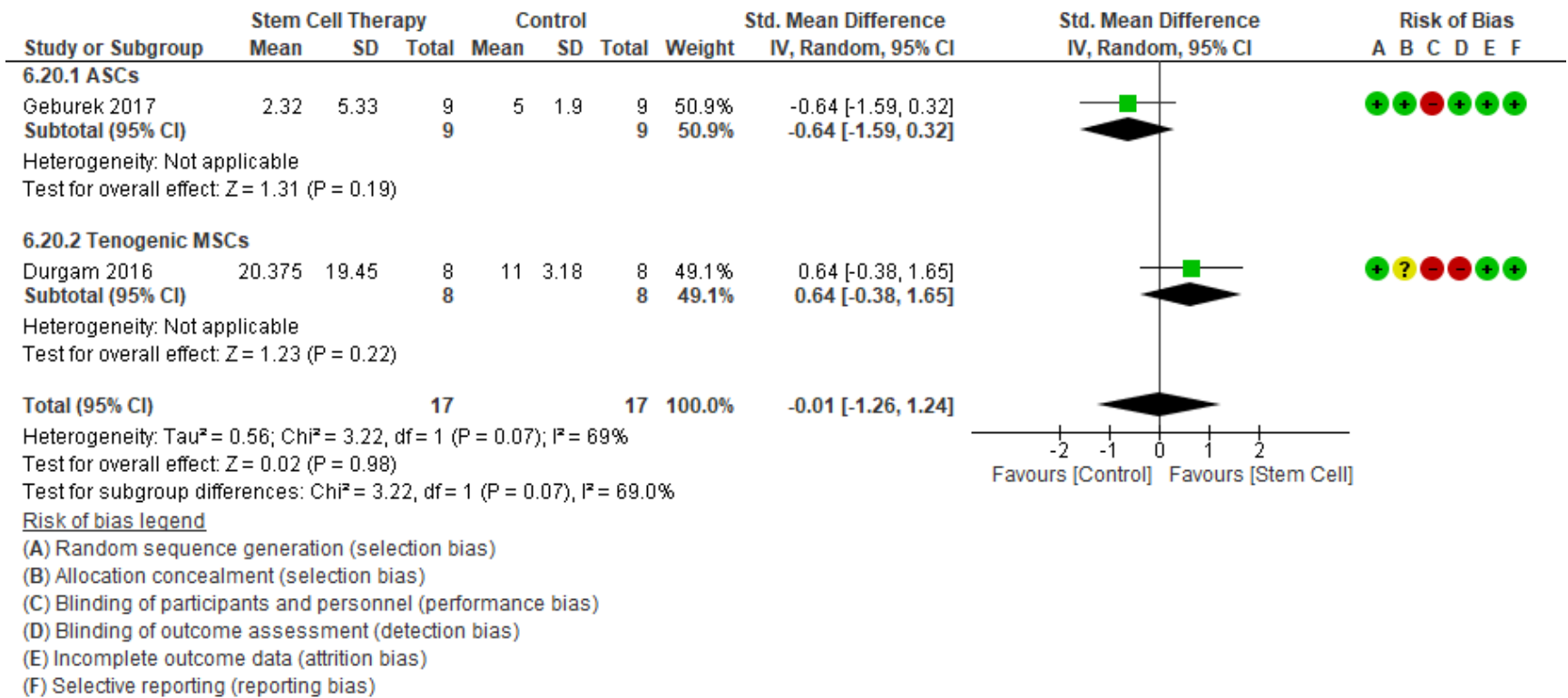

Figure 1.32. Forest plot for mechanical maximum stress. The lower value indicates lower stress at failure, while the higher value indicates higher stress at failure.

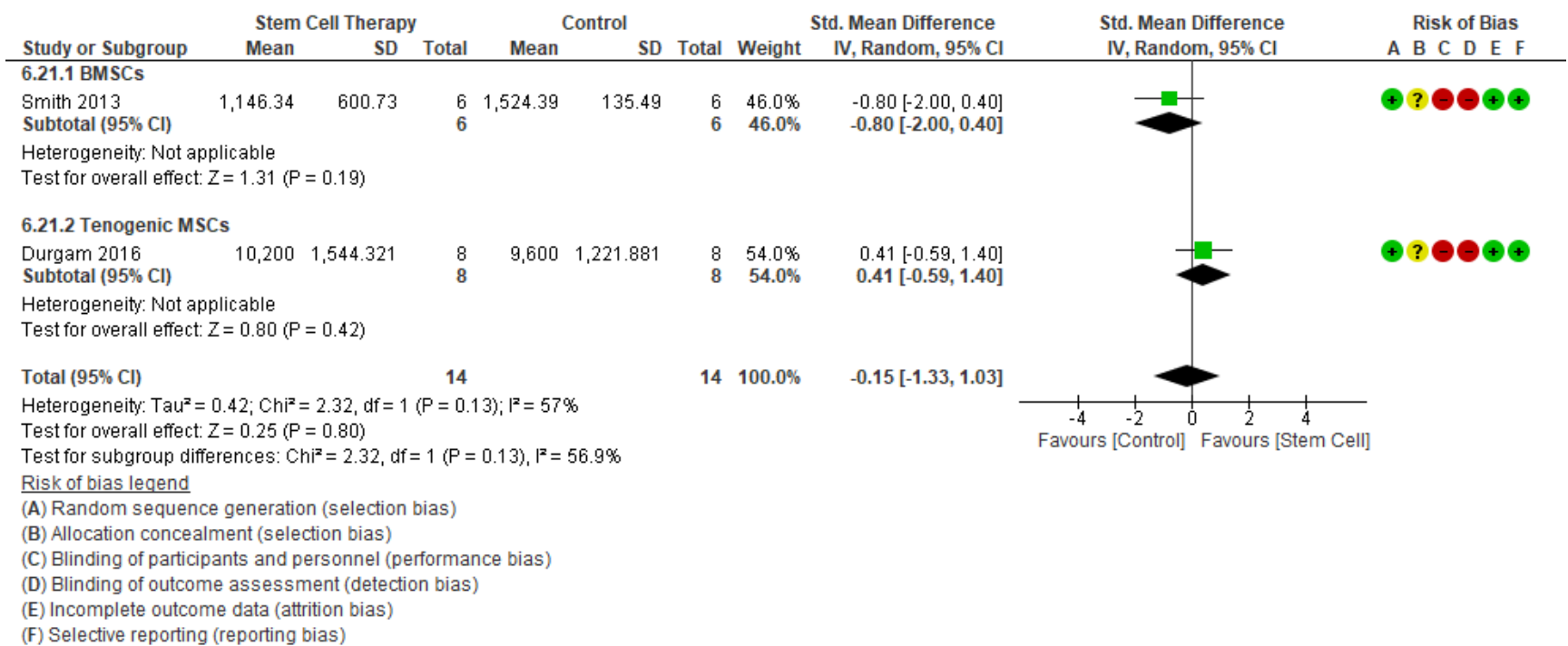

Figure 1.33. Forest plot for mechanical stiffness.

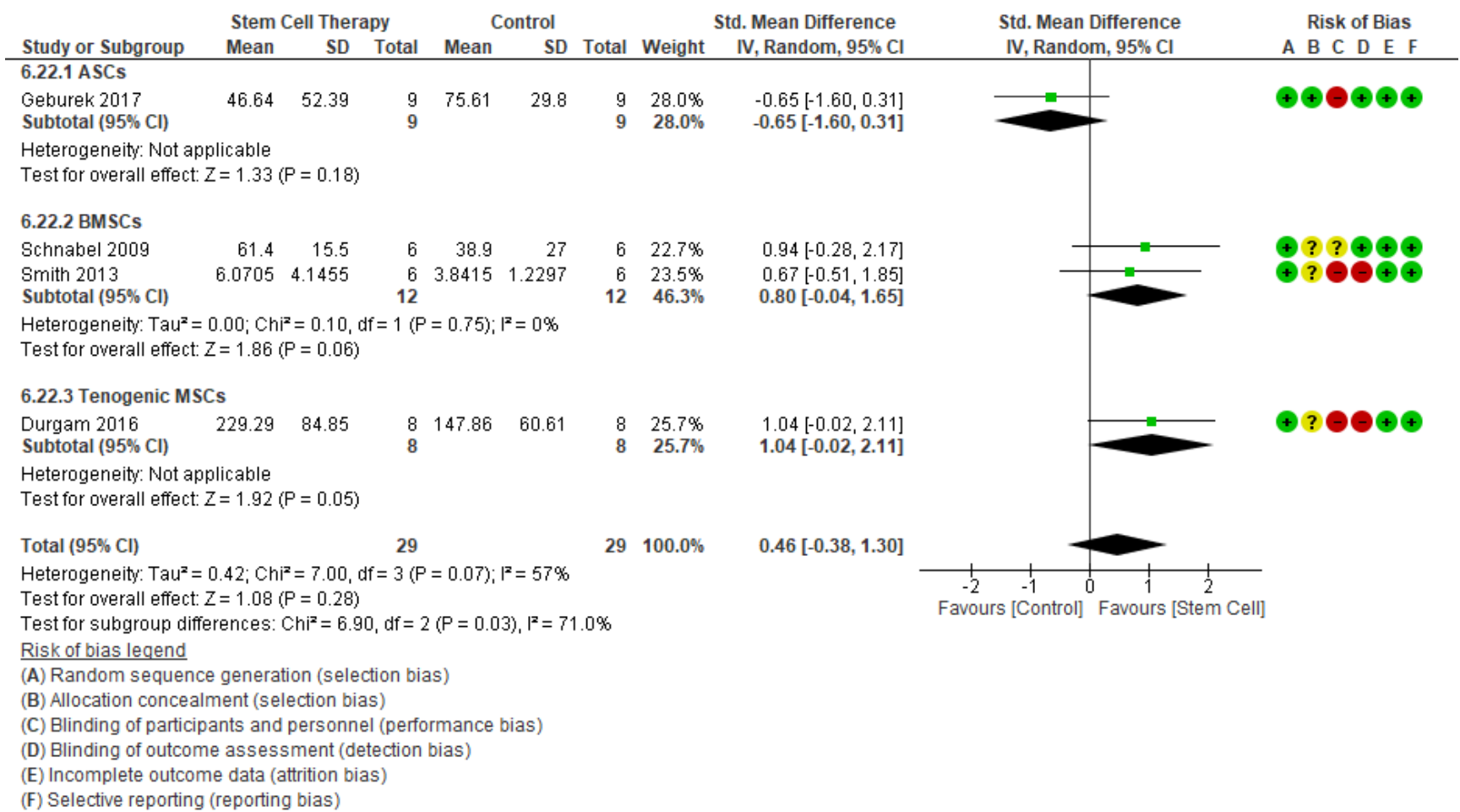

Figure 1.34. Forest plot for mechanical elastic modulus.

## 1.4. Discussion

### 1.4.1. Main Findings

The findings from our meta-analysis took advantages of accumulated clinical trials using equine tendon/ligament injuries that recapitulate human tendon/ligament injury scenarios most closely while allowing extensive investigation on healed tendon/ligament under controlled conditions. The analysis revealed the effects of cellular therapies being beneficial in facilitating vascularization at earlier phase and restoring microstructures of healed tissue. Specifically, when cellular therapy was used to treat both naturally-occurring and experimentally induced equine tendon/ligament injuries there was: 1) an increased rate of returning to performance; 2) a facilitated recovery of ultrasonographic echogenicity and increased vascularity at earlier healing phase evaluated by color doppler; and 3) an improved microstructure of healed tissue by reduced cellularity and improved fiber alignment. Our study provides potential mechanistic insights into improved healing of tendon/ligament injuries in both human and equine.

### 1.4.2. Rate of Returning to Soundness after Injury

Return to performance after tendon/ligament injuries following cellular therapies is a major milestone for not only equine athletes but also human. Our study indicated intralesional injection of MSCs increases the rate of return to performance in the horse. In humans, the return to performance is often measured indirectly by pain evaluation using established scoring rubrics. In a meta-analysis of human clinical trials, it was reported MSCs injection into commonly injured tendons in humans improved the overall outcome pool that includes pain, functional scores, MRI or ultrasonography parameters, and arthroscopic findings after 6 to 12 months.[32] As the cited study's outcomes include functional scores, the findings were largely comparable to our study. Additionally, bone marrow stimulation, an alternative method of delivery of MSCs to the site of injury without *ex vivo* processing, was evaluated for its efficacy on treatment of RC repair in a meta-analysis. And it was reported that bone marrow stimulation performed during arthroscopic RC repair reduced the re-tear rate and improved the constant pain score but not range of motion (ROM) of the joint or other pain scores evaluated at after 12 months post-operatively,[62] of which the lower re-tear rate was in agreement with our findings. Non-cellular injection, autologous blood injection, was found to improve pain scores of human patients with epicondylitis as early as 3 months and as long as 1 year post-injection compared to corticosteroid injection in meta-analysis.[63] The rate of return to performance in horses cannot be directly compared to pain scores, functionality, or re-tear rate evaluated by imaging modalities in human. However, combined effects from reduced pain and improved functionality may have led to the higher rate of return to performance in horses. One limitation of our findings with regard to the rate of return to performance is that the analysis was made from 1 PCS (prospective cohort study) and 5 RCS (retrospective case series) without RCT (randomized controlled trial), indicating re-evaluation with RCT is necessary to further validate our findings. Moreover, many of the studies included in return rate analysis were unbalanced towards the majority of subjects assigned to cellular therapy groups. It is known that unbalanced trials tend to favor interventions of interest,[64] indicating that conclusions from this particular finding need to be carefully drawn.

### 1.4.3. Ultrasound Evaluation of Injured Tendon and Ligament

Facilitated ultrasonographic echogenicity recovery and increased early phase vascularity by MSCs that were non-differentiated or tenogenically differentiated suggested healing of tendon injuries were facilitated by cellular injection. These outcomes have not been evaluated in meta-analysis of human clinical trials, yet they were consistent with findings from several RCTs. For example, BMSCs improved tendon structure as evaluated with ultrasound and MRI at 6 months post-injection among human patients affected by chronic patellar tendinopathy.[65] In human ultrasonographic structure evaluation is conducted often with ultrasound tissue characterization (UTC), yet it is not commonly used in equine studies.[66] The only structural evaluation using ultrasound and statistically significant in the present meta-analysis was echogenicity of lesions. However, a previous study found a weak correlation between tendon echogenicity and stress,[67] indicating mechanical properties may not be evaluated via ultrasonographic echogenicity alone. Therefore, although our analysis indicated facilitated recovery of echogenicity by cellular therapies, it may not necessarily represent improved healing of injured tissue. The possible association between tendon echogenicity and mechanical properties was suggested to be echogenicity change during fatigue-induced failure, as larger echogenicity change during fatigue loading was associated with high rate of failure using human cadaveric calcaneal tendons.[68] A useful and interesting application of these findings would be to investigate echogenicity of lesions both at weight bearing and non-weight bearing.

Additionally, another outcome that was improved by cellular therapies in our meta-analysis, improved vascularization during early phase, may have an important clinical implication, since neovascularization plays an essential role in tendon/ligament healing.[69] In this regard, our findings from clinical trials were uniquely valuable in terms of mechanistic insights on MSCs' effects on tendon/ligament healing. For example, neovascularization of tendon at the early healing phase has not been reported in human clinical study. The only available information on early neovascularization by MSCs are mostly pre-clinical studies. In a recent study using a mice calcaneal tendon injury model, MSCs induced early angiogenesis of injured tendon via polarization of macrophages into anti-inflammatory M2 phenotype by their extracellular vesicles (EVs).[70] It was also confirmed that MSCs primed to exert immunomodulatory properties increased endothelialization and presence of M2 macrophages at injured sites, especially in the early phase of healing at 4 days post-operatively in a rat medial collateral ligament injury model.[71] Therefore, our findings on improved vascularity of injured equine tendons by MSCs injection may be partly explained by an increased M2 phenotype polarization and accumulation of M2 macrophages at the injury site during early healing.

In this study, cellular therapies did not reduce either lesion CSA ratio or tendon CSA. In humans, a RCT reported BMSCs reduced ultrasonographic lesion size of chronic patellar tendinopathy 6 months post-injection, whereas PRP failed to reduce.[65] Additionally, BMSCs had better UTC outcomes than PRP. On the other hand, ASCs-derived exosomes applied to the partial-thickness RC tear of rabbits delivered by fibrin gel led to increased tendon thickness of healed tendon.[72] Therefore, it is likely that the effect of cellular therapies are to reduce lesion size, whereas tendon size tends to increase. The lack of difference in both lesion CSA and tendon CSA in our meta-analysis may be due to less accurate CSA measurement using ultrasonography than MRI.[73] Furthermore, relatively small lesion size often seen as core lesion in horses compared to human tendon/ligament lesions with which their size almost double the original size,[74] and small number of trials included in our meta-analysis may have led to type II error. Nonetheless, the structural improvement was consistently found using MRI without functional and pain score improvement in human clinical trial of RC tear using MSCs, and may not exactly represent improved healing. In that study, retear rate evaluated with MRI was 28.5% in arthroscopic treatment and 14.3% in ASCs treatment following arthroscopic RC tear repair, while no difference in visual analog scale (VAS) or range of motion (ROM) in 3 positions was found.[75] Therefore, the effects of cellular therapies may be more notable in imaging outcomes than in functional outcomes.

### 1.4.4. Gene Expression of Injured Tendon and Ligament

Expression of tenogenic genes was not different between cellular therapies and controls in our analysis. The reason for lack of treatment effects may be that the analysis focused on the late phase of healing as opposed to an early phase. In general, although expression patterns differ among genes, the tenogenic genes compared in the studies including transcription factors that upregulate during embryonic development or early phase of healing and gradually downregulate during postnatal development or late phase of healing. For example, in healing rat tendons from chemically-induced injury, *Scx* expression peaked at 8 days post-injury and returned to baseline by 21 days; same trend was observed in *procollagen III*, *LOX*, *Tnc*. On the contrary, maturation marker *Tnmd* continuously upregulated throughout the 21 day post-injury monitoring period.[76] In another study, mice received punch biopsy defects in the patellar tendon, and it was observed *Scx*, *Tnmd*, and *Col1a1* all upregulated gradually up to 8 weeks post-injury and downregulated to near baseline by 12 weeks.[77] The discrepancy on the patterns of gene expressions is potentially multifactorial, including defect type and size as well as model species. Nonetheless, it can be generally considered that upregulation of tenogenic genes is an early response to the injury that is followed by protein deposition or improvement in clinical outcomes and mechanical properties. The majority of studies included in our analysis had gene expression measurement after 12 weeks post-injury. Therefore, it is possible the treatment effects by cellular therapies on gene expression receded by the time of evaluation.

### 1.4.5. Compositional Analysis of Injured Tendon and Ligament

There was also no difference in compositional analysis of healed tendons between cellular therapies and controls in the present meta-analysis. However, there was a trend of reducing DNA and GAG contents by cellular therapies. In fact, the trend of lower DNA by cellular therapies was consistent with reduced cellularity evaluated histologically from our analysis. Since compositional analysis normally evaluate bulk composition of both lesion and intact tissues at macrostructure level as opposed to histological scoring at microstructure level, it is possible cellular therapies may reduce cellularity at the microstructure level yet not at the macrostructure level. Or it may be due to a smaller size of core lesions as compared to the sizes of injured tendons/ligaments. Nonetheless, DNA content measurement is not routinely performed in healed tendon except in horses, which makes outcome from the present meta-analysis unique and insightful. With regard to GAG contents, it is known GAG deposition is a maladaptation of healing tendon. Clinically, GAG deposition is also debilitating, as greater GAG content of the patellar tendon was correlated with greater tendon dysfunction and pain in humans.[78] Although there has not been a report on GAG contents from injured tendons/ligaments treated by cellular therapies in human, it was reported human umbilical cord-derived MSCs nearly completely prevented GAG deposition on RC defect created by punch biopsy in the rat.[79] The exact mechanisms of reduced GAG in injured tendon is yet to be elucidated, however, evidence suggests multifactorial effects. For example, in the horse, decellularized SDFT scaffold from chronic tendinopathy induced seeded MSCs' downregulation of tenogenic genes and increased release of GAG compared to scaffold from a healthy horse.[80] In a same study, it was also observed that cellular alignment was severely compromised on scaffold that has disrupted collagen fiber alignment from a horse with tendinopathy, indicating the critical role collagen fiber alignment plays on residing cellular phenotypes.[80] Therefore, it is possible improved collagen fiber alignment evaluated with histological analysis, evident from our meta-analysis, may indirectly lead to a trend of lower GAG deposition.

Similar to DNA and GAG contents, collagen and hydroxyproline contents were not different between cellular therapies and controls in our analysis, although there was a trend of increase amounts caused by the treatments. Collagen and hydroxyproline contents both measure synthesis of collagen contents directly or indirectly.[81] Yet, collagen content quantification itself may not be a reliable measure of tendon healing, as there is discrepancy of healing responses among types of collagen. In particular, quantification of collagen type I and III, as well as their ratio, are more indicative of functionality. Structurally, intact tendons/ligaments comprise higher content of collagen type I and lower collagen type III, which leads to improved mechanical properties such as modulus.[82] And it was reported MSCs reduce the ratio of collagen type III over I when applied to rat calcaneal tendon defect along with allograft than with allograft only.[83] It was also observed that MSCs seeded onto decellularized human flexor tendon prevented significant deposition of collagen type III upon subcutaneous implantation in immunocompromised rats.[84] Therefore, it is possible cellular therapies increased collagen type I synthesis while reducing type III deposition in horses, which then led to lack of difference in collagen contents among treatments. It may be more informative to compare the ratio of collagen type I and III among studies to draw a conclusion in future equine clinical trials. As an indirect quantification of collagen contents, hydroxyproline content was also reported to be an important indicator of collagen synthesis in healed tendon of rats, and higher contents of hydroxyproline were positively associated with higher failure load. In a previous study, MSCs were shown to increase hydroxyproline in healed tendon. For example, allogenic SDFT graft implanted into the SDFT defect of lambs led to higher hydroxyproline content when MSCs were injected simultaneously compared to no injection or PRP injection.[85]

### 1.4.6. Histological Analysis of Injured Tendon and Ligament

The effects of cellular therapies were more clearly demonstrated at the microstructure level evaluated by histology. Our meta-analysis indicates cellular therapies decreased cellularity and improved fiber alignment of healed tendons in horses. The decreased cellularity of healed tendon indicated cellular therapies restored more normal structures according to the rubrics used in the studies. Although there have been no reports of decreased cellularity by cellular injection into tendon or ligament using meta-analysis, several preclinical studies demonstrated this approach is effective in reducing cellularity of healed tissue. For example, MSCs had a trend of reducing histological cellularity in healed rat calcaneal tendon when combined with polyglycolic acid mesh as early as 6 days post-operatively.[86] On the other hand, another commonly applied regenerative therapy, PRP, had increased cellularity measured as DNA contents of healed SDFT in equine surgical injury model after 24 weeks.[61] The increased cellularity as well as GAG content was also observed in human calcaneal tendon after acute rupture and administration of PRP. At the same time, healed tendons had a higher deposition of collagen type I and lower collagen type III/I ratio as well as lower vascularization.[87] Therefore, as opposed to PRP injection, our analysis indicates cellular therapies are effective in obtaining more matured tendon structure in terms of cellularity. With regard to improved fiber alignment of healed tissue, there also has not been any report from meta-analysis. Nonetheless, a study of a rat calcaneal tendon model reported contradictory findings of fiber alignment not being improved with application of BMSCs but, rather by EVs.[88] Interestingly, similar findings from a study of a rat calcaneal tendon injury model indicated superior fiber alignment restored by exosomes, EVs of smaller size, over ectosomes, EVs of larger size, at 5 weeks post-injury.[89] Although not EVs, fibroblast growth factor (FGF)-2, a trophic factor secreted from MSCs in the form of exosome,[90] containing hydrogel also improved collagen fiber orientation of primary repair of RC tear in rats as early as 4 weeks post-operatively.[91] This evidence suggests cellular therapies improve fiber alignment of healed tendons/ligaments via paracrine effects.

While there were statistical significances in reducing microstructural cellularity and improving fiber alignment score by cellular therapies based on our analysis, there was no effect of cellular therapies on tenocyte morphology score, variation in cell density score, collagen type I distribution, vascularity, inflammatory cell infiltrate score, GAG, and total score. Therefore, cellular therapies may not exert consistent effects on these microstructural properties. Yet, these are the outcomes often improved by cellular therapies under controlled experimental conditions. For example, it is often observed that the addition of cells to tendon healing prevents ectopic chondroid formation. A recent study also observed in a tendon after injury GAG deposition is dramatically decreased by administration of 2-deoxy-D-glucose (2-DG) that inhibits conversion of glucose to glucose-6-phosphate.[92] MSCs are highly glycolytic cells whose growth largely relies on availability of glucose,[93] and their ATP synthesis is virtually abolished by 2-DG treatment. Therefore, a trend of lower GAG (standard mean difference in score: - 0.75) in MSCs-treated tendons in the present meta-analysis may partly be explained by deprivation of glucose from MSCs uptake at the healing site. In fact, there was a trend of lower GAG deposition in our analysis and GAG contents in our analysis, although neither reached statistical significance. Interestingly, it was also reported MSCs-derived exosomes were effective in preventing fat infiltration into the healing tissue of rabbit RC upon primary repair augmented with exosome-loaded polyurethane patch.[94] Collectively, this evidence suggests cellular therapies potentially prevent ectopic tissue formation at the sites of administration. On the other hand, total histological scores may not have been different between treatments and controls because of inconsistency among varieties of scoring systems. These scoring systems are often based on a few established rubrics, yet each scoring system is also commonly modified in each newer investigation. Notably, almost all studies with histological outcome measures included in our meta-analysis employed unidentical scoring systems based on a few systems and modified by each investigator. This potentially increased risk of type II error. Another potential reason for the lack of difference in total score by cellular therapies may be the late timing of evaluation relative to the whole healing process. An example of larger treatment effects at an early phase of healing was reported in a study demonstrating human ESCs-derived MSCs with upregulated *Scx* cultured on collagen/silk scaffold; this treatment resulted in improved total histological scores in non-*Scx* upregulated cell construct 4 weeks after implantation into the rat calcaneal tendon defects but not 8 weeks.[95] Therefore, it is of interest to assess healing at an earlier phase than that commonly selected in equine tendon/ligament injury clinical trials.

**1.4.7. Mechanical Properties of Injured Tendon and Ligament**

Mechanical properties compared in the present meta-analysis were also not different between treatment and control groups. The lack of treatment effects by cellular therapies may be due to multiple factors that affect mechanical properties or early evaluation timing in comparison to lengthy remodeling process of tendon and ligament to recover full strength that can span multiple years.[96] In general, tendinopathy or ligamentopathy increases CSA, decreases stiffness, and lowers Young's modulus among middle-aged individuals.[97] However, a recent systematic meta-analysis confirmed tendinopathic calcaneal tendon had a lower global stiffness, lower global modulus and lower local modulus than a normal calcaneal tendon in humans,[98] while tendinopathy did not affect the mechanical properties of the patellar tendon, suggesting the effects of tendinopathy can vary depending on the anatomical location and function of each tendon that is affected. Other contributors to the mechanical properties of tendons and ligaments are an individual's age and gender. For example, age negatively affects shear modulus of uninjured calcaneal tendon particularly among women, although the same age effect was not observed among men.[99] Collectively, these multiple factors that variably affect mechanical properties of tendons and ligaments may make it difficult to detect treatment effects by cellular therapies on tendon/ligament healing using *in vitro* uniaxial tensile testing. Partly due to this reason, in humans, tendon/ligament tissue elasticity is often estimated by ultrasonography and elastography,[100] which may increase the sensitivity to detect treatment effects.[101] Yet, there have not been reports of cellular effects on mechanical properties in human meta-analysis. The high sensitivity of elastography was also confirmed in equine SDFT,[102] hence the application of this modality is warranted in future investigations.

As to an ideal timing for evaluating mechanical properties on healed equine tendons, there is no consensus. And it may take even longer than multiple years for the treatment to be effective due to progressing tendinopathy/ligamentopathy for chronic cases. For example, human tendinopathy/ligamentopathy is a chronic degenerative condition from which affected individuals rarely recover to full function, especially among adults. However, in rodents, up to 12 weeks post-injury recovery seems to be a reasonable time frame to detect improved mechanical properties by cellular therapies. Examples of improved mechanical properties were reported in a rat calcaneal tendinopathy treated by exosomes[89] and a rat RC tear model treated by MSCs conditioned media.[103] Although not exosome, FGF-2, a component of MSCs-derived exosome,[90] also increased ultimate failure load and stiffness of rat chronic RC tear both at 6 and 12 weeks after injury.[104] Also, PRP combined with collagen scaffold seeded with tenocyte was shown to improve ultimate failure load of healed rat calcaneal tendon compared to collagen-tenocyte construct as early as 1 week post-operatively.[105] At the same time, differences in failure load and stiffness were much higher after 8 weeks than 4 weeks of post-anterior cruciate ligament (ACL) reconstruction and injection of MSCs-derived exosome than PBS or sham injection in rats.[106] And similar findings were observed in rat RC transection and primary repair combined with FGF-2 containing gelatin hydrogel, which resulted in a larger difference in improvement of ultimate load to failure at 12 weeks than 6 weeks post-operatively,[91] again emphasizing the lengthy remodeling phase to clearly demonstrate treatment effects on mechanical properties.

### 1.4.8. Limitations

The outcomes evaluated in this study that were not different between cellular therapy interventions and control included all gene expression levels, compositional quantities, and mechanical properties. All these outcomes are macrostructural analysis of healing tissue rather than microstructural analysis performed by histological evaluation. One possible explanation for the lack of cellular therapy effects at the macro level may be due to the small lesion typically in the form of longitudinal core lesion within affected tendons, which inevitably leads to inclusion of both healing and healthy tissue in a same sample specimen. Another potential reason for lack of significance, and indeed a study limitation, is controls included in the analyses that were not consistent among studies compared. Often cellular therapies were compared to PBS, serum, PRP, conventional therapies, or untreated. Therefore, the outcome of this study can be interpreted as the efficacy of cellular therapies in comparison to the other non-cellular therapies. Nonetheless, it is often common that cellular therapies fail to improve healing of human tendon/ligament chronic injuries at the functional or macrostructural level. For example, human umbilical cord blood-derived MSCs did not improve tunnel enlargement or functional and pain scored after application onto reconstructed ACL of human.[107]

### 1.5. Conclusion

In conclusion, our meta-analysis indicates that cellular therapies facilitate recovery of ultrasonographic echogenicity and increased vascularity at an earlier healing phase and improves microstructure of healed tissue by reduced cellularity and improved fiber alignment in horse tendon/ligament injuries. However, in terms of the rate of return to performance, the efficacies of cellular therapies remain inconclusive due to scarcity of RCTs. Hence, our study supports the beneficial effects of cellular therapies for equine tendon/ligament injuries in improving microstructural organization, and provides potential mechanistic insights into improved healing of tendon/ligament injuries in equine.

# Chapter 2. De Novo Tendon Neotissue from Equine Adult Stem Cells

## 2.1. Introduction

Tendinopathy and desmitis comprise a large majority of musculoskeletal injuries that are responsible for up to 72% of lost training days and 14% of early retirements by equine athletes.[12-14] Superficial digital flexor tendinopathy and suspensory ligament (SL) desmitis are the most common, comprising 46% of all limb injuries.[15,16] The predominant type of tendon and ligament injury varies among disciplines, but virtually all equine companions can be impacted. Strain induced injuries are common in the equine suspensory apparatus including the SL, superficial digital flexor tendon (SDFT), and deep digital flexor tendon (DDFT).[17] Many acute and chronic tendon and ligament lesions are thought to result from focal accumulation of microtrauma and poorly organized repair tissue that can coalesce into large lesions and predispose to spontaneous rupture in many species.[18]

Diagnosis is usually made with a combination of physical examination and ultrasound imaging.[108] Treatments vary widely and can range from rest with anti-inflammatory drugs, cold therapy and pressure bandaging to intralesional injections of various therapeutics and extracorporeal shock wave therapy.[19] Additionally, intralesional regenerative therapies such as platelet rich plasma, stem cells, and genetic material have been applied with variable success.[22] Short-term outcomes of these treatments are favorable. However, poor or abnormal tissue repair contributes to a reinjury rate in horses as high as 67% within 2 years.[20,21] To date, there is no single, gold standard to promote healing of ligament and tendon lesions.

There are four recognized stages of tendon and ligament healing: an acute inflammatory phase, a subacute reparative phase, a collagen phase, and a chronic remodeling phase. Low cell numbers and metabolic activity, limited blood supply, and failure of endogenous tenocytes and ligamentocytes to migrate to the injury site affect all stages of healing and contribute to poor tissue healing capacity in adult animals.[109,110] However, recent research confirms enhanced healing capacity of neonatal tendon over that of adults owing to migration of endogenous tenocytes to the site of injury and replacing early fibrous scar tissue with normal tendon.[111]

Autologous tenocyte implantation is one mechanism to deliver endogenous cells to the site of tendon or ligament injury in adult animals and humans,[23-25] however the therapy is limited by few harvest sites and harvest morbidity, and it is not practical in horses. Administration of exogenous adult multipotent stromal cells (MSCs) is reported to augment natural healing in naturally-occurring and experimentally-induced equine tendon and ligament injuries.[29-31] Results are mixed, in part due to differences among cell isolates, lesions, individual healing capacity, and low engraftment of exogenous cells (< 0.001%).[30,33] Further, there is evidence that an inflammatory environment may impede differentiation of MSCs, and the cells may assume an abnormal phenotype leading to unwanted side effects.[34,35]

The delivery of cells on scaffold matrix, often made of both natural and synthetic polymers, improves cellular retention at the site of implantation. Collagen is the most abundant natural polymer in the body and a common material for tissue engineering templates due to inherent biocompatibility.[112] Collagen type I (COLI) comprises 60 – 80% of tendon and ligament structure, and there are numerous commercially available FDA-approved formulations.[113,114] Scaffolds composed of COLI are routinely used for delivery and retention of stem cells in tendon and ligament tissue,[115] and published information confirms that COLI matrix supports differentiation of equine MSCs into diverse tissue lineages.[112,116,117] Additionally, evidence suggests differentiation of MSCs into the specific lineages orthogonal to the site of injection prior to implantation is essential, as pre-implantation of cells into tenocytes was reported to minimize the risk of ectopic bone and cartilage formation as well as tumor formation at the site of injection.[118] A recent study also reported the efficacy of pre-differentiated tenocytes' injection into naturally-occuring tendinopathy, as injection of tenogenically differentiated allogeneic equine MSCs into naturally-occurring tendon lesions resulted in a lower reinjury rate (18%) than conventional treatments (44%) 24 months post-treatment.[119] Regardless of specific target tissue, current knowledge supports that MSCs that are induced to assume characteristics of native tissue lineage and embedded in scaffold matrix prior to implantation have better engraftment and promote more robust tissue healing than undifferentiated primary cell isolates.[120]

To further advance therapeutic capacity of pre-differentiated tenocytes, application of physical stimulus such as strain is known to play an essential role. Strain can be applied both statically and dynamically. When applied statically, elevated matrix tension instigates fibroblast-to-myofibroblast activation eliciting scar-like phenotypes *in vitro*.[121] On the other hand, tenocytes maintained under no strain for 24hr resulted in a marked upregulation of matrix degradation protein, matrix metalloproteinase-1 (MMP-1) expression.[122] In the same study, the importance of dynamic strain was demonstrated, as MMP-1 expression was completely eliminated at 3% and 6% strain with frequency of 0.017 Hz or 1% strain with frequency of 1.0 Hz. Several other studies also supported the efficacy of dynamic strain to promote tenogenic differentiation, cellular viability, and to regulate catabolic/anabolic state.[123-126]

In this study, tendon neotissue was created by culturing equine adipose-derived MSCs (ASCs) on COLI templates in specifically formulated tenogenic medium maintained in custom-designed perfusion bioreactor both under static and dynamic strain. The hypotheses tested were: 1) ASCs assume a tenoblast-like morphology, express tendon-specific genes, and produce more organized extracellular matrix (ECM) in tenogenic versus stromal medium perfusion culture under static strain; and 2) cell tendon-specific gene expression and ECM deposition increase with dynamic versus static strain in tenogenic medium perfusion culture. The objectives were: 1) to evaluate neotissue constructed by culturing ASCs on COLI templates in stromal and tenogenic medium for up to 21 days under static strain for distribution and morphology of viable cells, cell number, tendon-specific gene expression, and micro- and ultra-structure; and 2) to evaluate neotissue constructed by culturing ASCs on COLI templates in tenogenic medium for 21 days under dynamic strain for distribution and morphology of viable cells, cell number, and micro- and ultra-structure.

## 2.2. Materials and Methods

### 2.2.1. Study Design

Supragluteal ASCs, passage 2, from 8 adult horses (4 geldings, 4 mares) were seeded at $1.0 \times 10^6$ cells/cm$^3$ onto bovine COLI sponge (n = 48), each sizing 6.0 x 4.0 x 1.0 cm$^3$, and were then rolled into a cylinder and wrapped with a finger trap suture. The constructs were cultured in stromal (n = 24) or tenogenic (n = 24) medium using custom-designed perfusion bioreactors for 7 (n = 8), 14 (n = 8), and 21 (n = 8) days under 10% static strain (Fig 2.1).

Gross appearance of constructs was documented prior to specimen harvest at each time point for molecular evaluations. Qualitative and quantitative cell growth kinetics, tendon-specific gene expression of constructs were evaluated with calcein-AM/EthD-1 staining and resazurin reduction, and RT-PCR, respectively. Microstructure of constructs was evaluated histologically with H & E staining and immunohistochemical staining against fibromodulin deposition. Ultrastructure of constructs was evaluated with SEM and TEM.

Additionally, the effect of dynamic strain on construct culture was evaluated. To apply dynamic strain, one end of each construct was continuously displaced at 10% strain in a sine wave at 1 Hz frequency during culture in tenogenic medium for 21 days. Specimens were harvested and evaluated histologically with H & E staining and ultrastructurally with SEM.

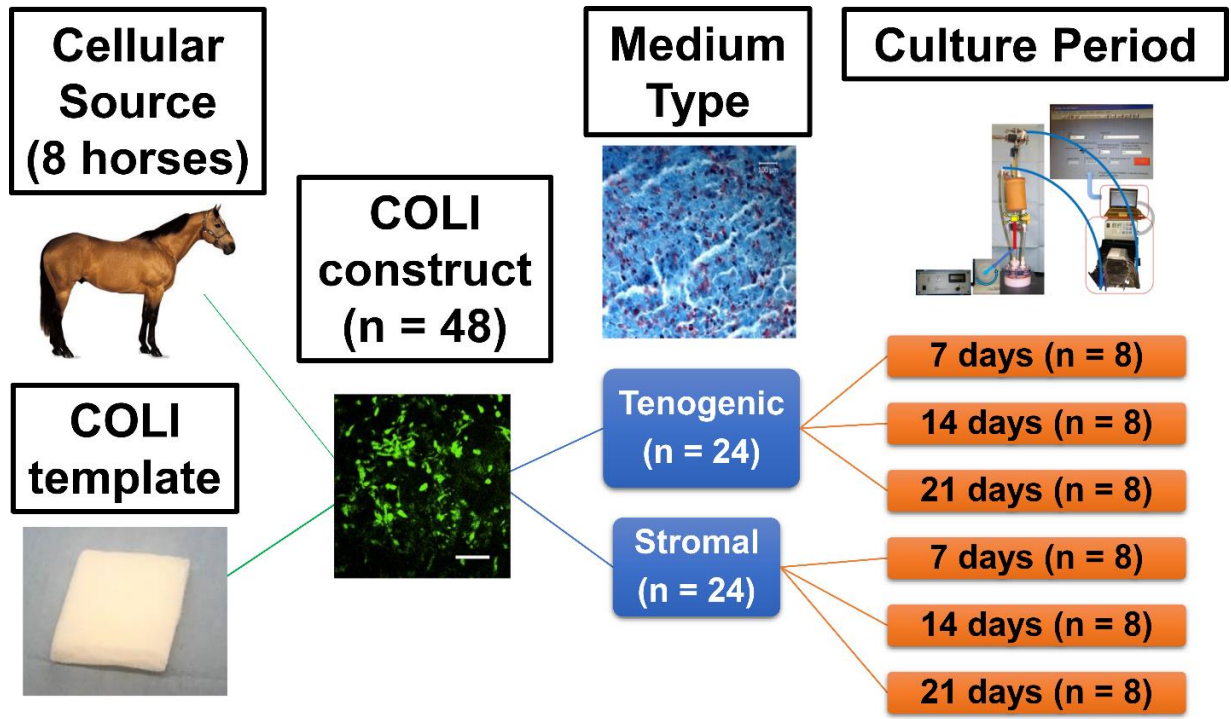

Figure 2.1. Neotendon culture using ASCs and COLI template. ASCs were seeded onto COLI template and cultured in tenogenic or stromal medium for up to 21 days under 10% static strain.

### 2.2.2. Equine Subcutaneous Adipose Tissue-derived Multipotent Stromal Cells (ASCs)

Equine subcutaneous adipose tissue was aseptically harvested from the supragluteal region

of 4 adult geldings and 4 mares euthanized for reasons unrelated to this study immediately post-

mortem. Harvested adipose tissues were minced and digested at 37 °C in 0.1% type I collagenase

(Worthington Biochemical, Lakewood, NJ) in PBS for 3 hours. ASCs were maintained in stromal

medium (DMEM-Ham's F12, 10% FBS, 1% antibiotic/antimycotic) until 80% confluence followed by cryopreservation at passage 0 (P0).

Cryopreserved ASCs were revitalized and expanded to P1 in stromal medium prior to construct culture in tenogenic (DMEM-high glucose, 1% FBS, 10 ng/ml transforming growth factor (TGF)-β1, 50 mM L-ascorbic acid 2-phosphate sesquimagnesium salt hydrate, 0.5 mg/ml insulin, 1% antibiotic/antimycotic) or stromal medium at P2.

### 2.2.3. Perfusion Bioreactor System

For each template, a COLI sponge section (Avitene™ Ultrafoam™ Collagen Sponge, Davol Inc., Warwick, RI), 6.0 x 4.0 x 1.0 $cm^3$, was rolled into a column with a diameter of 1.0 cm and length of 6.0 cm. The column was wrapped by a finger trap composed of #0 polydioxanone suture (PDS$^{®}$ II, Ethicon, Somerville, NJ) with 1 cm long loops on each end to secure the constructs to one immobile and one adjustable bar within a bioreactor chamber (Fig 2.2).

The bioreactor consisted of a top lid, core frame, and base chamber. The immobile horizontal bar is located inside the core frame and the adjustable horizontal bar is attached to the top lid via vertical threaded bar (Fig 2.3 – 2.5). The distance between the bottom immobile horizonal bar and top adjustable horizontal bar was set at 6.6 cm to apply 10% strain on the COLI construct during the culture period.

The bioreactor chamber was connected to a perfusion system that consists of a medium reservoir, peristaltic pump, and medium stirrer (Fig 2.6). The lowest port of each bioreactor was attached to a port of a 10 ml medium reservoir (Synthecon, Houston, TX) with tubing (4.8 mm inner diameter: Tygon$^{®}$, Compagnie de Saint-Gobain, Courbevoie, France). The reservoir had an oxygenator membrane at the bottom side for gas exchange. The port on the bioreactor cap and the other port of the medium reservoir were connected to a computer-controlled peristaltic pump (ISM404b, Ismatec, Wertheim, Germany) via 3-way stopcocks attached to 0.22 µm microfilters to which tubing (1.0 mm inner diameter: Compagnie de Saint-Gobain) between the bioreactor and the pump was attached.

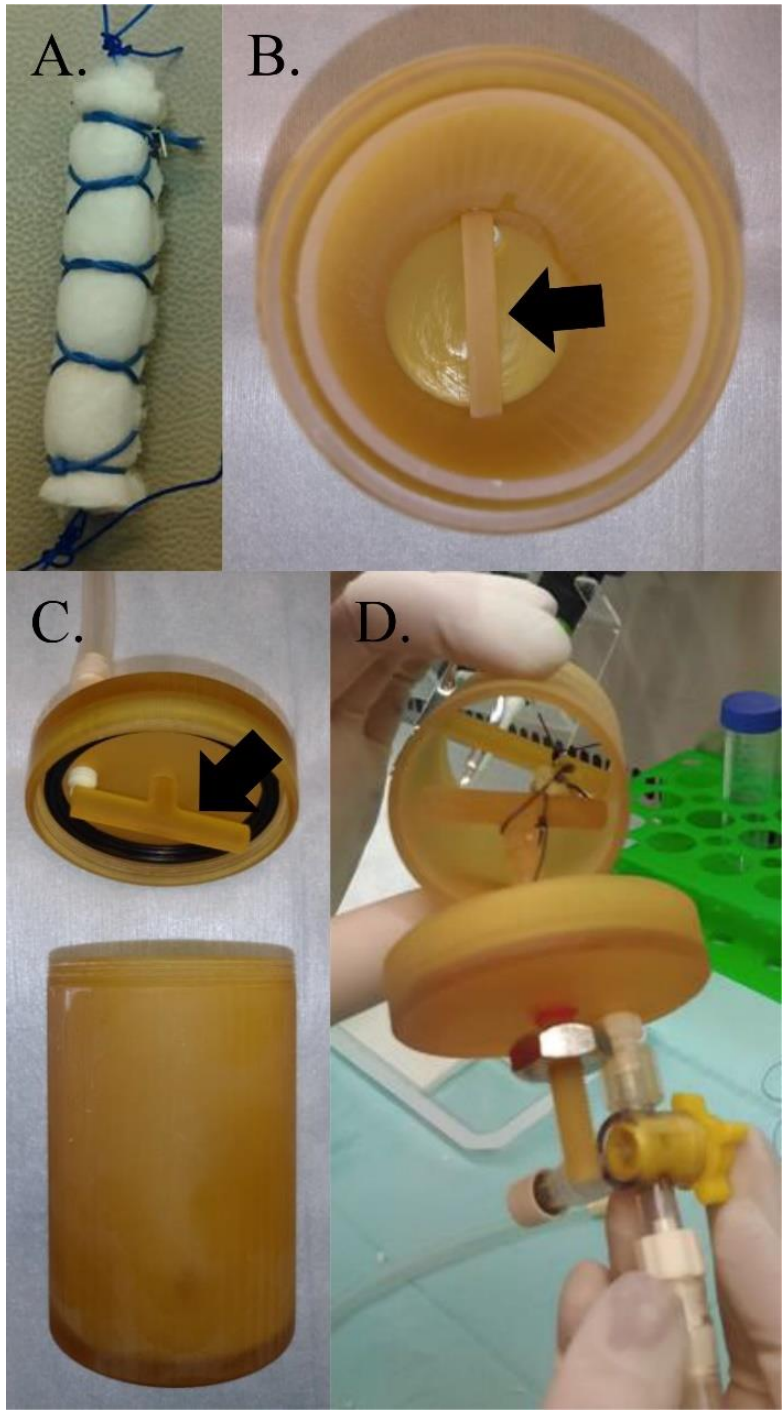

Figure 2.2. COLI template preparation and placement inside bioreactor chamber. COLI template is rolled into a column with a diameter of 10 mm and length of 60 mm and surround by a finger trap composed of # 0 PDS® II with 1 cm long loops on each end (A). One end of suture loop is tied to the immobile horizontal bar (black arrow) at the bottom of bioreactor chamber (B). The other end is tied to an adjustable horizontal bar (black arrow) at the top of bioreactor chamber to apply 10% static strain to construct during culture period (C). Constructs secured to the both horizontal bars (D).

Figure 2.3. Assembly of bioreactor. Bioreactor is composed of top lid (#1), core frame (#3), and base chamber (#4). Immobile horizontal bar locates inside core frame and adjustable horizontal bar (#2) is attached to top lid via vertical threaded bar.

System fluid flow rate was computer controlled (LabView™, National Instruments, Austin, TX) at 10 ml/minute, and the direction reversed before the medium reached one side of microfilter. Additional medium perfusion was rendered via stirrer (2.5 x 0.7 cm) at 300 rpm. All bioreactor system parts were sterilized with ethylene oxide prior to assembly and use. The perfusion system was maintained in a $CO_2$ incubator (5% $CO_2$, 37 °C) for the duration of the culture period. P2 ASCs were infused into COLI templates at 1.0 x $10^6$ ASCs/$cm^3$ template in stromal or tenogenic medium through the 3-way stopcocks and maintained for 21 days. Medium was exchanged every 7 days.

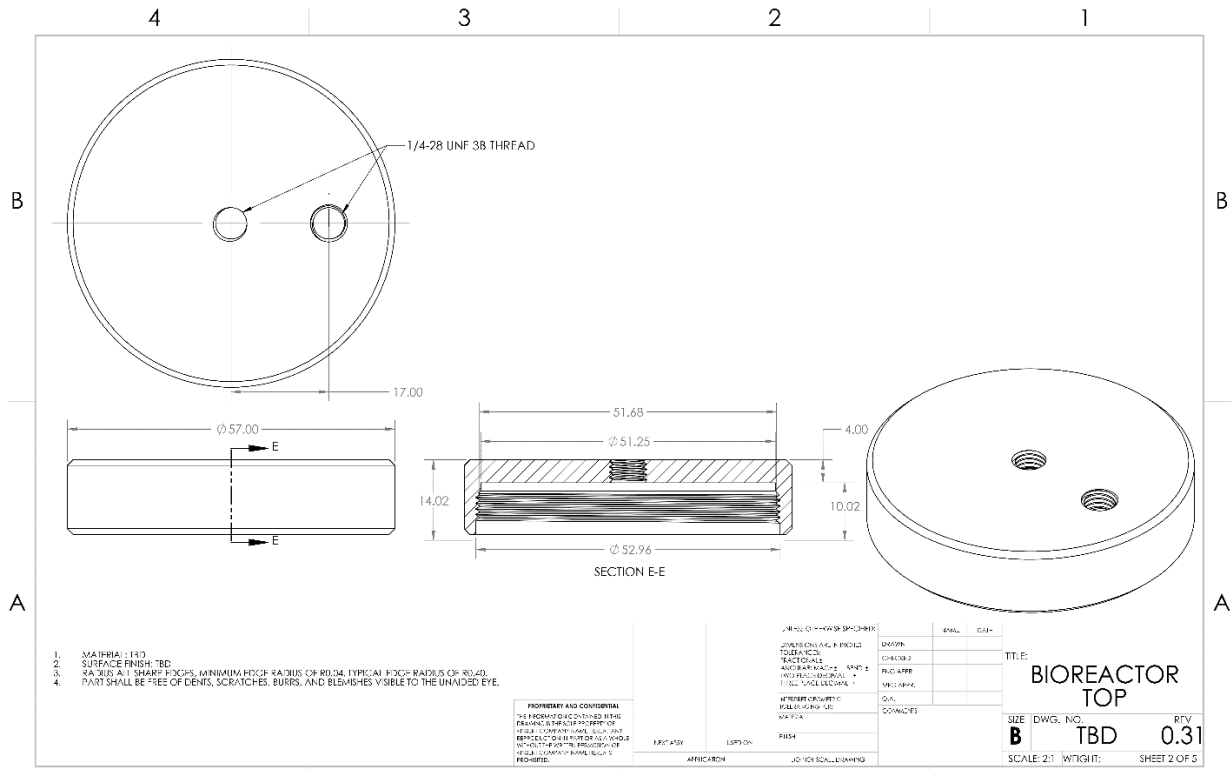

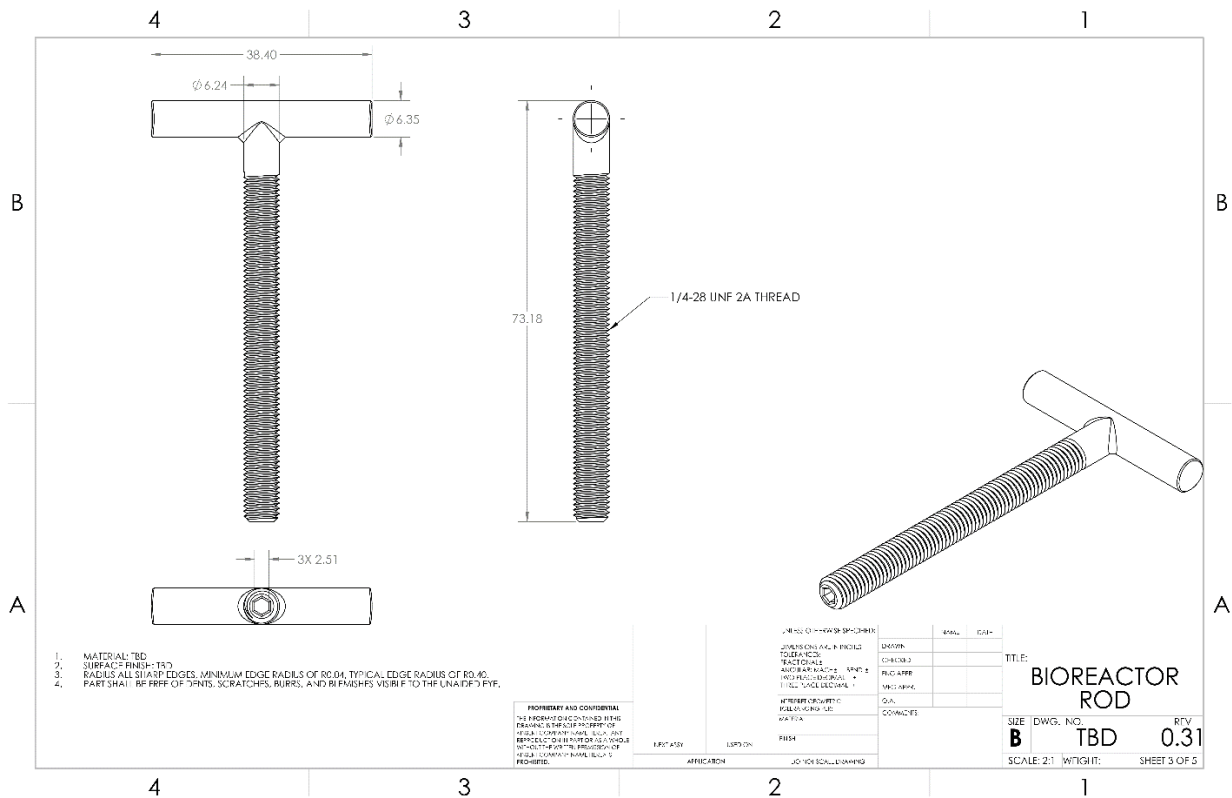

Figure 2.4. Dimensions of top lid (upper panel) and adjustable horizontal bar (lower panel).

Figure 2.5. Dimensions of core frame (upper panel) and base chamber (lower panel).

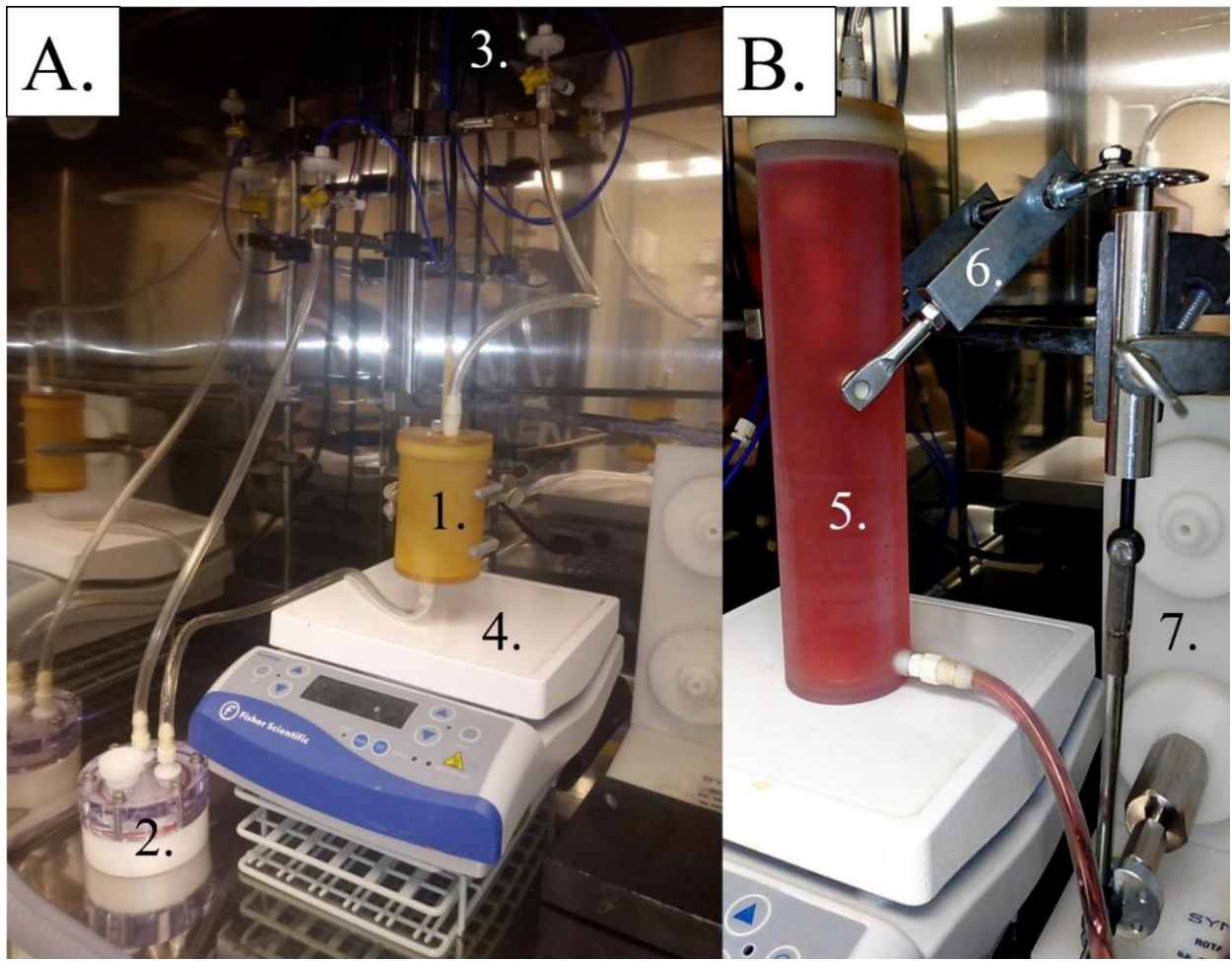

Figure 2.6. Perfusion bioreactor system and cyclic tensioner. For construct culture under static strain (A), a bioreactor (1) was attached to a 10 ml medium reservoir for gas exchange (2), both of which were connected to a computer-controlled peristaltic pump via 3-way stopcocks (3) through 0.22 µm microfilters between the bioreactor and the pump. Additional medium perfusion was rendered via stirrer (4). For construct culture under dynamic strain (B), a taller bioreactor (5) allowed vertical motion of stirrer attached to the bottom of construct by cyclic tensioner (6) externally applying magnetic field, while stirrer at the bottom of the bioreactor chamber was stirred. Cyclic tensioner was driven by vessel rotator.

### 2.2.4. COLI Construct Specimen Harvest

Once each culture period was reached, the COLI construct was harvested from the bioreactor. Suture was removed, construct unrolled, and full thickness cylindrical specimens were collected with a biopsy punch (diameter 4.0 mm) from the top, middle and bottom regions of the long axis of each construct. A total of three specimens from each region was allocated to each outcome. The measure (Fig 2.7) remaining of each COLI construct sample after collecting cylindrical specimens for other outcome measures was used as a single representative specimen combining all 3 top, middle, and bottom regions for tenogenic gene expression measurement.

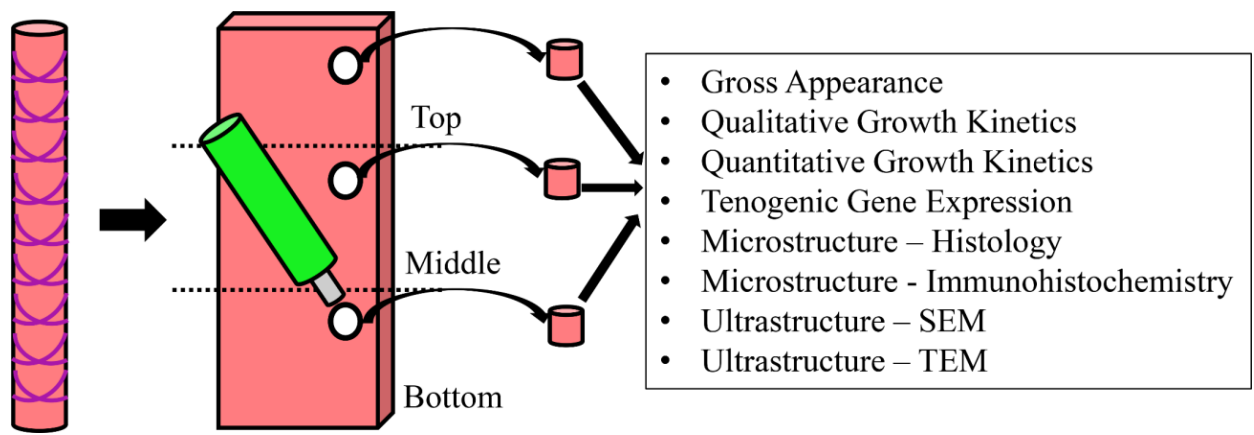

Figure 2.7. COLI construct specimen harvest scheme. After each culture period, COLI construct was removed from the bioreactor, suture unwrapped, construct unrolled, and full-thickness cylindrical cut was made to obtain specimen. Three specimens were obtained from each of the top, middle, and bottom region of construct for each outcome measure.

### 2.2.5. Qualitative and Quantitative Growth Kinetics

To evaluate qualitative growth kinetics of cells within constructs, specimens were stained with calcein acetoxymethyl (calcein-AM: Thermo Fisher, Waltham, MA) for viable cells and ethidium homodimer-1 (EthD-1: Thermo Fisher) for nonviable cells both at the concentrations of 4.0 μM in PBS at 37 °C for 30 minutes. After incubation with calcian-AM and EthD-1, specimens were briefly washed with PBS, placed on a glass slide with a cover glass. The microstructural distribution of viable and nonviable cells as well as cellular morphology within entire specimen were imaged with a confocal laser microscope (TCS SP8: Leica, Wetzlar, Germany) at multiple focal planes across the full-thickness of sample.

To evaluate quantitative growth kinetics of cells within a construct, the number of viable cells in each specimen was indirectly quantified using cellular metabolic activity measured by incubating with 100 μl of 50 μM resazurin reduction (Thermo Fisher) at 37 °C for 3 hours. After incubation, resazurin mixture was collected, of which 50 μl was mixed with 50 μl of PBS, and resorufin fluorescence measured at an excitation wavelength of 540 nm and an emission wavelength of 590 nm using a microplate reader (SPARK® Multimode Microplate Reader: TECAN, Männedorf, Switzerland). Resazurin solution incubated without specimen was used as a negative control to subtract fluorescence from the sample values.

### 2.2.6. Gene Expression - Reverse Transcription Polymerase Chain Reaction (RT-PCR)

The remaining portion of each COLI construct sample after collecting cylindrical specimens for other outcome measures was used as a single representative specimen for each COLI construct sample combining the 3 top, middle, and bottom regions. Specimens were digested at 37 °C in 0.1% type I collagenase in PBS for 1 hour, spun down at 300 x g for 10 minutes, and supernatant removed. One milliliter of TRI reagent® (Sigma, St. Louis, MO) was added to the precipitate, and homogenized by passing mixture through 18-gauge needle 30 times. Homogenate was spun down at 21,000 x g for 15 minutes at 4 °C. Total RNA was extracted from supernatant by phenol-chloroform extraction according to the manufacturer's instructions. Isolated RNA was cleaned up by RNeasy® Mini Kit (QIAGEN, Hilden, Germany). One microgram of total RNA was used for cDNA synthesis (QuantiTect® Reverse Transcription Kit, QIAGEN).

Equine-specific primers for tendon-specific genes, *scleraxis* (*Scx*), *mohawk* (*Mkx*), *early growth response 1* (*Egr1*), *connective tissue growth factor* (*CTGF*), *lysyl oxidase* (*LOX*), *collagen 1a1* (*Col1a1*), *collagen 3a1* (*Col3a1*), *decorin* (*Dcn*), *elastin* (*Eln*), *tenascin-c* (*Tnc*), *biglycan* (*Bgn*), *fibromodulin* (*Fbmd*), *collagen 14a1* (*Col14a1*), and *truncated hemoglobin 4* (*THBS4*) were quantified using primers previously validated[127-130] or designed with Primer-BLAST (National Center for Biotechnology Information, Bethesda, MD). PCR was performed with the denaturation step at 95°C for 15 minutes, followed by 40 cycles of denaturation at 94 °C for 15 seconds, annealing at 52 °C for 30 seconds, and elongation at 72°C for 30 seconds using SYBR Green system (QuantiTect® SYBR® Green PCR Kits, QIAGEN). Relative gene fold change was determined by standard means ($2^{-\Delta\Delta Ct}$). *Glyceraldehyde 3-phosphate dehydrogenase* (*GAPDH*) was used as the reference gene.

### 2.2.7. Histological Microstructure

Specimens were fixed in 4% paraformaldehyde (PFA) overnight at 4°C, serially dehydrated in an increasing concentration of ethanol and xylene, paraffin embedded, and sectioned (5 μm). Sections were deparaffinized in xylene, and serially rehydrated in decreasing concentrations of ethanol, followed by staining with haematoxylin and eosin (H & E). H & E staining was performed with incubation of sections with hematoxylin at room temperature for 3 minutes followed by washing with deionized water and tap water. Sections were then incubated with eosin at room temperature for 30 seconds, serially dehydrated with increasing concentrations of ethanol and xylene, then mounted with mounting medium (Permount™ Mounting Medium: Thermo Fisher) and cover glass.

Cellular morphology, distribution, and extra cellular matrix deposition were evaluated after digital images generated with a slide scanner (NanoZoomer, Hamamatsu Photonics K.K, Hamamatsu City, Japan) or a light microscope (DM4500B, Leica, Wetzlar, Germany) fitted with a digital camera (DFC480, Leica).

Table 2.1. Equine-specific Primer Sequences

| Gene | | Sequence (5' – 3') | Accession Number |
|------|---------|--------------------|------------------|
| *Scx* | Forward | TCTGCCTCAGCAACCAGAGA | NM_001105150.1 |
| | Reverse | AAAGTTCCAGTGGGTCTGGG | |
| *Mkx* | Forward | AGTGGCTTTACAAGCACCGT | XM_023632371.1 |
| | Reverse | ACACTAAGCCGCTCAGCATT | |
| *Egr1* | Forward | CCTACGAGCACCTGACCTCAG | XM_001502553.5 |
| | Reverse | GATGGTGCTGAAGATGAAGTGG | |
| *CTGF* | Forward | ACCCGCGTTACCAATGACAA | XM_023651101.1 |
| | Reverse | GGCTTGGAGATTTTGGGGGT | |
| *LOX* | Forward | CAGGCGATTTGCGTGTACTG | XM_023617821.1 |
| | Reverse | ACTTCAGAACACCAGGCACT | |
| *Col1a1* | Forward | CAAGAGGAGGGCCAAGAAGA | XM_023652710.1 |
| | Reverse | TCCTGTGGTTTGGTCGTCTG | |
| *Col3a1* | Forward | TCCTGGGGCTAGTGGTAGTC | XM_008508902.1 |
| | Reverse | GGCGAACCATCTTTGCCATC | |
| *Dcn* | Forward | TTATCAAAGTGCCTGGTG | XM_005606467.3 |
| | Reverse | CATAGACACATCGGAAGG | |
| *Eln* | Forward | CTATGGTGTCGGTGTCGGAG | XM_023655466.1 |
| | Reverse | GGGGGCTAACCCAAACTGAG | |
| *Tnc* | Forward | TACTGATGGGGCCTTCGAGA | XM_023628745.1 |
| | Reverse | AGCAGCTTCCCAGAATCCAC | |
| *Bgn* | Forward | TGATTGAGAACGGGAGCCTGAG | XM_023633175.1 |
| | Reverse | TTTGGTGATGTTGTTGGTGTGC | |
| *Fbmd* | Forward | GCTTCTGCTGAGGGACAC | NM_001081777.1 |
| | Reverse | GATTTCTGGGGTTGGGAC | |
| *Col14a1* | Forward | CTGGACGATGGAAGTGAG | XM_005613197.3 |
| | Reverse | GTGACCCTGAACTGCTGC | |
| *THBS4* | Forward | ACGTAAACACCCAGACGGAC | XM_023618094.1 |
| | Reverse | CACCAACTCGGAGCCTTCAT | |
| *GAPDH* | Forward | GTGTCCCCACCCCTAACG | NM_001163856.1 |
| | Reverse | AGTGTAGCCCAGGATGCC | |

## 2.2.8. Immunohistochemical Microstructure

The same paraffin blocks prepared for histological microstructural analysis were sectioned

(5 µm), deparaffinized in xylene, and serially rehydrated in decreasing concentrations of ethanol,

followed by incubation in PBST (0.1% Triton X-100 in PBS) at room temperature for 10 minutes.

Antigen-retrieval was performed in antigen retrieval buffer (100 mM Tris, 5% Urea, pH 9.5) at 121 °C for 30 minutes using an autoclave. Specimens were incubated in blocking buffer (1% BSA and 22.52 mg/ml glycine in PBST) at room temperature for 30 minutes.

Sections were stained with rabbit anti-human fibromodulin (PA5-26250: Invitrogen, Waltham, MA) polyclonal antibody at a concentration of 1:100 in incubation buffer (1% BSA in PBST) overnight at 4°C. Sections were washed with PBS at room temperature for 15 minutes each 3 times, then stained with goat anti-rabbit IgG conjugated with Alexa Fluor™ 488 (A11070: Molecular Probes, Eugene, OR) at concentration of 1:200 in incubation buffer for 1 hour at room temperature. Following washing with PBS 3 times, nuclei were counter-stained with 4',6-diamidino-2-phenylindole (DAPI: Thermo Fisher) at a concentration of 10 μM in PBS at room temperature for 10 minutes. Sections were washed with PBS once and mounted with mounting medium (Vectashield® Antifade Mounting Medium: Vector Laboratories, Newark, CA) and cover glass.

Images were obtained at an excitation wavelength of 490 nm and an emission wavelength of 525 nm using confocal microscope (TCS SP8: Leica). Sections stained with only secondary antibody were used as negative control, and sections of equine DDFT were used as positive control.

### 2.2.9. Scanning Electron Microscopy (SEM)

Specimens were fixed in 2% PFA and 1.25% glutaraldehyde in 0.1 M sodium cacodylate (CAC) buffer (pH 7.4) for 1 hour at room temperature, and transferred to buffer (3% glutaraldehyde in 0.1 M CAC buffer, pH 7.4) for 30 minutes. They were rinsed with washing buffer (5% sucrose in 0.1 M CAC buffer, pH 7.4), post-fixative buffer (1% osmium tetroxide in 0.1 M CAC buffer, pH 7.4), and water. Specimens were serially dehydrated, critical point dried, and sputter coated with gold. Digital images were created with a scanning electron microscope and camera at 15 kVp (Quanta 200, FEI Company, Hillsboro, OR).

### 2.2.10. Transmission Electron Microscopy (TEM)

Specimens were collected from constructs cultured in stromal and tenogenic medium both for 21 days. Specimens were fixed in 2% PFA and 2% glutaraldehyde in 0.1 M PBS (pH 7.4) at 4 °C overnight. They were washed 3 times in 0.1 M PBS for 30 min each, and post-fixed in 2% osmium tetroxide ($OsO_4$) in 0.1 M PBS at 4 °C for 3 hours. Specimens were dehydrated in grading ethanol, infiltrated with propylene oxide twice for 30 minutes each, and were placed in a 70:30 mixture of propylene oxide and resin for 1 hour, followed by polymerization in 100% resin at 60 °C for 48 hours.

The polymerized resins were sectioned at 70 nm with a diamond knife using an ultramicrotome (Ultratome Leica EM UC7: Leica), mounted on copper grids, and stained with 2% uranyl acetate at room temperature for 15 minutes. They were stained with lead stain solution at room temperature for 3 minutes. Images were obtained with a transmission electron microscope (JEM-1011, JEOL Ltd., Tokyo, Japan) at an acceleration voltage of 80 kV.

### 2.2.11. Statistical Analysis

Results are presented as mean ± standard error of the mean (SEM). Normality of data was examined with the Kolmogorov–Smirnov test. Outcome measures were compared with ANOVA. When overall difference was detected, pairwise comparisons between groups were performed using Tukey's post-hoc test. Fold changes of tenogenic gene expression of COLI constructs cultured in tenogenic medium normalized to those cultured in stromal medium for identical periods were compared to 1 using a one sample t-test for normally distributed results and Wilcoxon signed rank test for non-normally distributed results. All analyses were conducted using Prism (GraphPad Software Inc., San Diego, CA) with significance considered at $p < 0.05$.

## 2.3. Results

### 2.3.1. Gross Appearance

After each culture medium and period, the COLI construct was harvested from the bioreactor and its gross appearance imaged prior to the specimen collection. In stromal medium, constructs were initially slightly swollen and the majority of the COLI template appeared to be intact without noticeable degradation at day 7 (Fig 2.8.A). By day 14, COLI constructs still appeared to be swollen and the area from the middle to bottom started to degrade with fibrous debris separating from the surface of constructs, forming rough surface. Moreover, there was a fissure developing in some areas of the COLI template (Fig 2.8.B). By day 21, constructs remained swollen, and degradation further progressed especially in the bottom area. This was evident from discoloration and decomposition of construct around suture nots (Fig 2.8.C).

On the contrary in tenogenic medium, constructs started to contract by day 7 as was evident from space between the suture and COLI template, and the smooth surface (Fig 2.8.D). The contraction of constructs further advanced by day 14, and the surface remained smooth throughout all construct areas (Fig 2.8.E). Contraction of constructs were most significant at day 21 with the majority of areas separated from the suture, and the surface of constructs becoming smooth and with a continuous texture rather than the rough and porous original COLI template texture (Fig 2.8.F). Constructs were much firmer than those cultured in stromal medium or the original COLI template. Of note was themmore significant contraction of constructs in the top area. Unrolling constructs was also more difficult after 21 days of culture in tenogenic medium due to the strong adhesion inside constructs. Additionally, the outermost end of the COLI template overlapping the inner area appeared to have merged, forming a nearly complete cylindrical structure by the end of 21 day culture in tenogenic medium.

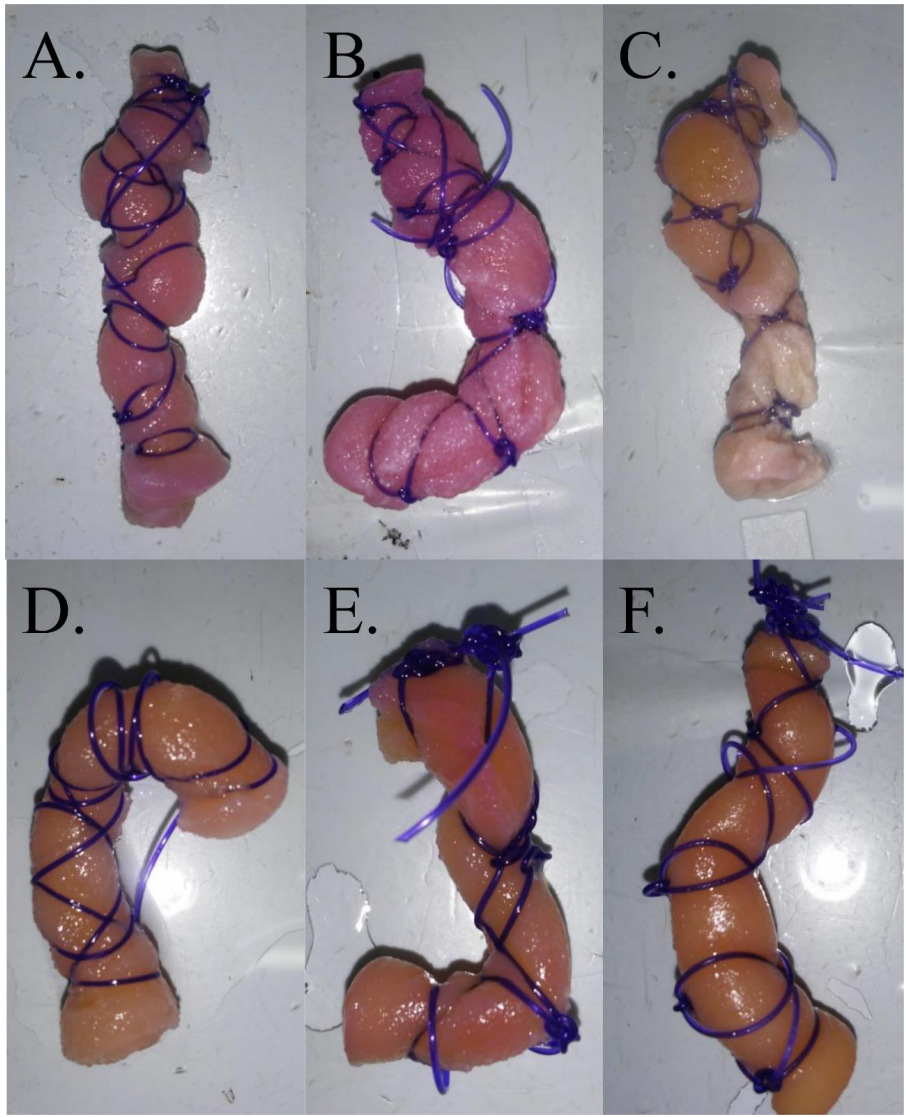

Figure 2.8. Gross appearance of COLI constructs cultured in stromal and tenogenic medium up to 21 days. COLI constructs were cultured in stromal (A – C) and tenogenic (D – F) medium for 7 (A and D), 14 (B and E), and 21 (C and F) days. Gross images were obtained after harvesting constructs from bioreactor and washing with PBS to remove coloration from medium. Coloration of constructs stems from each medium color. The upper area of each image corresponds to the top in the vertical axis of construct during culture.

### 2.3.2. Qualitative Growth Kinetics

Most cells remained viable in both stromal and tenogenic medium throughout the culture period with the majority of cells stained with calcein-AM and a few stained with EthD-1 (Fig 2.9). The majority of cells were spherical in stromal medium for all culture periods and in tenogenic medium at day 7. Cells assumed a spindle-shape and parallel alignment in tenogenic medium at days 14 and 21.

Cell numbers appeared to have been the same in stromal medium throughout the culture period, whereas cells appeared to have proliferated in tenogenic medium from day 7 to 14, and slightly decreased from day 14 to 21.

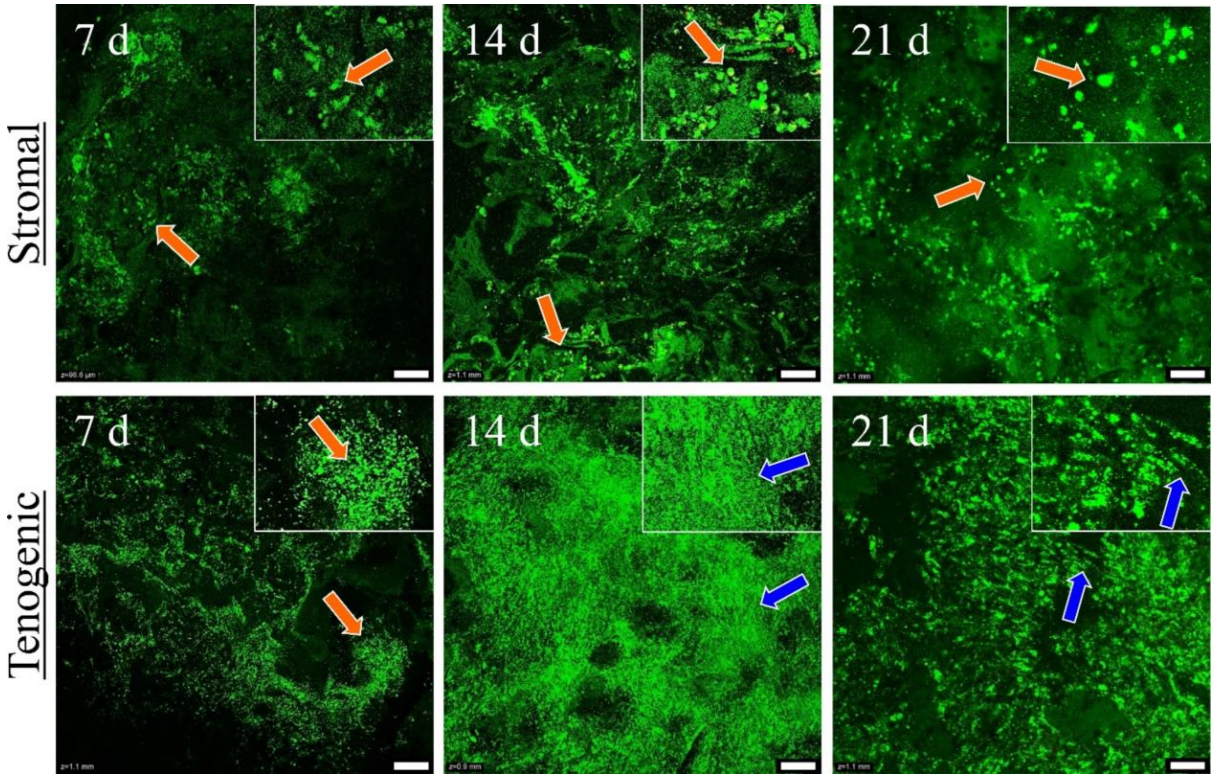

Figure 2.9. Cellular morphology and distribution in COLI templates under static strain. Constructs were cultured in stromal (top row) or tenogenic (bottom row) medium for 7 (left column), 14 (middle column), and 21 (right column) days. Viable cells were stained with calcein-AM (green) and non-viable cells were stained with ethidium homodimer-1 (EthD-1, red). Cells were spherical (orange arrows) or elongated (blue arrows). Each area indicated with arrow was enlarged in an inset of top right corner. Scale bars = 100 μm.

### 2.3.3. Quantitative Growth Kinetics

There appeared to have been more cells in the bottom region of constructs regardless of medium type or culture period. The second most populated region of constructs appeared to be the top, followed by the middle. In fact, there was large disparity between specimens heavily populated and those without cells. Therefore, cellular numbers from each region were combined to represent the entire constructs' cell numbers.

Combining all 3 regions of constructs, cell numbers did not change in stromal medium throughout the culture period, while cell numbers of constructs cultured in tenogenic medium increased from day 7 to 14 and remained unchanged since then (Fig 2.10). Additionally, cell numbers were higher in tenogenic medium compared to those in stromal medium throughout the culture period. Of note was mitigated cellular number disparity among the specimens in tenogenic medium at day 21, represented by a smaller standard error.

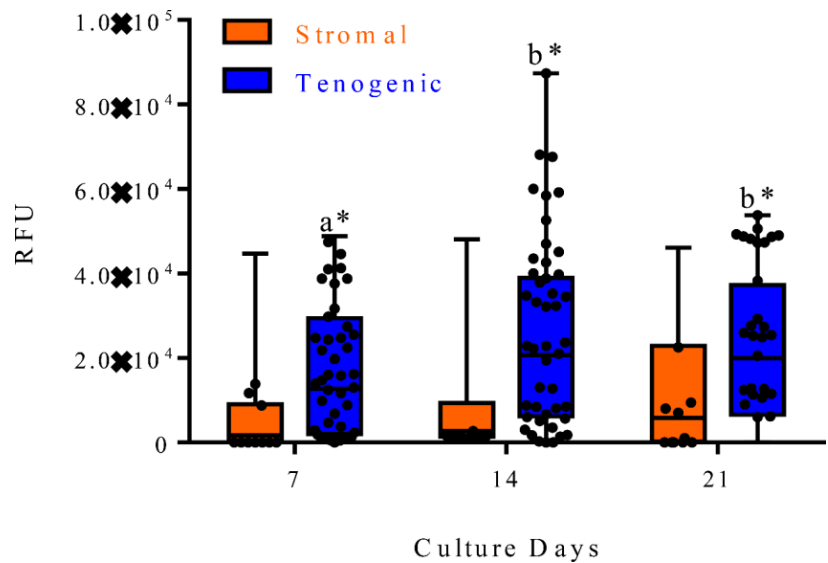

Figure 2.10. Quantitative growth kinetics of cells in constructs under static strain. Constructs were cultured in stromal (orange bars) or tenogenic (blue bars) medium for 7, 14, and 21 days. Fluorescent intensity (RFU) of reduced resorufin by viable cells were measured with microplate reader. Data presented as mean ± SEM. Distinct lowercase letters indicate differences among culture days within each medium type. Asterisks indicate differences between medium types within each culture day.

### 2.3.4. Gene Expression - Reverse Transcription Polymerase Chain Reaction (RT-PCR)

Throughout the culture period, all tenogenic genes tested had a trend of upregulation in COLI constructs cultured in tenogenic medium compared to those cultured in stromal medium.

In particular, tenogenic transcription factors (Fig 2.11A) appeared to have more strongly upregulated at day 7 and gradually returned to the baseline. Among those, *Mkx* downregulated (0.25 ± 0.05-fold) at day 14 and upregulated (2.28 ± 0.46-fold) at day 21 in constructs cultured in tenogenic medium compared to those cultured in stromal medium. Similarly, *Egr1* upregulated (7.31 ± 1.71-fold) at day 7; *CTGF* upregulated both at day 7 (5.48 ± 2.47-fold) and 21 (7.69 ± 2.50-fold); and *LOX* upregulated at day 7 (86.96 ± 78.50 -fold).

The trend of stronger upregulation at an early culture period followed by gradual decrease was also noted in the expressions of tenogenic ECM genes (Fig 2.11.B). *Col1a1* upregulated at day 7 (2781 ± 2627-fold). *Col3a1* upregulated both at day 7 (29.81 ± 20.84-fold) and 14 (26.20 ± 15.94-fold). *Eln* upregulated both at day 7 (303.3 ± 144.4-fold) and 14 (462.9 ± 365.4-fold). *Tnc* upregulated both at day 14 (7.29 ± 2.40-fold) and 21 (5.12 ± 1.35-fold). *Bgn* upregulated at day 7 (8.52 ± 3.03-fold).

For mature tendon markers, although non-significant, there appeared to be a trend of initial upregulation followed by further strengthened upregulation towards the end of the culture period (Fig 2.11.C). *Fbmd* upregulated at day 7 (9.75 ± 4.68-fold). *Col14a1* upregulated at day 7 (20.98 ± 11.05-fold). *THBS4* upregulated at day 14 (58.48 ± 35.51-fold).

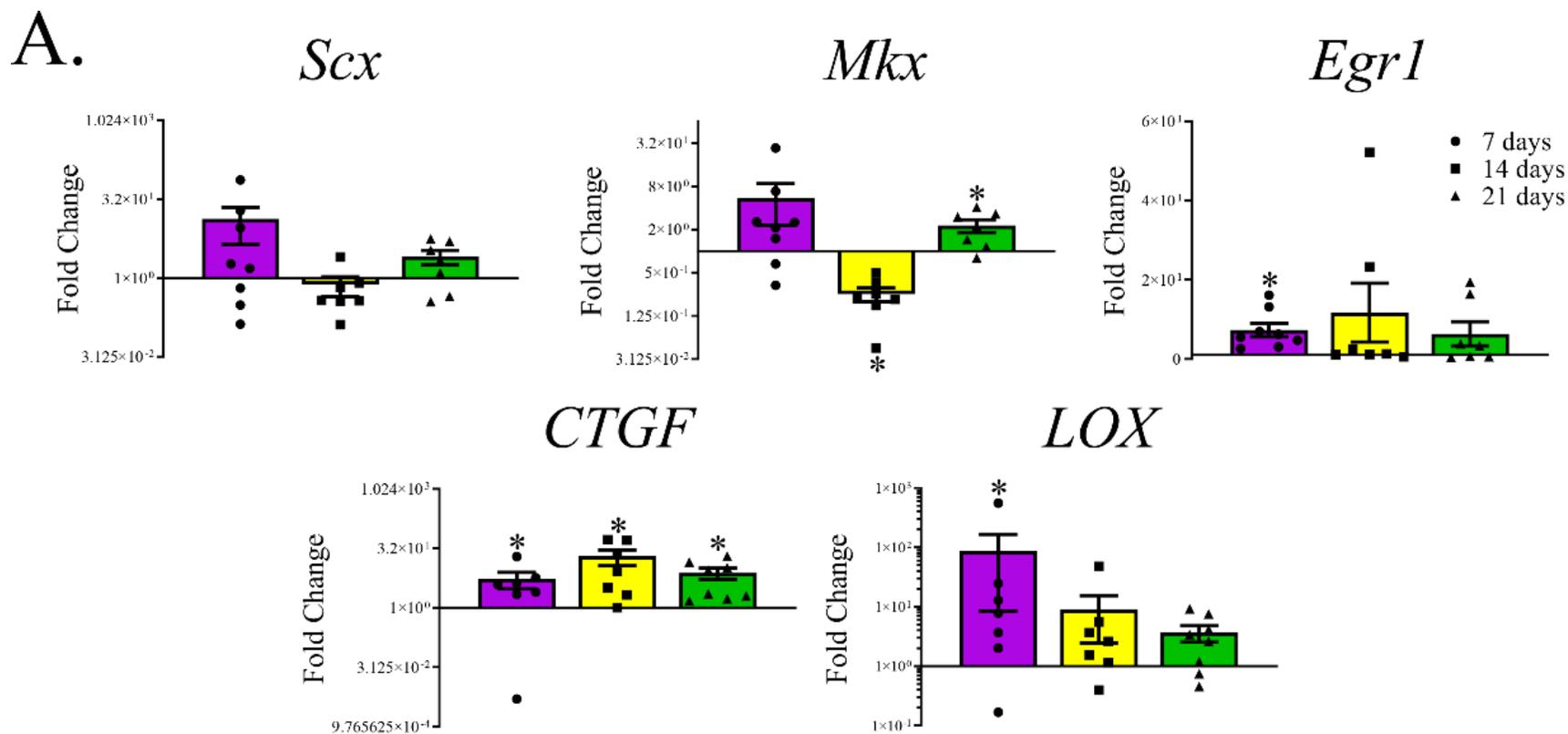

Figure 2.11. Tendon-specific gene expression of cells in constructs cultured under static strain. Constructs were cultured in stromal and tenogenic medium for 7 (purple bars), 14 (yellow bars), and 21 (green bars) days, and gene expression of cells cultured in tenogenic medium was normalized to that of stromal medium and expressed as fold change at each culture period. Expression of tenogenic transcription factors, *Scx*, *Mkx*, *Egr1*, *CTGF*, and *LOX* (A). Expression of tenogenic ECM genes, *Col1a1*, *Col3a1*, *Dcn*, *Eln*, *Tnc*, and *Bgn* (B). Expression of mature tendon markers, *Fbmd*, *Col14a1*, and *THBS4* (C). Data presented as mean ± SEM. Asterisks indicate difference from 1-fold within each culture period. Different lowercase letters indicate difference among culture periods.

(fig. cont'd)

# B.

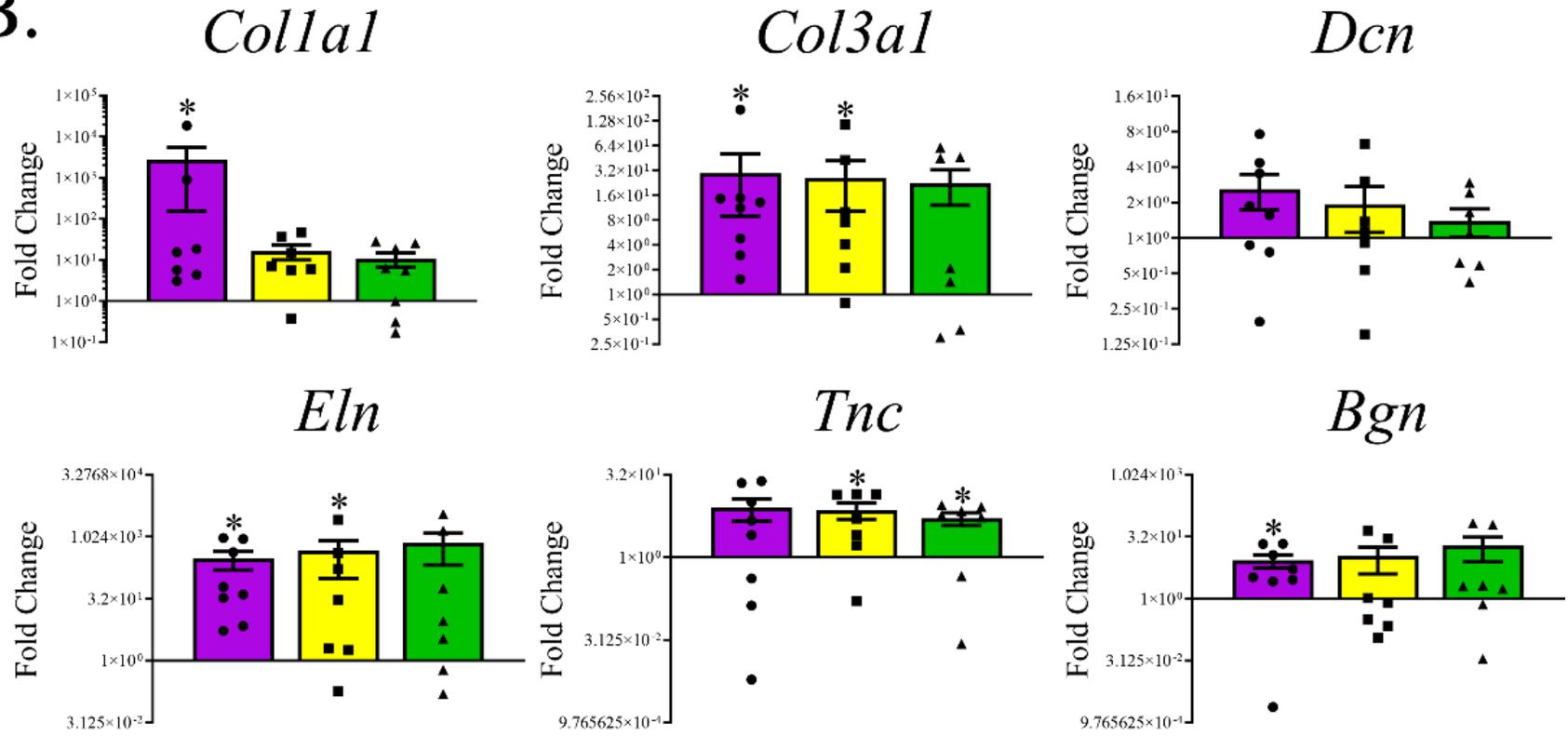

(fig. cont'd)

C.

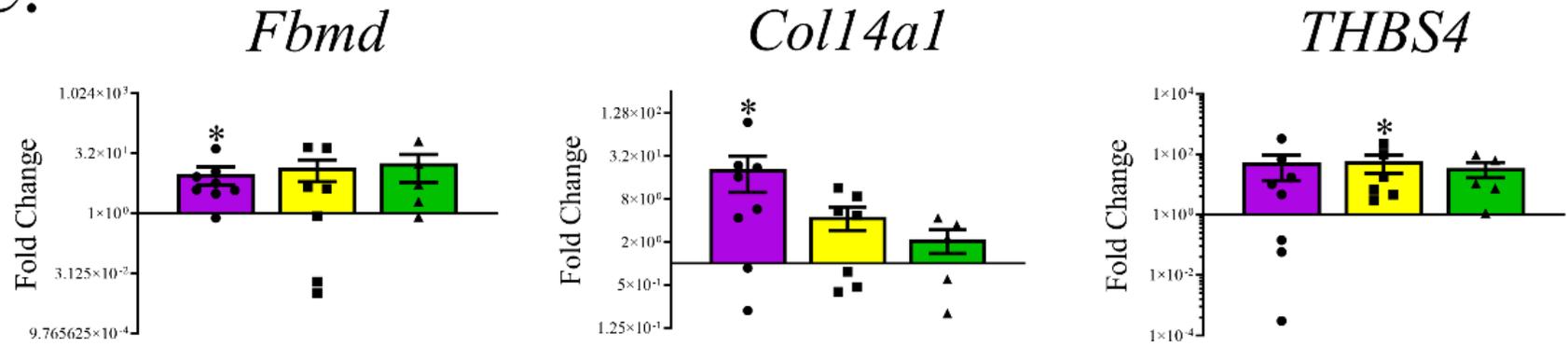

### 2.3.5. Histological Microstructure

Histologically, all cells in both stromal and tenogenic medium throughout the culture period appeared to be viable with nuclei that had good integrity without any signs of fragmentation (Fig 2.12). Additionally, cells appeared to have homogenously stained cytosol without extensive vacuole formation. Similar to the findings of qualitative and quantitative growth kinetics, there was a disparity among specimens that were heavily populated by cells and those that had sparse cellular distribution, especially in constructs cultured in stromal medium.

Morphologically, spherical to rhomboid cells were sparsely distributed throughout the constructs in stromal medium regardless of culture periods. They appeared to loosely attach to the collagen fibers of the template via a limited number of filopodium, or be entrapped by the web of collagen fibers within the template at day 7. Cell morphology and density were largely unchanged in stromal medium for the remaining culture periods. However, cells appeared to have increased the numbers of filopodium they project, and formed mesh-like ECM around themselves to attach to collagen fibers. At the same time, there still were cells loosely attached to or entrapped by collagen fibers. Of note was minimum amount of ECM deposited by the cells in stromal medium throughout the culture periods, preserving porous structure of original COLI template.

In contrast to cells in constructs cultured in stromal medium, those in tenogenic medium were more elongated and assumed a spindle shape by day 7. Cells tended to form clusters that contain numerous cells embedded in a newly deposited fibrous ECM network. Moreover, there appeared to be a larger number of cells within constructs. This characteristic was further progressed by day 14 with larger clusters formed within constructs at higher frequency. The deposition of newly synthesized ECM also progressed, resulting in filling of the original COLI template pores with cells and ECM. The other significant change from day 7 to 14 was increased parallel alignment among cells, many of which had elongated rod-like nuclei. By day 21 in tenogenic medium, most cells appeared to have extensively elongated rod-like nuclei and coalesced with surrounding cells, aligning parallel to each other. The deposition of ECM also advanced and many areas of the COLI template had its pore filled with tissue-like structure. The clusters of cells also tended to form parallel alignment along adjacent collagen fibers.

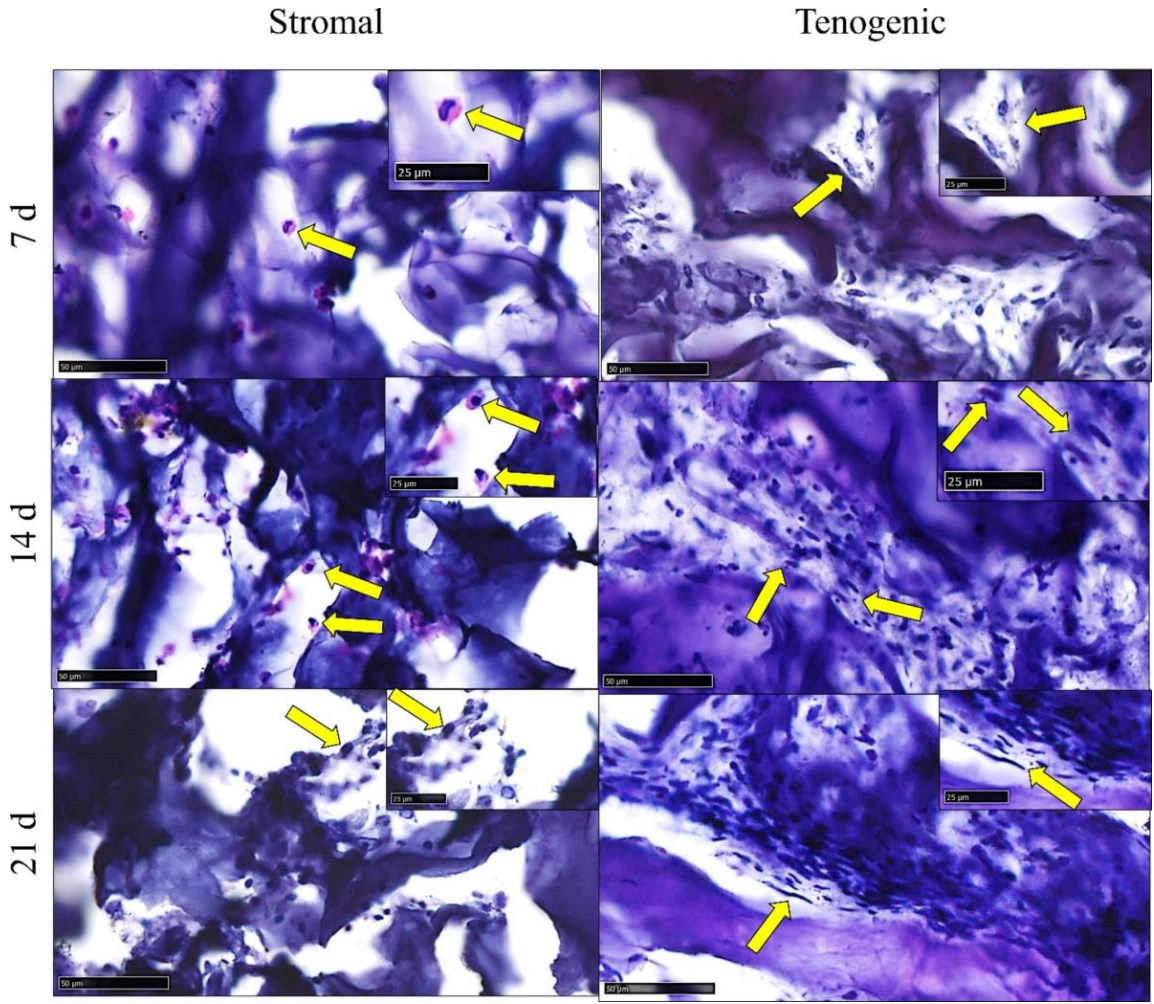

Figure 2.12. Histological microstructure of constructs. Constructs were cultured in stromal (A, C, and E) or tenogenic (B, D, and F) medium for 7 (A and B), 14 (C and D), and 21 (E and F) days. Formalin fixed paraffin embedded constructs were sectioned at 5 μm, followed by H & E staining. Cells within constructs are indicated with yellow arrows. Representative area with cells in each image was enlarged in an inset of top right corner. Scale bars = 50 μm (A – C, and E) and 100 μm (D and F).

### 2.3.6. Immunohistochemical Microstructure

The deposition of tenogenic maturation marker protein fibromodulin appeared to be absent in constructs cultured in stromal medium both at day 7 and 14 of culture (Fig 2.13). After 21 days of culture in stromal medium, however, there appeared to be minimum amounts of fibromodulin deposition throughout the constructs. Of note was fibromodulin distribution that was not consistent with cellular distribution, suggesting fibromodulin was deposited onto COLI template itself rather than the protein existed in the cytosol of cells. Consistent with the other outcome measures, the distribution of cells represented by nuclei stained with DAPI was not homogeneous, and they formed several clusters within constructs.

Similar to constructs cultured in stromal medium at day 7 and 14, those cultured in tenogenic medium for 7 days had no deposition of fibromodulin onto COLI template. On the contrary, after 14 days of culture in tenogenic medium, there was progressively increasing deposition of fibromodulin onto COLI constructs. Fibromodulin deposition was moderate and distributed throughout the constructs regardless of cellular distribution. Cells were non-uniformly distributed within constructs, forming clusters represented by aggregates of nuclei. The deposition of fibromodulin further increased from day 14 to day 21 in tenogenic medium, resulting in extensive amounts of fibromodulin deposited throughout the constructs by day 21. The distribution pattern of fibromodulin appeared to be more abundant in the areas cells did not form clusters than those immediately adjacent to the cell clusters. And there appeared to be a parallelly aligned fibrous structure of fibromodulin deposited areas in the constructs at 21 days, suggesting cells deposited fibromodulin directly onto collagen fibers of the COLI template. Interestingly, the deposition of fibromodulin was even greater in the constructs cultured in tenogenic medium for 21 days than native equine DDFT tissue.

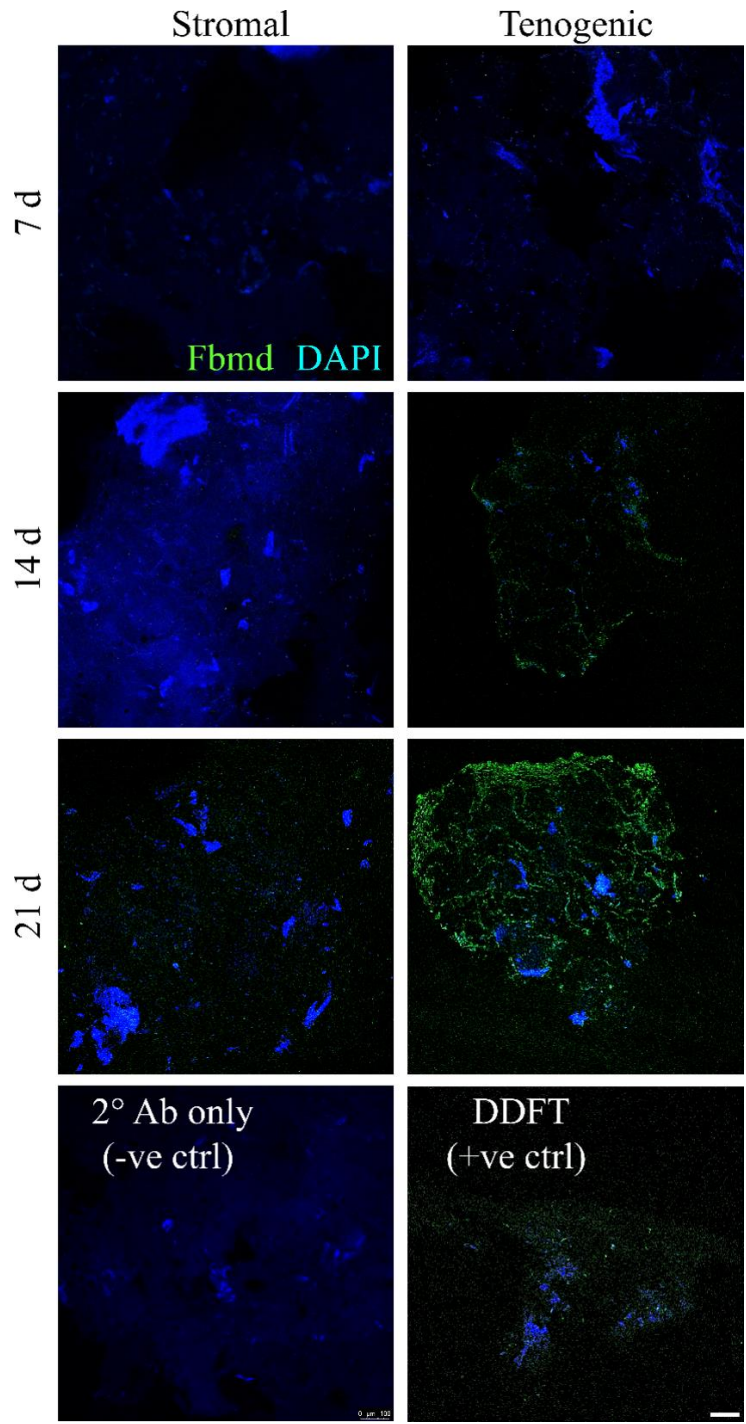

Figure 2.13. Immunohistochemical microstructure of constructs. Constructs were cultured in stromal (left panels) or tenogenic (right panels) medium for 7 (top panels), 14 (second top panels), and 21 (second bottom panels) days. Secondary antibody only section (2° Ab only) was used for negative control (-ve ctrl) and DDFT section was used for positive control (+ve ctrl) both at the bottom panels. Green color represents fibromodulin (Fbmd) staining and blue color represents nucleus (DAPI) staining. Scale bars = 100 μm.

### 2.3.7. Scanning Electron Microscopy (SEM)

Ultrastructural characteristics of cells using SEM indicated cells were spherical to rhomboid within stromal medium cultured constructs throughout the culture period (Fig 2.14.A – C). Additionally, they deposited minimum amounts of ECM onto the COLI template. Consistent with histological microstructure, cells attached to the COLI template with mesh-like ECM deposited around themselves (Fig 2.14.B). On the contrary, cells were much more elongated and assumed a spindle shape in tenogenic medium cultured constructs (Fig 2.14.E and F). The deposition of ECM appeared to be limited until 14 days of culture in tenogenic medium, whereas parallel, large spindle-shaped cells were tightly adhered and surrounded by well organized, fibrous ECM, all attached to the COLI template in tenogenic medium at day 21 (Fig 2.14.F).

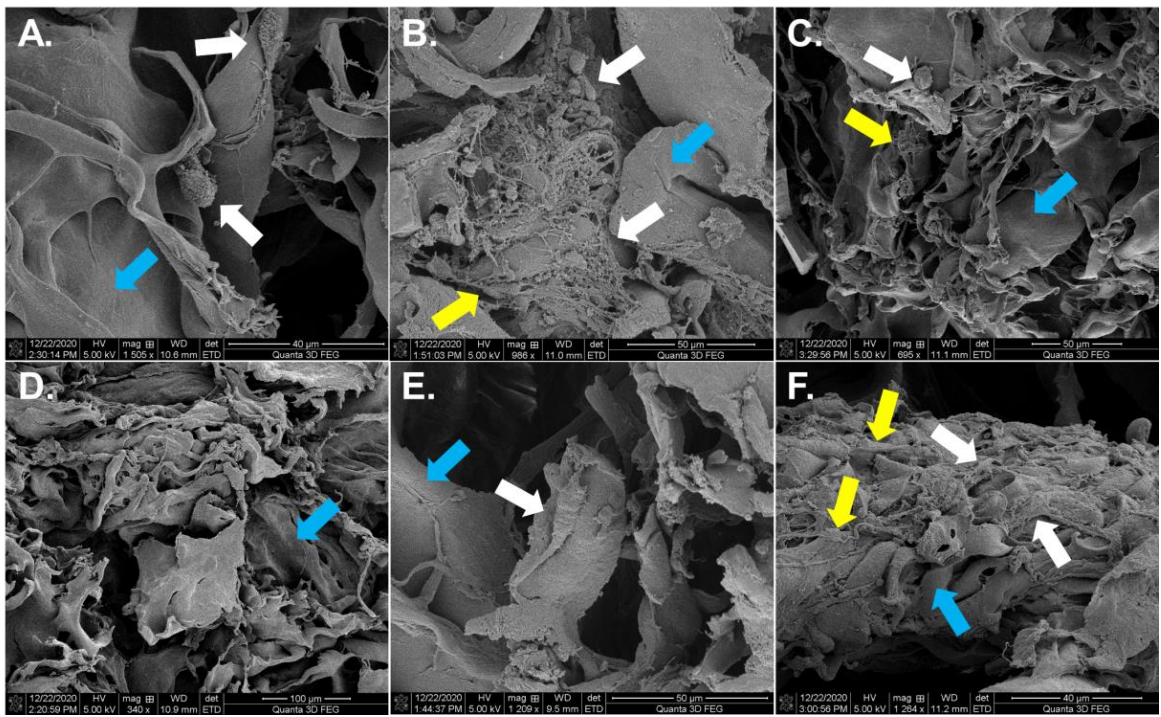

Figure 2.14. Ultrastructure of constructs. Constructs were cultured in stromal (A - C) or tenogenic (D - F) medium for 7 (A and D), 14 (B and E), and 21 (C and F) days and images obtained with SEM. Cells (white arrows) were found surrounded by newly deposited ECM (yellow arrows) on collagen fibers of original template (blue arrows). Scale bars = 40 μm (A and F), 50 μm (B, C, and E), 100 μm (D).

### 2.3.8. Transmission Electron Microscopy (TEM)

Ultrastructural characteristics imaged with TEM indicated cells were spherical and loosely attached to collagen fibers of the COLI template in stromal medium cultured constructs (Fig 2.15.A). Cells maintained integrity of both the plasma nucleus membranes, suggesting they were viable. They were rich in mitochondria and golgi apparatus-like structures. Nucleus to cytosol ratio was relatively high. Lack of ECM-like structures around the cells indicated cells did not actively deposit ECM on the existing original COLI template fibers.

In tenogenic medium cultured constructs, cells assumed more elongated morphology represented by the elongated rod-like shape of the nuclei (Fig 2.15.B). Nucleus membranes appeared to have maintained integrity, suggesting viability of cells. The most significant difference of cells in tenogenic medium cultured constructs compared to those in stromal medium cultured constructs was the presence of a collagen fiber like structure present in the cytosol of the cells (Fig 2.15.B. yellow arrows). The presence of collagenous fibers within cells was a tenocytic characteristic. The areas of heterochromatin was more abundant than in stromal medium, indicating transcription activity was relatively less active in cells from tenogenic medium cultured constructs.

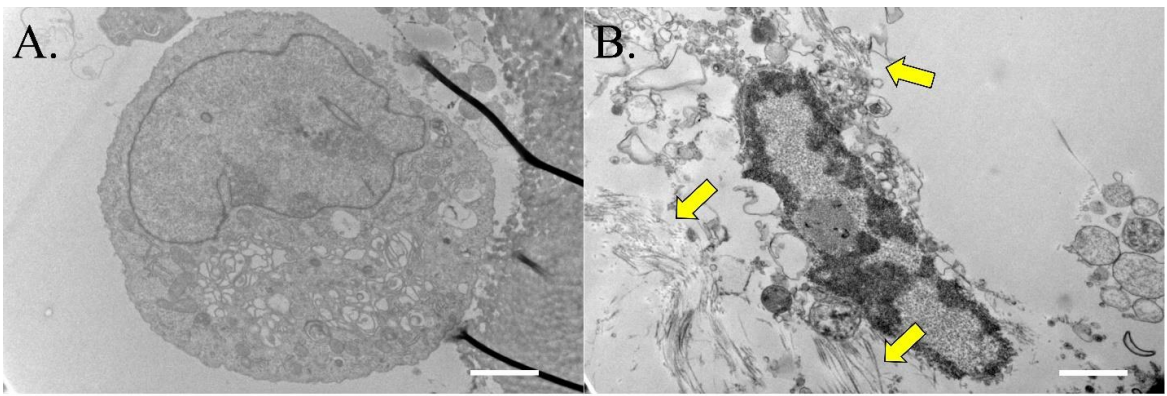

Figure 2.15. Ultrastructure of constructs. Constructs were cultured in stromal (A) or tenogenic (B) medium for 21 days and images obtained with TEM. Cells contained collagenous fibers (yellow arrows) within cells. Scale bars = 2 μm (A) and 1 μm (B).

### 2.3.9. Dynamic Strain Culture

Both qualitative and quantitative growth kinetics revealed significantly diminished cellularity within constructs, which resulted in a non-detectable level of cellular number by resazurin reduction. Microstructurally, cells were round and formed clusters with surrounding granulofilamentous ECM in tenogenic medium at day 21 (Fig 2.16). Compared to the constructs cultured under static strain, cell cluster numbers were much lower, indicating interruption of cellular attachment to the COLI template. Ultrastructurally, spherical cells surrounded by abundant granulofilamentous ECM formed clusters in tenogenic medium at day 21 (Fig 2.17). Although cluster numbers were fewer than those cultured under static strain, cellular density within clusters was high and deposition of ECM appeared to be abundant, entrapping cells between pillars of COLI template fibers.

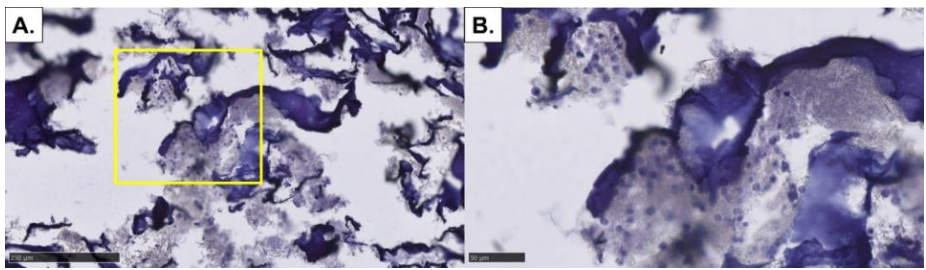

Figure 2.16. Histological microstructure of constructs. Constructs were cultured in tenogenic for 21 days under dynamic strain. Sections were stained with H & E. Yellow inset in A is enlarged in B. Scale bars = 250 µm (A) and 50 µm (B).

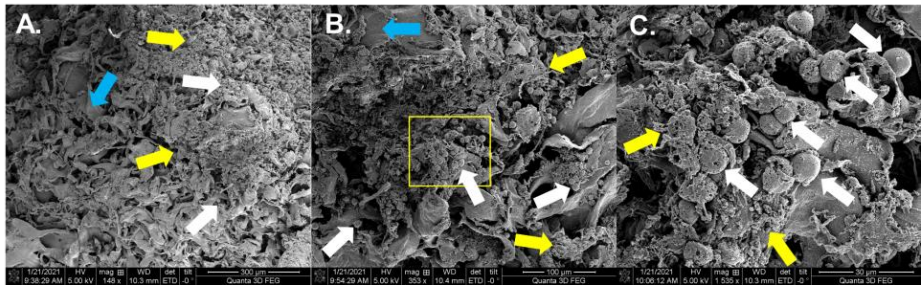

Figure 2.17. Ultrastructure of constructs. Constructs were cultured in tenogenic medium for 21 days under dynamic strain. Cells (white arrows) were found surrounded by newly deposited ECM (yellow arrows) on collagen fibers of original template (blue arrows). Yellow inset in A is enlarged in B. Scale bars = 300 µm (A), 100 µm (B), and 50 µm (C).

## 2.4. Discussion

### 2.4.1. Main Findings

The major findings of this study are: 1) equine ASCs remained viable during the 21 day culture period both in stromal and tengenic media with static strain; 2) equine ASCs proliferated from day 7 to 14 in tenogenic medium, but did not proliferate in stromal medium throughout the culture period with static strain; 3) equine ASCs assumed a tenocyte-like morphology and organized into neotendon-like tissue on collagen templates by depositing newly synthesized ECM and embedding themselves in tenogenic medium, but remained spherical morphology and distributed sparsely and loosely attaching to collagen templates that contain scant ECM deposited by cells in stromal medium with static strain; 4) the morphological transformation of collagen templates to neotendons in tenogenic medium with static strain corroborated with earlier upregulation of tenogenic transcription factors and later upregulation of tenogenic maturation markers; 5) the maturation of neotendons during 21 day culture period in tenogenic medium with static strain coincided with progressive accumulation of tenomodulin expression on constructs; 6) equine ASCs formed clusters of cells aligned parallel to collagen fibers and cells produced collagn fibrils around themselves after 21 days of culture in tenogenic medium with static strain, all of which were ultrastructurally evident; and 7) application of dynamic strain during 21 day culture in tenogenic medium decreased cellular number that attached to templates, maintained spherical morphology of cells that were loosely attached to templates, and cells deposited fibrous ECM around themselves. Collectively, our study supports equine ASCs cultured on collagen templates in tenogenic medium with static strain form premature neotendon that recapitulate embryonic tendon development *in vitro*.

## 2.4.2. Growth Kinetics

To date, there have been numerous attempts to culture MSCs onto collagenous scaffold for differentiation into tenocytes. And the most commonly used collageneous scaffold for this purpose is hydrogel. Cells or cellular aggregates can be embedded in hydrogels during solidification, after which viable cells embedded in hydrogels and assume elongated morphology, especially in tenogenic medium.[131,132] Although there have been reports of cells assuming spindle morphology in stromal medium and maintaining viability in collagenous hydrogels, often cells fail to assume tenocyte-like morphology in stromal medium. For example, it was reported both equine MSCs and tendon-derived MSCs retained non-spindle morphology and instead assembled into cell clusters in a collagen/synthetic peptide hydrogel in stromal medium culture,[133] which was consistent with our constructs cultured in stromal medium. This evidence indicates tenogenic medium culture is an essential component of neotendon formulation. The novel and significant advantage of the culture system developed in this study compared to existing neotendon engineering methods is the robust proliferation potential of cells on constructs in tenogenic medium. To date, there have been no been reports on MSCs' maintaining robust proliferative properties in 3D cultures, especially on collagenous scaffolds regardless of medium type. Current consensus on poor proliferative properties of MSCs in 3D cultures was consistent only with cells cultured in stromal medium in the present study. Additionally, robust cellular proliferation in tenogenic medium contradicted anti-prolifeative properties of tenogenic medium. For example, MSCs adhered to decellularized tendon scaffolds,[134] yet cells did not proliferate during the culture period. And these anti-proliferative properties of tenogenic medium were attributed to the effects of TGF-β.[135] One of the reasons for robust proliferative capacity of cells in tenogenic medium observed in our study may be highly expressed *CTGF* gene consistently upregulated throughout the culture period. In a previous study, mouse ASCs cultured with CTGF not only upregulated tenognic genes but also increased proliferation in a dose-dependent mannar.[136]

As opposed to hydrogel embedding of MSCs, culturing of MSCs onto collagen sponge similar to that used in this study has shown limited success. The reason for low cellular adherance and uniform distribution within sponge was reported to be the closure of pores upon hydration of the sponge, which prevented migration of cells into the scaffold and exchange of both oxygen and nutrients inside the construct. To mitigate this disadvantage, reinforcement of mechanical properties of the sponge with synthetic polymer and application of a stirring culture method have been demonstrated to be effective.[137] In particular, a stirring culture method was shown to increase proliferation of MSCs as well at as low as 50 rpm in poly-ethylene terephthalate (PET) reinforced collagen sponge but not in non-PET reinforced sponge.[137] This was consistent with our findings of non-proliferative cells in stromal medium. Our findings, however, were unique in achieving cellular proliferation in tenogenic medium without reinforcement of the scaffold with synthetic polymers. The reason for highly proliferative cells within our neotendon might stem from high shear stress produced by a higher stirring rate employed in our culture condition. Normally, the higher stirring rate leads to more degradation of collagen sponge during culture, and it was demonstrated collagen sponge lost as much as 50% of its size and the surface of the sponge became rough after 7 days of stirring the culture at 150 rpm without synthetic polymer reinforcement,[137] consistent with constructs cultured in stromal medium in our study. In contrast, our neotendon cultured in tenogenic medium contracted in size and its surface became smoother overtime, and it did not show apparent degradation despite stirring the culture at 300 rpm for 21 days. The reason for this was potentially the reinforcement of the collagen sponge with abundant ECM progressively deposited by tenogenically-induced

ASCs. This was consistent with previous reports on improved differentiation capacity in ESCs by medium flow,[138] and improved chondrogenesis by dynamic culture.[139] Our findings are one of the novel applications of high shear stress to cells to promote tenogenic differentiation. And it was consistent with previous report in that shear stresses induced by collagen fiber sliding during mechanical loading was adequate to maintain tendon mechanical properties via activation of mechanosensitive ion channel PIEZO1.[140]

### 2.4.3. Gene Expression

In terms of tenogenic transcription factors' expression patterns, the expression of *Scx* had a trend of upregulation at the early culture period followed by a gradual return to the baseline. This was consistent with the fact that *Scx* is an essential transcription factor for initiating tenogenic differentiation. Its effect in tendon healing is to directe progenitor cells into a tenogenic lineage by preventing progenitor cells from differentiating into chondrogenic cells under the influence of TGF-β.[141] Indeed, *Scx* depletion had a negative effect on tendon healing at day 28 post-injury but not at day 56 in mice.[142] A similar trend was also evident in the expression of *Mkx* with early upregulation followed by a gradual decline. *Mkx* is also another important tenogenic transcription factor, and its function was reported as preventing progenitor cells from differentiating into myofibroblasts at the onset of the healing process and resulting angiofibrosis.[143] However, in the present study, the *Mkx* remained upregulated in tenogenic medium compared to stromal medium at day 21. Therefore, it was likely that cells maintained certain levels of active differentiation to tenocytes throughout the culture period. With regard to the expression of *Egr1*, upregulation was statistically significant only at day 7 in the present study. The expression pattern corroborated that of healing tendon in a rabbit model, demonstrating the highest expression of *Egr1* at the injury site after 7 days post-injury, followed by decline to the original level by day 28.[144] In contrast to other transcription factors evaluated in this study, the expression of *CTGF* was significantly higher in tenogenic medium than stromal medium throughout the culture period consistently without clear distinction between early and late periods. This may be explained partly by *CTGF*'s funtion as an essential molecule to prevent aging of tenogenic progenitors.[145] It was also noted CTGF is among the highest expressed constitutive tenogenic transcription factors.[146] However, *CTGF* had been upregulated only at day 3 post-injury in the chicken compared to the baseline.[146] Therefore, it is possible our tenogenic differentiation condition recapitulates the embryonic development of tendon tissue more closely than the healing process. In terms of early upregulation followed by gradual decline, *LOX* had the most apparent trend, although the expression levels were higher in tenogenic medium compared to stromal medium both at 7 and 21 days of culture. LOX is an enzyme required for cross-linking collagen fibils in tendons and plays an essential role in translating embryonic movements into development of mechanical properties during gestation.[147] And inhibition of LOX during embryonic development was demonstrated to severely impair collagen crosslinking and mechanical properties of chicken tendon.[148] Therefore, the higher expression of *LOX* followed by its decline suggests the neotendon constructed in our study is again likely to recapitulate embryonic tendon development. And it had been reported that application of ASCs to transected calcaneal tendon of rat upregulated *LOX* than non-treated or ASCs combined with growth differentiation factor (GDF) - 5 at day 14 post-injury.[149] In that study, the application of ASCs was most effective in restoring mechanical properties of injured tendon, suggesting the upregulation of *LOX* is an important indicator of healed tendon functionality.

One of the ECM producing genes, *Col1a1*, a subunit of collagen type I, had an expression pattern that followed a similar trend with the transcription factors. In equine tendons, *Col1a2*, another subunit of collagen type I, expressed constitutively in both adult healthy tendons and acutely injured tendons, whereas expression was lower in tendons with chronic tendinopathy.[150] Therefore, cells cultured in tenogenic medium in this study were likely in a healthy state or actively healing state rather than a dormant state of chronic tendinopathy. As opposed to *Col1a1*, the expression pattern of *Col3a1* was relatively stable and upregulated as compared to those in stromal medium throughout the culture period with consistent upregulation. Regardless of expression patterns, both genes are essential in maintaining mechanical properties of tendons. The expression of both genes can be abolished by lack of mechanical stimulus, as demonstrated in a study with dogs whose digital flexor tendon was unloaded by limb suspension and all collagen type I, II, and III expressions were severely impaired by day 42.[151] At the same time, *Col1a1* and *Col3a1* were shown to upregulate in the mouse digital flexor tendon upon transection up to day 21 by gradual increase, yet expression of both were negligible by day 28.[152] So that early upregulation of *Col1a1* at day 7 observed in our study may be similar to the early response to injury. As to *Tnc*, it is normally expressed at a negligible level in tendon, and upregulates upon injury.[153] Indeed, *Tnc* is most likely to be the essential gene for tendon development and healing in horses as well. By comparing expression levels of *Tnc* among equine fetal, yearling, and adult tendons, the expression increased from fetal to yearling and returned to fetal level at adult.[150] Similarly, its expression was high in acutely injured tendons, while it was low in healthy tendons or those affected with chronic tendinopathy. Combined, expression patterns of ECM producing genes also suggests neotendon cultured in tenogenic medium recapitulates maturing embryonic tendon or early healing tendon.

Tenogenic maturation marker, *Fbmd* was upregulated throughout the culture period in tenogenic medium in the present study. It is a highly and constitutively expressed gene in mature healthy tendons. It transiently decreases expression upon injury and gradually upregulates over the healing period.[153] On the other hand, it has a monotonic increase in expression throughout gestation and up to 14 days postnatallyin mice patellar tendon,[154] suggesting its crucial role during embryonic development of tendons. In contrast, *Col14a1* had a clearer trend of early strong upregulation followed by a gradual return to baseline. This was in agreement with normal development of the flexor digitorum longus tendon in the mouse that showed strong expression of *Col14a1* both in gene and protein levels until postnatal 10 days but diminished by postnatal 30 days.[155] The functional importance of *Col14a1* at an early stage of tendon development was evident from the fact that knockout of *Col14a1* negatively affected mechanical strength of the mouse tendon at an early stage but not at a late stage. Additionally, it was reported that another tenogenic construct made of hydrogel and tenocytes showed higher expression of *Col14a1* at day 7 of culture and later downregulated by day 45. Therefore, it is likely *Col14a1* is an early regulatory marker for neotendon maturation.[156] In the present study, *THBS4* was highly upregulated consistently in constructs cultured in tenogenic medium, although statistically significant difference when compared to stromal medium was achieved only at day 7 of culture in tenogenic medium. *THBS4* was reported to be a tendon ECM composition modifying protein, and knockout mice demonstrated compromized limb strength due to weakened muscle.[157] In the ovine calcaneal tendon, *THBS4* is also highly expressed in an earlier gestation period and gradually decreases its expression leading to low expression in adult tendon.[158] Relatively high expression of *THBS4* in the present study suggests again that our neotendon may more closely resemble developing tendons.

### 2.4.4. Microstructure

Microstructures of constructs revealed progressive maturation of neotendon in tenogenic medium over the 21 day culture period, presented by an increasingly elongated morphology of cells that aligned in parallel with surrounding cells and abundant ECMs deposited around the clusters of cells. The findings were unique in that cells appeared to have proliferated and formed premature tendon-like tissue using collagen type I sponge *in vitro*. There have been reports on MSCs cultured on collagen type I sponge to treat tendon injury,[95,159,160] all of which led to beneficial effects of constructs to improve healing. However, microstructural evaluation demonstrating progressive cellular proliferation and maturation in the scaffold has not been reported. A report using similar methods to culture constructs made of collagen type I sponge with ASCs cultured in tenogenic medium demonstrated progressive deposition of ECM onto the scaffold without numerous cells.[126] Although cells did not demonstrate robust proliferation within constructs, numerous MSCs' colonization with parallel alignment along the direction of mechanical stimulation evident from histological evaluation of construct was reported for hydrogel-based neotendon that embedded MSCs.[161] Similarly, numerous cell colonies within a collagen-based scaffold were achieved by formulating cartilagenous constructs made of human tendon-derived progenitor cells and culturing them on collagen type II sponge cultured in chondrogenic medium for 21 days.[162] In that study, numerous spindle-shape cells that were homogenously distributed were observed in constructs. The discrepancy between numerous studies using collagen type I or type II sponges reporting conflicting findings with regard to cellular proliferation when MSCs were seeded may stem from simply the structural difference between properties of collagen type I and II. For example, it was shown that higher ratio of type I to type II collagen in hydrogel led to lower void space in comparison to a 1:1 ratio, although a higher percentage of type I increased superior mechanical properties.[163] Another potential reason for the numerous cell colonies observed in the study using collagen type II sponge may be the additional cross-linking process of type II collagen sponge needed to reinforce mechanical properties. Collectively, our neotendon formulation protocol presents a novel method to construct highly cellular neotissue with organized structure that has not been attained for collagen sponge-based constructs. Our findings on microstructural transformation over 21 days not only demonstrate detailed morphologies of cells within neotendon, but also provide a novel insight on morphological transformation during progressive maturation of neotendon.

The progressive increase of fibromodulin staining in tenogenic medium-cultured constructs during the culture period presented possible maturation of tenoblasts at an earlier phase as compared to tenocytes at later stages. Fibromodulin is collagen-binding leucine-rich proteoglycan, and functions to stabilize small-diameter collagen fibril-intermediates.[164] Deficiency in fibromodulin and its same family proteoglycan lumican in mice led to misaligned knee patella, severe knee dysmorphogenesis, and extreme tendon weakness, characteristic features of Ehlers-Danlos syndrome.[165] On the other hand, enhanced expression of fibromodulin via liposome-based transfection or delivery of protein in the form of hydrogel led to increased biomechanical properties in injured rat calcaneal tendon and improved gait performance[166],[167] indicating its essential role in the healing process. Therefore, implantation of constructs that contain abundant fibromodulin, such as those cultured in tenogenic medium in the present study, by itself may be therapeutic. In our study, fibromodulin was present in a large amount at 21 days of culture in tenogenic medium after a gradual increase, while the expression of tenomodulin in the healthy DDFT, the positive control, was relatively low. Although fibromodulin is generally considered a maturation marker, the expression level is not monotonically increasing with age, and the trend varies depending on the area of the tendon. For example, its expression monotonically increases during gestation, reaching a peak 6 days postnatally and start declining at 21 days postnatally in mice tendon.[168] To date, there has been no report on tenomodulin expression in the equine DDFT. However, our findings were consistent with the expression pattern of the mouse neonate tendon, as highly proliferative neotendon at 21 days of culture in tenogenic medium had a much more abundant presence of tenomodulin than healthy DDFT from a mature horse. Collectively, these indicate our neotendon closely resembles developing tendon before fully maturing, and may augment tendon healing partly by the mechanism of fibromodulin delivery.

### 2.4.5. Ultrastructure

Ultrastructure of constructs further confirmed the progressive maturation of neotendon cultured in tenogenic medium. The celluarity was low and cells were spherical in constructs cultured in stromal medium. Although collagen sponge contains both fibrous and sheet-like structures that often mimic cellular morphology,[169] the presence of cells loosely attaching to collagen fibrils was clearly demonstrated. In tenogenic medium, however, cells assumed a spindle-shape and aligned in parallel with surrounding cells, especially after 21 days of culture. The spindle-shape of cells was similar to that of *Scx*-overexpressing human ESCs-derived MSCs (hESC-MSCs) on silk-collagen scaffold, although cells did not obtain parallel alignment in that study.[95] In another study using silk-collagen scaffold with cultured hESC-MSCs, highly elongated cells with parallel alignment in the direction of dynamic mechanical stimulation along with highly upregulated *Scx* was achieved despite them being cultured in stromal medium,[170] demonstrating that dynamic strain is an effective method to promote tenogenesis. In our study, there was parallel alignment of cells on collagen fibrils, and the number of cells found in the construct cultured in tenogenic medium after 21 days was much higher and they formed dense cell clusters without dynamic tensioning. It is possible the higher shear stress of fluid flow in the swirling motion created an optimum mechanical stimulus to orient cells in parallel alignment in the tenogenic medium without the use of a dynamic tensioning apparatus. Further, TEM images show the collagen fibrils within and around cells in constructs cultured in tenogenic medium, which was charcteristic to tenocytes.[171]

## 2.4.6. Dynamic Strain Culture

Cyclic tensioning applied to constructs in our study resulted in severely reduced cellularity. To date, it is widely accepted that cyclic tensioning improves the viability of cells within the scaffold and promotes tenogenic differentiation under tenogenic conditions,[126] which in turn results in improved healing of injured tendon.[160] The low cellularity of construct cultured under cyclic tensioning may be due to lower attachment of cells to collagen fibrils evident from the spherical morphology of cells observed both microstructurally and ultrastructurally. The reason for the lower attachment of cells to fibrils can be exceeding strain duration and length. For example, 2% strain for 30 minuts per day at 0.5 cycles per minute frequency was adequate to promote cellular proliferation and expression of tenogenic genes as well as ECM deposition in MSCs cultured on decellularized human umbilical vein using tenocytes extracts.[172] Moreover, exceeding strain length was reported to interfere with cellular attachment when applied also statically. In that study, 20% static strain applied to a collagen-based scaffold cultured with MSCs resulted in fewer cells compared to 0% strain, while 20% strain enhanced cell alignment in the direction of strain on the scaffold with opposing gradients of mineral content and structural alignment.[173] The spherical morphology of cells on the collagen sponge network was similar to that of embryonic neural stem cells (NSCs) cultured on collagen-glycosaminoglycan sponge ultrastructurally. Although the cell type and scaffold were different from those used in our study,

NSCs-collagen construct improved locomotion of mice affected by spinal cord injury.[174] Therefore, it is possible, despite the lower cellularity and weak attachment to the scaffold, that constructs cultured under dynamic tensioning may augment healing of tendon injuries.

## 2.5. Conclusion

In conclusion, the results of this study support *de novo* tendon neotissue formation with adult equine ASCs on a collagen template by culturing in tenogenic medium under high shear stress. Distinct differences in neotissue characteristics with static versus dynamic strain confirms the value of the *in vitro* study environment to evaluate neotendon formation to guide therapeutic interventions. In the future, implantation of proliferation phase neotissue may promote tendon healing in equines and beyond.

# Chapter 3. Healing Capacity of Implantable Collagen Constructs for Equine Tendon Regeneration in an Elongation-induced Rat calcaneal Tendinopathy Model

## 3.1. Introduction

Musculoskeletal injury is responsible for up to 72% of lost training days and 14% of retirements by equine athletes,[12,14] and tendon injuries account for the large majority. Recovery from tendon injuries is prolonged due to a slow healing response attributable to low tendon tissue cellularity, metabolic activity and blood supply[109] that contribute to permanent scar formation and high re-injury rates.[21,110] Research shows that neonatal tendon heals without scar tissue by a mechanism recently identified in a murine model.[111] Specifically, following tendon injury in neonates, a transient fibrous patch (or scar) is formed by extrinsic fibroblasts. Subsequently, tenocytes migrate to and replace the fibrous tissue with neotendon. In adults, tenocytes in undamaged tissue are activated but don't migrate to the injury, so persisting fibroblasts form a permanent scar.[111]

To overcome this recognized challenge, adult multipotent stromal cells (MSCs) have been directly injected into naturally-occurring and experimentally-induced tendon injuries with mixed and unpredictable results.[175] The variability has been attributed to an engraftment efficiency of less than 0.001%.[175] Additionally, injection of undifferentiated MSCs into tendon can have serious side effects including chondroid and ectopic bone formation at the injection site.[34] To improve retention and engraftment, collagen type I (COLI), an FDA-approved, biocompatible, and biodegradable biomaterial,[114] has been used as template material for MSCs to grow in cardiac neotissue formation.[176] It serves as a cellular delivery/retention platform that is compatible with injectable therapies. Additionally, tenogenically differentiated adipose-derived MSCs (ASCs) obviate abnormal tissue formation from administration of undifferentiated MSCs. By combining these strategies, we have successfully generated tendon neotissue from ASCs that are tenogenically differentiated and grown on COLI scaffold templates to circumvent these limitations.

Neotissue generated by MSCs only or combined with COLI templates requires testing in an appropriate rodent tendinopathy model prior to preclinical equine testing. To date, the treatment effects of tendon neotissue has been most commonly investigated in rodent acute tendon injury models created by surgical resection of portion of tendon. They were treated with MSCs alone,[177] COLI scaffolds populated with MSCs,[126] and tendon stem/progenitor cell (TSC) sheets,[178] all of which resulted in improved mechanical strength and histological scores. However, caution needs to be taken with interpretation of these outcomes, since rodents possess stronger tendon regenerative potential[179] and surgically created tendon defect models does not simulate clinical scenario characterized by chronic tendinopathy and subsequent rupture. Moreover, surgical implantation of neotissue at the time of injury is not a feasible approach for clinical application. More realistically, the optimum time to deliver exogenous cells to an injured tendon or ligament in horses is around the sixth day after injury at the transition between the inflammatory and subacute reparative phases.[180]

In this regard, tendinopathy models that resemble clinical tendinopathy scenario both etiologically and pathologically, as well as administration strategies that are clinically feasible are essential to test efficacy and therapeutic value of novel treatment modalities. One of the most commonly used tendon for tendinopathy model creation is rat calcaneal tendon composed of insertions of the plantaris, gastrocnemius and soleus muscles due to its accessibility for controlled injury and subsequent treatment administration.[181] The types of tendinopathy creation mechanisms range from chemical injury induction with collagenase, cytokines, prostaglandins and fluoroquinolone, or mechanical injury induction with electrical muscle stimulation and downhill/uphill running have been used across multiple animal species.[182] Among these, downhill/uphill running can most closely recapitulate the overuse nature of tendinopathy which account for majority of naturally-occurring tendon injuries in equine athletes. And rats underwent strenuous uphill treadmill running for 12 weeks were shown to have developed tendinopathy resembling naturally occurring human tendinopathy characterized by decreased collagen fiber organization, increased cellularity with endothelial cells and fibroblasts, while causing minimum inflammatory response.[183]

However, strain injuries induced by these methods often can be inconsistent and subject to inherent individual variation. It is believed that mechanism of overuse tendinopathy is excessive loading and tensile strain beyond 4% of its length where the collagen fibers start to slide past one another as the intermolecular cross-links fail, and, at approximately 8% of elongation, a macroscopic rupture occurs because of tensile failure of the fibers and interfibrillar shear failure.[184] Indeed, the elongation of superficial digital flexor tendon (SDFT) is a sign of tendinopathy presented as medial or lateral displacement,[185] or sometimes as bowed tendon in horses.[186] As such, surgical elongation of tendon using a tool devised to create strain injuries in canine ligaments[187] is an appealing tendinopathy model creation modality, as it can consistently control strain level to minimize variations while simulating etiology of naturally-occurring tendinopathy.

Another important consideration in animal tendinopathy model creation is the role immune system plays in tendon healing.[188] This is especially important, because varieties in expression level and haplotype of major histocompatibility complex (MHC) II within horse population had been reported to elicit immune responses despite of MSCs' general status as hypoimmunogenic to non-immunogenic from lack of MHC II expression.[189] Moreover, haplotype-mismatched allogenic MSCs are known to be short-lived *in vivo* upon transplantation, which contributes to failure of implants.[190] Combining difficulties in identifying MHC II haplotype-matched donors among 50 known haplotypes[191] and superior accessibility of allogenic versus autologous MSCs in horses, it is important to develop neotissue from allogenic MSCs that does not elicit immune response and augment healing in immnocompetent recipient individuals. Therefore, the use of both immunocompromised and immunocompetent rats to create tendinopathy model can elucidate the role immune system play in tendon healing by neotissue and leads to development of readily accessible immunocompatible neotissue.

In this study, tendon neotissue created from equine ASCs cultured on COLI template in tenogenic or stromal medium for 21 days under static strain, or phosphate buffered saline as no treatment were implanted via needle in randomly selected limb of each rat 6 days after bilateral elongation-induced calcaneal tendon injury creation. Hypotheses tested were: 1) tenogenic neotissue implanted is immunocompatible and regenerative without eliciting immune response and form tendon-like tissue, while stromal neotissue implant is immunocompatible yet non-regenerative without forming tendon-like tissue; and 2) the immune system plays an role in a mechanism of non-regenerative tissue formation by stromal neotissue. The objectives of the study were: 1) to compare microstructure of implanted both tenogenic and stromal neotissues in immunocompetent rat calcaneal tendon; and 2) to compare microstructure of implanted stromal neotissues in both immunocompetent and immnocompromised rat calcaneal tendon.

## 3.2. Materials and Methods

### 3.2.1. Study Design

Protocols (#21-022 and #21-051) were approved by the Louisiana State University Institutional Animal Care and Use committee prior to study initiation. calcaneal tendon elongation injury was made bilaterally to both hindlimbs of each Sprague Dawley (SD) rat (n = 18) and RNU

Nude (RNU) rat (n = 12). Bovine corium COLI (Avitene™ Ultrafoam™ Collagen Sponge, Davol Inc., Warwick, RI) templates were infused with equine ASCs from 1 adult gelding, and cultured in tenogenic (n = 30) or stromal (n = 15) medium for 21 days.

Six days after calcaneal tendon elongation injury, tenogenic neotissue (SD rat: n = 18; RNU rat: n = 12) was percutaneously injected into tendon lesion of one randomly selected hindlimb in each rat via needle. The opposite tendon lesion in each animal was injected with stromal neotissue (SD rat: n = 9; RNU rat: n = 5) or PBS (SD rat: n = 9; RNU rat: n = 5). Limb use was evaluated with an established rubric for all rats 1 week before surgery, daily up to 14 days after surgery and then weekly until harvest 6 weeks after surgery.

Tendons from 15 rats (SD rat: n = 9; RNU rat: n = 5) were evaluated for tensile mechanical properties. Tendons from the other 15 rats (SD rat: n = 9; RNU rat: n = 5) were divided sagitally to give four quadrants, medial axial, medial abaxial, lateral axial, and lateral abaxial. One quadrant from each rat was assigned to one of four outcome measures, gene expression, collagen composition, and micro- and ultrastructure (Fig 4.1, 4.2), based on a randomized block design.

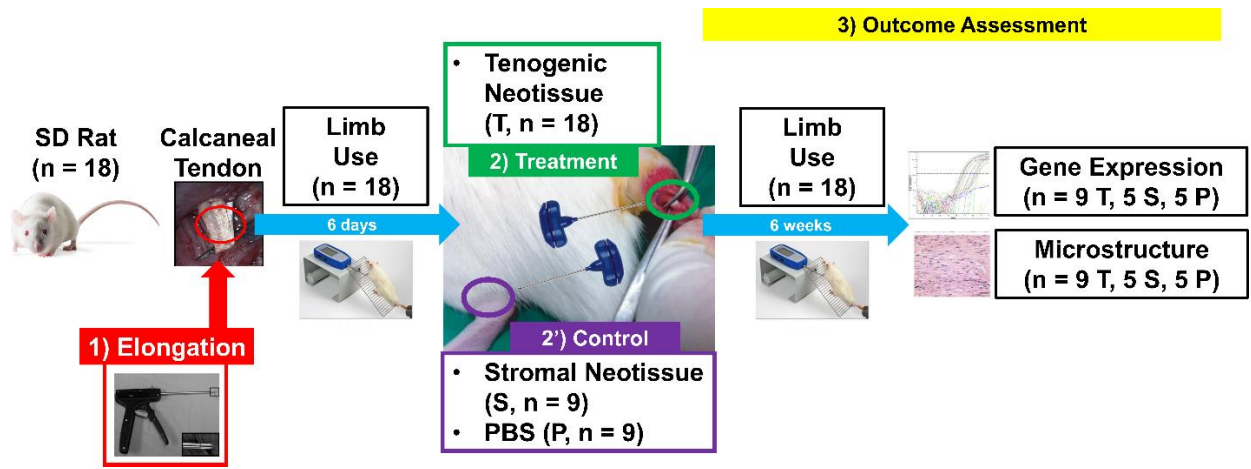

Figure 3.1. Study design of SD rat bilateral calcaneal tendon elongation injury model creation and neotissue treatment.

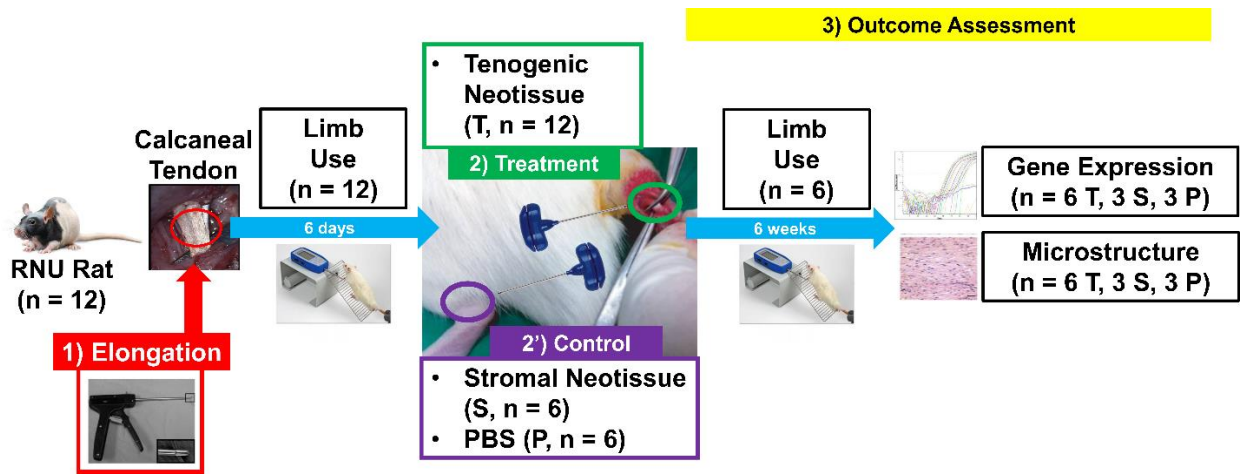

Figure 3.2. Study design of RNU rat bilateral calcaneal tendon elongation injury model creation and neotissue treatment.

### 3.2.2. Perfusion Bioreactor System

COLI scaffold templates were secured to a movable horizontal bar under continuous uniaxial static tension and maintained inside a custom-designed bioreactor chamber that is connected to a perfusion system. The perfusion system consists of a 10 ml medium reservoir (Synthecon, Houston, TX) that permits gas exchange while a computer-controlled peristaltic pump (Ismatec 404b, Glattbrugg, Switzerland) maintains a bidirectional fluid flow rate of 1 ml/min. The cells were infused into COLI scaffold templates ($6.0 \times 4.0 \times 1.0$ $cm^3$) at $1.0 \times 10^6$ cells/$cm^3$ scaffold template through the injection port of each chamber and maintained in tenogenic (DMEM-high glucose, 1% fetal bovine serum (FBS), 10 ng/ml transforming growth factor (TGF)-β1, 50 mM L-ascorbic acid 2-phosphate sesquimagnesium salt hydrate, 0.5 mg/ml insulin, 1% antibiotics) or stromal (DMEM-F12, 10% FBS, 1% antibiotics) medium. The perfusion system was maintained in 5% $CO_2$ at 37°C for 21 days with medium changes every 7 days.

### 3.2.3. Elongated calcaneal Tendinopathy Models

Both Sprague Dawley (SD) and RNU nude (RNU) rats aging 8 – 10 weeks old and weighing 200 – 250 g were obtained from a vendor (Charles River Laboratories, Wilmington,

MA). Elongation-induced calcaneal tendinopathy was created in both hindlimbs of rats using an adaptation of an elongation device validated for canine cranial cruciate ligament elongation[187].

Briefly, rats were induced and maintained with isofluorane (2.0%) in 100% oxygen at a flow rate of 1L/min inside an induction chamber and then a facemask. Premedication was given by subcutaneous injection of glycopyrrolate (0.02 mg/kg) and butorphanol (0.5 mg/kg). Following aseptic preparation, L-shaped skin incisions were made superior and lateral to each calcaneal tendon to elevate the skin. After carefully excising the paratenon of each calcaneal tendon, fine grit stainless steel nail file was placed underneath the tendon to prevent slippage during force application.

Single-interrupted # 4-0 Vicryl® sutures (Ethicon, Somerville, NJ) were placed through the calcaneal tendon with the distal suture just proximal to the calcaneus and the proximal suture at the junction of the calcaneal tendon and gastrocnemius muscle. A spring scale was used to apply 2.5 N of tension to the proximal aspect of the calcaneal tendon via stainless steel S-shape hook and # 0 PDS® II suture (Ethicon). The distance between the distal edge of the proximal suture and the proximal edge of the distal suture was measured during force application to determine the pre-elongation length (PrE) with a vernier caliper.

The tendon elongation device has a hollow cylinder (7.5 mm inner diameter; 10 mm outer diameter) with a 2.5 mm-wide textured (80 grit) rim. Within the cylinder, a removable 10-gauge stainless steel hook (5 mm diameter, 2 mm depth) is attached to a retractable ratchet system that locks every 1.5 mm. The hook was placed beneath calcaneal tendon midway between sutures. The textured surface of the cylinder engaged with the tendon surface by pressing the device firmly to prevent slippage during hook retraction and limit elongated tissue to the area within the inner diameter of cylinder. The hook was retracted 1 ratchet step (1.5 mm) to elongate the isolated section by 8 ratchet steps (12 mm) and held for 1 minute until release.

Post-elongation length (PoE) was determined identically to the PrE, and percent elongation (PE) was calculated as PE = (PoE−PrE)/PrE × 100. Skin was approximated with # 4-0 Vicryl® sutures in a simple continuous subcuticular suture pattern. The suture line was be sealed with tissue adhesive. Post-medication was given by subcutaneous injection of carprofen (5 mg/kg), enrofloxacin (10 mg/kg), and 5 ml of warmed 0.9% saline.

### 3.2.4. COLI Construct Implantation

Six days after tendon elongation surgeries, ASC-COLI constructs cultured in tenogenic or stromal medium, or PBS were implanted via needle under general anesthesia identically as for surgery.

A half of cylindrical piece (4 mm diameter, 10 mm thickness) collected from construct using biopsy punch was placed into the tip of a 18-gauge 1.5 inch needle (Exel international, Salaberry-de-Valleyfield, Canada), 20-gauge 1 inch needle (Becton Dickinson, Franklin Lakes, NJ) inserted into 18-gauge needle from the tip side, then construct pushed into 20 gauge needle using stylet of 22-gauge 1 inch IV catheter (Terumo Medical Corporation, Southaven, MS).

After aseptic preparation of hindlimb, 20-gauge needle was inserted transdermally into calcaneal tendon to proximal end of tendon, and construct was implanted by advancing 22-gauge IV catheter stylet inside 20-gauge needle while withdrawing needle.

The procedure was performed on both limbs, with one randomly assigned to receive a tenogenic ASC-COLI construct, while the other received a stromal ASC-COLI construct or PBS (0.5 ml). Rats were maintained in cages without movement restriction for 6 weeks after which they were humanely sacrificed according to current AVMA standards and calcaneal tendons harvested.

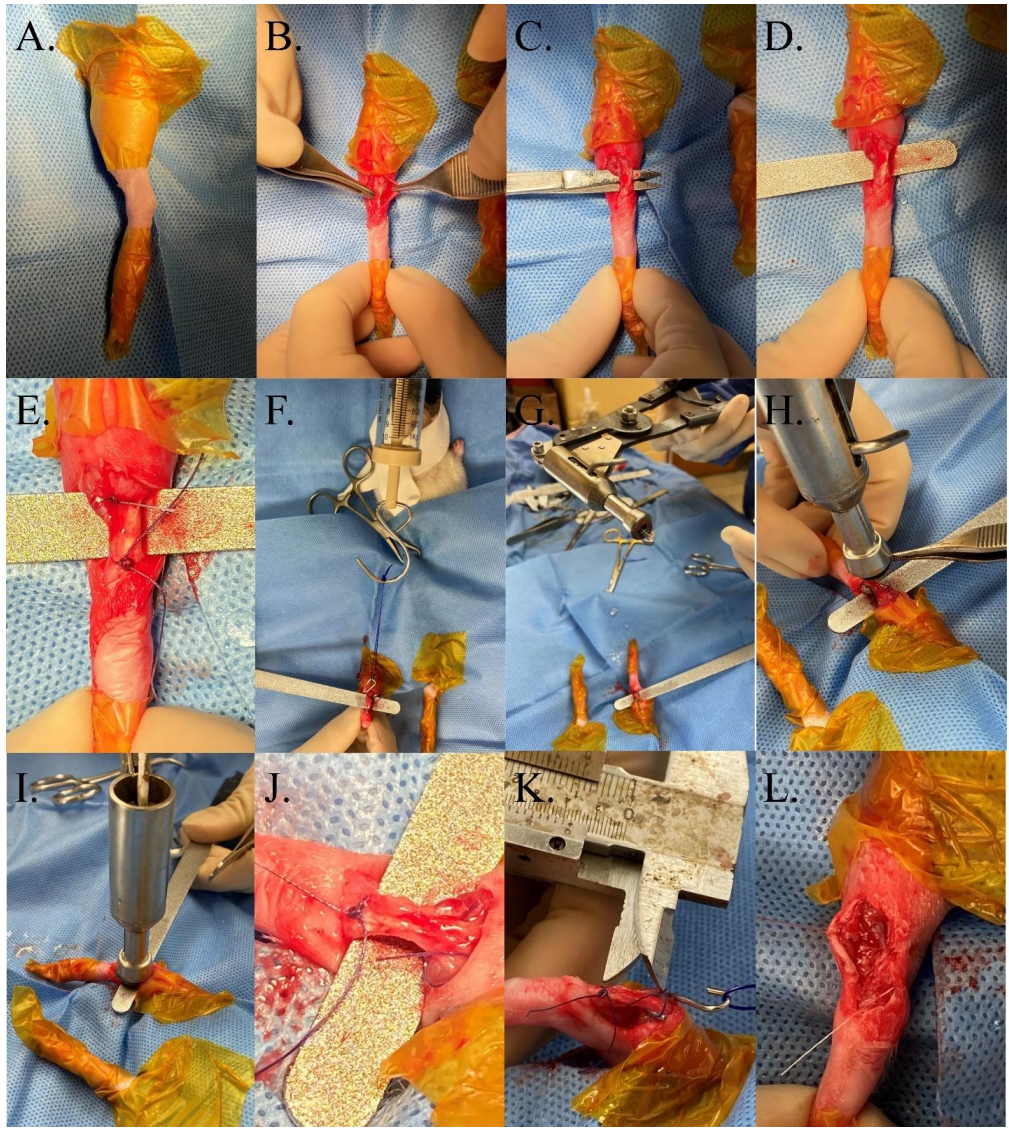

Figure 3.3. Elongation-induced calcaneal tendinopathy model creation. Hindlimb was passed through fenestrated drape and both proximal and distal limb covered with iodine-impregnated incision drape (A), L-shaped skin incisions was made superior and lateral to each calcaneal tendon to elevate the skin (B), isolate and elevate calcaneal tendon (C), flat stainless steel stick with etched surface was placed underneath tendon (D), single-interrupted 5-0 Vicryl® sutures were placed through the calcaneal tendon with the distal suture just proximal to the calcaneus and the proximal suture at the junction of the calcaneal tendon and gastrocnemius muscle (E), a bar type tension gauge applied 2.5 N of tension to the calcaneal tendon (F), tendon elongation device (G), lift tendon with the hook inside at midway between proximal and distal sutures (H), textured surface of the cylinder will be engaged with the tendon surface by pressing the device firmly to prevent slippage during hook retraction and limit elongated tissue to the area within the inner diameter of cylinder (I) and retract tendon inside cylinder. The distance between proximal and distal suture ob elongated tendon (J) was measured with a vernier caliper. Subcutaneous tissue and skin were closed in routine manner (L).

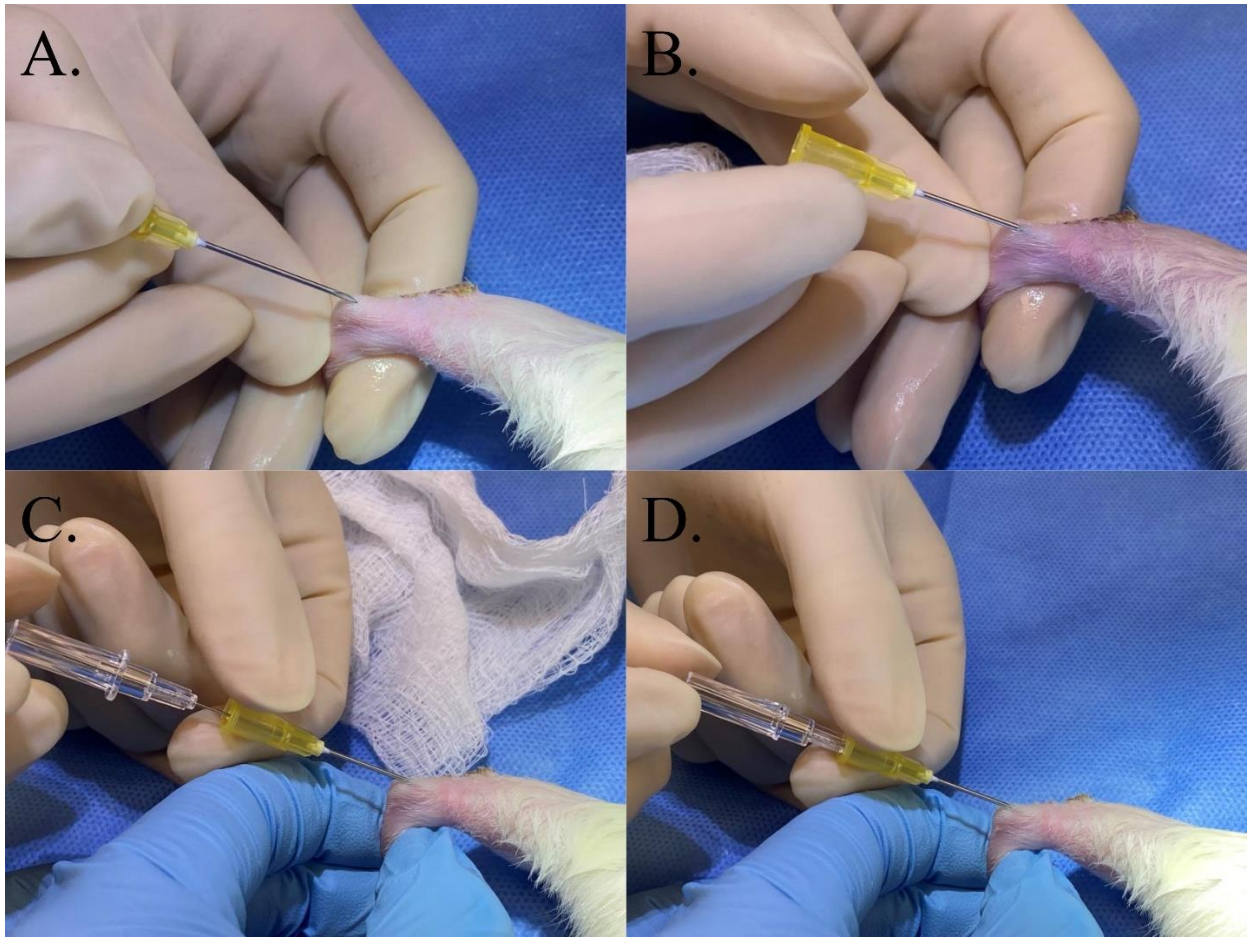

Figure 3.4. Needle implantation of ASC-COLI construct into rat calcaneal tendon. A half of cylindrical piece (4 mm diameter, 10 mm thickness) placed into the tip of 20-gauge 1 inch needle was inserted into calcaneal tendon transdermally from distal (A) to proximal (B) end of tendon. Stylet of 22-gauge 1 inch IV catheter was inserted into 20-gauge needle (C) and advanced while withdrawing needle to implant construct.

### 3.2.5. Functional Outcome

All rats were monitored once daily for signs of pain according to the Rat Grimace Scale (RGS)[192,193], and for excessive swelling, redness or drainage at the surgery site. A scoring system previously developed to evaluate rat limb use[194] was used to evaluate limb use one week before surgery, daily up to 14 days after surgery, and then weekly up to 6 weeks.

### 3.2.6. Microstructure

A quadrant of calcaneal tendon was fixed in 4 % paraformaldehyde (PFA), paraffin embedded, sectioned (5 μm) and stained with haematoxylin and eosin or Masson's trichrome. Digital images generated with a slide scanner (NanoZoomer, Hamamatsu Photonics K.K, Hamamatsu City, Japan) or a light microscope (DM4500B, Leica, Wetzlar, Germany) fitted with a digital camera (DFC480, Leica) and cellular morphology, distribution, and extra cellular matrix organization were be evaluated.

### 3.2.7. Gene Expression - Reverse Transcription Polymerase Chain Reaction (RT-PCR)

A quadrant of calcaneal tendon was snap frozen in liquid nitrogen and transferred to a cooled grinding cylinder of BioPulverizer (BioSpec, Bartlesville, OK) and pulverized.[195] Ground samples were collected by rinsing the cylinder with 1 ml of TRI reagent® (Sigma, St. Louis, MO) and transferred to 1.5 ml tube. Ground sample was homogenized by passing mixture through 18-gauge needle 30 times. Total RNA was extracted from supernatant by phenol-chloroform extraction according to the manufacturer's instructions. Isolated RNA was cleaned up by RNeasy® Mini Kit (QIAGEN, Hilden, Germany). One microgram of total RNA was used for cDNA synthesis (QuantiTect® Reverse Transcription Kit, QIAGEN). Rat-specific primers for tendon-specific genes, *collagen 1a1* (*Col1a1*), *collagen 3a1* (*Col3a1*), *tenascin-c* (*Tnc*), *scleraxis* (*Scx*), and *tenomodulin* (*Tnmd*) were quantified using primers previously validated.[126] PCR was performed with denaturation step at 95°C for 15 minutes, followed by 40 cycles of denaturation at 94 °C for 15 seconds, annealing at 52 °C for 30 seconds, and elongation at 72°C for 30 seconds using SYBR Green system (QuantiTect® SYBR® Green PCR Kits, QIAGEN). Relative gene fold change was determined by standard means ($2^{-\Delta\Delta Ct}$). *Glyceraldehyde 3-phosphate dehydrogenase* (*GAPDH*) was used as the reference gene.

Table 3.1. Rat-specific Primer Sequences

| Gene | | Sequence (5' – 3') | Accession Number |
|------|------|--------------------|------------------|
| *Col1a1* | Forward | GGAGAGAGTGCCAACTCCAG | NM_053304.1 |
| | Reverse | GTGCTTTGGAAAATGGTGCT | |
| *Col3a1* | Forward | TCCCAGAACATTACATACCACT | NM_032085.1 |
| | Reverse | GCTATTTCCTTCAGCCTTGA | |
| *Tnc* | Forward | AGATGCTACTCCAGACGGTTTC | NM_053861.1 |
| | Reverse | CACGGCTTATTCCATAGAGTTCA | |
| *Scx* | Forward | AACACGGCCTTCACTGCGCTG | NM_001130508.1 |
| | Reverse | CAGTAGCACGTTGCCCAGGTG | |
| *Tnmd* | Forward | GTGGTCCCACAAGTGAAGGT | NM_022290.1 |
| | Reverse | GTCTTCCTCGCTTGCTTGTC | |
| *GAPDH* | Forward | AAGTTCAACGGCACAGTCAAGG | NM_017008.4 |
| | Reverse | CGCCAGTAGACTCCACGACATA | |

### 3.2.8. Statistical Analysis

Results are presented as mean ± standard error of the mean (SEM). Normality of data was examined with the Kolmogorov–Smirnov test. Measures of functional outcome, post-elongation percentages, and post-elongation tendon width were compared with ANOVA among groups. When overall difference was detected, pairwise comparisons between groups were performed using Tukey's post-hoc test. Fold changes of tenogenic gene expression were compared: to 1 using one sample t-test and between groups using two-sample t-test for normally distributed results; and to 1 using Wilcoxon signed rank test and between groups using Mann Whitney test for non-normally distributed results. All analyses were conducted using Prism (GraphPad Software Inc., San Diego, CA) with significance considered at $p < 0.05$.

### 3.3. Results

### 3.3.1. Functional Outcome

Following elongation injuries, functional scores dramatically decreased despite rats were ambulatory all the time. Additionally, minimum swelling was noted at the surgery site without signs of infections or excessive inflammation for all rats. Functional scores gradually recovered over 6-week period for both SD and RNU rats (Fig 3.5). However, SD rats appeared to have better recovery of limb use than RNU rats by the end of post-operative period. After 6 weeks of post-operative period, neither SD rats nor RNU rats fully regained original limb function.

In SD rats, functional scores had a trend of low values in PBS treatment group compared to tenogenic construct or stromal construct groups after 10 days post-operatively. There appeared to be no difference in functional scores between tenogenic construct and stromal construct groups (Fig 3.5.A). On the contrary, there was no trend of difference in functional scores among treatment groups for RNU rats all the time (Fig 3.5.B). Compared to SD rats, functional recovery of limbs in RNU rats appeared to be biphasic rather than continuous.

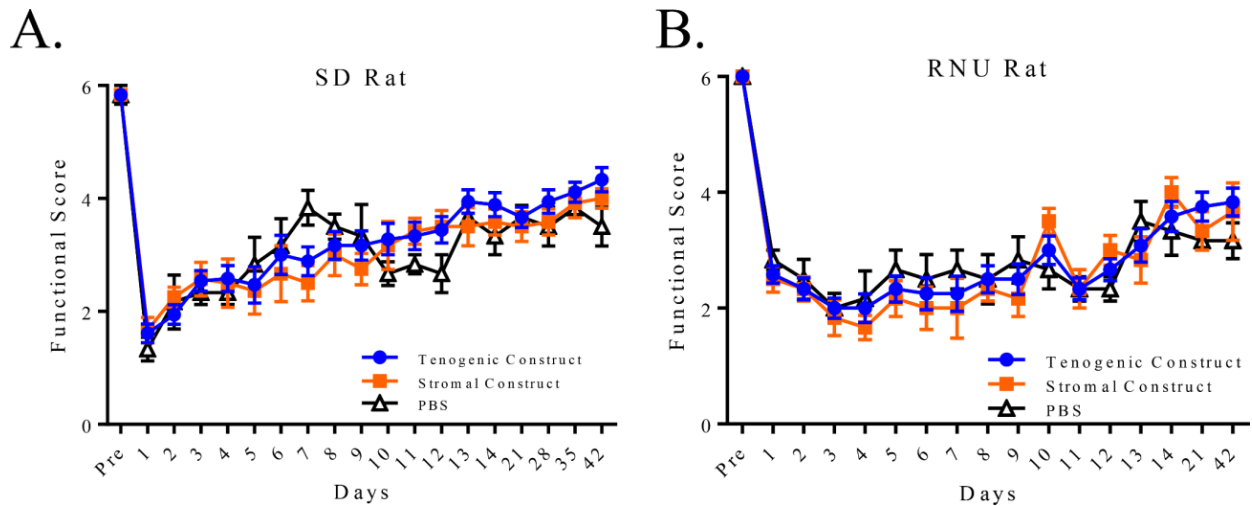

Figure 3.5. Functional score of hindlimbs in SD (A) and RNU (B) rats from pre-injury to 6 weeks post-injury. Functional scores were obtained for limbs that received tenogenic constructs (blue), stromal constructs (orange), or PBS (black). Each treatment was administered at day 6.

### 3.3.2. Post-Elongation Percentage and Tendon Width

Percentage of elongation did not differ among treatment groups for SD rats (Fig 3.6.A). Overall elongation percentage was $54.24 \pm 3.756\%$. Elongation percentages were generally higher than those that may lead to acute rupture. As a result, loosened midportion of calcaneal tendon following elongation between sutures was noted (Fig 3.3.J). Individually, elongation percentages were 60.12 ± 6.666% for tenogenic construct group, 50.65 ± 3.889% for stromal construct group, and 43.79 ± 4.984% for PBS group. Variation in elongation percentages was higher in tenogenic construct group, and as high as 144% of elongation was achieved without causing complete rupture.

Six weeks post-injury, the widths of healed calcaneal tendon at midportion was higher in tenogenic construct group (2.444 ± 0.1175 mm) compared to PBS group (1.903 ± 0.08724 mm). Widths of calcaneal tendon in stromal construct group (2.164 ± 0.1088 mm) were not different from tenogenic construct group or PBS group (Fig 3.6.B). There was no visible partial or complete rupture noted in gross appearance of healed tendons among all treatment groups.

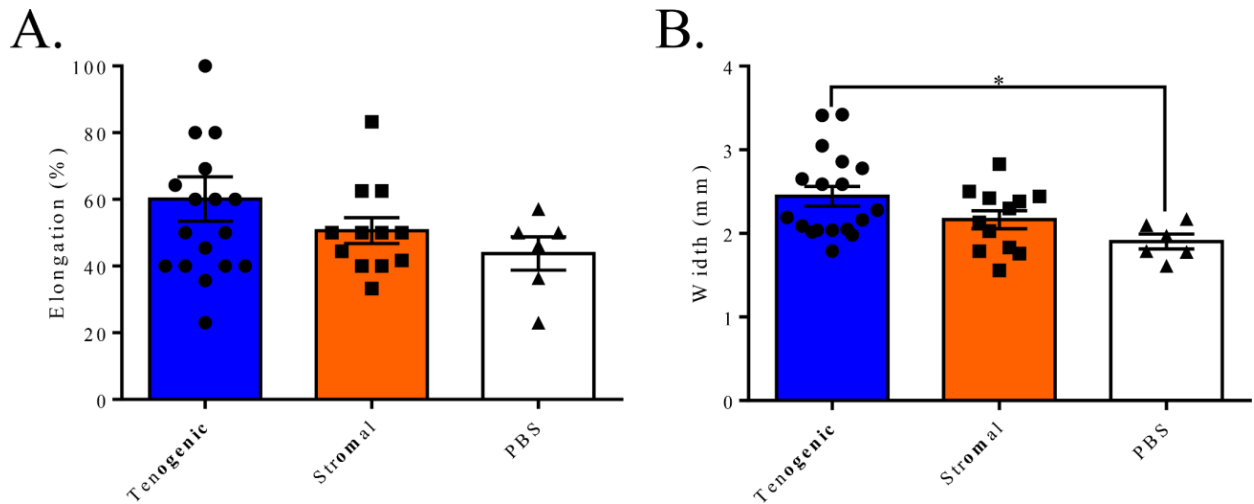

Figure 3.6. Post elongation percentage (A) and tendon width at harvest (B) of SD rats. Post elongation percentage was determined at the time of surgery and width measured at tissue harvest for each tendon of all rats.

Similar to SD rats, percentage of elongation did not differ among treatment groups for RNU rats (Fig 3.7.A). Overall elongation percentage was 70.17 ± 6.297% and appeared to be higher than that for SD rats. Individually, elongation percentages were 68.91 ± 10.64% for tenogenic construct group, 55.72 ± 6.789% for stromal construct group, and 87.16 ± 9.224% for PBS group. Variation in elongation percentages was again higher in tenogenic construct group, and as high as 150% of elongation was achieved without causing complete rupture. High variability demonstrated less control in size of injury compared to transection of tendon.

Unlike SD rats, the widths of healed calcaneal tendon at midportion in RNU rats 6 weeks after injury were not different among treatment groups (Fig 3.7.B). Individually, widths were 2.082 ± 0.04139 mm for tenogenic construct group, 2.012 ± 0.06183 mm for stromal construct group, and 1.992 ± 0.1262 mm for PBS group. Although non-significant, there was a trend of widest tendon widths by tenogenic construct treatment, followed by stromal construct and PBS treatment. There was also no visible partial or complete rupture noted in gross appearance of healed tendons among all treatment groups in RNU rats.

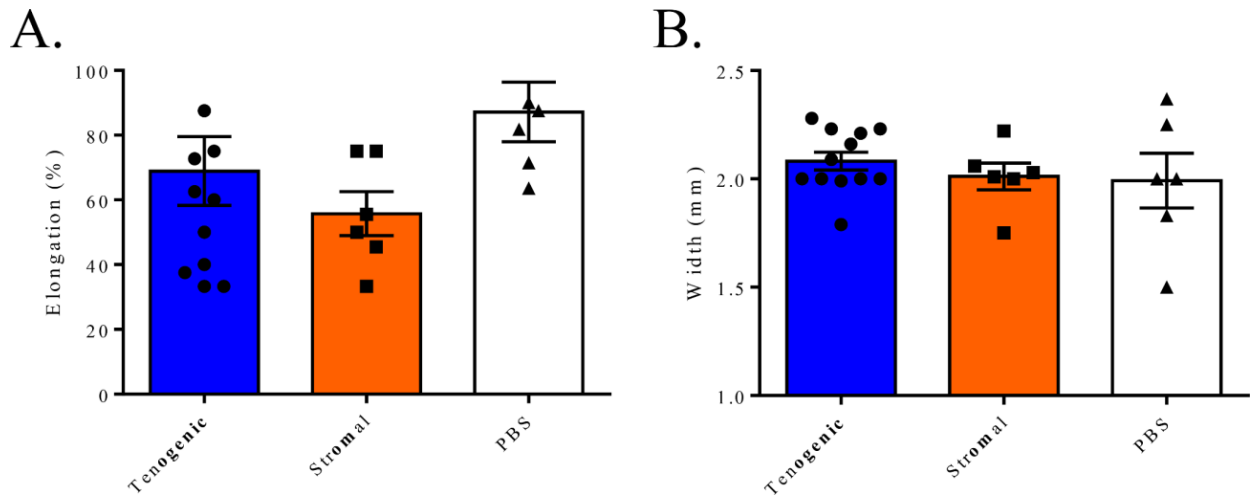

Figure 3.7. Post elongation percentage (A) and tendon width at harvest (B) of RNU rats. Post elongation percentage was determined at the time of surgery and width measured at tissue harvest for each tendon of all rats.

### 3.3.3. Microstructure

In calcaneal tendons of SD rats who received tenogenic constructs, there were no obvious regions of distorted cellular alignment or ECM orientation at the midportion where elongation injury was created. Implanted tenogenic constructs were clearly identified as cylinder-like structure with basophilic in H & E staining and light blue in trichrome staining within implanted calcaneal tendon of rats at the time of harvest (Fig 3.8). Although implants were ejected from needle while withdrawing from calcaneal tendon along the entire length, majority of portion of implants appeared to have been deposited at the proximal end of tendons (Fig 3.8.A). However, lesser amounts of implants were deposited along the entire length of tendons. Implants were relatively acellular and mostly consisted of fibrous ECM and diffusely residing fibroblastic cells. Cellularity within implants were relatively higher than surrounding native tendon, yet low cellularity of implants resembled native tendon tissue. In terms of immune response to the implant, there were no signs of mononuclear cells' infiltration, indicating lack of immune response to implants. Of note was the presence of red blood cells within empty spaces of implants, which suggested neovasculization was also induced (Fig 3.8.F).

Majority of cells assumed tenocyte-like morphology with elongated rhomboid nucleus. There were areas where cells aligned randomly inside the implant, while there were areas where cells obtained parallel alignment with surrounding native tendon (Fig 3.8.D). In the areas where cells aligned randomly, cells tended to form clusters with relatively higher density. Around those areas, fibrous ECM also maintained random alignment. In the areas where cells aligned parallelly, cellularity was lower and cells aligned parallel to the surrounding tendon tissue. Additionally in those areas, fibrous ECM within gained parallel alignment as well to the surrounding native tendon. Additionally, these ECM developed crimp pattern characteristic of tendon tissue (Fig 3.8.D).

The immediate adjacent areas of implants did not accumulate fibrotic tissue to sequester implant. This again indicated lack of foreign body reaction to implant. Moreover, there were apparent integration of implants with surrounding native tendons characterized by gradual transition of implants towards native tendons merging ECMs from both areas (Fig 3.8.E). Cells within merging areas also appeared to be more elongated mature tenocyte-like cells.

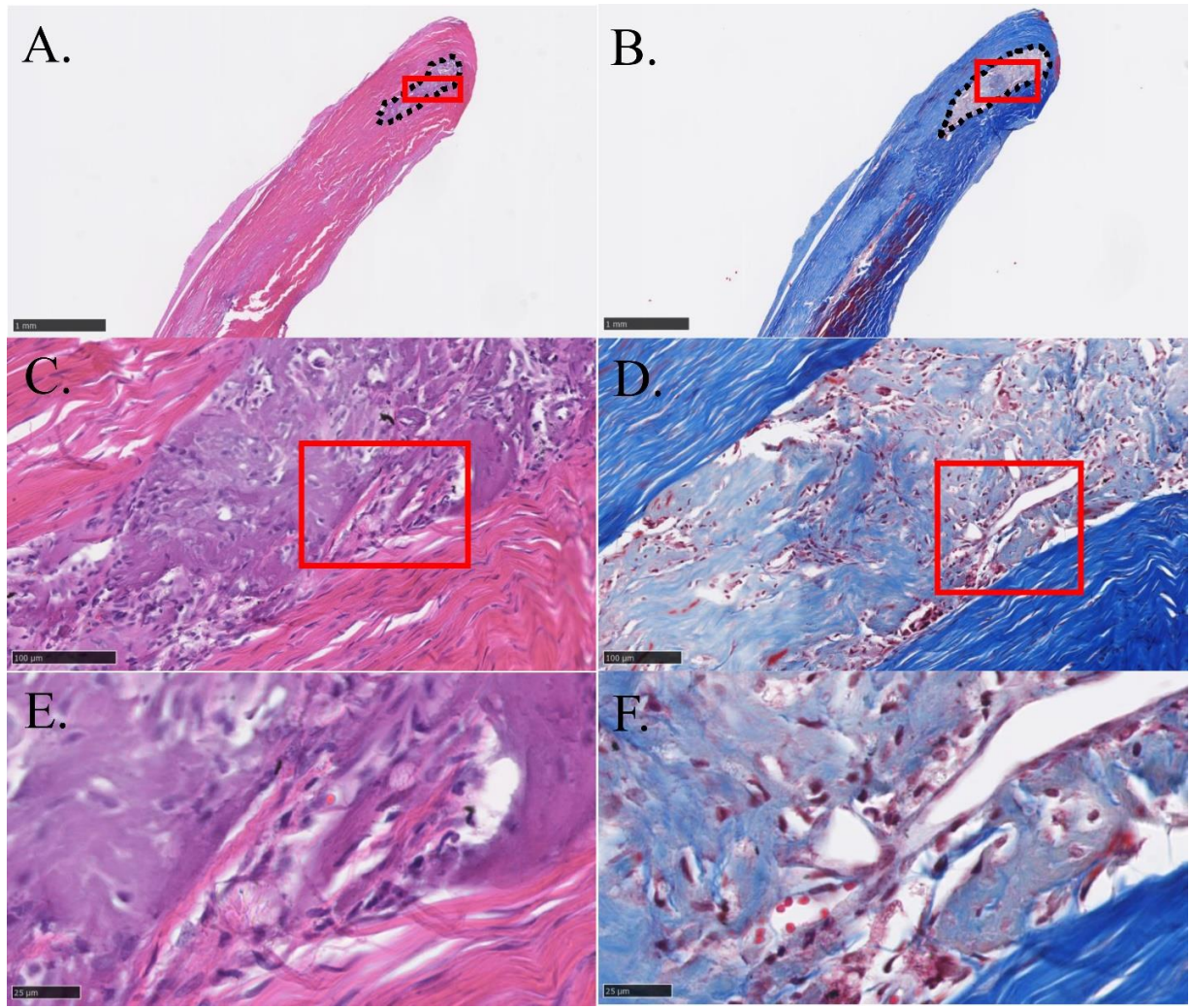

Figure 3.8. Microstructure of calcaneal tendon treated with tenogenic construct in SD rats. Harvested calcaneal tendon was stained with hematoxylin and eosin (A, C, E) or Masson's trichrome (B, D, F). Implanted construct was indicated with black dotted line. Panels E, F, C, and D, are enlarged images represented by red rectangles in panels C, D, A, and B, respectively. Scale bars = 1 mm (A, B), 100 μm (C, D), and 25 μm (E, F).

Similar to tendons that received tenogenic construct administration, those with stromal constructs also lacked obvious disorganized area in the midportion. The implants were also clearly identified as cylindrical structure within tendon demarcated from surrounding tissue. The implants did not have infiltrating monocytes or surrounding fibrous encapsulation. Cells within implants appeared to be more round to rhomboid with oval nucleus. Implants were high in cellularity and contained less amounts of immature ECM that was stained red by trichrome (Fig 3.9).

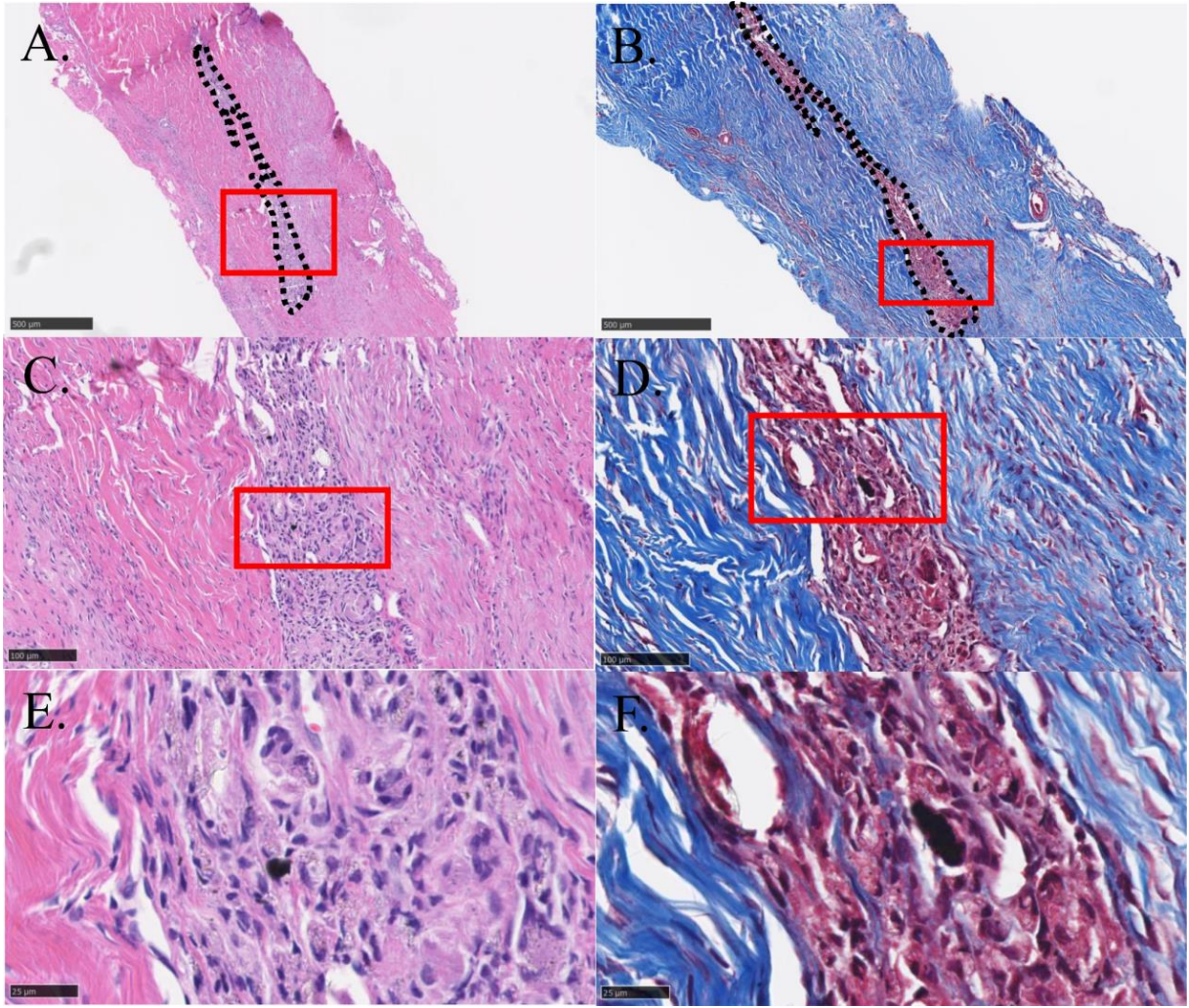

Figure 3.9. Microstructure of calcaneal tendon treated with stromal construct in SD rats. Harvested calcaneal tendon was stained with hematoxylin and eosin (A, C, E) or Masson's trichrome (B, D, F). Implanted construct was indicated with black dotted line. Panels E, F, C, and D, are enlarged images represented by red rectangles in panels C, D, A, and B, respectively. Scale bars = 500 μm (A, B), 100 μm (C, D), and 25 μm (E, F).

Tendons that received PBS treatment had apparent disorganized areas at the midportion characterized by high cellularity and non-parallel alignment of both cells and ECM. However, majority of areas within tendon maintained low cellularity with parallel alignment along with longitudinal direction (Fig 3.10). There were no signs of hematoma formation at the time of harvest. Additionally, abnormal cellular differentiation such as chondrocyte-like or osteocyte-like cell formation was not evident as well.

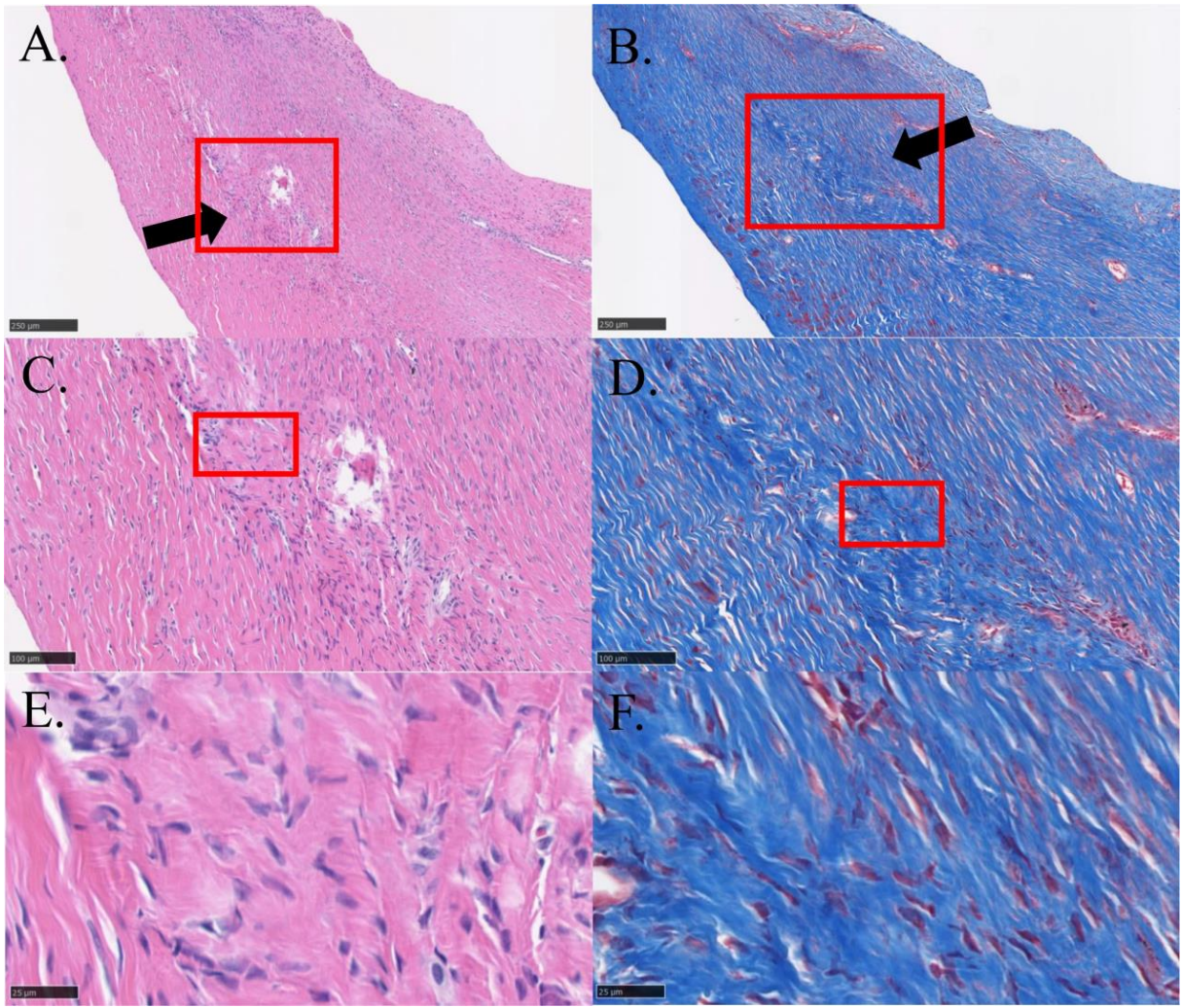

Figure 3.10. Microstructure of calcaneal tendon treated with PBS in SD rats. Harvested calcaneal tendon was stained with hematoxylin and eosin (A, C, E) or Masson's trichrome (B, D, F). Transverse lesion in the mid-substance of tendon was indicated with black arrow. Panels E, F, C, and D, are enlarged images represented by red rectangles in panels C, D, A, and B, respectively. Scale bars = 250 μm (A, B), 100 μm (C, D), and 25 μm (E, F).

Similar to tendons of SD rats that received implant treatments, obvious elongation-induced lesions were not found in tendons of RNU rats into which tenogenic constructs were implanted. Implants in the tendons were clearly identified as cylindrical structures (Fig 3.11). They were basophilic in H & E staining and light blue in trichrome staining. The areas stained red by trichrome staining were observed in multiple portions, and it was not consistent finding with tenogenic constructs implanted into SD rats' tendons. The organization of cells and ECM appeared to be less organized than tenogenic constructs implanted in calcaneal tendons of SD rats.

Although cellularity was low homogenously, much less cells aligned parallelly to each other and to the surrounding native tendons (Fig 3.11.D). Distinct forms of ECM were observed within implant, which were fibrous and caseous appearance. Among these, fibrous ECM obtained parallel to oblique alignment with longitudinal direction of native tendon (Fig 3.11.F). Of note was stronger blue coloration of fibrous ECM than tenogenic constructs implanted tendons of SD rats by trichrome staining. The degree of staining with blue color was almost indistinguishable between fibrous ECM within implants and collagen fibers of native tendons, potentially more advanced maturity of fibrous ECM. The same areas with fibrous ECM also had close eosinophilic coloration to native collagen fibers. Cells had oval to spindle shape nucleus and populated more frequently in the areas where fibrous ECM existed. Cellular nucleus was also intact and no cells had apoptotic or necrotic appearances. Cells were also not assuming chondrocyte-like morphology characterized by round cells embedded in hyaline-like ECM, or osteocyte-like morphology characterized by calcium deposition around them. Also, there was no cells that contain vacuoles resembling adipocyte-like cells or macrophages.

As expected from immunocompromised status of RNU rats, there was no sign of immune reactions characterized by monocyte infiltration or encapsulation of implants by fibrous tissues (Fig 3.11.C). Not only within implants themselves, but also within native tendons around implants lacked infiltration of inflammatory cells. Unlike tenogenic constructs implanted into calcaneal tendons of SD rats, presence of red blood cells within implants was not noted, indicating lack of neovascularization. In terms of integration with surrounding native tendon structure, the boundary between implants and native tendons was clear and lacked insertion of collagenous fibers from native tendons to implants, suggesting limited integration of implants (Fig 3.11.C and D).

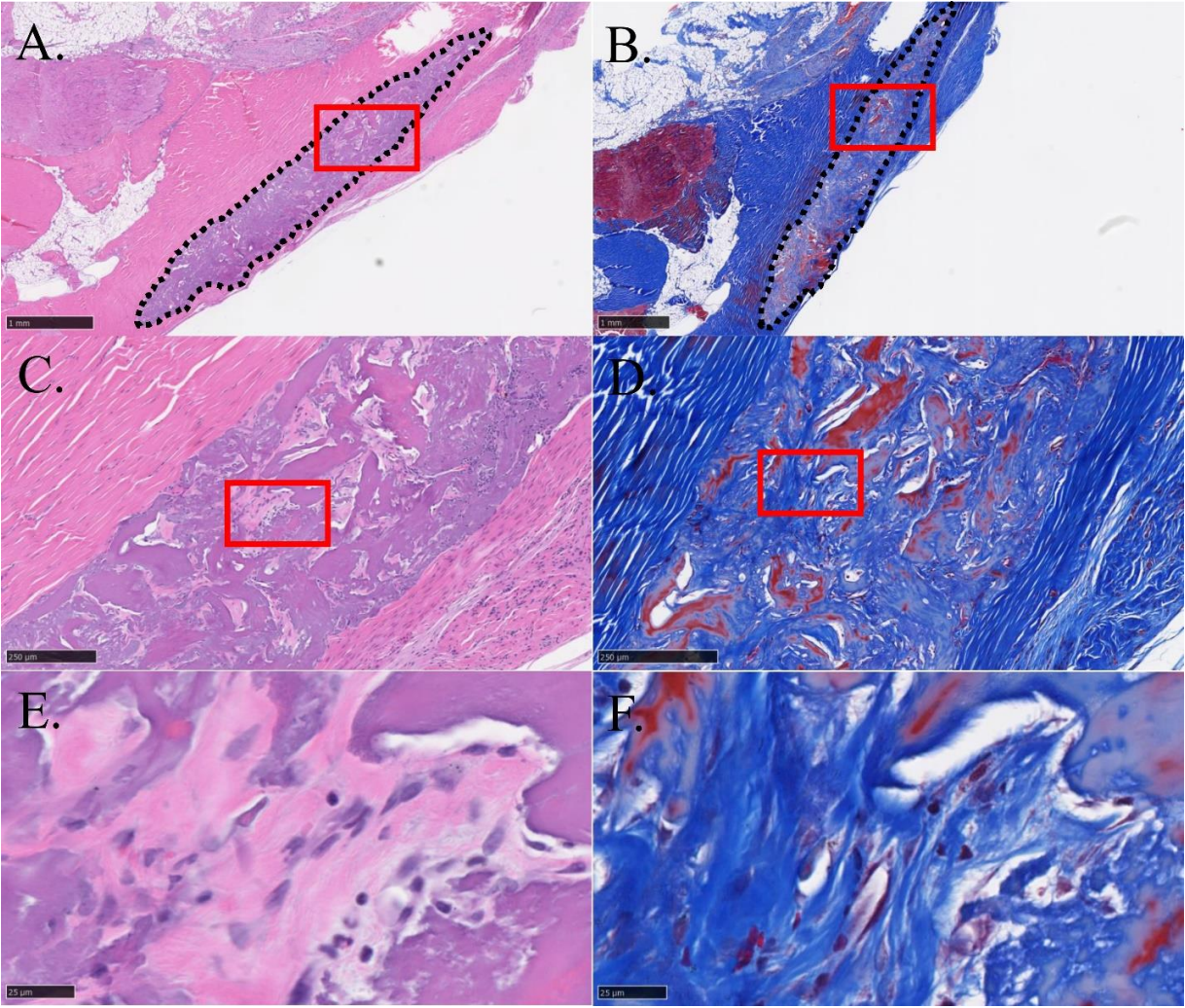

Figure 3.11. Microstructure of calcaneal tendon treated with tenogenic construct in RNU rats. Harvested calcaneal tendon was stained with hematoxylin and eosin (A, C, E) or Masson's trichrome (B, D, F). Implanted construct was indicated with black dotted line. Panels E, F, C, and D, are enlarged images represented by red rectangles in panels C, D, A, and B, respectively. Scale bars = 1 mm (A, B), 250 μm (C, D), and 25 μm (E, F).

Clear elongation lesions were not observed in tendons of RNU rats that received stromal constructs. Implants appeared similar to those of tenogenic constructs (Fig 3.12). Cells were mostly maintaining viable morphology with intact nucleus. Like tenogenic constructs, both fibrous and caceous ECM existed with more cellularity within areas of fibrous ECM (Fig 3.12.C). The organization of fibrous ECM was random, and lesser blue coloration with trichrome staining suggested less mature fibers (Fig 3.12.F). Immune response was not observed.

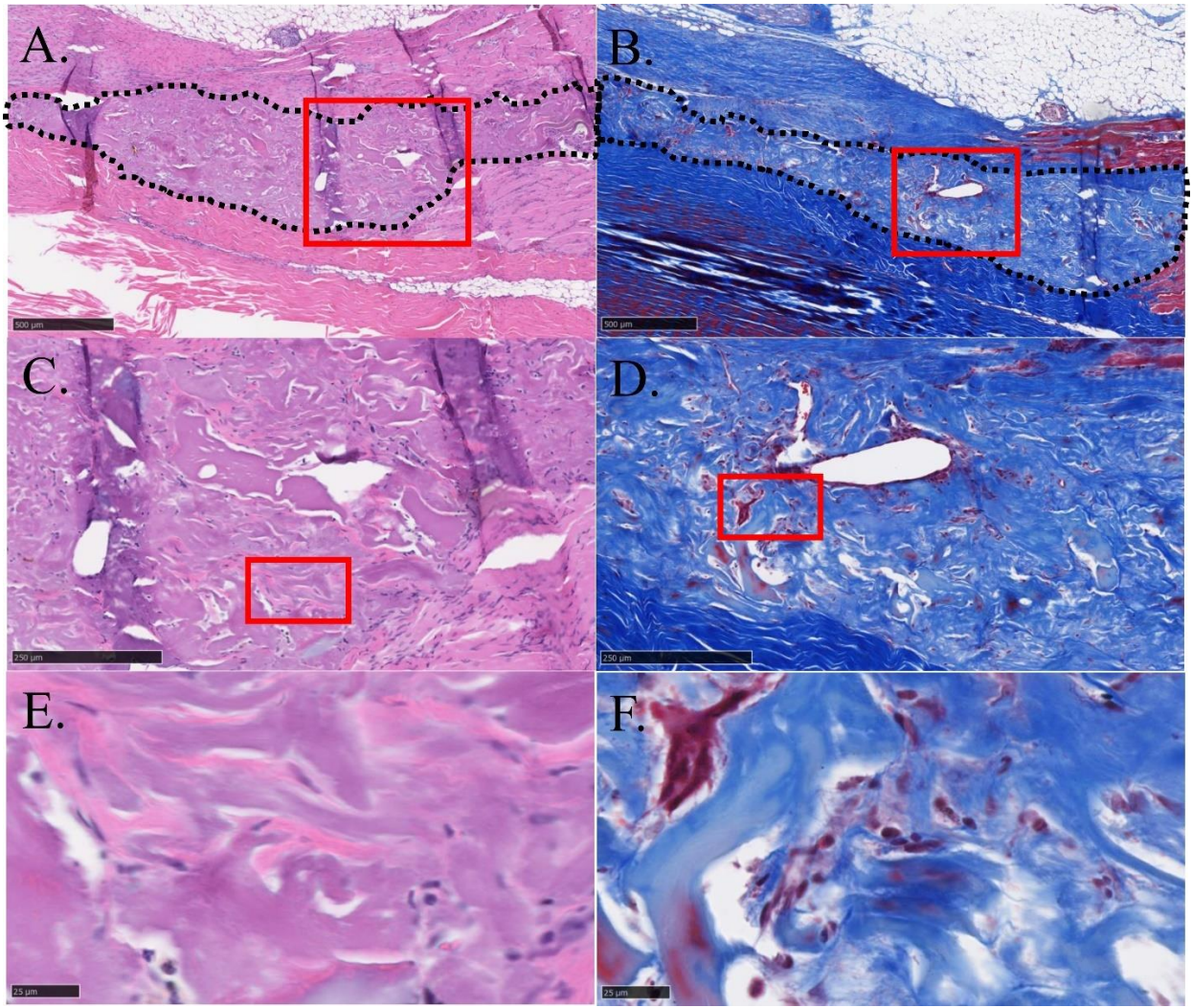

Figure 3.12. Microstructure of calcaneal tendon treated with stromal construct in RNU rats. Harvested calcaneal tendon was stained with hematoxylin and eosin (A, C, E) or Masson's trichrome (B, D, F). Implanted construct was indicated with black dotted line. Panels E, F, C, and D, are enlarged images represented by red rectangles in panels C, D, A, and B, respectively. Scale bars = 500 μm (A, B), 250 μm (C, D), and 25 μm (E, F).

In the tendons of RNU rats that received PBS treatment, elongation-induced lesion characterized by higher cellularity and disorganized cell-ECM orientation was observed (Fig 3.13.A and B, black arrows). However, no areas contained apoptotic- or necrotic-like cells, and had excessive inflammatory changes, indicating the tendons were in the phases of both proliferative and remodeling (Fig 3.13.E and F). Some areas had apparent chondroid formation with circular non-stained proteinaceous deposition that contain round cells (Fig 3.13.C and D).

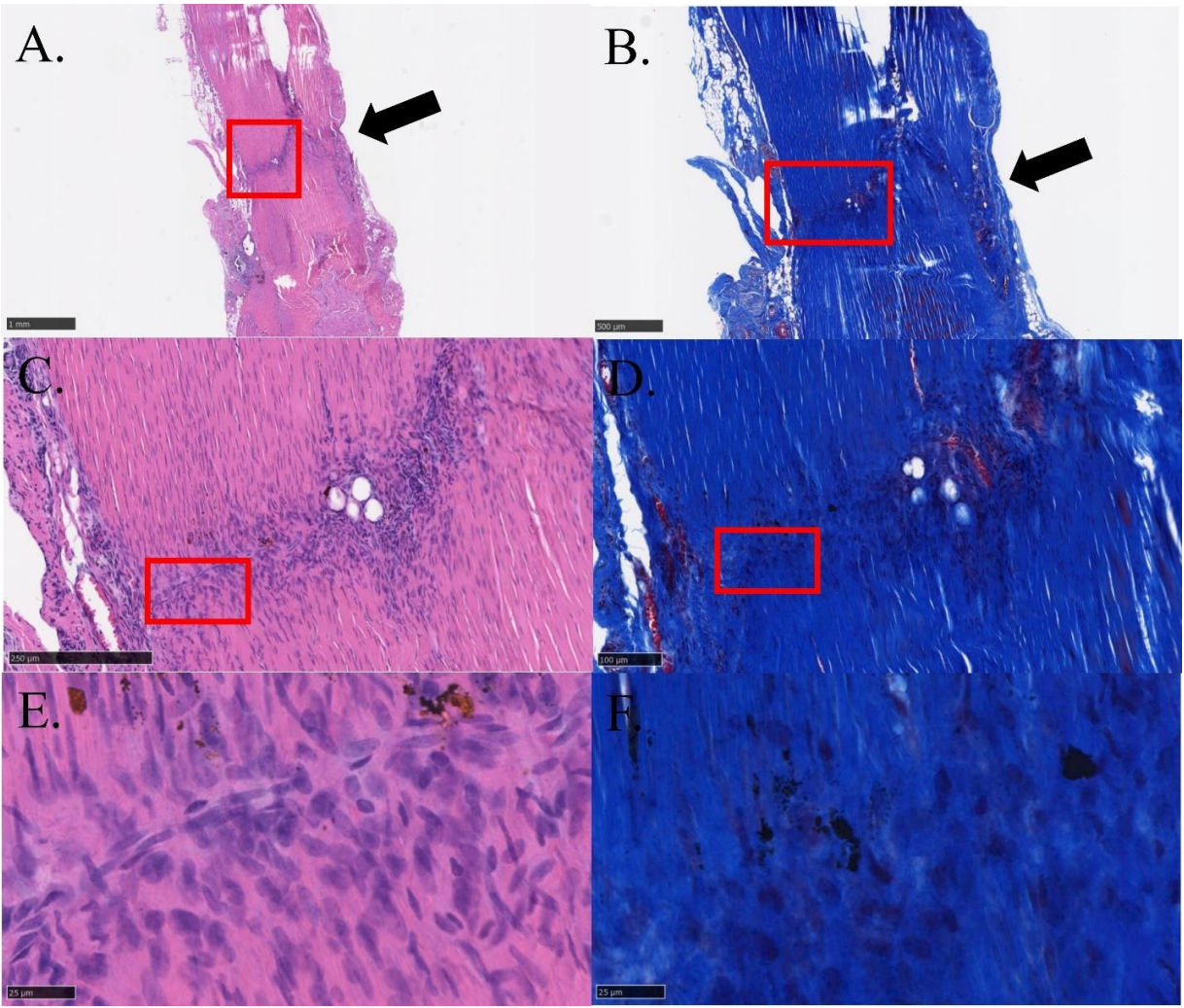

Figure 3.13. Microstructure of calcaneal tendon treated with PBS in RNU rats. Harvested calcaneal tendon was stained with hematoxylin and eosin (A, C, E) or Masson's trichrome (B, D, F). Transverse lesion in the mid-substance of tendon was indicated with black arrow. Panels E, F, C, and D, are enlarged images represented by red rectangles in panels C, D, A, and B, respectively. Scale bars = 1 mm (A), 500 μm (B), 250 μm (C), 100 μm (D), and 25 μm (E, F).

### 3.3.4. Gene Expression - Reverse Transcription Polymerase Chain Reaction (RT-PCR)

Regardless of control treatments on contralateral calcaneal tendons, those treated with tenogenic constructs had a trend of overall higher expressions of tendon-specific genes within each SD rat (Fig 3.14). Specifically, *Col1a1* had a trend of upregulation in calcaneal tendons treated with tenogenic constructs compared to those treated with stromal constructs (1.516 ± 0.5197-fold) and upregulated compared to those treated with PBS (2.612 ± 0.5515-fold), and the difference between tenogenic constructs and treatments was larger with PBS treatments than stromal constructs. Similarly, *Col3a1* had a trend of upregulation in calcaneal tendons treated with tenogenic constructs compared to those treated with stromal constructs (1.488 ± 0.3843-fold) and those treated with PBS (2.240 ± 0.4108-fold). *Tnc* also had a trend of upregulation in calcaneal tendons treated with tenogenic constructs compared to those treated with stromal constructs (1.503 ± 0.5312-fold) and those treated with PBS (2.824 ± 0.8847-fold).

Although the trend of larger difference between tenogenic constructs and PBS than stromal constructs remained same, expression of *Scx* in calcaneal tendons treated with tenogenic constructs appeared to have downregulated compared to those treated with stromal constructs (0.7536 ± 0.1462-fold) or unchanged compared to those treated with PBS (1.035 ± 0.04775-fold). And the similar trend was observed also with the expression of *Tnmd*. The expression of *Tnmd* in calcaneal tendons treated with tenogenic constructs appeared to have been unchanged compared to those treated with stromal constructs (1.137 ± 0.2151-fold) or slightly upregulated compared to those treated with PBS (1.311 ± 0.2681-fold).The same trend of higher tendon-specific gene expressions in calcaneal tendons treated with tenogenic constructs compared to contralateral tendons treated with controls was not consistently found in RNU rats (Fig 3.15).

However, *Col1a1* had a trend of upregulation in calcaneal tendons treated with tenogenic constructs compared to those treated with stromal constructs (1.783 ± 0.07086-fold) and those treated with PBS (1.924 ± 0.6127-fold). In contrast to the trend observed in SD rats, the differences between tenogenic constructs and stromal constructs appeared to be larger than PBS in *Col3a1*, *Tnc*, and *Tnmd*.

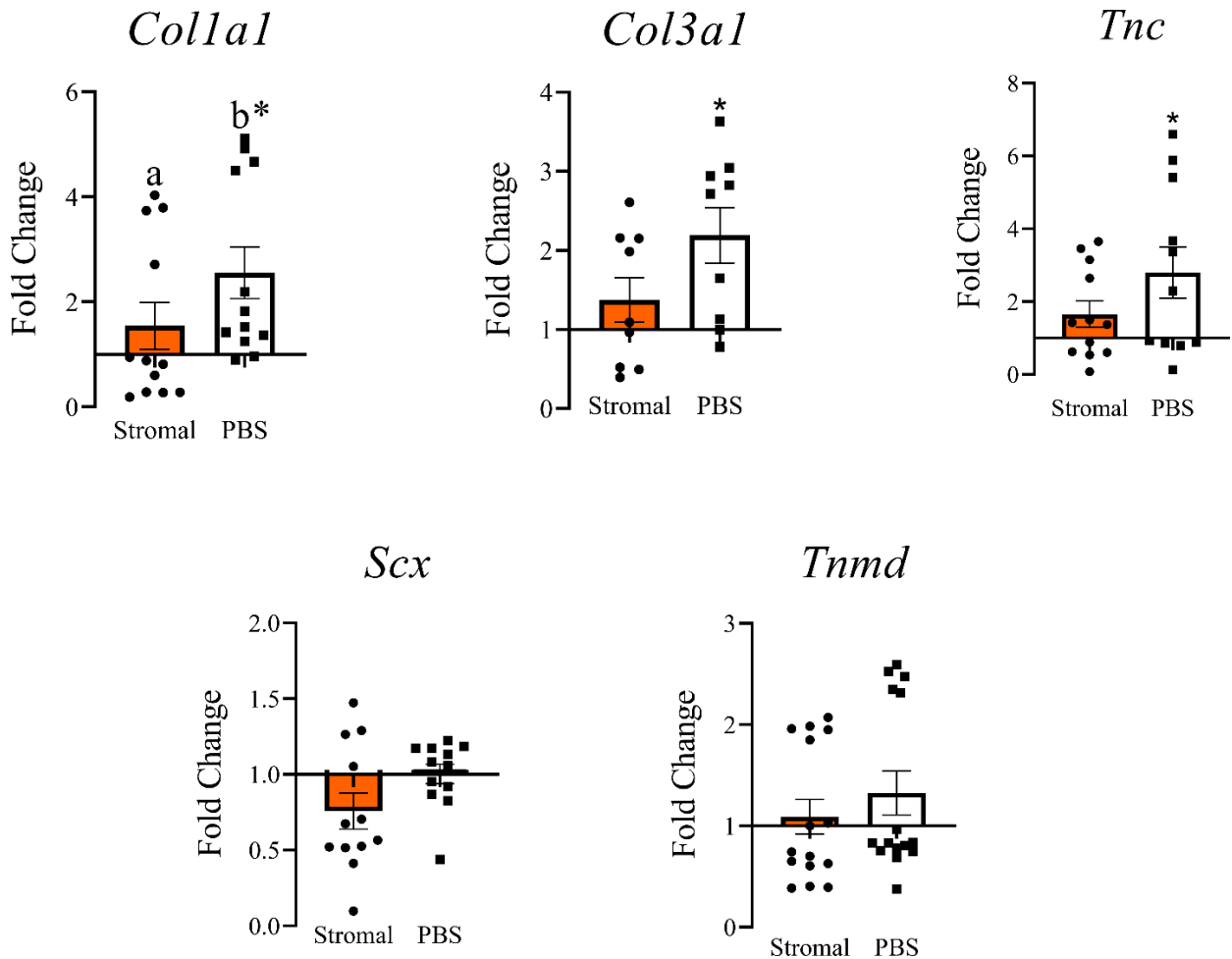

Figure 3.14. Tendon-specific gene expression of SD rat calcaneal tendons treated by tenogenic constructs, stromal constructs (Stromal), or PBS 6 weeks post-injury. Gene expressions of calcaneal tendons treated with tenogenic constructs were normalized to those treated with stromal constructs (orange) or PBS (white) and expressed as fold change. Data presented as mean ± SEM. Asterisks indicate difference from 1-fold within each stromal construct and PBS group (difference from tenogenic construct group). Different lowercase letters indicate difference between stromal construct and PBS groups.

For example, *Col3a1* had a trend of upregulation in calcaneal tendons treated with tenogenic constructs compared to those treated with stromal constructs (4.010 ± 0.1538-fold) and those treated with PBS (1.308 ± 0.2276-fold). *Tnc* also had a trend of upregulation in calcaneal tendons treated with tenogenic constructs compared to those treated with stromal constructs (2.674 ± 0.8937-fold) and those treated with PBS (1.755 ± 0.5960-fold). And the similar trend was observed also with the expression of *Tnmd*. The expression of *Tnmd* in calcaneal tendons treated with tenogenic constructs had a trend of upregulation compared to those treated with stromal constructs (1.722 ± 0.1576-fold) and those treated with PBS (1.095 ± 0.05383-fold). The expression of *Scx* in calcaneal tendons treated with tenogenic constructs appeared to be unchanged compared to those treated with stromal constructs (0.8518 ± 0.1174-fold) or upregulated compared to those treated with PBS (1.779 ± 0.5199-fold).

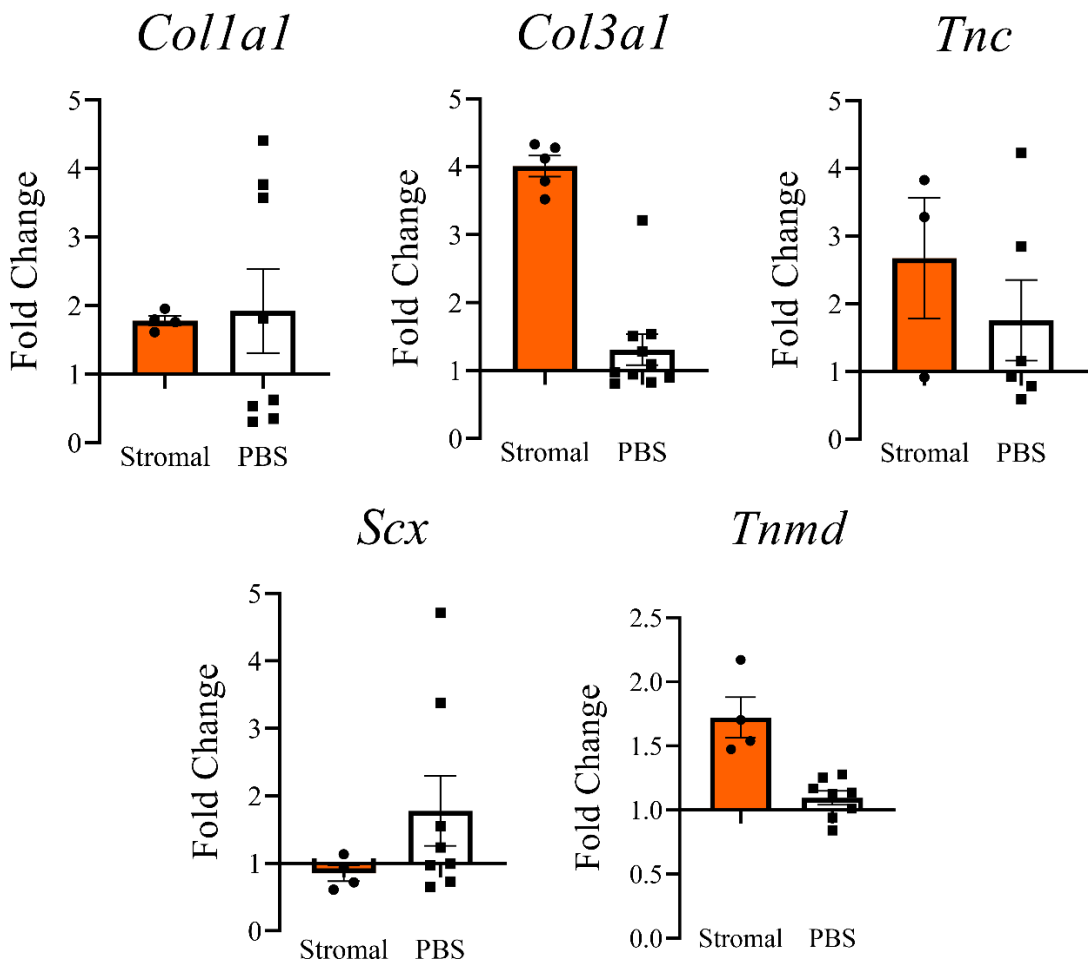

Figure 3.15. Tendon-specific gene expression of RNU rat calcaneal tendons treated by tenogenic constructs, stromal constructs (Stromal), or PBS 6 weeks post-injury. Gene expressions of calcaneal tendons treated with tenogenic constructs were normalized to those treated with stromal constructs (orange) or PBS (white) and expressed as fold change. Data presented as mean ± SEM. Asterisks indicate difference from 1-fold within each stromal construct and PBS group (difference from tenogenic construct group). Different lowercase letters indicate difference between stromal construct and PBS groups.

## 3.4. Discussion

### 3.4.1. Main Findings

The major findings of this study were: 1) elongation-induced calcaneal tendon of both immunocompetent and immunocompromised rats had minimum morbidity on general conditions, yet affected limb use was detectable using established rubric; 2) the recovery of impaired limb use corroborated with healed calcaneal tendon with minimum microstructural derangement; 3) the constructs implanted via needle into tendon remained withing implantation cites during 6 weeks post-operative period without clear signs of immune response; 4) however, the tenogenic constructs formed tendon-like neotissue with elongated cells aligned in parallel to longitudinal axis of tendon, while stromal constructs transformed amorphous tissue with randomly aligned cells that contain numerous vacuoles in SD rats; and 5) tenogenic and stromal constructs both formed tendon-like neotissue with elongated parallel cells with abundant ECM that appeared to have integrated with native orthotopic tendon tissue. Thus, our study suggested neotendon tissue, especially the one cultured in tenogenic medium, had a potential to be integrated into native tendon tissue and augment healing both in immunocompetent and immunocompromised animals.

### 3.4.2. Functional Outcome

The elongation-induced injury of rat calcaneal tendons rendered minimum morbidity to the general condition of animals, while eliciting detectable effects on limb use. The injury created in our study by elongation device was uniquely representative of acute overstrain calcaneal tendon injury. In human, acute calcaneal tendon rupture is the most common tendon rupture in the lower extremity, and occur typically in individuals who are active only intermittently,[196] a different etiology from overuse calcaneal rupture. Therefore, although calcaneal tendon rupture was not caused in animals, our elongation-induced injury was likely to represent the most common type of calcaneal tendon injury in human. A similar strain-induced tendon injury model was reported previously using cyclic strain of rat patellar tendon to attain fatigue damage accumulation.[197] And this type of injury represents chronic overuse-induced tendinopathy rather than acute tendon injury.

The limb use of rats after elongation injury in the present study showed immediate impairment of limb use, yet limb use gradually recovered in both immunocompetent and immunocompromised rats. However, impaired limb use was detectable only with the use of established rubric, and rats were ambulatory throughout the study period. Limb use evaluated in our study using subjective scoring showed no difference among treatments regardless of rat phenotypes. The reasons for no difference in limb use among treatments may be subjective scoring system or lower morbidity rendered to the limb use by elongation injury. It has be reported more invasive injury creation using enzymatic degradation or surgical transection of tendon led to the detection of impaired limb use only by the gait analysis using force measuring platform. For example, a study evaluating limb use after collagenase-induced patellar tendon injury in rats observed rather subtle impairment of gait parameters by injuries, and parameter that was associated with pain was double stance duration.[198] Additionally, step length and ground reaction forces were the only affected parameters after RC tear and repair in rats, which was also not different from uninjured limb by day 28,[199] indicating objective gait analysis system is necessary to detect changes in gait after tendon injury and treatment. Another reason for no difference among treatments maybe the needle implantation itself might have rendered limb use impairment that potentially had larger effects on limb use than treatments themselves. For example, it was reported a simple needle injection of saline into tendon severely impaired many of gait parameters evaluated in rats.[198] Therefore, the injury type and treatment administration modality employed in the present study might require objective gait analysis system to detect treatment effects.

Of note was that although not statistically different, the recovery of limb use in immunocompromised rats appeared to be slower than that of immunocompetent rats in the present study. This indicated immune system plays an important role in healing of tendon injury. As athymic RNU rats used in the present study were depleted of T cells, it was suggested T cells may be a crucial part of healing process upon elongation injury. Indeed, in a previous study, it was reported M1/M2 macrophage ratio peaked at day 3, while T helper and Treg cells increased over time during healing of transected calcaneal tendon.[200] The importance of T cells in tendon healing was also demonstrated in a study reporting expression of chemokine receptor 2 (CCR2), an important surface marker normally expressed in circulating bone-marrow monocytes and a mediator of macrophage recruitment, in both macrophage and T cells of healthy tendon. And knockout of CCR2 resulted in reduced numbers of myofibroblasts and reduced functional recovery during late healing.[201]

### 3.4.3. Post-Elongation Percentage and Tendon Width

With regard to tendon width, it is commonly seen the size of healing tendon assessed by cross-sectional area (CSA) or thickness increase after injury. Indeed, ratio of CSA of ruptured calcaneal tendons to that of contralateral healthy tendons increased gradually upon primary surgical repair up to 6 months and decreased by 12 months in human.[202] And several reports indicated cellular treatments decrease size of healed tendon. For example, injection of ASCs was reported to decrease thickness of recalcitrant patellar tendinopathy along with decreased pain scores in 6 months period.[203] On the other hand, other studies reported increased CSA coincided with increased maximum load and stiffness of rat calcaneal tendon both after 2 and 4 weeks of mid-substance defect creation followed by tendon stem/progenitor cell sheet wrapping compared to non-treatment.[178] In the present study, the width of the healed calcaneal tendons treated by tenogenic constructs were larger than those treated by PBS, which was consistent with the later reports. The difference between those studies reporting contradicting effects of cellular therapies on size of healed tendon may largely stem from differences in phases of tendons healing. And it may be case that cellular therapies including neotendon in the present study increase tendon size during early phase of healing after acute injury, whereas treatments decrease size of chronically injured tendon. Combined, increase of healed tendon width at earlier stages and decrease at later stages both may represent a better healing.

In the present study, average 50 – 70% elongation of calcaneal tendon immediately after strain application was achieved for both immunocompetent and immunocompromised animals. It is an extensive elongation which may result in acute rupture based on established stress-strain curve of tendons, since macroscopic rupture starts at above 8% elongation in human.[204] In a previous study, the failure strain of human calcaneal tendon was on average 12.8% for the bone-tendon complex and 7.5% for the tendon substance at the 1% per second rate strain application, whereas mean failure strain was 16.1% for the bone-tendon complex and 9.9% for the tendon substance at 10% rate.[205] In uninjured calcaneal tendon of rat, 2 bundles that consist calcaneal tendon, plantaris longus and gastrocnemius tendons, both reached approximately 20% strain at failure.[206] Although the strain at failure of rat calcaneal tendon was reported to be lower than that of our study potentially due to different mechanisms of strain application, evidence suggests rat calcaneal tendon tends to withstand higher strain without rupture than human, and it may have led to high elongation percentage achieved in the present study. This discrepancy may be attributed to extensive rotation of 2 calcaneal tendon bundles in rat,[207] which potentially allows more elongation under stress.

In human, elongation of tendon itself occurs normally after healing of acute rupture. For example, human calcaneal tendons were reported to elongate 0.15 to 3.1 cm after acute injury and healing,[208] approximately 0.8 – 16% elongation of original length, given average length of calcaneal tendon is approximately 18 cm.[209] In rats, strain at failure of transected and healed calcaneal was reported to be approximately 1.0 cm,[210] which is close to 70% elongation for non-repaired calcaneal tendon. It was also reported primary repair of transected rat calcaneal tendon maintain original length of approximately 1.0 cm post-operatively, whereas non-repaired tendon resulted in callus formation and approximately 60% increase of original length to 1.6 cm after 8 weeks.[211] Although elongation achieved in the present study is not directly comparable to post-injury calcaneal tendon elongation commonly observed in human after acute rupture, our model may recapitulate a clinical scenario of strain-induced acute calcaneal tendon rupture in human most closely among currently available tendon injury models.

### 3.4.4. Microstructure

Regardless of recipients' immune status, the present study revealed implanted constructs, especially those cultured in tenogenic medium, maintained morphologically viable and assuming orthotopic tendon characteristics. When placed to fill calcaneal tendon defect of rat, COLI sponge itself increased ultimate failure load, shortened healed tendon, and decreased stiffness than empty defect.[212] Therefore, COLI sponge has tenoconductive properties. Using this tenoconductive properties, COLI sponge's most prominent clinical use is to augment primary repair as a patch most often combined with RC tear repair.[213] The advantage of COLI augmentation was reported to be reduced radiographic retear rate, increased complete healing rate, and improved functions.[213,214] Indeed, cells infiltrate and form immature tendon like neotissue 6 month after implantation in human, yet COLI implant itself was found to be absorbed and replaced by newly formed tissue.[214] Given its relatively small size compared to repaired RC and non-parallel collagenous fibers of neotissue, it was unclear whether neotissue formed contributed directly to the increased function. Or it may be possible the residual cells within neotissue exerted paracrine effects to promote natural healing of repaired RC.

This is true especially because addition of cells to collagenous patch led to increased RC healing and mechanical property in rat, indicating augmentation of healed tendon size may not be the primary benefits of the treatment. In that study, bovine decellularized pericardium colonized with ASCs led to better fibrocartilage and tidemark formation at the bone-to-tendon interface, increased collagen fiber density, improved fiber orientation, and increased load-to-failure than patch only treatment.[215] And survival of neotissue implant into tendon may not depend on the viability of recipient tissue. For example, it had been shown that collagen sponge seeded with MSCs maintain high cellularity and viability within construct under controlled cyclic tensioning after implanted into ovine tendon *ex vivo*. Although the implants were placed non-viable tissue ex vivo in that study, cells maintained viability and uniform distribution within construct most likely due the pumping of sponge from cyclic tensioning action that leads to exchange of oxygen and nutrients between inside and outside of construct.[216] Therefore, survival of cells within implanted neotissue in the present study might be aided by the stress-relaxation cycles applied to recipient tendons during animals' ambulation.

The constructs implanted into calcaneal tendons of both immunocompetent and immunocompromised rats had dense ECM structure with few void spaces populated by numerous viable cells. The much denser structure of constructs found in our study may be due to the presence of tenogenically-induced MSCs or longer post-implantation period after which constructs were examined histologically. On the contrary, it was reported implanted COLI sponge without cells into the gap of transected calcaneal tendon of rats formed much less denser crisscross meshwork of collagen fiber bundles that was infiltrated by endogenous fibroblastic cells at 2 weeks.[212] Nonetheless, the less dense structure observed in the COLI implanted calcaneal tendon did not lead to inferior mechanical properties compared to the untreated gap that had much dense cellular and ECM structure.[212] Therefore, it is important to evaluate the mechanical properties of healed tendons treated by our constructs in the future. In terms of the difference between the appearance of constructs cultured in tenogenic and stromal medium, those cultured in stromal medium were more disorganized with randomly oriented round to oval cells surrounded by much fewer collagenous ECMs stained blue with trichrome staining[217] after implanted into immunocompetent rats. This contradicted the similar composition of constructs cultured in both stromal and tenogenic media after implanted into immunocompromised rats. This discrepancy revealed the roles that immunology plays to maintain phenotype of implanted constructs and direct maturation of implanted neotissue towards orthotopic tissue.

In terms of immunogenicity, pre-implantation differentiation may increase immunogenicity of cells. For example, it was reported that undifferentiated xenogenic human MSCs had better retention in subcutaneous space of immunocompetent mice with less lymphocyte and macrophage infiltration than osteogenically differentiated MSCs.[218] Similarly, allogenic MSCs differentiated to myocytes increased their immunogenicity by upregulating major histocompatibility complex (MHC) -Ia and -II, resulting in earlier elimination from infarcted heart of rats and lost functional improvement by 5 months.[219] Therefore, pre-implantation differentiation may elicit a stronger immune response to both xenogenic and allogenic MSCs, which was in contrast to our findings. On the contrary, pre-implantation differentiation is likely to be beneficial in terms of maintaining orthotopic phenotype. Previously, MSCs delivered in collagen hydrogel implanted into the central defects of rabbit patellar tendons resulting in ectopic bone formation up to 28% of treated tendons.[34,220] This indicates that undifferentiated MSCs do not tend to maintain undifferentiated state upon implantation and may differentiate into ectopic lineages. Our study added an insight on mechanisms of this ectopic tissue formation of MSCs to be partly immune-related. And the importance of immune status as well as scaffold types was also previously demonstrated clearly in a study investigating sarcoma formation by implanted MSCs. In that study, host-derived sarcomas developed, when MSC/collagen sponge constructs were implanted subcutaneously into syngeneic and immunocompromised mice, but not when allogeneic MSCs were used or MSCs were injected as cell suspensions.[221]

It is of noteworthy that there was no clear evidence of immune response to construct cultured in tenogenic medium after implantation into immunocompetent animals despite of xenogenic materials used for neotendon formulations. The signs of foreign body reaction including mononuclear cell infiltration and fibrotic encapsulation of implant were observed in a previous study using human decellularized dermis implanted into rats.[222] Those signs were not evident in either of constructs cultured in tenogenic or stromal medium in our study, suggesting that the roles immune played in formation of amorphous tissue in immunocompetent animals was not likely foreign body reaction to eliminate implants. An example of interplay between immune system and differentiation of MSCs was reported in a study that demonstrated myelodysplastic syndrome (MDS) cells, a type of aberrant hematopoietic stem cells (HSCs), impair osteogenesis of MSCs in HSCs' niche that resulted in reduction of hematopoiesis. Interestingly, enforced osteogenic differentiation of MSCs rescued normal hematopoiesis of MDS mice.[223] Therefore, orthotopically differentiated MSCs are essential components to maintain homeostasis of MSCs-immune interactions at each specific niche. Specific to tenocytes, it was reported activated tenocytes by pro-inflammatory cytokines induced partial M1 phenotype polarization by upregulating CD80 surface expression and downregulating HLA-DR expression, both M1 phenotype markers, of macrophage.[224] And this M1 polarization of macrophage appeared to be an important component of normal tendon healing, as exosomes from M2 macrophage promote peritendinous fibrosis,[225] a common complication following tendon injuries. In human, the involvement of macrophages during tendinopathic state was reported by increased presence of macrophages in tendinopathic calcaneal tendons.[226]

In our study, calcaneal tendons had mild signs of tendinopathy such as increased cellularity or loss of parallel cell alignment along longitudinal axis regardless of treatment in both immunocompetent and immunocompromised rats, and lacked signs of severe tendinopathy such as calcified deposits or chondroid formation that can be formed by collagenase injection.[227] Additionally, the lesion was not limited to the mid-substance and found inconsistently throughout the length of the tendons. This may be attributed to the large size of hook used to lift tendons in our study and hollow cylinder that was used to prevent slippage of tendon while applying strain. As a result, the stress might have concentrated at both proximal and distal contacts between the cylinder and tendon. Moreover, robust healing capacity of rat calcaneal tendon was evident from the study reporting reattachment and histological characteristics of only increased cellularity and lost parallel fiber orientation as early as 21 days after transection and saline treatment.[228] Robust healing of tendon was further prominent especially in young rats compared to old rats. For example, 20 months old SD rats showed lower histological scores and more adipocyte accumulation combined with decreased synthesis of tendon-related proteins compared to 2 months old rats[229] which are close age to rats used in our study. Therefore, the effects of each treatment on elongation-induced calcaneal tendon injury was not clearly delineated between treatments in the present study.

## 3.5. Conclusion

Collectively, our study indicated a novel elongation-induced calcaneal tendon bilateral injury model developed for both immunocompetent and immunocompromised rats closely recapitulated acute human calcaneal tendon strain or rupture etiopathologically, yet rendered minimum morbidity to general conditions of animals. The model demonstrated robust capability to clearly represent treatment responses that can be depend both on neotissue types but also on immunological status of recipient animals. Neotissues implantation techniques developed in the present study via needle and viable implanted neotissues in orthotopic tendon recipient tissue established neotissue delivery system without compromising neotissue viability and structural integrity. This was a major advancement in applicability of neotissue without the need of surgical implantation. Additionally, tenogenic constructs demonstrated it maintained tendon-like phenotype throughout the healing process in immunocompetent animals, which was not attainable by stromal constructs. The neotendon cultured in tenogenic medium may maintain tendon-like phenotype upon implantation into injured tendon and augment healing by both integrating into native tissue and facilitating endogenous healing by paracrine effects.

# Chapter 4. Therapeutic Effects of De Novo Tendon Neotissue on Accessory Ligament of Deep Digital Flexor Tendon Core Injuries

## 4.1. Introduction

Tendinopathy and desmitis comprise a large majority of musculoskeletal injuries that are responsible for up to 72% of lost training days and 14% of early retirements by equine athletes.[12-14] Many acute and chronic tendon and ligament lesions are thought to result from focal accumulation of microtrauma and poorly organized repair tissue that can coalesce into large lesions and predispose to spontaneous rupture across species.[18] Low cell numbers and metabolic activity, limited blood supply, and failure of endogenous tenocytes and ligamentocytes to migrate to the injury site contribute to poor tissue healing capacity.[109,110] Poor or abnormal tissue repair contributes to a reinjury rate in horses as high as 67% within 2 years.[20,21]

Administration of exogenous adult multipotent stromal cells (MSCs) is reported to augment natural healing in tendon and ligament injuries.[29-31] However, treatment efficacy of undifferentiated MSCs is limited by low retention rate[30,33] and ectopic formation of unwanted tissue types.[34,35] Evidence suggests delivery of MSCs cultured on scaffold template and differentiated into specific lineages prior to implantation improve retention and function of target organ.[120] By differentiating equine adipose-derived MSCs (ASCs) into tenocytes on biocompatible[112] commercially available FDA-approved collagen type I (COLI)[113,114] that is known to support retention of tendon stem cells[115] and multiple differentiation of MSCs,[112,116,117] we have generated *de novo* tendon neotissue that resemble proliferation-phase healing tendon. Tendon neotissue was biocompatible when orthotopically implanted into immunocompetent rat calcaneal tendon without eliciting immune response, and formed immature tendon-like tissue with excellent integration into surrounding native tendon tissue. To further advance this technique towards clinical application, evaluation of tendon neotissue in validated equine preclinical tendinopathy model is necessary.

Equine tendinopathy, especially naturally-occurring tendon injury, resembles human tendinopathy etiologically and histologically most closely among many species. Tendinopathy in both species occurs as a result of accumulated damage after prolonged, high-intensity exercise, and similar extracellular (ECM) changes observed at microstructure level.[230] Ultrasonographically, tendon lesions in both species appear as hypoechoic region where teared tendon retract and void space filled with fluid.[231,232] Majority of equine tendinopathy occurs in forelims,[233] and the prevalence of specific tendons affected by tendinopathy was reported to be 11.1% for forelimb superficial digital flexor tendon (SDFT), 3.61% for forelimb suspensory ligament (SL), 0.06% for hindlimb SDFT, and 0.14% for hindlimb SL.[234] Core lesions are among the most common type of lesions seen in acute equine tendon and ligament injuries.[235] The several mechanisms were suggested as the reason for lesion formation at the core of tendon among horses. One of them is central fibrils' tendency to reach the end of the stress-strain curve toe region earlier than fibrils at the periphery, resulting in stress concentration at the central core region.[236] Another mechanism of core lesion formation is the thermal damage. It was reported core of equine SDFT reaches temperatures as high as 45°C during high-speed locomotion,[237] and local temperature rise above 42.5°C was shown to negatively affect viability of fibroblast cell.[238]

To replicate the clinical scenario for testing of novel treatment options, lesions can be created in the SDFT by chemical[47] and surgical[30,239,240] means. Among chemical induction of tendinopathy models in horses, injection of collagenase is most commonly used method due to its relative ease in creating lesions. The method was originally reported as a model for naturally-occurring tendinopathy due to its similarity in clinical and histological course following injury.[241] To date, the model has been extensively utilized to evaluate the efficacies of multiple modalities including pharmaceutical, cellular, and gene therapies.[242-245] However, the lesions created by collagenase does not morphologically resemble naturally-occurring core lesions both ultrasonographically and histologically,[242,246] and is often associated with uncontrollable inflammation in surrounding area. Moreover, it was reported that tendinopathy induced by collagenase results in poor correlation between clinical presentation and microscopical characteristics of lesion.[247] Elastography, a ultrasound-based modality to measure tissue elasticity, is reliably applied to assess naturally-occurring core lesion of equine tendinopathy,[248] yet outcomes of elastography assessment in collagenase-induced tendinopathy was also reported to be poorly correlated with biochemical changes of tendinopathy,[249] further demonstrating the limitations of collagenase-induced tendinopathy model.

On the contrary, mechanical disruption with surgical instrumentation reduces the risk of excessive inflammation and allows more control of lesion size than chemical injury.[240] Surgical creation of core lesion in equine tendon is most commonly performed by the use of arthroscopic burr to make tunnel at the center of SDFT.[30,240,250] The approach has advantages of not only being able to control the size of lesion without causing excessive and uncontrollable inflammation, but also it creates pathologically similar lesion to naturally-occurring tendinopathy. Those characteristics included disrupted collagen fibers, increased glycosaminoglycan content, increased presence of tenocytes with plump nuclei, the scarcity of inflammatory cells, increased matrix metalloproteinase (MMP) activity, and neovascularization.[250] The model has been widely used to evaluate the efficacy of cellular therapies, with which tenogenically differentiated MSCs and undifferentiated ASCs improved regeneration while minimizing scar formation characterized by increased collagen type I deposition and collagen crosslinking with minimum deposition of collagen type III and smooth muscle actin.[30,46]

Core lesion creation in SDFT simulates clinical scenario most closely. Additionally, equine

SDFT, an equivalent of human calcaneal tendon, shares functionality both as critical energy-storage systems during high-speed locomotion and pathophysiology of tendinopathy from accumulated exercise- and age-related microdamage preceding rupture.[251] This makes equine SDFT tendinopathy model also an ideal model for translational preclinical model for human tendinopathy research. However, despite of the robust similarity to clinical scenario and translatability to human tendinopathy application, SDFT tendinopathy requires euthanasia for evaluation of tissue regeneration to harvest injured tendon. In this regard, tendinopathy model creation in accessory ligament of the deep digital flexor tendon (ALDDFT) is an appealing alternative, as desmotomy of ALDDFT does not impair horse's performance and euthanasia is not necessary for tissue harvest.[252] Despite ligament and tendon are anatomically distinguished, genetic lineage tracing revealed that both tenocytes and ligamentocytes arise from the same progenitor pool,[253,254] suggesting potential interchangeability of both lesion model and treatment between tendon and ligament. Due to lower incidence of ALDDFT injuries in horses, it is also often considered non-load bearing ligament, yet more than 2-fold higher strain experienced on ALDDFT than SDFT, deep digital flexor tendon (DDFT), SL at walk was reported.[255] Additionally, naturally-occurring tendinopathy of ALDDFT is frequently presented as hypoechoic core lesion-like area similar to that of SDFT ultrasonographically.[256] Using naturally-occurring ALDDFT tendinopathy, single injection of tenogenically induced MSCs injection was found to improve ultrasonographical appearance in 80% of animals.[257] Therefore, it is likely that experimentally-induced ALDDFT tendon injury model provide robust yet humane preclinical model to assess novel regenerative therapies.

In this work, a novel robust yet nonterminal equine ALDDFT desmitis model by surgical creation of core lesion was established. The model consisted mechanical disruption of ALDDFT under controlled surgical conditions, similar to the SDFT model, and then subsequent harvest by minor modification of the standard surgical technique for ALDDFT desmotomy.[258] The established model was then validated for its robustness to evaluate therapeutic potential of tendon neotissue. The hypotheses tested in this study were: 1) core lesion creation in ALDDFT does not cause significant morbidity; 2) core lesion created in ALDDFT resembles that of SDFT ultrasonographically, allowing morphological evaluation of lesion and implant; and 3) tendon neotissue augment healing of lesion than COLI template or ASCs alone treatments microstructurally. The aims of this study were to evaluate horses pre-operatively and 11 weeks post-operatively for: 1) objective gait analysis; 2) ultrasonographic morphology of lesion and implant; and 3) histological characteristics of lesion and implant upon treatment by tendon neotissue, COLI template, and ASCs.

## 4.2. Materials and Methods

### 4.2.1. Study Design

Protocol (#21-076) was approved by the Louisiana State University Institutional Animal Care and Use committee prior to study initiation. Three adult mares (body condition 4 - 7, 5 - 15 years, 425-500 kg) without evidence of forelimb lameness or systemic illness based on physical exam were be selected from the Louisiana State University research herd. A core lesion was created in right forelimb ALDDFT using a technique modified from validated lesion creation in SDFT.[30,240] Based on existing knowledge, the optimum time to deliver exogenous cells to an injured tendon or ligament horses is around the sixth day after injury at the transition between the inflammatory and subacute reparative phases.[180] Hence, 6 days after surgery, ligament neotissue generated from male donor ASCs was percutaneously administered to ALDDFT core lesion in a horse via 18-gauge needle under ultrasonographic guidance. Using the same administration technique, ASCs from same donor in PBS; or COLI template were administered to the ALDDFT core lesions of the other horses. The total volume and cell number were identical between treatment and controls. Twelve weeks after lesion creation, ALDDFTs were harvested from each horse. Gait was evaluated at 5, 7, 9, and 11 weeks both after lesion creation and ALDDFT harvest. Lesion was evaluated ultrasonographically at 2, 5, 7, 9, and 11 weeks both after lesion creation and ALDDFT harvest. Harvested each ALDDFT was equally divided along the midsagittal axis. One half was tested for tensile mechanical properties while the other was assessed for ligament-specific gene expression and extracellular matrix composition as well as microstructure (Fig 5.1).

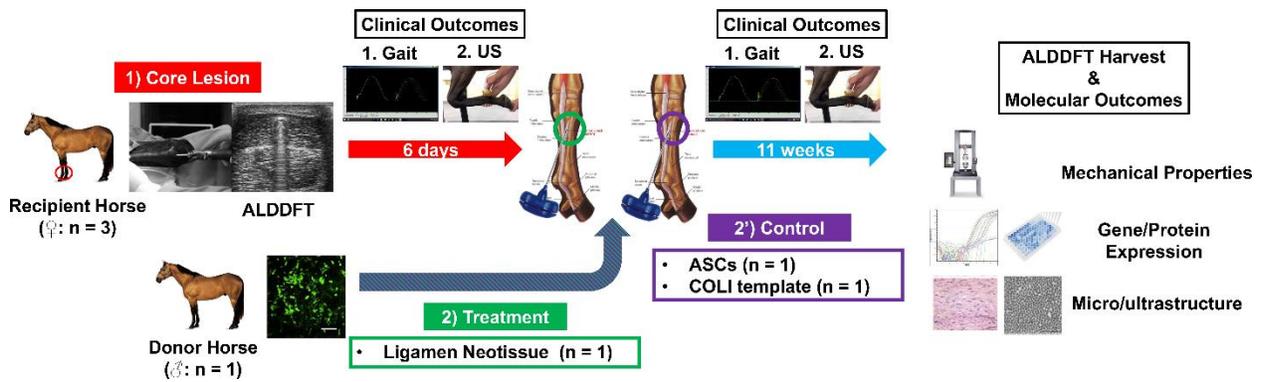

Figure 4.1. Study design of horse accessory ligament of the deep digital flexor tendon core lesion model creation and neotissue treatment.

## 4.2.2. Accessory Ligament of the Deep Digital Flexor Tendon (ALDDFT) Core Lesion Model Creation

ALDDFT core lesions were surgically created with a validated adaptation of an established SDFT surgical core lesion model[30,240]. With patient on left lateral recumbency, right forelimb was clipped with #40 blade and aseptically prepared. After appropriate drapping of the body, limb was rested on the mayo stand and table height was adjusted according to the surgeon preference. An iodine impregnated drape was applied to the limb due to its inadherent nature, and a window was cut through it to make a surgical incision centering over the proximal third of the palmarolateral area where proximal and middle thirds of the metacarpal bone III meet using #10 blade approximately of 6 cm length. A 1 cm incision was made through carpal retinaculum 3 cm distal to the origin of ALDDFT origin using a new #10 blade. Paratenon was bluntly dissected with metzenbaum scissors to isolate ALDDFT from deep digital flexor tendon (DDFT). ALDDFT was elevated using curved crile forceps, and a 2.8-mm trocar intramedullary (IM) pin (Jorgensen Laboratories, Inc., Loveland, CO) was inserted from proximal end of isolated ALDDFT, approximately 8 cm distal from accessory carpal bone.

IM pin was advanced distally inside the ligament core for 4 cm by marking IM pin with predetermined distance using sterile marker pen. Subsequently, a 2.0-mm small-joint hooded abrasion burr (2.0 mm x 8.0 mm: stryker®, 0375-641-000) was inserted slowly into the tunnel to the end of probed tunnel. Once reached the distal end of tunnel, burr was activated while gradually removing burr from tunnel. Tunnel was further widened with 4.0-mm tomcat cutter (stryker®, 375-545-000) using the same motion. Paratenon and subcuataneous tissues were sutured in a simple continuous pattern with poliglecaprone 25 (MONOCRYL® 2-0 USP; Ethicon, Norderstedt, Germany). Skin was closed using skin stapler (35 wide, Midwest Veterinary Supply, Inc., Lakeville, MN). Both longitudinal and transverse images were obtained ultrasonographically to document the size and shape of core lesion.

### 4.2.3. Neotendon Implantation

Six days after creation of the ALDDFT core lesion, 3 cylinders, 4.0 x 10.0 mm (diameter x length), of neotendon placed at the tip of an 18-gauge 1.5 inch needle (Exel international) were injected into core lesion of a horse using ultrasound guidance with the horse under standing sedation and local cutaneous/subcutaneous anesthesia with carbocaine (2%, 5 – 10 ml)[259].

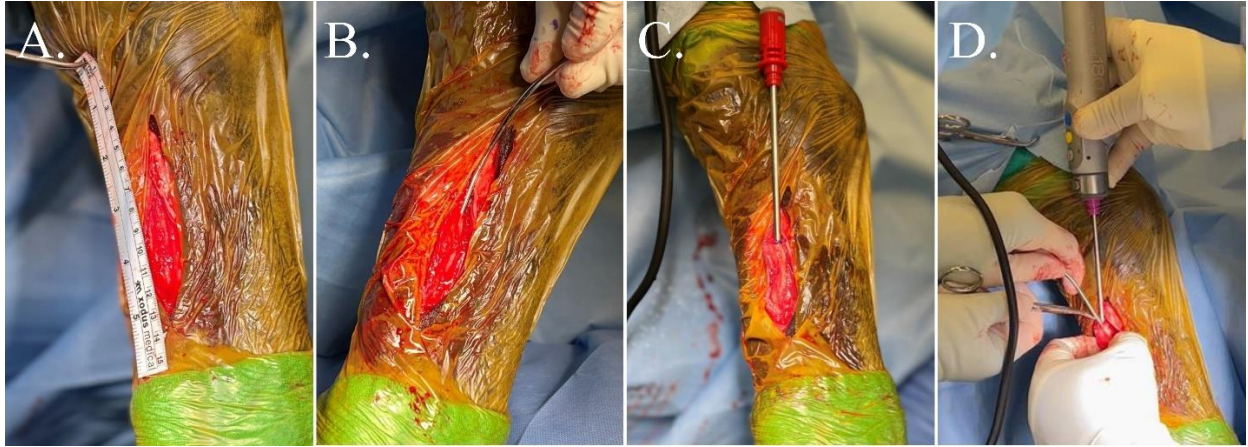

Figure 4.2. Surgical ALDDFT core lesion creation. The ALDDFT is exposed through a 6 cm incision (A). An IM pin is inserted from proximal end of isolated ALDDFT, approximately 8 cm distal from accessory carpal bone. IM pin was advanced distally inside the ligament core for 4 cm by marking IM pin with predetermined distance using sterile marker pen to create a tunnel (B). A 2.0-mm small-joint hooded abrasion burr was inserted slowly into the tunnel to the end of probed tunnel. Once reached the distal end of tunnel, burr was activated while gradually removing burr from tunnel (C). Tunnel was further widened with 4.0-mm tomcat cutter using the same motion (D).

To place neotendon into the needle tip, cylinders were collected from construct cultured for 21 days in ligamentogenic medium using 4.0 mm diameter biopsy punch after finger trap suture removal and unwrapping of construct. Neotendon cylinders were inserted from the tip of needle using stylet of 18-gauge intravenous catheter. Needle was inserted at the proximal level of core lesion under ultrasound guidance perpendicular to the limb to the center of lesion. Neotendon was administered into the core lesion by gradually advancing stylet of 18-gauge intravenous catheter from the side of hub to the predetermined length of 55.0 mm (Fig 5.3).

The other horses received ASCs in PBS or COLI template. ASCs ($3.77 \times 10^5$ cells) from same donor at P2 were resuspended in identical volume ($377 \text{ mm}^3$) and injected via needle. Three cylinders of COLI template, 4.0 x 10.0 mm (diameter x length), were injected identically as neotendon.

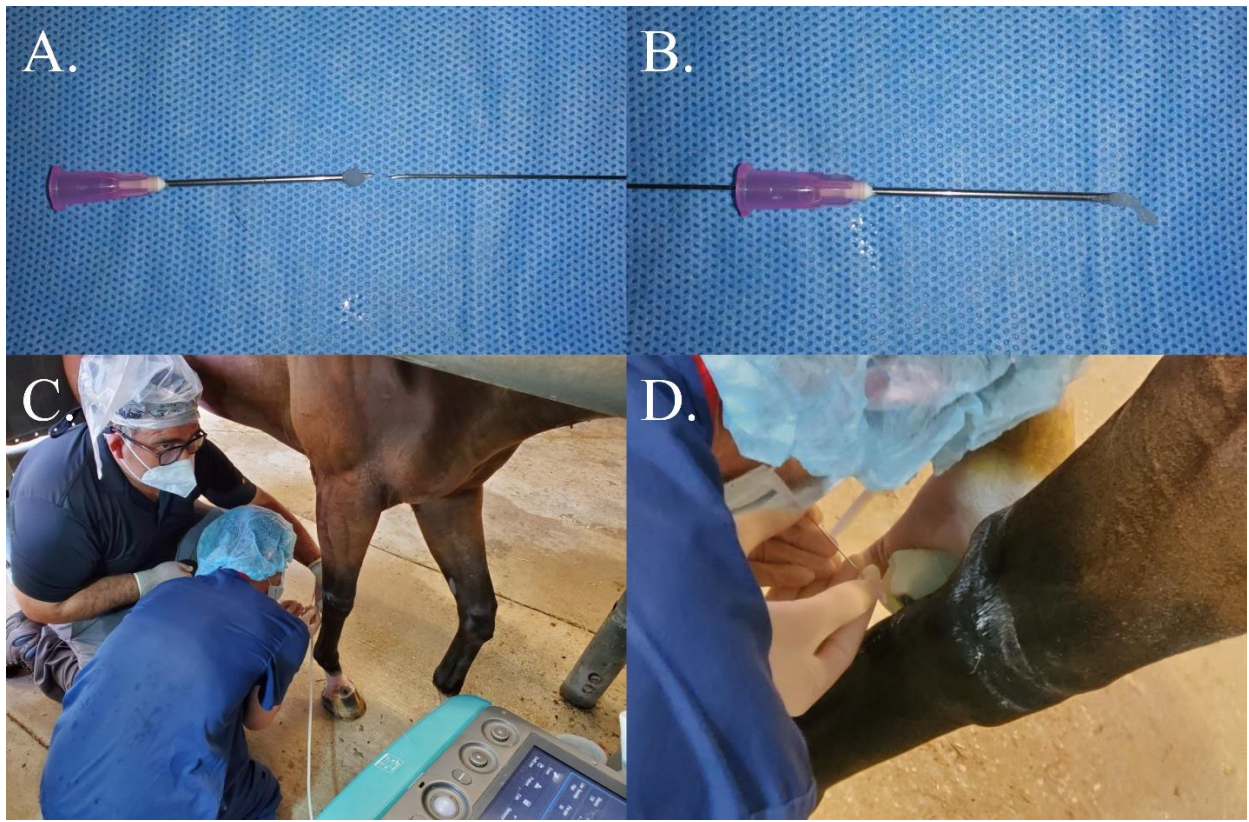

Figure 4.3. Neotendon implantation into the core lesion of ALDDFT. Neotendon cylinders were placed at the tip of 18-gauge needle and pushed inside using 18-gauge intravenous catheter stylet (A). Neotendon was ejected from the tip of 18-gauge needle by passing stylet through the needle (B). Needle was inserted at the proximal level of core lesion under ultrasound guidance perpendicular to the limb to the center of lesion (C). Neotendon was administered into the core lesion by gradually advancing stylet of 18-gauge intravenous catheter from the side of hub to the predetermined length of 55.0 mm (D).

### 4.2.4. Post-operative Care and Rehabilitation

Horses were housed in stalls from 1 day before to 5 weeks after each surgical procedure of lesion creation and ALDDFT harvest. Horses received phenylbutazone (2g, IV) the day of surgery and for 10 consecutive days (1g, PO, q12h). All horses also received sulfamethoxazole/trimethoprim (21 mg/kg, PO, q12h) for 10 days. The limb was supported with a Robert Jones bandage that was changed every other day for 14 days. Horses underwent a standard rehabilitation regimen for ALDDFT desmotomy[260] after each post-operative procedure. Specifically, horses were restricted to stall confinement without exercises for 14 days.

Subsequently, they were walked for 5, 10, and 15 minutes per day, at 3, 4, and 5 weeks post-operatively. Following completion of a 5-week rehabilitation, horses were housed in a small paddock. After 90 days in a small paddock, horses were allowed to gradually increase exercise over 18 weeks, after which they were allowed to resume full exercise.

### 4.2.5. Objective Gait Analysis

Gait analysis was conducted 1 day before and 5, 7, 9, and 11 weeks after each surgical procedure of core lesion creation and ALDDFT harvest. Horses were trotted on a 900 x 900 mm force platform embedded in the center of a 40 m concrete runway (Model #BP900900, Advanced Mechanical Technology, Inc., Watertown, MA) by experienced handlers. A minimum of 5 trials per side (left and right) were collected at a velocity of 2.0 – 3.8 m/s and acceleration of - 0.9 to 0.9 m/s$^2$. Ground reaction forces were recorded at a rate of 1000 Hz and processed with commercially available software (Acquire V7.3, Sharon Software Systems, Bengaluru, India). Measured forces included y (craniocaudal, braking and propulsion) and z (vertical) peak force and impulse. Percent change in each GRF measure was calculated as [(GRF after treatment – GRF at baseline)/(GRF at baseline)] × 100 at each time point.

### 4.2.6. Ultrasonographic Evaluation

Evaluation was conducted to obtain pre-operative, immediately post-operative, and at 2, 5, 7, 9, 11 weeks post-operative images of each surgical procedure. A real-time B-mode ultrasound machine with a 12 mHz linear array probe was used. Both longitudinal and cross-sectional images were captured to evaluate fiber parallelism and ligament versus lesion area, respectively. Images were evaluated by investigators unaware of treatments.

### 4.2.7. ALDDFT Harvest

Twelve weeks after creation of the lesion, the ALDDFT was harvested through the same surgical approach used to create the lesion. Following digital document of the ALDDFT gross appearance, the excised tissue was divided along the sagittal direction into longitudinal medial and lateral halves through the center of the lesion. Randomly selected one half was assigned to microstructure analysis. Cubes ($1.0 \times 1.0 \times 1.0$ cm) were collected from both visible lesions filled and unfilled with implants.

### 4.2.8. Microstructure

Samples collected from ALDDFT were fixed in 4 % paraformaldehyde (PFA) at 4 °C overnight, paraffin embedded, sectioned (5 μm) and stained with haematoxylin and eosin or Masson's trichrome. Digital images were generated with a slide scanner (NanoZoomer, Hamamatsu Photonics K.K, Hamamatsu City, Japan) or a light microscope (DM4500B, Leica, Wetzlar, Germany) fitted with a digital camera (DFC480, Leica) and cellular morphology, distribution, and extra cellular matrix organization were be evaluated.

### 4.2.9. Statistical Analysis

Results are presented as mean ± standard error of the mean (SEM). Normality of data was examined with the Kolmogorov–Smirnov test. Percent changes of ground reaction at each time period was compared to baseline (0%) using one-sample t-test for normally distributed data and one-sample Wilcoxon signed rank test for non-normally distributed data. Outcome measures were compared with ANOVA. When overall difference was detected, pairwise comparisons between groups were performed using Tukey's post-hoc test. All analyses were conducted using Prism (GraphPad Software Inc., San Diego, CA) with significance considered at $p < 0.05$.

## 4.3. Results

### 4.3.1. Objective Gait Analysis

The objective gait analysis of horse that received COLI template treatment showed increase in PVF on z-axis at 9 (4.99 ± 0.62%) and 11 (2.7 ± 0.32%) weeks post-injury compared to pre-injury baseline (Fig 4.4 and Table 4.1). Similarly, Imp on z-axis increased at 7 (8.55 ± 1.86%) and 11 (11.06 ± 0.21%) weeks. PVF on y(b)-axis increased at 9 (18.09 ± 2.38%) weeks, and Imp on y(b)-axis increased both at 9 (25.93 ± 4.72%) and 11 (25.93 ± 6.93%) weeks. There was no difference in PVF and Imp of y(p)-axis at all post-injury time points.

In the horse treated with tenogenic construct, PVF on z-axis increased at 5 (27.23 ± 2.109%), 7 (16.29 ± 0.8142%), 9 (23.04 ± 0.6371%), and 11 (14.46 ± 0.7545%) weeks (Fig 4.5 and Table 4.2). Imp on z-axis increased at 5 (18.48 ± 3.387%), 7 (17.34 ± 1.032%), 9 (19.37 ± 1.177%), and 11 (13.67 ± 1.291%) weeks. PVF on y(b)-axis increased at 7 (111.5 ± 16.10%), 9 (88.89 ± 10.52%), and 11 (86.83 ± 14.94%) weeks, and Imp on y(b)-axis increased both at 9 (127.8 ± 10.39%) and 11 (133.3 ± 24.22%) weeks. PVF on y(p)-axis decreased at 7 (-10.34 ± 2.729%) and 11 (-21.71 ± 4.218%) weeks, and Imp on y(p)-axis decreased at 11 (-22.58 ± 6.035%) weeks.

In contrast to those with COLI template or tenogenic construct treated horses, objective gait analysis of horse treated with undifferentiated ASCs had overall decreased parameters throughout the post injury period. For example, PVF on z-axis decreased at 5 (-3.875 ± 0.7615%), 7 (-15.95 ± 0.5830%), and 9 (-4.895 ± 0.8403%) weeks (Fig 4.6 and Table 4.3). Similarly, Imp on z-axis decreased at 5 (-16.76 ± 1.313%), 7 (-23.93 ± 0.8107%), 9 (-11.53 ± 1.669%), and 11 (-13.76 ± 0.6678%) weeks. PVF on y(b)-axis decreased at 7 (-15.74 ± 5.141%), yet increased at 11 (37.32 ± 6.718%) weeks. Imp on y(b)-axis increased at 11 (53.33 ± 9.718%) weeks. Imp on y(p)-axis increased at 9 (30.77 ± 7.195%) weeks.

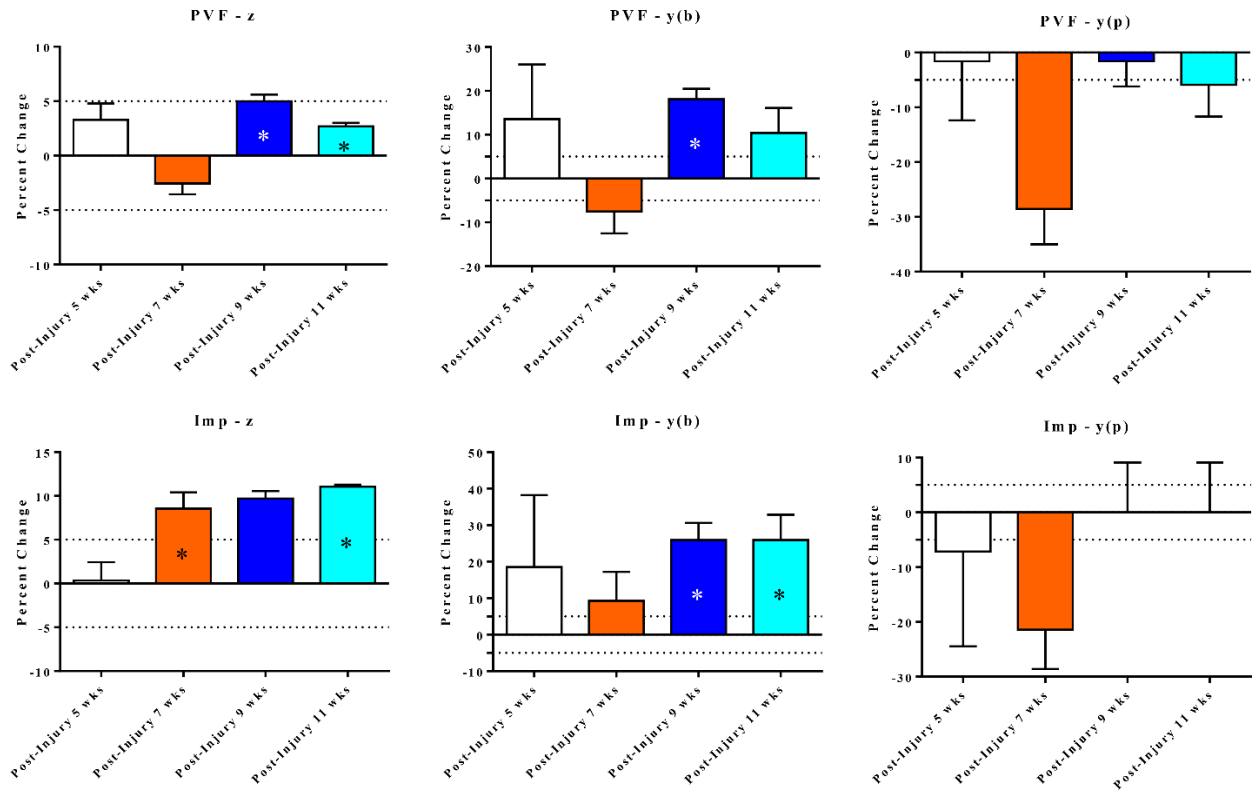

Figure 4.4. Objective gait analysis performed on a horse treated with COLI template. Percent change (mean ± SEM) in right forelimb ground reaction forces over baseline (1 day preoperative). Peak force (PVF) and impulse (Imp) in vertical (z) and craniocaudal braking (y(b)) and propulsion (y(p)) after 5 (white), 7 (orange), 9 (blue), and 11 (light blue) weeks from lesion creation. Asterisks indicate differences from baseline (0%). A 5% change from baseline (dotted lines) is indicated.

Table 4.1. Objective gait analysis parameters of horse treated with COLI template.

|  | Post-Injury 5 wks | Post-Injury 7 wks | Post-Injury 9 wks | Post-Injury 11 wks |
|---|---|---|---|---|
| PVF - z | 3.29 ± 1.5 | -2.56 ± 0.98 | 4.99 ± 0.62 | 2.7 ± 0.32 |
| Imp - z | 0.34 ± 2.08 | 8.55 ± 1.86 | 9.69 ± 0.86 | 11.06 ± 0.21 |
| PVF - y(b) | 13.57 ± 12.45 | -7.54 ± 4.99 | 18.09 ± 2.38 | 10.39 ± 5.73 |
| Imp - y(b) | 18.52 ± 19.73 | 9.26 ± 7.97 | 25.93 ± 4.72 | 25.93 ± 6.93 |
| PVF - y(p) | -1.6 ± 10.79 | -28.53 ± 6.5 | -1.6 ± 4.59 | -5.87 ± 5.82 |
| Imp - y(p) | -7.14 ± 17.31 | -21.43 ± 7.14 | 0 ± 9.11 | 0 ± 9.11 |

Percent changes of peak vertical force (PVF) and impulse (Imp) for vertical (z), craniocaudal braking [y(b)], and craniocaudal propulsion [y(p)] normalized to baseline (pre-injury) in right forelimb were measured at 5, 7, 9, and 11 weeks after lesion creation. Data presented as mean ± SEM.

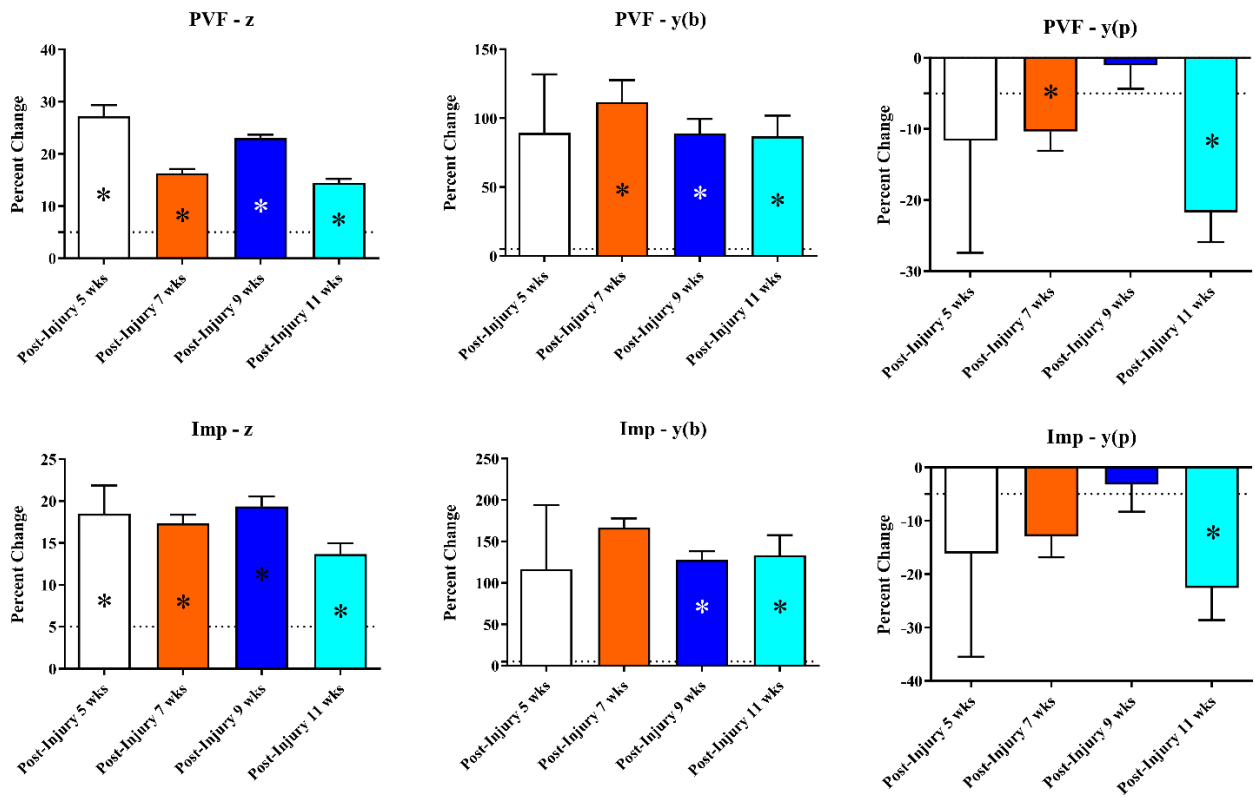

Figure 4.5. Objective gait analysis performed on a horse treated with tenogenic construct. Percent change (mean ± SEM) in right forelimb ground reaction forces over baseline (1 day preoperative). Peak force (PVF) and impulse (Imp) in vertical (z) and craniocaudal braking (y(b)) and propulsion (y(p)) after 5 (white), 7 (orange), 9 (blue), and 11 (light blue) weeks from lesion creation. Asterisks indicate differences from baseline (0%). A 5% change from baseline (dotted lines) is indicated.

Table 4.2. Objective gait analysis parameters of horse treated with tenogenic construct.

|              | Post-Injury 5 wks | Post-Injury 7 wks | Post-Injury 9 wks | Post-Injury 11 wks |
|--------------|-------------------|-------------------|-------------------|--------------------|
| PVF - z      | 27.23 ± 2.11      | 16.29 ± 0.81      | 23.04 ± 0.64      | 14.46 ± 0.75       |
| Imp - z      | 18.48 ± 3.39      | 17.34 ± 1.03      | 19.37 ± 1.18      | 13.67 ± 1.29       |
| PVF - y(b)   | 89.3 ± 42.54      | 111.5 ± 16.1      | 88.89 ± 10.52     | 86.83 ± 14.94      |
| Imp - y(b)   | 116.7 ± 77.28     | 166.7 ± 11.11     | 127.8 ± 10.39     | 133.3 ± 24.22      |
| PVF - y(p)   | -11.63 ± 15.78    | -10.34 ± 2.73     | -1.03 ± 3.31      | -21.71 ± 4.22      |
| Imp - y(p)   | -16.13 ± 19.35    | -12.9 ± 3.95      | -3.23 ± 5.1       | -22.58 ± 6.04      |

Percent changes of peak vertical force (PVF) and impulse (Imp) for vertical (z), craniocaudal braking [y(b)], and craniocaudal propulsion [y(p)] normalized to baseline (pre-injury) in right forelimb were measured at 5, 7, 9, and 11 weeks after lesion creation. Data presented as mean ± SEM.

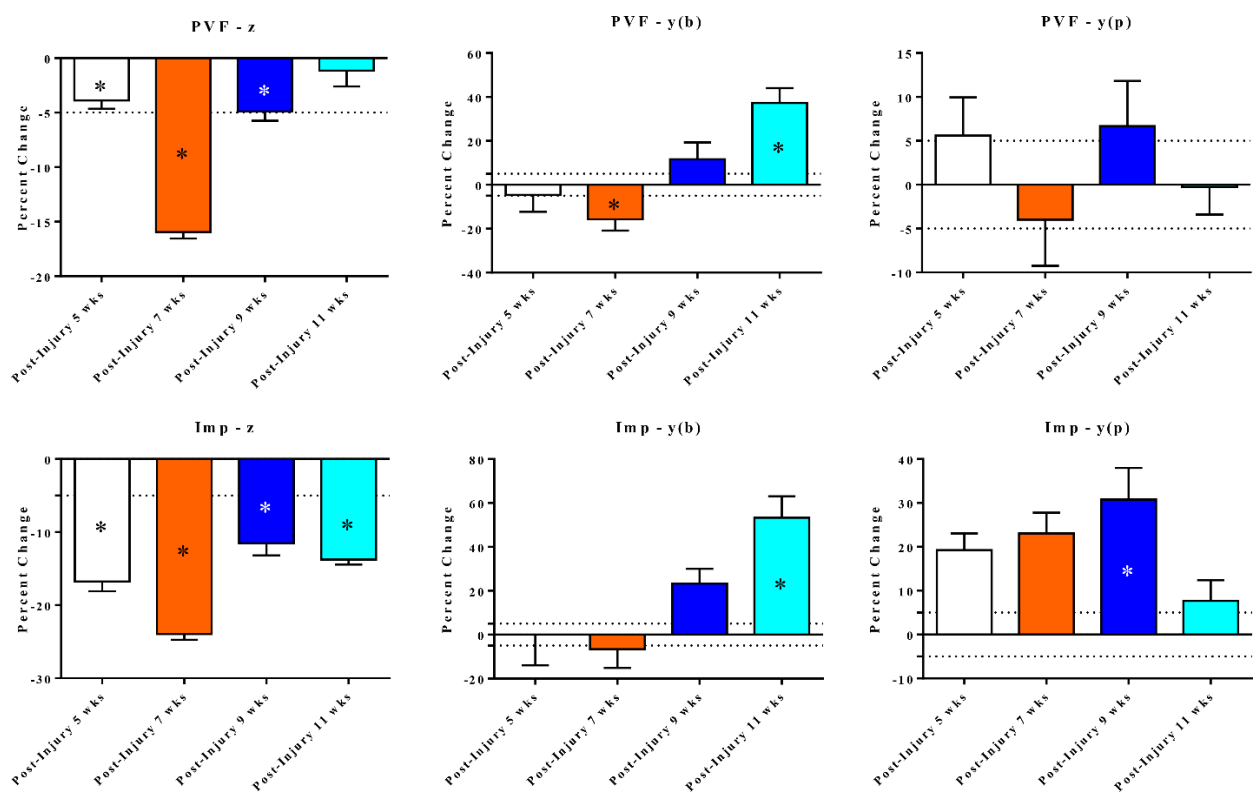

Figure 4.6. Objective gait analysis performed on a horse treated with undifferentiated ASCs. Percent change (mean ± SEM) in right forelimb ground reaction forces over baseline (1 day preoperative). Peak force (PVF) and impulse (Imp) in vertical (z) and craniocaudal braking (y(b)) and propulsion (y(p)) after 5 (white), 7 (orange), 9 (blue), and 11 (light blue) weeks from lesion creation. Asterisks indicate differences from baseline (0%). A 5% change from baseline (dotted lines) is indicated.

Table 4.3. Objective gait analysis parameters of horse treated with undifferentiated ASCs.

|  | Post-Injury 5 wks | Post-Injury 7 wks | Post-Injury 9 wks | Post-Injury 11 wks |
|---|---|---|---|---|
| PVF - z | -3.88 ± 0.76 | -15.95 ± 0.58 | -4.9 ± 0.84 | -1.13 ± 1.46 |
| Imp - z | -16.76 ± 1.31 | -23.93 ± 0.81 | -11.53 ± 1.67 | -13.76 ± 0.67 |
| PVF - y(b) | -4.67 ± 7.64 | -15.74 ± 5.14 | 11.66 ± 7.64 | 37.32 ± 6.72 |
| Imp - y(b) | 0 ± 13.94 | -6.67 ± 8.5 | 23.33 ± 6.67 | 53.33 ± 9.72 |
| PVF - y(p) | 5.6 ± 4.35 | -4 ± 5.23 | 6.67 ± 5.15 | -0.27 ± 3.14 |
| Imp - y(p) | 19.23 ± 3.85 | 23.08 ± 4.71 | 30.77 ± 7.2 | 7.69 ± 4.71 |

Percent changes of peak vertical force (PVF) and impulse (Imp) for vertical (z), craniocaudal braking [y(b)], and craniocaudal propulsion [y(p)] normalized to baseline (pre-injury) in right forelimb were measured at 5, 7, 9, and 11 weeks after lesion creation. Data presented as mean ± SEM.

### 4.3.2. Ultrasound Analysis

Core lesions initially appeared as hypoechoic round to oval areas with smooth boundaries in transverse plane, and gradually increased in echogenicity with less discernable boundaries by the end of the evaluation period regardless of treatment (Fig 4.7.A). Additionally, there was a trend of increased irregularities in morphology, and lesions conformed stellar morphology from which multiple fissures appeared to have propagated since 5 weeks post-injury.

The core lesion CSA percentage (Fig 4.7.B) increased to 250.4% of the pre-treatment CSA 4 weeks after COLI administration, after which it decreased to less than the pre-treatment CSA. After neotendon treatment, the core lesion CSA remained stable and did not exceed 122.2% of original size throughout the post-injury period. With ASC treatment, the core lesion CSA increased to 198.9% of the pre-treatment CSA by 6 weeks post-treatment, decreased to 80.5% at 8 weeks, and increased to 228.9% by 11 weeks.

The total CSA of ALDDFT itself was subjectively increased from pretreatment 1 week post-injury, immediately prior to treatment, and remained enlarged regardless of treatment. On the other hand, there was no clear effect of injury to surrounding tissue, such as enlarged neighboring tendons or induced tendon sheath effusion. Increased size of the structure was also detectable by palpation during routine physical examination. There were no pathologic changes noted in the SL, DDFT or SDFT.

The COLI and neotenton were evident in the core lesion with ultrasound immediately after administration. By 11 weeks post-treatment, echogenicity of the core lesion subjectively increased up to ≥ 90% of native tissue echogenicity, respectively, at the level of the largest CSA. Further, the fissures from lesions appeared to have closed by 11 weeks post-injury.

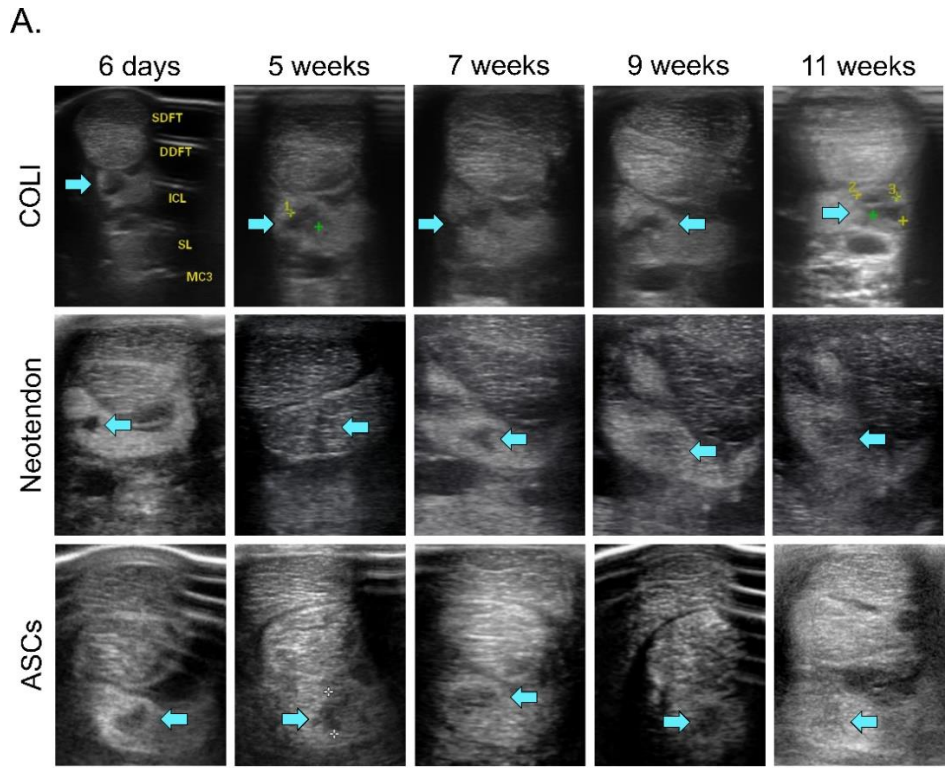

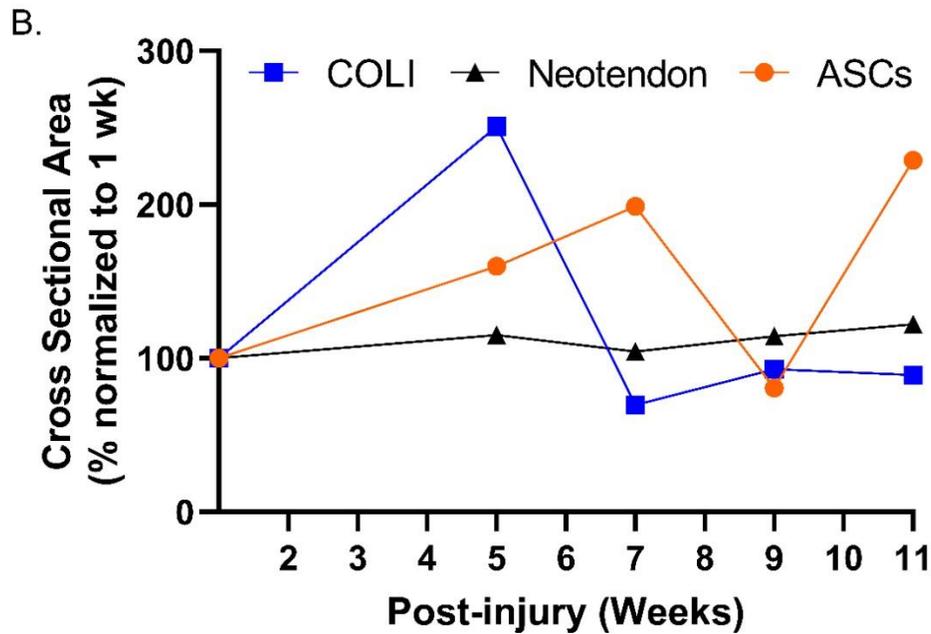

Figure 4.7. Ultrasound images of ALDDFT core lesions (blue arrows) immediately after treatment administration (6 days) and up to 1 week before harvest (11 weeks) (A) and lesion cross sectional area (CSA) percentage normalized to 6 days post-injury at each post-injury week of evaluation (B). CSA percentage was calculated as (CSA before treatment 6 days post-injury)/(CSA at subsequent post-injury evaluations) x 100. ICL: accessory ligament of the deep digital flexor tendon; MC3: metacarpal bone III.

### 4.3.3. Microstructure

In the tendon that received COLI template treatment, a distinct gross difference existed between the core lesion filled with COLI template (Fig 4.8.a and b) and empty core lesion (Fig 4.8.c and d). Implant was smooth and shiny in appearance and yellow in color. It was firmly adhered to the wall of core lesion. Implant filled almost all inner space of lesion spanning proximal half. The inner space of empty core lesion had narrower diameter, indicating contraction of surrounding native tendon tissue. The core lesion filled with COLI template had fibrotic scar tissue around the lesion characterized by white coloration (Fig 4.8.A upper rectangle). The surrounding areas lacked signs of necrosis or hematoma, and had minimum cellularity mostly consisted of fibrous ECM. In contrast, COLI template implanted (Fig 4.8.a and b, dotted lines) had numerous cellular infiltration with moderate amounts of randomly organized fibrous ECM. Cells contained round and rhomboid nucleus, representing both immature fibroblast-like cells. Although there was no clear infiltration of mononuclear cells, notable amounts of cells that contain eosinophilic granules were found inside the implant. Those cells had unsegmented nucleus, indicating they were unlikely to be eosinophils. Of note was no ectopic chondroid or bone formation observed based on the absence of chondrocyte-like cell or osteocyte-like cell formation within implant. The empty core lesion had gross dark red coloration, indicating the presence of hematoma or necrosis (Fig 4.8.A, lower rectangle). Microscopically, the surrounding areas of empty core lesion had necrosis characterized by severely degraded short fibers of disorganized ECM populated with scant cells (Fig 4.8.c and d). The cells within this region had non-fragmented intact rhomboid nucleus, indicating the area of tendon was in proliferation state infiltrated by fibroblastic cells. There was almost no red blood cells found within lesion and surrounding areas.

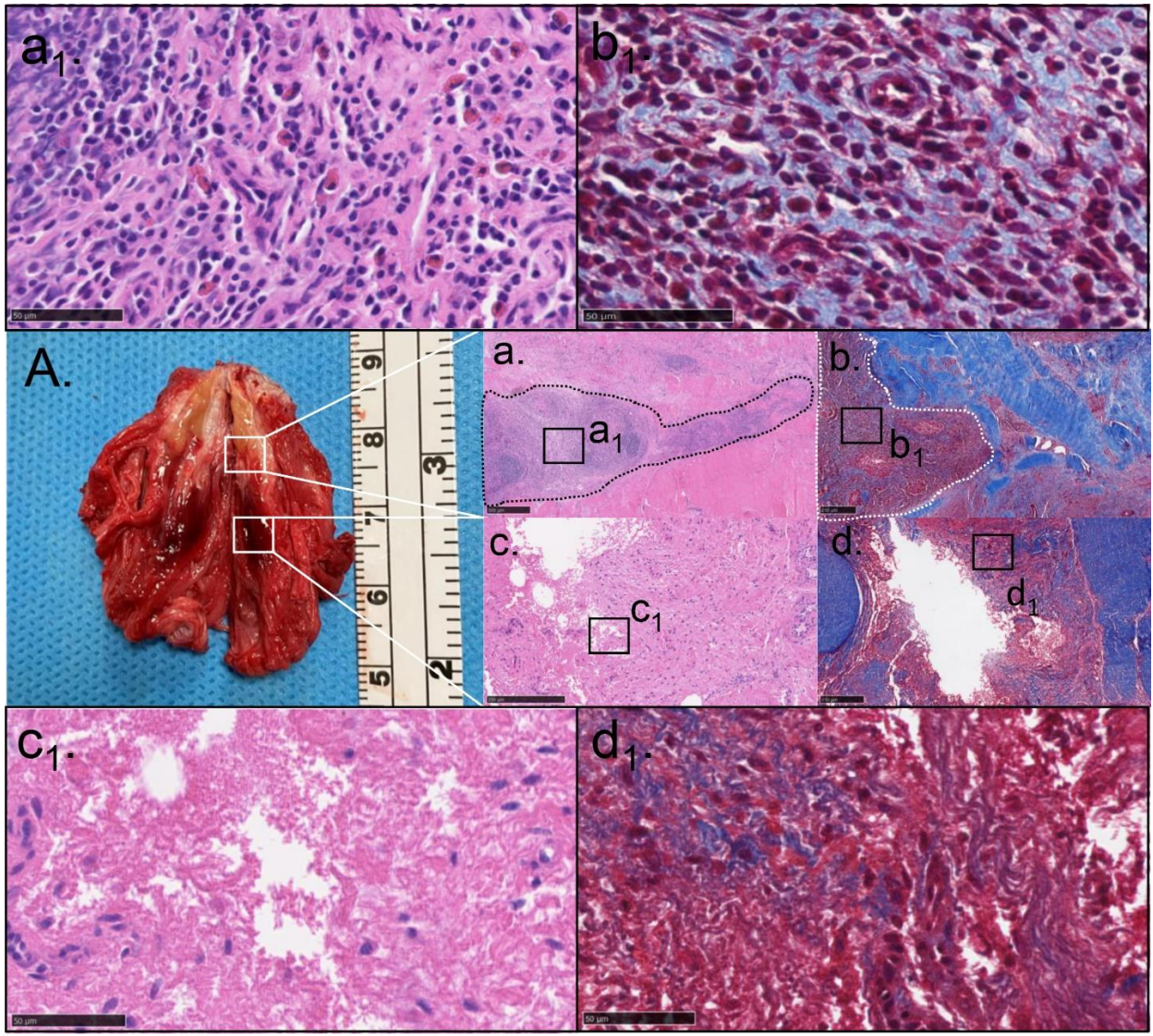

Figure 4.8. Histological images of ALDDFT harvested 12 weeks after injury creation for COLI template only treatment horse. Gross appearance of ALDDFT and core lesion in longitudinally transected ALDDFT (A). Core lesion area filled with COLI template was enlarged in a and b. Core lesion area without filling by COLI template was enlarged in c and d. Insets in a – d (a1 – d1) are enlarged. Formalin fixed paraffin embedded ALDDFT specimen was sectioned at 5 μm, followed by H & E (a, c, a1, and c1) and Masson's trichrome (b, d, b1, and d1) staining. COLI template filling core lesion was indicated by dashed lines. Scale bars = 100 μm (a – d) and 50 μm (a1 – d1).

The core lesion was grossly filled with neotissue that had yellow to brown color and gelatinous appearance 12 week post-operatively, and obvious necrosis was not found in surrounding region of core lesion (Fig 4.9.A and B, white arrows). Area surrounding core lesion had indistinguishable color and fibrous appearance, indicating no clear fibrotic areas existed. Microscopically, core lesion was not clearly discernable from surrounding healthy tissue except an area with the remaining implanted neotendon (Fig 4.9.C and D, red rectangles) that had characteristic high cellularity and more random cellular alignment compared to healthy ligament tissue. Since, there was nearly no remaining void space within healed ligament, no areas that contain serous or hematous fluids accumulation. The implanted neotendon had distinct characteristics of high cellularity with relative random alignment than native tissue (Fig 4.9.E and F), yet cells at the peripheral area of implanted neotendon attained oval to oblong nucleus aligning along the longitudinal axis of recipient ligament (Fig 4.9. G and H). Additionally, some areas in neotendon had an extensive deposition of fibrous *de novo* ECM within which cells were aligned in parallel and gained wavy cell-ECM organization. Of note was the size of cells in the core region of neotendon tended to be larger and appeared to have aligned in concentric. Although there was no clear signs of mononuclear cells infiltration or giant cell formation within neotendon, smaller area of healed ligament contained chondrocyte-like cells characterized by lacunae around themselves and hyaline-like ECM deposited in surrounding area (Fig 4.9 G and H). This indicated a certain degree of incomplete healing was present.

The core lesion remained unfilled throughout 12 week post-operatively in the ALDDFT of horse treated with ASCs, and hollow core lesion was grossly evident (Fig 4.10.A, white arrow). Area surrounding core lesion had white coloration compared to native red ligament, indicating fibrotic scar tissue formation. Microscopically, unfilled core lesion was evident in the center of ALDDFT (Fig 4.10.B and C), and there was no obvious necrotic region characterized by cells with fragmented nucleus or debris of ECM surrounding the core lesion (Fig 4.10.C). Although surrounding areas of core lesion were not necrotic, ECM was short and randomly aligned.

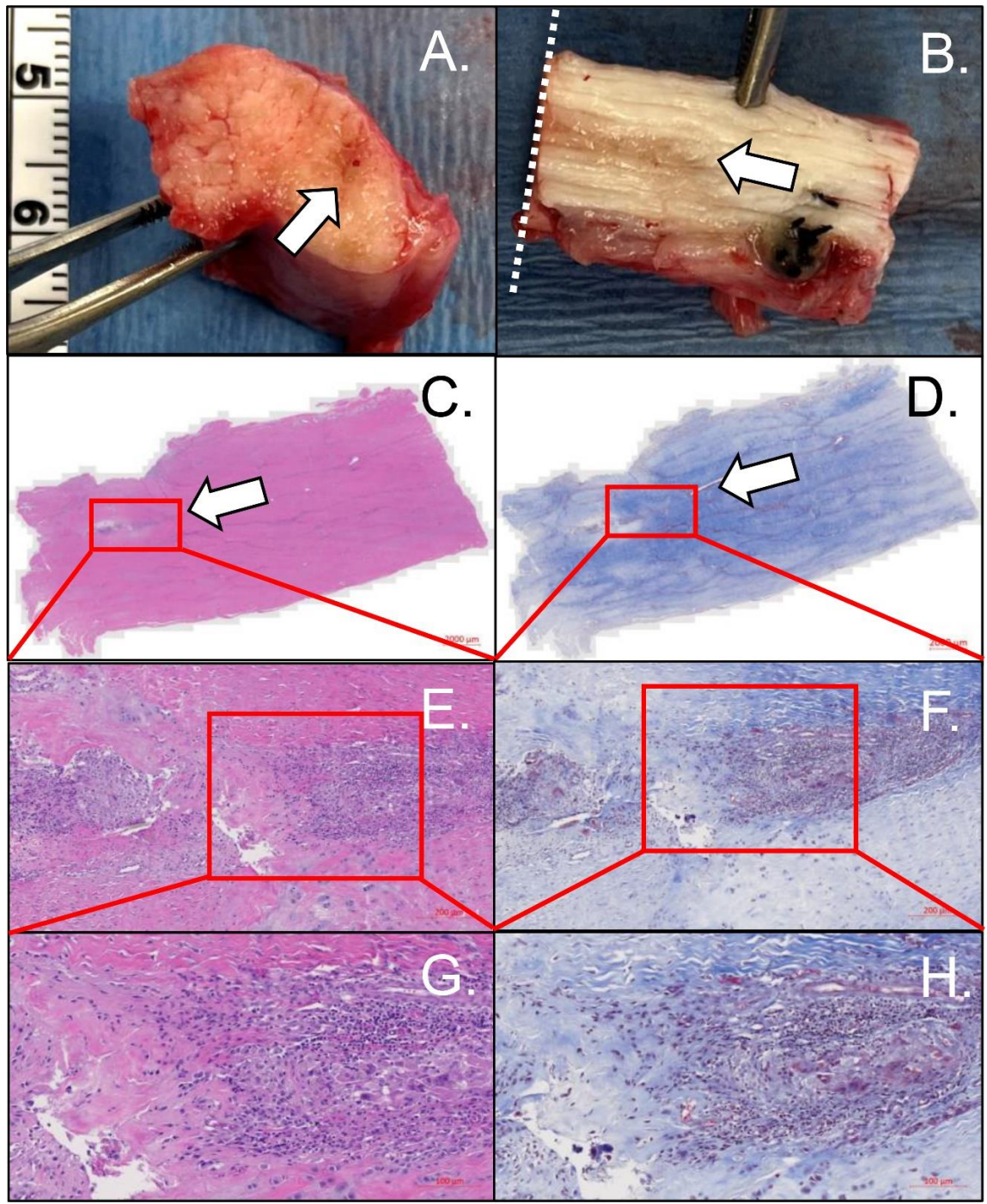

Figure 4.9. ALDDFT harvested 12 weeks after injury from the horse treated with neotendon. Gross appearance of core lesion (white arrows) at the middle transverse plane (A). Gross appearance of core lesion at the middle coronal plane within proximal half of harvested ALDDFT (B). The plane indicated by white dotted line in B is facing up in A. Histological images of specimen in B are shown in C and D. Areas within red rectangles are enlarged to panels below. Specimens were stained by H & E (A, C, E, and G) and Masson's trichrome (B, D, F, and H). Scale bars = 2000 μm (C and D), 200 μm (E and F), 100 μm (G and H).

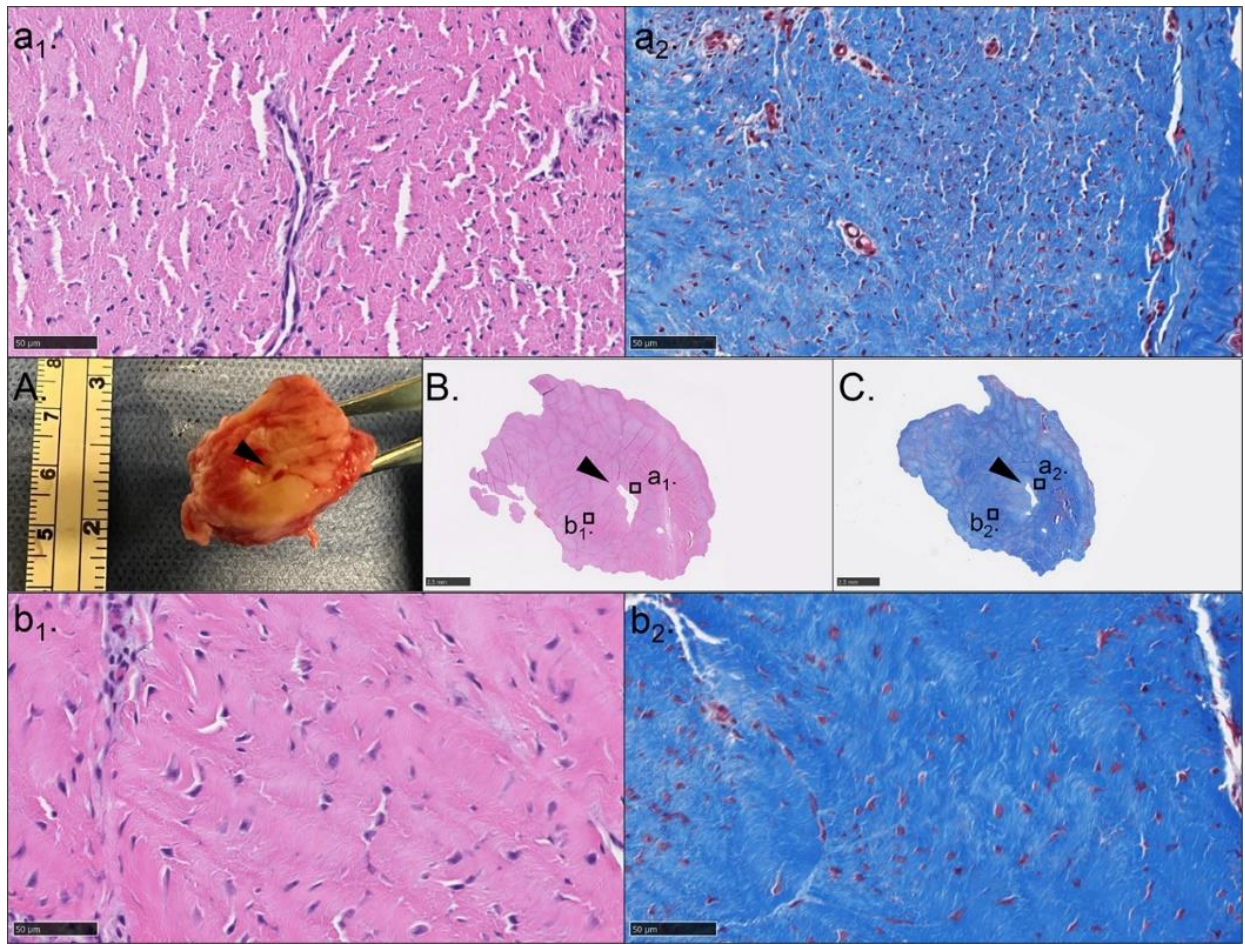

Figure 4.10. Histological images of ALDDFT harvested 12 weeks after injury creation for ASCs only treatment horse. Gross appearance of ALDDFT and core lesion in transversally transected ALDDFT (A). Formalin fixed paraffin embedded ALDDFT specimen was sectioned at 5 μm, followed by H & E staining (B – D). Red rectangles in B and C were enlarged in C and D, respectively. Scale bars = 2.5 mm (B), 1 mm (C), and 100 μm (D).

## 4.4. Discussion

### 4.4.1. Main Findings

The main findings of this study were: 1) surgically-induced core lesion in equine ALDDFT caused minimum and transient impairment in limb use; 2) removal of ALDDFT also did not cause limb use impairment or systemic adverse effects; 3) core lesion was identified as cylindrical hypoechoic lesion in the center of ALDDFT post-injury similar to that of naturally-occurring core lesion; 4) CSA of core lesion transiently increased and echogenicity increased overtime again similar to the trend normally observed in naturally-occurring core lesion; 5) core lesion remained void surrounded by necrotic area when lesion was not filled with implant 12 weeks post-injury; 6) COLI implantation filled core lesion with surrounding fibrotic scar tissue without necrosis; 7) neotendon implantation led to filled core lesion with premature to mature tendon-like tissue and neotendon itself formed premature tendon-like tissue that integrated to surrounding non-fibrotic native tissue; and 8) ASCs injection did not fill the core lesion yet led to premature healing tendon tissue at the peripheral of lesion characterized by high cellularity and fragmented ECM fibers. Combined, the surgically-induced core lesion in equine ALDDFT elicit similar clinical presentation to that of naturally-occurring ALDDFT injury, and has robust potential to evaluate distinct responses to different treatments.

### 4.4.2. ALDDFT Core Lesion Model Creation

Horse is an ideal animal model for tendon and ligament injury because of the similarity with human in terms of relative size and function of tendon and ligament. Additionally, horse is one of the few animal species that naturally develop tendon/ligament injury beside human. These make horse the most suitable for translational study of chronic tendinopathy/ligamentopathy treatment. Similar size of horse tendon/ligament with those of human allows application of similar diagnostic tools such as US and magnetic resonance imaging (MRI) as well as surgical tools for reconstruction.[30,261] Due to natural tendency to develop injury, an established rubric for evaluating ultrasonographic images and pain associated with tendon/ligament injury are also available.[46,262,263] Functionally, it was suggested human tendons have lower energy-storing capacity measured as higher hysteresis than most animals including horses.[264,265] However, it was reported ALDDFT of equidae experience approximately 5% strain during walking,[255,266] which is within a physiological range of human tendons.[267-269] Moreover, tendons of equidae are likely to be under physiological load of approximately 1,000 – 4000 N,[255] higher than that of ovine tendons which is below 310 N.[270] Additionally, horse flexor tendons have closer strength measured as maximum stress to human Achilles tendon than sheep Achilles tendon.[271-273] Therefore, among quadrupedal animals, horse is the closest to human in terms of mechanical properties, since human tendons such as calcaneal tendon can be exposed to as high as 9,000 N during physiological motions.[274] As a result, horse tendons have been historically compared to those of human mechanically and histologically.[275]

However, the use of horse as an animal model has disadvantages of ethical concern due to horse's status as companion animal and the higher cost associated with raising and maintaining skeletally mature horse, as well as specific husbandry requirements. Thus, the non-terminal ALDDFT injury model developed in this study addresses both disadvantages of ethical concern and higher cost by limiting morbidity to minimum and allowing reuse of horses in multiple investigation, which ultimately leads to significant improvement of 3R concept (Replacement, Reduction, Refinement) and cost performance.[276] Historically, horses have been reused in multiple investigations with appropriate interval and careful assessment,[277] because they are ideal animal for a wide variety of fields such as reproductive research as well as educational purpose.[278] Therefore, this versatility of horse as both research and educational animal makes this model successful in addressing existing limitations. To date, non-terminal equine musculoskeletal injury model developed includes post-traumatic osteoarthritis (PTOA) model that recapitulate human PTOA pathology has been previously established using horses,[279] and used in many investigations. Another advantage of the model developed in this study is the availability of multi-sampling. Multi-sampling allows the application of repeated measures on study outcomes and increase statistical power.[280] Multi-sampling is available from large tendon such as SDFT in the form of biopsy without the need for sacrifice,[43] yet this prevents mechanical testing of healed tendon which normally requires entire section of tendons. On the contrary, the model developed in this study allows removal of maximum 4 whole ALDDFTs without significant morbidity. Moreover, it also allows long-term monitoring of horses after ALDDFT removal. This is particularly relevant with cell or implant therapies often associated with tumorigenesis,[281] that can also be systemic such as implant-associated anaplastic large cell lymphoma.[282]

### 4.4.3. Objective Gait Analysis

Clinical presentation of the model developed in this study showed temporal impairment of limb use for up to 7 weeks post-operatively as similar to that of naturally-occurring ALDDFT injuries. In naturally-occurring ALDDFT or SDFT injuries, lameness is often subjectively absent or mild, although swelling of metacarpal region is a consistent findings.[283] In this study, limb use was evaluated objectively using GRF measurement to detect mild lameness, as any decrease in PVF is associated with increased pain in that limb of horse.[284] As a result, transient limb use impairment observed in this study was longer than that previously reported in SDFT surgical core lesion for up to 2 weeks post-operatively using subjective lameness assessment.[46] Potential reason for longer limb use impairment was due to higher sensitivity of objective gait analysis than subjective lameness evaluation. This was also evident in a study that reported significant decrease in PVF after LPS injection into equine joint, despite subjective lameness scores indicated no impaired limb use.[284] Since 9 weeks post-injury, horses returned to original performance level, which was consistent with normal clinical presentation of horses both after ALDDFT injury or excision. For example, desmotomy of ALDDFT has been traditionally performed to treat flexural deformity of lower extremity and is known to cause minimum effects on to limb use.[285] Thus, not only similar clinical presentation of the model developed in this study can recapitulate natural course of response to injury and treatment, but also minimally invasive injury is advantageous due to less medication is required to control post-operative pain that might affect outcomes of testing treatment.

### 4.4.4. Ultrasound Analysis

Ultrasonographically, core lesion was identified as hypoechoic area in the center of ALDDFT that maintained circular- to oval-shape transversely and contained short fibers longitudinally. These characteristics were consistent with those of naturally-occurring tendon/ligament injury in both horse and human, since reduction of echogenicity and disruption of linear fibers are criteria of tendon/ligament injury for both species.[286] As expected, core lesion created in the present study also resembled ultrasonographic presentation of surgically-induced SDFT core lesion.[240] Compared to naturally-occurring ALDDFT desmitis, ultrasonographic presentation of core lesion created in this study may represent approximately half of clinical cases, as the other half was presented as diffusely hypoechoic ligament.[283] Morphologically, surgically-induced core lesion in the present study had relatively uniform lesion size and echogenicity clearly distinguishable from normal tissue. On the contrary, chemically-induced lesion tends to lead to non-homogenous size, morphology, and echogenicity.[287] This is an important advantage of surgically-induced core lesion for both limiting variations and clinical assessment, because the lesion size evaluation is an essential part of tendon/ligament healing due to association with local pain in equine.[41] In the present study, the lesion CSA increased transiently by 5 week post-injury and gradually decreased, while echogenicity of lesion increased overtime. This was consistent with same trends observed in both chemically and surgically induced SDFT core lesion.[41,43,50] Of note is that several factors have effects on core lesion size. For example, the transient increase of lesion CSA and associated pain was reported only after MSCs treatment in a study of surgically-induced equine SDFT core lesion.[41] Hence, it is possible ASCs treatment administered in the present study might have contributed to the lesion CSA increase in one of the horses. Additionally, when measuring lesion CSA, it is important to consider post-injury management, since surgically induced SDFT core lesion CSA varies between bandaging or casting.[288] Therefore, it is important also to administer a consistent post-injury management to all horses throughout the study period to minimize variation.

In human, tendon/ligament lesion often presents itself as diffuse hypoechoic area rather than confined central core lesion. Therefore, CSA of injured tendon or ligament itself rather than lesion CSA is used for evaluation. Indeed, increased tendon/ligament CSA is a common change associated with injury in human, which leads to decreased mechanical properties and decreased functionality.[289,290] Clinically, tendon/ligament CSA increases gradually over 12 weeks after injury in human, and CSA is an important early healing phase that affect long-term recovery.[291] The same measure is applicable to our ALDDFT core lesion model as well, since overall ligament CSA subjectively increased overtime and accompanied by persistent swelling of limb circumference at the surgical site throughout the study period. This was consistent with previous reports of surgical and chemical SDFT core lesion model where increased tendon/ligament CSA persisted as long as 6 months post-injury.[29,30,46] Tendon CSA increase upon injury can be replicated also in small animals such as rabbit,[292] yet evaluation of ultrasonographic images is much less established than that in horse. Therefore, our ALDDFT core lesion model closely recapitulates not only equine tendon/ligament core lesions but also share characteristics with human tendon/ligament injury ultrasonographically.

Core lesion created in the present study gained echogenicity over 3 months post-injury period. This is consistent with naturally-occurring tendon injury that regains echogenicity over several months period.[293] Prolonged presence of hypoechoic core lesion despite of earlier clinical improvement observed in the present study was also consistent with naturally-occurring tendon/ligament injury.[283] In terms of the period to regain original echogenicity, our ALDDFT model is close to naturally-occurring ALDDFT desmitis, since it was reported half of the horses had normal appearance of ALDDFT ultrasonographically after 4 months from initial injury.[283] Another important property of ALDDFT core lesion established in the present study is the lack of lesion propagation into DDFT, since it can introduce a confounding factor. This was the case for surgically-induced SDFT, propagating itself vertically and increased its lesion length overtime.[288] Regardless of tendon/ligament types, the limitation of surgical core lesion may be the creation of cylindrical empty space inside tendon, not truly recapitulating pathologic changes of naturally-occurring injuries. For example, hypoechoic core lesion of naturally-occurring tendon/ligament injury is often characterized by hemorrhage and effusion.[293] Yet, same limitation can be considered an unique advantage, since empty core lesion creates partial defect that can test regenerative capacity of implants such as hydrogel.[294-296]

### 4.4.5. Microstructure

Macrostructurally, part of core lesion that was not filled with COLI was characterized with void space and necrotic surrounding area. Microstructurally, necrotic area surrounding void core lesion contained degrading short ECM fibers and fewer round to rhomboid cell infiltration, representing transitional phase between inflammatory and proliferative stages of healing. Similarly, lesion that received ASCs administration remained void, although the surrounding area was not necrotic. Therefore, ALDDFT core lesion created in the present study likely to have established critical-sized defect (CSD), an essential criteria to establish in each musculoskeletal tissue defect model to delineate treatment effects from spontaneous healing.[297] Establishment of CSD in equine ligament is a significant achievement for tendon/ligament regeneration strategies, since it evaluates regenerative capacities of interventions to treat large defects that are otherwise incurable.[298-301] To date, non-critical-sized partial defect such as central window defect, punch biopsy defect, or complete resection models have been created in small animals such as rat and rabbit or large animals such as dog or sheep.[302,303] In this regard, creation of CSD in equine ligament is novel and establishes a foundation for testing of truly regenerative therapeutics. The potential reason for remained core lesion without being filled with COLI or neotendon in equine ligament in contrast to small animals' tendon/ligament may be robust regenerative capacities of small animals that are not always representative to those of human.[126,304] Interestingly, creation of CSD has not been reported for surgically-induced equine SDFT core lesion,[30,46] potentially due to the use of cutter to remove tendon tissue in the present study rather than burr to simply disrupt tissue in most of previous reports. This also may relate to slower healing of ALDDFT core lesion in the present study compared to naturally-occurring SDFT injury that can reach remodeling phase by 3 months post-injury.[305]

In contrast to empty core lesion area, the area filled with COLI had non-necrotic surrounding area that was characterized with white and firm scar tissue-like gross appearance. Microscopically, enormous round to oval cells infiltrated and randomly aligned inside implanted COLI, and surrounding area was characterized with less organized ECM. The amount of ECM within COLI implant was rather scares and small amount of fibrous ECMs separated cells within densely populated COLI implant. Although mononuclear cell infiltration, giant cell, or signs of granulomatous inflammation were not observed inside or surrounding area of implanted COLI, cells with eosinophilic granules were found inside COLI, indicating potential immune response. Infiltration of round to oval fibroblastic cells into both synthetic and organic scaffolds used for tendon repair augmentation is a commonly observed outcome in human.[306] In particular, COLI was used in primary repair augmentation of rotator cuff tendon in human, and it formed premature tendon-like tissue over repair site without rejection in 3 to 6 months post-operatively.[214,307] On the contrary, the presence of eosinophilic cells inside COLI was noted in the current study. Eosinophil accumulation is a prominent feature of multiple allergic diseases in horse.[308] With regards to immune response to implants, accumulation of eosinophil around implant is a common signs of reaction to some metal implants.[309-311] Therefore, our finding indicated horse may be more sensitive to foreign body than human, and the model developed in the present study is an ideal pre-clinical animal model for investigation of immunogenicity associated with implants.

The morphology of implanted neotendon was similar to that of implanted COLI. It was populated by numerous cells with oval to oblong nucleus. The cells, however, attained better parallel alignment with longitudinal axis of ALDDFT subjectively. Moreover, there appeared to have better integration of implanted neotendon to surrounding native tissue, since the cellular alignment was more organized at peripheral area of implanted ligament than central region and there was more extensive communication of ECM fibers between neotendon implant and surrounding native tissue. Interestingly, there was an area of apparent chondrogenic differentiation of endogenous cells in surrounding area of neotendon potentially as a form of aberrant healing response. This was often common healing response to tendon/ligament injuries, since granulation tissue that form after natural injury often contains ectopic chondroid and largely fibrotic.[293] In terms of immune response to neotendon, the signs observed were largely similar to those observed for COLI implant, including lack of mononuclear cell infiltration or giant cell, lack of fibrotic capsule formation around implant, yet there was presence of eosinophilic cells within implant. Subjectively, however, there appeared to have been less eosinophilic cells within neotendon compared to COLI implant. Of note was surrounding areas of both implanted COLI and neotendon as well as that of core lesion treated by ASCs were non-necrotic. The potential reason for this finding is cells, potentially MSCs, that infiltrated COLI or those cultured in neotendon as well as ASCs injected into the core lesion might have exerted anti-necrotic paracrine effects, as MSCs have anti-necrotic effects to treat osteonecrosis.[312-314] ASCs are also reported to improve structural organization of chemically-induced SDFT core lesion,[51] which might have contributed to the lack of necrotic surrounding area.

## 4.5. Conclusion

Surgically-induced equine ALDDFT core lesion model established in the present study presents a novel non-terminal equine ligament injury model that closely resembles naturally-occurring equine tendon/ligament injuries both clinically and etiopathologically. It is characterized with mild and transient limb use impairment as well as ultrasonographic hypoechoic lesion that follows size, morphology, and echogenicity changes typically seen in naturally-occurring injuries. Additionally, core lesion does not heal without any intervention and leads to void space surrounded by largely necrotic tissue, demonstrating robust testing capability of regenerative therapies. Collectively, the model is an effective mechanism with improved accessibility that facilitates clinical translation of novel tendon/ligament treatment strategies.

# Vita

Takashi Taguchi, originally from Sapporo, Hokkaido, Japan, received bachelor's degree in veterinary medicine in 2008 at Osaka Prefecture University. In the same year, he passed board examination of veterinary medicine and pursued a career as a small animal veterinary clinician. After 6 years of small animal clinical trainings including internship at Tokyo University of Agriculture and Technology, he started graduate research study in 2014 and received master's degree in veterinary medicine in 2017 from Western University of Health Sciences. He continued graduate research at the School of Veterinary Medicine at Louisiana State University in Baton Rouge to pursue doctoral degree.